\documentclass[traditabstract]{aa}
\usepackage{array,multirow,pstricks,rotating,graphics,amssymb,amsmath,threeparttable}
\usepackage{aas_macros}
\usepackage{natbib}
\bibpunct{(}{)}{;}{a}{}{,} 
\usepackage{txfonts}
\usepackage{longtable}

\newcommand{\cm}{\textup{ cm}}

\newcommand{\K}{\textup{ K}}

\newcommand{\HH}{\textup{H}}

\newcommand{\He}{\textup{He}}
\newcommand{\e}{\textup{e}^-}

\newcommand{\kb}{k_{\rm B}}
\newcommand{\rmd}{\,{\rm d}}

\begin{document}
\title{Low-temperature gas opacity}

\subtitle{\AE SOPUS: a versatile and quick computational tool}

\author{Paola Marigo$^1$ \and Bernhard Aringer$^{2,3}$}

\institute{
 Department of Astronomy, University of Padova,
        Vicolo dell'Osservatorio 3, I-35122 Padova, Italy \and
Astronomical Observatory of Padova -- INAF,
        Vicolo dell'Osservatorio 5, I-35122 Padova, Italy \and
Department of Astronomy, University of Vienna, T\"urkenschanzstra\ss e 17, A-1180 Wien, Austria}

\date{Received 29 May 2009 / Accepted 27 October 2009}
 
\abstract{
We introduce a new tool -- \AE SOPUS: Accurate Equation of State and
OPacity Utility Software -- for computing the equation of state 
and the Rosseland mean (RM) opacities of matter in the ideal gas phase. 
Results are given as a function of one pair of state
variables, (i.e. temperature $T$ in the range
$3.2~\le~\log(T)~\le~4.5$, and parameter $R=~\rho/(T/10^6\,{\rm K})^3$
in the range $-8~\le~\log(R)~\le~1$), 
and arbitrary chemical mixture. The chemistry is
presently solved for about 800 species, consisting of
almost 300 atomic and 500 molecular species. The gas opacities 
account for many continuum and discrete sources, including atomic
opacities, molecular absorption bands, and collision-induced absorption. 
Several tests made on 
\AE SOPUS have proved that the new opacity tool is accurate in the results,
flexible in the management of the input prescriptions, and  
agile in terms of computational time requirement.   
Purpose of this work is to greatly expand the public availability
of Rosseland mean opacity data in the low-temperature regime.
We set up a web-interface (http://stev.oapd.inaf.it/aesopus) which 
enables the user to compute and shortly retrieve 
RM opacity tables according to his/her specific needs, allowing a full
degree of freedom in specifying the chemical composition
of the gas. As discussed in the paper, useful 
applications may regard, for instance,  RM opacities of
gas mixtures with 
i) scaled-solar abundances of metals, choosing among various solar
mixture compilations available in the literature;
ii) varying CNO abundances, suitable for evolutionary models of red
and asymptotic giant branch stars and massive stars 
in the Wolf-Rayet stages;
iii) various degrees of enhancement in $\alpha$-elements, and
C-N, O-Na, and Mg-Al abundance anti-correlations, necessary to
properly describe the properties of stars in early-type galaxies 
and Galactic globular clusters; 
iv) zero-metal abundances appropriate for studies of gas opacity in
primordial conditions. 
 }
\authorrunning{Marigo \& Aringer}
\titlerunning{\AE SOPUS, a computational tool for low-temperature gas opacity}
\keywords{
Equation of state --
Atomic processes  --
Molecular processes --
Stars: abundances --
Stars: atmospheres --
Stars: AGB and post-AGB
              }
\maketitle

\section{Introduction}
\label{sec_intro}
In a gas under conditions of local thermodynamical equilibrium (LTE) and in the limit
of the diffusion approximation (DA), the solution to the radiation transfer 
equation simplifies and the total flux of radiation $F$ as a
function of radius $r$ is given by:
\begin{equation}
F(r)=-\displaystyle\frac{4\pi}{3}\frac{1}{\kappa_{\rm
R}(\rho,T)}\displaystyle\frac{\partial B(r,\,T)}{\partial r}
\end{equation}
where $T$ is the gas temperature, 
$\rho$ denotes the density, $B(r,\,T)$ is the integral of the
Planck function over frequency, and the relation
\begin{equation}
  \frac{1}{\kappa_{\rm R}(\rho,T)} = \displaystyle\frac{\displaystyle\int_0^\infty \displaystyle\frac{1}{\kappa(\nu)}
\frac{\partial B_\nu}{\partial T} d\nu}{\displaystyle\int_0^\infty \displaystyle\frac{\partial B_\nu}{\partial T} d\nu}\,,
\label{eq_rosseland}
\end{equation}
first introduced by Rosseland (1924), defines the Rosseland mean opacity $\kappa_{\rm R}(\rho,T)$.
Being a harmonic average over frequency, $\kappa_{\rm R}$ emphasises  
spectral regions of weak absorption, across which the energy flux
is most efficiently transported.

Both LTE and DA  conditions are usually met in the stellar interiors, where 
collisions dominate the thermodynamic state of matter, 
the photon mean free-path is  much shorter than 
the typical scale length of the temperature gradient, and the
Kirchoff's law applies with the source function being the Planckian.
However, in the outermost layers of a star the photon mean-free path may become
so long that the DA conditions break down, thus invalidating the use of the RM opacity.
In these circumstances,   
a straight arithmetic 
average of the monochromatic absorption coefficient (Eddington 1922), designated with $\kappa_{\rm P}$,
Planck mean (PM) opacity:
\begin{equation}
{\kappa_{\rm P}(\rho,T)} =
\displaystyle\frac{\displaystyle\int_0^\infty \displaystyle
\kappa(\nu)B_\nu d\nu}{\displaystyle\int_0^\infty B_\nu d\nu}
\label{eq_planck}
\end{equation}
may be more suitable to represent the absorption 
properties of the gas in a simplified version of the radiation transport
equation (e.g. Helling et al. 2000).

Both RM and PM opacities are frequency-integrated averages, so that they
only depend on two independent state variables, e.g. temperature $T$
and density $\rho$ (or pressure $P$), and the chemical composition of the gas.

In stellar evolution models it is common practise to describe the 
absorption properties of matter with the RM opacity formalism, adopting
pre-computed static tables of $\kappa_{\rm R}$ which should encompass a region of the
bi-dimensional space $T$-$\rho$ wide enough to cover all possible
values met across the stellar structure during the evolution, from the 
atmosphere down to the central core.
The chemical composition is usually specified by a set of abundances,
e.g.: the total metallicity $Z$, the hydrogen abundance $X$, and the
partitions $\{X_i/Z\}$ of heavy elements in the mixture, which 
depend on the specific case under
consideration. Frequent choices are assuming solar partitions
$\{X_i/Z\}=\{X_i/Z\}_{\odot}$, or deriving
$\{X_i/Z\}$ from
other constraints such as the enhancement in $\alpha$-elements
(expressed by the ratio $[\alpha/{\rm Fe}]$), 
or the over-abundances in C and O necessary to describe the hydrogen-free
chemical profile in He-burning regions.

In the literature several authors have calculated $\kappa_{\rm
R}(\rho,T)$ for different combinations of the state variables and 
chemical composition. 
Let us limit here to briefly recall the most relevant efforts, i.e. those
mainly designed for supplying the scientific community with extended and 
continuously updated RM opacity databases.

In the high-temperature regime, 
i.e. $10^4 {\rm K} \la T \la 10^9 {\rm K}$, calculations of RM opacities 
are mainly provided by two independent teams, namely:
the Opacity Project (OP) international collaboration coordinated
by Seaton (Seaton 2005, and references therein); and the
Opacity Project at Livermore (OPAL) being carried on by Iglesias, Rogers
and collaborators (see Iglesias \& Rogers 1996, and references
therein). Both groups have set up a free 
web-access to their RM opacity
calculations, via either a repository
of static tables and/or source routines, 
or an interactive web mask where the user can specify 
the input parameters and run the calculations in real time.

In the low-temperature regime,
$10^3{\rm K} \la T \la 10^4{\rm K}$,
widely-used RM opacity tables are those provided by the
research group of the Wichita State University 
(Ferguson et al. 2005 and references therein).
A web page hosts an archive of static RM opacity tables, for both
scaled-solar and $\alpha$-enhanced mixtures, which cover a wide 
range  of metallicities including the $Z=0$ case.
It should be acknowledged the large body of work made by Kurucz, who
provides, via web or CD-ROMs, all necessary atomic and molecular data as well as 
FORTRAN codes to calculate $\kappa_{\rm R}$ (see Kurucz 1993abc), 
in the temperature interval $10^3{\rm K} \la T \la 10^5{\rm K}$, for 
scaled-solar and $\alpha$-enhanced mixtures.
More recently, Lederer \& Aringer (2009) have calculated and made available
via the VizieR Service a large catalogue of RM opacity tables for C- and
N- rich compositions, with the purpose to supply  
RM opacity data suitable for the modelling of asymptotic giant branch 
(AGB) stars.
Helling \& Lucas (2009) have produced a set of gas-phase  
Rosseland and Planck mean opacity tables for various metallicities,  
C/O and N/O ratios.      
It is due mentioning also the recent paper by Sharp \& Burrows (2007), who 
provide an exhaustive and useful review on the thermochemistry, techniques, 
and databases needed to calculate atomic and molecular opacities at
low temperatures.

Despite the undeniable merit of all these works, 
the public access to low-temperature RM opacities
still needs to be widened to account for the miscellany of chemical patterns 
-- mostly relating to the photosphere
of stars -- that modern 
spectroscopy is bringing to our knowledge with an ever-growing richness of details, and also to allow
the exploration of possible opacity changes driven by any hypothetical chemical composition.
The peculiar abundance features in the atmospheres of AGB stars
 (e.g. McSaveney et al. 2007, Smith et al. 2002);
the $\alpha$-enhanced abundance pattern of stellar populations belonging to globular clusters
(e.g. Gratton et al. 2004) and 
elliptical galaxies (e.g. Clemens et al. 2006, 2009);
the large carbon overabundance and other chemical anomalies of the so-called
carbon-enhanced metal-poor stars in the Galaxy (e.g. Beers  \& Christlieb  2005);    
the striking C-N,  O-Na  and Mg-Al abundance anti-correlations exhibited by stars 
in Galactic globular clusters (e.g. Carretta et al. 2005); the chemical composition of 
the primordial gas after the Big Bang nucleosynthesis (e.g. Coc et al. 2004): 
these are a few among the most remarkable examples.

In this framework, {\em purpose of our work is to greatly expand the availability
of RM opacity data in the low-temperature regime, by offering the
scientific community an accurate and flexible computational
tool, able to deliver RM opacities tables on demand, and with a full
freedom in the specification of the chemical mixture.}

To this aim, we have developed the \AE SOPUS tool (Accurate Equation of State and
OPacity Utility Software), which consists of two fundamental  parts: 
one computes the equation of state (EOS)
of matter in the gas phase, and the other evaluates the total monochromatic
coefficient, $\kappa(\nu)$, as sum of several opacity sources, and then computes
the Rosseland mean.
The EOS is solved for $\approx 800$ chemical species, including
neutral atoms, ions, and molecules. The RM opacities take into account several true 
(continuum and discrete) absorption and scattering processes.
An interactive web-interface (http://stev.oapd.inaf.it/aesopus)
 allows the user to run \AE SOPUS according 
to his/her specific requirements just by setting the input parameters 
($T-R$ grid, reference solar composition, total metallicity, abundance of each chemical species)
on the web mask. 

The paper is organised as follows.
Section~\ref{sec_esopus} specifies the bi-modular structure of \AE SOPUS.
In Sect.~\ref{sec_eos} we illustrate the basic ingredients necessary
to set up and solve the equation of state. Numerical aspects are
detailed in Appendix~\ref{apx_method}.
Section~\ref{sec_opacso} indicates the opacity sources 
included in the evaluation of the
total monochromatic absorption coefficient.
The Rosseland mean is presented in
Sect.~\ref{sssec_rmk}, with details on the computing-time requirements 
provided in Sect.~\ref{sssec_fgrid}. Complementary information on the 
frequency integration is given in Appendix~\ref{apx_fgrid}.
The formalism introduced to describe the different ways the RM opacity
tables can be arranged, as a function of the state variables and 
chemical composition, is outlined in Sect.~\ref{sec_tables}.  
In Sect.~\ref{sec_results} we analyse five relevant cases 
of RM opacity calculations, characterised by different chemical patterns,
namely: scaled-solar elemental abundances (Sect.~\ref{ssec_ksun}), 
varying CNO abundances (Sect.~\ref{ssec_kcno}), $\alpha$-enhanced
mixtures (Sect.~\ref{ssec_alpha}), 
mixtures with peculiar C-N-O-Na-Al-Mg abundances 
(Sect.~\ref{ssec_anomal}),
 and metal-free compositions  (Sect.~\ref{ssec_kz0}).
Appendix~\ref{apx_afe} specifies the general scheme adopted to construct
non-scaled-solar mixtures.
Final remarks and indications of future developments of this work are expressed
in Sect.~\ref{sec_final}.

\section{The \AE SOPUS code}
\label{sec_esopus}

\subsection{Equation of state}
\label{sec_eos}
The equation of state quantifies the distribution of available
particles in the unit volume,  in the form of neutral and ionised atoms, electrons, 
and molecules. 
At low temperatures ($T \la 6\,000$ K)
and  sufficiently high densities, 
molecules can form in appreciable concentrations 
so as to dominate the equation of state at the coolest temperatures.
To this respect a 
seminal work was carried out by Tsuji (1964, 1973)  
who set up the theoretical foundation of most chemistry routines
still in use today.

In our computations the EOS is solved for atoms and molecules in the gas
phase, under the assumption of an ideal gas in both  thermodynamic equilibrium (TE)
and instantaneous chemical equilibrium (ICE).
This implies that the abundances of the various atomic and molecular 
species depend only on the local
values of temperature and density, regardless of the specific 
mechanisms of interaction among them. 

Solving a chemical equilibrium problem requires three general steps. 
First, one must explicitly define the
gas system in terms of its physical and thermodynamic nature. For example,
the classical problem in chemical
equilibrium computations is to calculate the state of a closed system
of specified elemental composition at fixed temperature $T$ and pressure $P$.
The nature of the physical-chemical model determines
the set of governing equations to be used in computations.
The second step is to manipulate this original set of equations into a
desirable form, to reduce the number of unknowns and/or to 
fulfil the format requirements of the adopted computation scheme.
The third step is to solve the remaining simultaneous equations,
usually be means of iterative techniques (see, for instance, Tsuji 1963).

Rather than solving sets of equations, the equilibrium computation can
be formulated as an optimisation problem, such as solving the so-called
classical problem by minimising the calculated free energy of the
system (Mihalas, D\"appen, \& Hummer 1988).
An alternative approach,
based on the neural network technique,  
has been recently proposed by Asensio Ramos \& Socas-Navarro (2005). 

In this study we adopt the Newton-Raphson iteration scheme to
solve the chemical equilibrium problem of a gas mixture
with assigned chemical composition, pressure $P$ (or density) 
and temperature $T$.
The adopted formalism and solution method are detailed below. 

\subsubsection{Equilibrium relations}
Under the ICE approximation, the gas species obey the equilibrium
conditions set by the dissociation-recombination and ionisation processes.
Generally speaking,  
the chemical interactions in the gas
between species $A$ and $B$
may involve the simple dissociation-recombination process
\begin{equation}
\label{eq_dissociative_equilibrium}
A  + B \leftrightarrows  AB 
\end{equation}
in which both {\em forward} and {\em reverse} reactions proceed at
the same rate. In the above equation $A$ or $B$ may be an atom,
molecule, ion or electron.
Of course one may postulate more complicated chemical interactions such as
\begin{equation*}
AB  + CD \leftrightarrows  AC + BD 
\end{equation*}
or
\begin{equation*}
A + B + C \leftrightarrows  ABC 
\end{equation*}
but these can ultimately be reduced to
Eq.~(\ref{eq_dissociative_equilibrium}), in the forms of simple 
dissociation-recombination reactions, i.e.
\begin{equation*}
(AB) + (CD)  \leftrightarrows (ABCD) \leftrightarrows (AC) + (BD) 
\end{equation*}
\begin{equation*}
(A + B) + C  \leftrightarrows  ABC\,.
\end{equation*}
 From statistical mechanics we know that for any species $A$ and $B$
 in equilibrium with their compound $AB$ (usually a molecule), 
the number densities $n_A$, $n_B$, and $n_{AB}$ are related by
the Guldberg-Waage law of mass action:
\begin{equation}
\label{eq_eqconst}
K_{AB}(T) = \frac{n_A n_B}{n_{AB}},
\end{equation}
where $K_{AB}(T)$ is the dissociation constant or equilibrium constant
of species $AB$,
which depends only on temperature.
It is expressed with
\begin{equation}
\label{eq_dissociation}
K_{AB}(T) = \displaystyle\left(\frac{2\pi \mu k T}{h^2}\right)^{3/2}  
\frac{Q_{{\rm int},A} Q_{{\rm int}, B}}{Q_{{\rm int},AB}} 
\exp{\left(-\frac{D_{AB}}{k_{\rm B} T}\right)}\, ,
\end{equation}
where $k_{\rm B}$ is the Boltzmann's constant; $h$ denotes the Planck's constant;  
$T$ is the local temperature; 
$\mu=\displaystyle\frac{m_A m_B}{m_{AB}}$ is the reduced
mass of the molecule; the $Q_{\rm int}$'s are the internal partition
functions; and 
$D_{AB}$ is the dissociation energy of the ($A$, $B$, $AB$) reaction
given by Eq.~(\ref{eq_dissociative_equilibrium}).
Species  $A$ and $B$ themselves can be either molecules or single atoms.

In the identical framework we can consider positive ionisation and
 recombination processes:
\begin{align*}
\nonumber
A^{+r} & \leftrightarrows A^{+r+1} + e^- 
\nonumber \, .
\end{align*}
Again, species $A$ is taken in
the general sense and can be either a molecule or a single atom, 
and the superscript $+r$ (or $+r+1$) denotes its ionisation stage.

These processes can be described through 
the corresponding equilibrium or ionisation constant:
\begin{equation}
\label{eq_sahaconst}
K^{\rm Saha}_{A^{+r+1}}(T) = \frac{n_{A^{+r+1}} n_{e^-}}{n_{A^{+r}}} 
\end{equation}
which is explicitly given in the form of the Saha equation
\begin{equation}
\label{eq_saha}
K^{\rm Saha}_{A^{+r+1}}(T)  =  \displaystyle\left(\frac{2\pi m_e k T}{h^2}\right)^{3/2}  
\frac{Q_{{\rm int},A^{+r+1}} \cdot 2 }{Q_{{\rm int},A^{+r}}} 
\exp{\left(-\frac{I_{A^{+r}}}{k_{\rm B} T}\right)}\, .
\end{equation}
Here $m_e$ is the mass of the electron; $I_{A^{+r}}$ is
the ionisation potential of species $A$ in the
$+r^{\rm th}$ ionisation stage; the $Q_{\rm int}$ are the internal
partition functions appropriate to the corresponding
species. The factor $2$ is the statistical weight $g_e$ for
free electrons, corresponding to two possible
spin states.

The same formalism with $r=-1$ can be applied to account for 
the electron-capture negative ionisation
\begin{align*}
\nonumber
A + e^- &\leftrightarrows A^{-}\, ,
\end{align*}
which is assigned the equilibrium constant
\begin{equation}
\label{eq_sahaconst_neg}
K^{\rm Saha}_{A^{-}}(T) = \frac{n_{A^{-}}}{n_{A} n_{e^-}}\,,
\end{equation}
and the Saha equation:
\begin{equation}
\label{eq_saha_neg}
K^{\rm Saha}_{A^{-}}(T)  =  \displaystyle\left(\frac{h^2}{2\pi m_e k_{\rm B} T}\right)^{3/2}  
\frac{Q_{{\rm int},A^{-}}} {Q_{{\rm int},A} \cdot 2 }
\exp{\left(\frac{I_{A^{-}}}{k_{\rm B} T}\right)}\,,
\end{equation}
where $I_{A^{-}}$ corresponds to the electron affinity, i.e. the energy released when an electron 
is attached to a neutral atom or molecule.

Where ionisation of diatomic and polyatomic
molecules is considered, there are at least three
energy-equivalent ways of forming positive molecular ions:
\begin{enumerate}
\item[1)] $A+B \rightarrow AB\,\,;\,\,\,\,\,\,\,\, AB-e^- \rightarrow AB^+$ 
\item[2)] $A-e^- \rightarrow A^+\,;\,\,\,\,\,\,\,\, A^+ +B \rightarrow AB^+$ 
\item[3)] $B-e^- \rightarrow B^+\,;\,\,\,\,\,\,\,\, A +B^+ \rightarrow AB^+\,.$ \end{enumerate}
Dissociation and ionisation equilibrium can be taken into account 
simultaneously  by choosing that dissociation path in which the atomic species 
that remains ionised is the one with the lowest ionisation potential.
For instance, for a ionised diatomic  molecule $AB^+$ with 
  $I_A<I_B$ the selected sequence is 2), so that 
the number density of the ionised molecule is calculated by combining
Eq.~(\ref{eq_eqconst}) and Eq.~(\ref{eq_saha}), obtaining: 
\begin{eqnarray}
\label{eq_kabp}
K_{AB^+}(T) & = & \frac{n_{A^+} n_B}{n_{AB^+}} 
= \displaystyle\frac{K_{AB}\,K^{\rm Saha}_{A^+}}{K_{AB}^{\rm Saha}}\\
\label{eq_ionmol}
& = & \displaystyle\left(\frac{2\pi \mu k_{\rm B} T}{h^2}\right)^{3/2}  
\frac{Q_{{\rm int},A^+} Q_{{\rm int}, B}}{Q_{{\rm int},AB^+}} 
\exp{\left(-\frac{D_{AB^+}}{k_{\rm B} T}\right)}\, ,
\end{eqnarray}
where the dissociation energy is given by $D_{AB^+}=D_{AB}+I_A-I_{AB}$
and $I_{AB}$ is the ionisation energy of the molecule $AB$.

In the case of negative molecular ions and assuming that dissociation of $AB$ 
produces $A^-$ and $B$ (hence
$I_{B^-} < I_{A^-}$), we can extend the same formalism of 
Eq.~(\ref{eq_ionmol}) to calculate the dissociation constant:
\begin{eqnarray}
\label{eq_kabm}
K_{AB^-}(T) & = & \frac{n_{A^-} n_B}{n_{AB^-}}
= \displaystyle\frac{K_{AB}\,K^{\rm Saha}_{A^-}}{K^{\rm Saha}_{AB^-}}\\
\label{eq_ionmol_neg}
& = & \displaystyle\left(\frac{2\pi \mu k_{\rm B} T}{h^2}\right)^{3/2}  
\frac{Q_{{\rm int},A^-} Q_{{\rm int}, B}}{Q_{{\rm int},AB^-}} 
\exp{\left(-\frac{D_{AB^-}}{k_{\rm B} T}\right)}\, ,
\end{eqnarray}
where the dissociation energy is now 
$D_{AB^-}=D_{AB}+I_{AB^-}-I_{A^-}$, and $I_{AB^-}$ denotes the electron
affinity  of $AB$, or equivalently,  the neutralisation energy of $AB^-$.

\subsubsection{Conservation relations}
\label{sssec_consrel}
In addition to the equilibrium relations (dissociation-recombination
and ionisation), 
there exist three additional types of equations that will
completely determine the concentrations of the
various species of the plasma, namely: 
i) conservation of atomic nuclei for each chemical species,
ii) charge neutrality, and
iii) conservation of the total number of nuclei.

Let us denote with $\mathcal{N}_{\rm el}$ the number of chemical
elements, $\mathcal{N}_{\rm
  mol}$ the number of molecules (neutral and ionised), and
 $\mathcal{N}_{\rm tot}$ the total number of species under
consideration  (neutral and ionised atoms and molecules).

Indicating with $N_\alpha$ the number density of nuclei of type
$\alpha$ (occurring in atoms, ions and molecules), and
$\varepsilon_{\alpha}=N_{\alpha}/N_{\rm a}$ its fractional abundance
with respect to the total number density of nuclei $N_{\rm a}$
(both in atoms and bound into molecules), then 
the conservation of nuclei requires that
each atomic species $\alpha$ (not a molecule)
fulfils the equation
\begin{equation}
\label{eq_atom}
\varepsilon_{\alpha}\, N_{\rm a} = N_{\alpha}=n_{\alpha} + \sum_{r=1}^{pz} n_{\alpha^{+r}} + n_{\alpha^{-}} +
\displaystyle\sum_{A=1}^{\mathcal{N}_{\rm mol}} \nu_{A,\alpha} n_{A}
\end{equation}
In the right-hand
side member, $n_{\alpha}$ is the number density of neutral atoms; 
the next two terms give the number density of ions  in all 
positive ionisation stages (up to the maximum stage $pz$), and in the negative ionisation stage; 
the last summation is performed
over all molecules (neutral and ionised) which contain the atom
$\alpha$. 
Here $\nu_{A,\alpha}$ corresponds to the stoichiometric coefficient, 
expressing the number of atoms $\alpha$ in molecule $A$.

Charge neutrality requires that
\begin{equation}
\label{eq_charge}
n_e =\displaystyle\sum_{i=1}^{\mathcal{N}_{\rm tot}}
\displaystyle\sum_{r=1}^{pz} r\, n_{A_i^{+r}}
- \displaystyle\sum_{j=1}^{\mathcal{N}_{\rm tot}} 
n_{A_j^{-}}
\end{equation}
where we include all appropriate atomic and molecular ions, with  both
positive and negative electric charges. For each species $A_i$, 
the total number of free electrons is evaluated by means 
of the second internal summation extended up to $pz$, which 
corresponds to the highest positive ionisation stage. 
Negative ionisation produces a loss of free electrons, which
explains the minus  preceding the last summation.

Finally, the necessary normalisation is given by the ideal gas
law, so that the total number density $n_{\rm tot}$ of all particles
obeys the relation:
\begin{equation}
\label{eq_ntot}
n_{\rm tot} = n_e + \displaystyle\sum_{A=1}^{\mathcal{N}_{\rm tot}} n_{A} 
= \displaystyle\frac{P}{k_{\rm B} T}
\end{equation}
where the summation includes all molecules and atoms (neutral and
ionised). 
The number density of each atomic species, $N_{\alpha}$, is then
obtained from Eq.~(\ref{eq_atom}) once the fractions
$\varepsilon_{\alpha}=N_{\alpha}/N_{\rm a}$
are given as a part of the problem specification.

The foregoing set of equations (\ref{eq_eqconst}) through 
(\ref{eq_ntot}) are sufficient
for problem solution, as illustrated in the following.

\subsubsection{Solution to the ICE problem}
The solution to the chemical equilibrium problem in \AE SOPUS is based
in large part on source code available under the GPL from the
SSynth project (http://sourceforge.net/projects/ssynth/) that is developed
by Alan W. Irwin and Ana M. Larson.
Basic thermodynamic data together with a few FORTRAN routines 
were adopted with the necessary modifications, as detailed below.

\subsubsection{Thermodynamic data}
From the SSynth package we make use, in particular, 
of the whole compilation of internal partition
functions, ionisation and dissociation energies. 
Each species (atomic and molecular) is assigned a set
of fitting coefficients of
the polynomial form
\begin{equation}
\ln Q =\displaystyle\sum_{i=0}^{\mathfrak{m}} a_i (\ln T)^i\,,
\label{eq_qint}
\end{equation}
based mostly  on the works by Irwin (1981, 1988) and Sauval \& Tatum
(1984). In most cases the degree of the polynomials is five ($\mathfrak{m}=6$).
The original compilation was partially modified and extended to include
additional ionisation stages for atoms, and two more molecules, 
H$_3^+$ and FeH, that may be relevant in the opacity computation.
We consider the ionisation
stages from I to V  for all elements from C to
Ni (up to VI for O and Ne), and from I to III for heavier atoms
from  Cu to U. 
Specifically, our revision/extension of the original Irwin's database
involve the following species.

The partition functions for the C to Ni group have been
re-calculated with the routine {\em pfsaha} of the ATLAS12 code (Kurucz
1993a), varying the temperature from $5000$ to $20000$ K in steps of
$100$ K. The partition functions of the $15$ rare earth elements
belonging to the Lanthanoid series, from La to Lu, have been re-computed
with the routine {\em pfword} from the UCLSYN spectrum synthesis code
(Smith \& Dworetsky 1988) incrementing the temperature from $6000$ to $30000$ K in steps of
$100$ K.  This revision was motivated by the substantial changes in the energy levels
of the earth-rare elements introduced in more recent years
(Alan Irwin, private communication; see e.g. Cowley et al. 1994). 
We have verified that,
 the UCLSYN partition functions for third spectra of the Lanthanides are in close
agreement with the data presented in Cowley et al. 1994, while the results from 
ATLAS12  or Irwin's (1981) compilation are usually lower, in some cases 
by up to a factor of two (e.g. for Ce$^{+3}$ and  Tb$^{+3}$).
The partition function for FeH is given from  Dulick et
al. (2003)  over a temperature range from $1000$ to $3500$ K in steps of
$100$ K.
Then, for all the revised species, we have obtained  the fitting coefficients of Eq.~(\ref{eq_qint}) by the method of least-squares fitting. 
In most cases the best fitting is achieved with a $\chi^2$ parameter 
lower than $10^{-4}$.
For H$_3^+$ we use the original fitting polynomial provided by Neale \& Tennyson (1995).
 
In total, our database of partition functions consists of
 $\mathcal{N}_{\rm tot}\approx 800$
species, including $\approx 300 $ atoms (neutral and ionised) 
from H to U,  and $\mathcal{N}_{\rm mol}\approx 500$ molecules.

\begin{table*}
\begin{center}
\begin{tabular}{l|l|l|l}
\hline
\hline
Process  & Symbol & Reaction & References and Comments \\
\hline
\multirow{3}*{\em Rayleigh} & $\sigma_{\rm Ray}$($\HH_2$) & $\HH_2 + h\nu
\rightarrow \HH_2 + h\nu'$ & Dalgarno \& Williams (1962) \\\cline{4-4}
& $\sigma_{\rm Ray}$($\HH$) & $\HH + h\nu \rightarrow \HH + h\nu'$&
\multirow{2}*{Dalgarno (1962)}\\
& $\sigma_{\rm Ray}$($\He$) & $\He + h\nu \rightarrow \He + h\nu'$& \\
\hline
{\em Thomson} & Th($\e$) &$\e + h\nu \rightarrow \e + h\nu'$ & NIST
(2006 CODATA recommended value)\\
\hline
\multirow{8}*{\em free-free} & $\sigma_{\rm ff}$($\HH^-$) & 
$\HH+\e +h\nu \rightarrow \HH + \e$& John (1988) \\
& \multirow{1}*{$\sigma_{\rm ff}$($\HH$)} & 
\multirow{1}*{$\HH^+ + \e + h\nu \rightarrow \HH^+ + \e $} & Method as in Kurucz (1970) based on  Karsas \& Latter (1961)\\
&$\sigma_{\rm ff}$($\HH_2^+$) & $\HH^+ + \HH + h\nu \rightarrow \HH^+ + \HH$ & Lebedev et al. (2003)\\
&$\sigma_{\rm ff}$($\HH_2^-$) & $ \HH_2 + \e +h\nu \rightarrow \HH_2 + \e$ & John (1975)\\
&$\sigma_{\rm ff}$($\HH_3$) & $\HH_3^+ + \e + h\nu \rightarrow \HH_3^+ + \e $& $\sigma_{\rm ff}(\HH_3)=\sigma_{\rm ff}(\HH)$ (assumed) \\
&$\sigma_{\rm ff}$($\He^-$) & $\He + \e + h\nu \rightarrow \He+ \e$ & Carbon et al. (1969) \\
&$\sigma_{\rm ff}$($\He$) &$\He^+ + \e + h\nu \rightarrow \He^+ + \e$& $\sigma_{\rm ff}(\He)=\sigma_{\rm ff}(\HH)$ (assumed)\\
&$\sigma_{\rm ff}$($\He^+$) &$\He^{++} + \e + h\nu \rightarrow \He^{++} + \e$ & $\sigma_{\rm ff}(\He^+)=\sigma_{\rm ff}(\HH)$ (assumed)\\
\hline
\multirow{7}*{\em bound-free} & $\sigma_{\rm bf}$($\HH^-$) & $\HH^- + h\nu \rightarrow \HH + \e$ & John (1998) \\\cline{4-4}
&  \multirow{2}*{$\sigma_{\rm bf}$($\HH$)} & \multirow{2}*{$\HH + h\nu \rightarrow \HH^+ + \e$} & Method as in Kurucz (1970) based on\\
& & & Gingerich (1969) and Karsas \& Latter (1961)\\\cline{4-4}
 & $\sigma_{\rm bf}$($\HH_2^+$) & $\HH_2^+ + h\nu \rightarrow \HH^+ + \HH $ & Lebedev et al. (2003)\\\cline{4-4}
&\multirow{2}*{$\sigma_{\rm bf}$($\He$)} &  \multirow{2}*{$\He + h\nu \rightarrow \He^+ + \e$} & Method as in Kurucz (1970) based on\\
& & & Gingerigh (1964) and Hunger \& Van Blerkom (1967) \\\cline{4-4}
& $\sigma_{\rm bf}$($\He^+$) & $\He^+ + h\nu \rightarrow \He^{++} + \e$ &  Hunger \& Van Blerkom (1967)\\
\hline
\multirow{1}*{\em bound-bound} & $\sigma_{\rm bb}$($\HH$) & \multirow{1}*{$\HH + h\nu \rightarrow \HH^*$}& Kurucz (1970) including Stark broadening\\
\hline
\multirow{6}{2cm}{\em Collision\\induced\\absorption} & \multirow{2}*{$\sigma_{\rm CIA}$($\HH_2/\HH_2$)} & \multirow{2}*{$\HH_2+\HH_2+h\nu \rightarrow \HH_2+\HH_2$} & $600\K<T<7000 \K$,  $20\cm^{-1}<\tilde{\nu}<20 000\cm^{-1}$ \\
& & & Borisow et al. (1997) \\\cline{4-4}
 & \multirow{2}*{$\sigma_{\rm CIA}$($\HH_2/\He$)} & \multirow{2}*{$\HH_2+\He+h\nu \rightarrow \HH_2+\He$} & $1000\K<T<7000\K$, $25\cm^{-1}y<\tilde{\nu}<20088\cm^{-1}$ \\
& & & J\o rgensen et al. (2000)\\\cline{4-4}
& \multirow{2}*{$\sigma_{\rm CIA}$($\HH/\He$)} & \multirow{2}*{$\HH+\He+h\nu \rightarrow \HH+\He$} & $1500\K <T<10000 \K$, $50\cm^{-1}<\tilde{\nu}<11000\cm^{-1}$\\
& & &  Borisow et al. (2001) \\
\hline
\end{tabular}
\end{center}
\caption{Scattering and absorption processes involving H and He
nuclei, considered in this work.} 
\label{tab_opacsource}
\end{table*}

\subsubsection{Method}
\label{sssec_method}
First we need to specify the list of atoms, ions and
molecules which should be considered, 
together with the values of  
gas pressure $P$,  temperature $T$ and chemical
abundances $\varepsilon_\alpha =N_{\alpha}/N_{\rm a}$.
Then, the code arranges a system consisting of $\mathcal{N}_{\rm el}+2$ 
non-linear equations 
for the number densities of neutral atoms $n_{\alpha}$, the total number 
density of atoms $N_{\rm a}$, and the electron density $n_{e}$.
Once these densities are known, the number densities of any other
ionised and/or molecular species are
calculated by solving for their concentrations 
in Eqs.~(\ref{eq_eqconst}), (\ref{eq_sahaconst}), 
(\ref{eq_sahaconst_neg}), (\ref{eq_kabp}), or (\ref{eq_kabm}) using the equilibrium/ionisation  
constant appropriate for each atom or molecule.
Given the non-linearity of the equations, the  system is 
conveniently solved by using 
a standard Newton-Raphson iterative method (Press et al. 1986).
Numerical details are given in Appendix~\ref{apx_method}.

It is worth remarking that the EOS in \AE SOPUS can easily deal with
{\em any} chemical mixture, including peculiar cases such as 
zero-metallicity $(Z=0)$ or hydrogen-free $(X=0)$ gas. In general, 
no convergence problem has been encountered within 
the assumed ranges of the state variables.

In place of the gas pressure $P$, 
it is also possible to specify 
the gas density $\rho$. 
In this case a second external iteration cycle is switched on   
according to a root-finding numerical scheme. At each $i^{\rm th}$ iteration 
a new value $P_i$ is assigned to the pressure and the 
EOS is solved yielding the corresponding 
$\rho_i=P_i\, (\mu m_{\rm u})/{k_{\rm B} T}$, where $\mu$ is the mean
molecular weight in units of atomic mass $m_{\rm u}$. 
The process is repeated until the 
difference $|\log(\rho_i) - \log(\rho)|$ 
decreases below a specified tolerance $\delta_{\rho}$. In our computations 
we adopt $\delta_{\rho}= 10^{-8}$, and convergence is reached typically 
after 3-4 iterations. 

%


\begin{table}
\begin{tabular}{ccl}
\hline
\hline
\multicolumn{2}{c}{Species} & Source \& Reference \\
\hline
\multirow{5}*{Atoms} & {C,N,O} &  \\
& { Ne,Na, Mg} & \\
& {Al,Si,S} &  OP: Seaton (2005)   for $\log(T) \ge 3.6$\\
& {Ar, Ca, Cr} &  \\
& {Mn,Fe,Ni} & \\
\hline
\multirow{3}*{Atoms} & \multirow{1}*{CI, NI} &  Method as in Kurucz (1970) based on\\
&    {OI, MgI} & Peach (1970) and Henry (1970)\\
&{AlI, SiI} & for $\log(T) < 3.6$\\
\hline
\multirow{20}*{Molecules} &  HF & LL: Uttenthaler et al. (2008) \\
 &HCl & LL: Rothman et al. (2005)\\
 &CH & LL: J\o rgensen (1997)\\ 
 &C$_2$ & LL: Querci et al. (1974) \\
 &CN &  LL: J\o rgensen (1997)\\
 &CO &  LL: Goorvitch \& Chackerian (1994) \\
 &OH &  LL: Schwenke (1997) \\
 &SiO &  LL: Rothman et al. (2005)\\
 &TiO &  LL: Schwenke (1998) \\
 &VO &  LL: Alvarez \& Plez (1998) \\
 &CrH & LL: Bauschlicher et al. (2001)\\
 &FeH & LL: Dulick et al. (2003) \\
 &YO & LL: Littleton (2007) (2001) \\
 &ZrO & LL: Plez (2007) \\ 
 &H$_2$O & LL: Barber et al. (2006) \\
 &HCN & LL: Harris et al. (2003) \\
 &C$_3$ &  OS: J\o rgensen et al. (1989) \\
 &CO$_2$ & LL: Rothman et al. (1995) \\
 &SO$_2$ &  LL: Rothman et al. (2005)\\
 &C$_2$H$_2$ &  OS: J\o rgensen (1997) \\
\hline
\end{tabular}
\caption{Data sources for the  atomic and molecular monochromatic absorption coefficients.
Atomic absorption coefficients (including both continuum and discrete
opacities) are  
from the Opacity Project (OP) database, while molecular absorption
coefficients are extracted from either line lists (LL) or opacity
sampling (OS) data.} 
\label{tab_opacmol}
\end{table}

\subsection{Opacity}
\label{sec_opacso}
In our computations  we consider the following {\em continuum} 
opacity processes
\begin{itemize}
\item Rayleigh scattering,
\item Thomson scattering,
\item Bound-free absorption due to photoionisation,
\item Free-free absorption,
\item Collision-induced absorption (CIA),
\end{itemize}
and {\em line} opacity processes
\begin{itemize} 
\item Atomic bound-bound absorption,
\item Molecular band absorption.
\end{itemize}

Denoting with $\sigma_j(\nu)$ the monochromatic cross section (in cm$^2$) of the $j^{th}$ 
absorption process (not scattering),  the monochromatic true absorption opacity and scattering 
opacity per unit mass (in cm$^2$ g$^{-1}$) are calculated with 
\begin{eqnarray}
\label{eq_kabs}
\kappa_j^{\rm abs}(\nu) & = & \displaystyle\frac{n_j}{\rho}\, \sigma_j^{\rm abs}(\nu)\, 
(1-e^{-h\nu/k_{\rm B}T})\\
\label{eq_kscat}
\kappa_j^{\rm scatt}(\nu) & = & \displaystyle\frac{n_j}{\rho}\, \sigma_j^{\rm scatt}(\nu)\,,
\end{eqnarray}
where $n_j$ is the number density of particles of type $j$,  $\rho$ is the gas density, and 
$(1-e^{-h\nu/k_{\rm B}T})$ is a correction factor for stimulated emission.

Tables~\ref{tab_opacsource} and \ref{tab_opacmol} detail 
the whole compilation of the scattering and absorption 
processes considered here. 

The monochromatic opacity cross sections for atoms (except for H and He), taken from the OP database, 
are interpolated in frequency, temperature and electron density, according to the formalism described 
in Seaton  et al. (1994) and Seaton (2005).
They include all radiative continuum and discrete opacity processes. 
Line broadening is taken into account as the
result of thermal Doppler effects, radiation damping and pressure
effects.
  
The monochromatic molecular absorption coefficient caused 
by each of the different
species included in our code is taken 
from opacity sampling
(OS) files produced for the selected frequency 
grid (see Sect.~\ref{sssec_fgrid} and
Appendix~\ref{apx_fgrid}),
that are in
most cases calculated directly from the corresponding line
lists (see Table~\ref{tab_opacmol}).
The only exceptions are C$_2$H$_2$ and
C$_3$ for which we use already existing pre-computed opacity sampling data.

Where line lists are adopted, the absorption cross section of a 
spectral line, involving the bound-bound transition from state $m$ to state $n$, 
is evaluated with the relation:
\begin{equation}
\sigma^{\rm abs}_{\rm bb}(\nu)=\frac{\pi e^2}{m_e c}\, \frac{gf}{Q(T)}\, e^{-E_0/k_{\rm B}T}\, 
\left(1-e^{-h\nu_0/k_{\rm B}T}\right)\,\phi(\nu)\,
\label{eq_sigmamol}
\end{equation}
with $e$ and $m_e$ the charge and mass of the electron, $c$ the speed of light, $h\nu_0$ the
energy of the corresponding radiation,
$Q(T)$ the total partition function 
(being the product, $Q_{\rm trans}~Q_{\rm int}$, of the translational and internal partition functions) 
of the molecular species under consideration, 
$E_0$ the excitation energy of the lower level $m$
of the transition, $g f$ the product of the statistical weight $g_{(m)}$ of the level times
the oscillator strength $f_{(m,n)}$ of the transition. The correction for stimulated emission 
is given by the term in brackets.
The normalised broadening function, $\phi(\nu)$,  for the line profile takes into account the effect of
thermal broadening and non thermal-contribution of microturbolent velocities, according to the equation:
\begin{equation}
\phi(\nu)=\frac{1}{\Delta_{\nu}\sqrt{\pi}}\,e^{-\left(\frac{\nu-\nu_0}{\Delta_{\nu}}\right)^2}\,
\label{eq_phi}
\end{equation}
with a Doppler width $\Delta_{\nu}$ given by
\begin{equation}
\Delta_{\nu}=\frac{\nu_{0}}{c}\sqrt{\frac{2 k_{\rm B}T}{m}+\xi^2}\,,
\label{eq_doppler}
\end{equation}
where $m$ is the mass of the molecule, and 
$\xi$ is the microturbolent velocity, which is assigned the value $2.5$~km/s.
More details about the input data and the treatment of molecular line opacities
can be found in Aringer (2000), Lederer \& Aringer~(2009), and Aringer et al.~(2009).

In summary, to generate the molecular OS files directly
from the line lists, prior to  the execution of \AE SOPUS, we proceed as follows.
For each value of a selected set of temperatures, 
($13$ values in the range $600$~ K~$\la~T~\la~10000$~K),
the monochromatic absorption coefficient of a molecular species
at a given wavelength point, $\sigma^{\rm abs}_{\rm mol}(\nu)$, is obtained by adding up 
the contributions of all the lines  
in the list with the corresponding broadening functions 
taken into account:
\begin{equation}
\sigma^{\rm abs}_{\rm mol}(\nu)=\displaystyle\sum_{\rm lines} \sigma^{\rm abs}_{\rm bb}(\nu)\,,
\end{equation} 
where each term $\sigma^{\rm abs}_{\rm bb}(\nu)$ is evaluated with Eqs.~(\ref{eq_sigmamol}) 
-- (\ref{eq_doppler}).
Then, during the computations with \AE SOPUS, 
we interpolate on the OS tables for any given temperature of the gas. 
We notice that the errors brought about by this
interpolation are marginal compared to all other sources of uncertainty
(e.g. molecular data, microturbolence velocity, solar abundances, etc.).
\subsubsection{The Rosseland mean}
\label{sssec_rmk}
Once the total monochromatic opacity coefficient is obtained by
summing up all the contributions of true absorption  and scattering
\begin{equation}
\kappa(\nu)=\displaystyle\sum_{j}\kappa_j^{\rm abs}(\nu) + \kappa_j^{\rm scatt}(\nu)\,\,\,\,,
\end{equation}
then the Rosseland mean opacity, classically defined by Eq.~(\ref{eq_rosseland})
is conveniently calculated with (see e.g. Seaton et al. 1994):
\begin{equation} 
\label{eq_krossu}
\frac{1}{\kappa_{\rm R}(\rho,T)}=\int_0^\infty\frac{F_R(u)}{\kappa(u)}\rmd  u 
\end{equation}
where
\begin{equation}
F_R(u)=\frac{15}{(4\pi^4)}\,u^4\exp(-u)/[1-\exp(-u)]^2.
\label{weight}
\end{equation}
In the above equations $\nu$ is the photon frequency, and
$u~=~h\nu/(\kb T)$ is the normalised photon energy.  In
our calculations $\kappa_{\rm R}$ denotes the absorption coefficient
per unit mass, and is always given in cm$^2$ g$^{-1}$.
Since the opacity coefficient $\kappa_{\nu}$ enters 
Eq.~(\ref{eq_rosseland}) as an inverse, the minima dominate the values of the Rosseland mean.
It follows that a large $\kappa_{\rm R}$  implies large absorption from the radiation
beam, while a small $\kappa_{\rm R}$ indicates that the energy losses from the beam remain little
as it propagates through the matter.

In practise, the numerical integration of Eq.~(\ref{eq_krossu}) requires to specify 
two finite (lower and upper) limits, $u_1$ and $u_2$, and the grid of frequency points.
The choice of the limits must guarantee the covering of the relevant wavelength region
for the weighting function ${\partial B_\nu}$/${\partial T}$, so as to include its maximum and
the declining wings. 

In this respect it useful to recall that, in analogy with  the Wien's displacement law 
for the Planck function, the wavelength $\lambda_{\rm max}$ of the 
the maximum of  ${\partial B_\nu}/{\partial T}$ is inversely proportional
to the temperature according to
\begin{equation}
\lambda_{\rm max}\,[\mu{\rm m}] = \displaystyle\frac{3756.56}{T[{\rm K}]}
\label{eq_lmax}
\end{equation}
It follows that the maximum of the function $F_{\rm R}$ is reached for $u_{\rm max} = 3.8300$.

In our calculations we adopt the integration limits
$u_1~\simeq~10^{-3}$ and $u_2\simeq~64$, corresponding to the 
wave numbers  $\tilde{\nu}_1~=~10 $ cm$^{-1}$ and
$\tilde{\nu}_2~=2~10^{5} $ cm$^{-1}$, and wavelengths $\lambda_1= 1000$
$\mu$m and $\lambda_2~=~0.04$ $\mu$m, respectively.
We have verified that these values largely satisfy the condition of
spectral coverage of the
weighting function over the entire temperature range, $3.2~\le~\log(T)
~\le~4.5$, here considered.

\subsubsection{The frequency grid and computing time}
\label{sssec_fgrid}
Since in our calculations a number of crucial opacity sources, i.e. molecular absorption bands,
 are included as OS data, it is convenient to specify, prior
of computations, a grid of frequency points, which should be common to both the OS treatment and 
the numerical integration of Eq.~(\ref{eq_rosseland}).
The frequency distribution will be determined as a compromise 
between the precision (and accuracy)  of the integration and the speed 
of calculations.

For this purpose we employ the algorithm by Helling \& J\o rgensen (1998),
that was developed to optimise the frequency distribution in the opacity
sampling technique when dealing with a small number of frequency points.
We performed a few tests adopting frequency grids with decreasing size, namely 
with $n_{\rm tot}= 5488,\,1799,\,944,\, 510,\,$ and $149$ frequency points. 
More details are given in Sect.~\ref{apx_fgrid}. 
The results discussed in the following sections refer to the grid with $n_{\rm tot}=944$ 
points, which has proved to yield reasonably accurate RM opacities.

Besides the quality of the results, another relevant aspect 
is the computing time.
With the present choice of the frequency grid , i.e. $n_{\rm tot}=944$ points,  
generating one table at fixed
chemical composition, arranged with the default grid 
of the state parameters ($T$ and $R$, see Sect.~\ref{ssec_tr}), 
i.e. containing $N_T \times N_R= 67\times 19 = 1273$ opacity values, 
takes $\tau\sim 45$ s with a 2.0 GHz processor.
Adopting other frequency grids would require shorter/longer computing times,
roughly  $\tau\sim 200$ s for $n_{\rm tot}=5488$; 
$\tau\sim 70$ s for $n_{\rm tot}=1799$;
$\tau\sim 30$ s for $n_{\rm tot}=510$; and
$\tau\sim 15$ s for $n_{\rm tot}=149$.
These values prove that \AE SOPUS is indeed a quick computational tool,
which has made it feasible, for the first time, 
the setup of a web-interface  (http://stev.oapd.inaf.it/aesopus)
to produce low-temperature RM opacity tables on demand and in short times.

The main reason of such a fast performance
mainly resides in the optimised use of the opacity sampling method 
to describe molecular line absorption, 
and the adoption of pre-tabulated absorption cross-sections for metals 
(available from the Opacity Project website). 
In this way the line-opacity data is extracted (e.g. from line lists and the OP database) 
and stored in a convenient format {\em before} the execution of {\AE}SOPUS,
thus avoiding to deal with huge line lists {\em during} the opacity computations. 
This latter approach is potentially more accurate, 
but extremely time-consuming (e.g. F05).

Moreover the  improvement in accuracy that would be achievable with the on-the-fly 
treatment of the line lists is in principle reduced  
when adopting a frequency grid for integration which is much sparser 
(e.g. $\sim 10^4$ frequency points as in F05) than the dimension of the line lists 
(up to $10^7 - 10^8$ line transitions).
On the other hand, as shown by our previous tests and also 
by F05, while the computing time scales
almost linearly with the number of frequency points,
the gain in precision does not, 
so that the RM opacities are found to vary just negligibly beyond a certain threshold
(see also Helling et al. 1998 and Appendix~\ref{apx_fgrid}).
All these arguments and the results discussed in 
Sects.~\ref{sssec_kcomp} support the indication  that the agile approach adopted in {\AE}SOPUS  
is suitable to produce RM opacities with a very favourable accuracy/computing-time ratio.

\section{Opacity tables: basic parameters}
\label{sec_tables}
Tables of RM opacities can be generated once a few input parameters are
specified, namely: 
the chemical composition of the gas, and the bi-dimensional space over
which one pair of independent state variables is made vary.

\begin{figure}
\resizebox{\hsize}{!}{\includegraphics{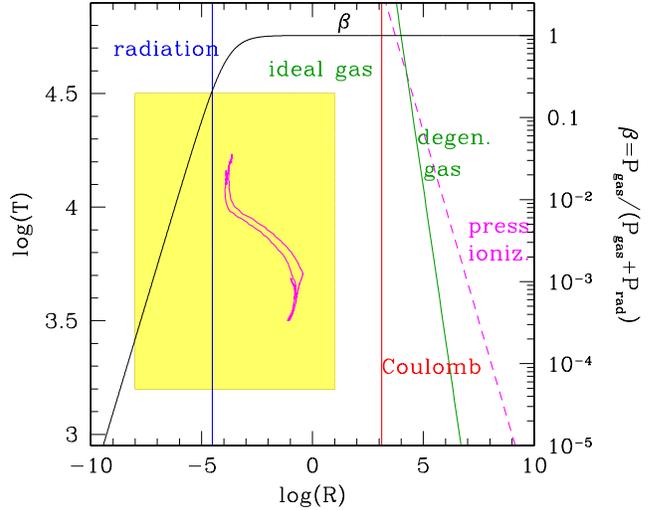}}
\caption{Location of our RM opacity tables in the 
$\log(T)-\log(R)$ diagram (shaded rectangular area), together with  
the approximate boundaries 
between regions where the total pressure is dominated by different
effects: radiation pressure, ideal gas, electron-degenerate gas,
Coulomb interactions, and pressure ionisation.
The vertical line to the left of which radiation dominates the  pressure
is given by $P_{\rm gas}=P_{\rm rad}$ with $\mu=0.5$.
Along the border line between the ideal and degenerate gas we equate 
the corresponding pressures of a non-relativistic electron gas with mean molecular weight $\mu_e=2$.
The vertical boundary at $\log(R)\sim 3$, beyond which the Coulomb coupling of
charged particles should become important, is
defined by the condition $\Gamma_{\rm C}=1$, where $\Gamma_{\rm C}=1.1\,10^{-5} T/\rho^{1/3}$ is
the Coulomb coupling parameter for an ionised-hydrogen plasma.
Pressure ionisation is assumed to become dominant at $\log(\rho)=0$
for $\log(T)\le 4.5$, a typical value according to the analysis developed by  Luo (1997).
The behaviour of the parameter $\beta$ (defined in the text)
is also shown. 
As an example, the evolution of the photospheric parameters 
($T_{\rm eff},\,R_{\rm phot}$) of a stellar model with initial mass
$M= 5\,M_{\odot}$, and metallicity 
$Z=Z_{\rm ref}=0.02$ is depicted by a magenta
line, covering the evolution from the pre-main sequence to the first pulses on
the TP-AGB (calculations performed with the Padova stellar evolution code).}
\label{fig_eos}
\end{figure}

\subsection{State variables}
\label{ssec_tr}
Under the assumption of ideal gas, described by the law
\begin{equation} 
P_{\rm gas}= \displaystyle\frac{ k_{\rm B}}{\mu m_{\rm u}} \rho T\,,
\end{equation}
one must specify  one pair of independent state variables.
Usual choices are, for instance,  $(P_{\rm gas}, T)$ or  $(\rho, T)$.
For practical and historical reasons, 
opacity tables are generally built as a function of the logarithm 
of the temperature $T$, and the logarithm of the $R$ variable, defined
as $R= \rho/(T_6)^3$, with $T_6=T/(10^6\,{\rm K})$.

An advantage of using the $R$ parameter, instead of $\rho$ or $P$,
is that the opacity tables can cover rectangular regions
of the $(R,\,T)$-plane, without the nasty voids over extended temperature
ranges that would come out if intervening changes in the EOS 
are not taken into account (e.g. transition from ideal to degenerate
gas). 

Interestingly, as pointed out by Mayer \& Duschl (2005; see their appendix D),
different $R$ values correspond to different gas/radiation
pressure ratios, $\delta = P_{\rm gas}/P_{\rm rad}$.
The relation between $\log(R)$ and $\log(\delta)$ is linear,
with larger $R$ values corresponding to larger $\delta$, i.e. an increasing
importance of $P_{\rm gas}$ against $P_{\rm rad}$.
Moreover, we  notice that the equality
$P_{\rm gas} = P_{\rm rad}$ takes place 
in the range at $-4.8 \la \log(R) \la -4.5$,
assuming a mean molecular weight varying in the interval $0.5 \la \mu \la 1$.
In Fig.~\ref{fig_eos} we also plot 
the quantity $\beta=P_{\rm gas}/(P_{\rm gas}+P_{\rm  rad})$, a 
parameter frequently used by stellar evolutionists.

In this respect Fig.~\ref{fig_eos} illustrates the rectangular region
covered by our RM opacity tables in the $\log(T)-\log(R)$ diagram,
defined by  the intervals $(3.2 \le \log(T) \le 4.5)$ and 
$(-8~\le~\log(R)~\le~1)$.
We note that the table area lies in the domain of the ideal gas, and it
extends into the region dominated by radiation pressure for  
$\log(R) \le -4.5$.
Non ideal effects related to electron degeneracy, Coulomb coupling of
charged particles, and pressure ionisation of atoms are expected to
become dominant outside the table boundaries, 
in the domain of high-density plasmas.

It is important to remark that our RM opacity tables can be
easily extended to higher temperatures, $\log(T) > 4.5$, with 
the RM opacity data provided by OPAL and OP. As a matter of fact the agreement between
our results and OPAL is good in the overlapping transition region, say
$3.9 \la \log(T) \la 4.5$ (see Sect.~\ref{sssec_kcomp} and panel c) of Fig~\ref{fig_kcomp_z02}).

Within the aforementioned limits of the state variables, 
the interactive web mask enables the user to freely
specify the effective ranges of $\log(T)$ and $\log(R)$ of interest as well as
the spacing of the grid points $\Delta\log(T)$ and $\Delta\log(R)$.
From our tests it turns out that a good sampling of the main opacity features
can be achieved with $\Delta\log(T)=0.05$ for $\log(T)>3.7$ and
$\Delta\log(T)=0.01$ for $\log(T)\le 3.7$, and $\Delta\log(R)=0.5$.
In any case, the choice should be driven by consideration of two aspects, i.e. 
maximum memory allocation, and accuracy of the adopted 
interpolation scheme.

\subsection{Chemical composition} 
\label{ssec_chem}
It is specified in terms of the following quantities:
\begin{itemize}
\item The reference solar mixture;
\item The reference metallicity $Z_{\rm ref}$;
\item The hydrogen abundance $X$;
\item The reference mixture;
\item The enhancement/depression factor $f_i$ of each element (heavier than helium), 
with respect to its reference abundance.  
\end{itemize}

The reference solar mixture can be chosen among various options, which are
referenced in Table~\ref{tab_sun}. For their relevance to the opacity issue,
the corresponding solar metallicity, $Z_{\odot}$, and the 
$({\rm C/O})_{\odot}$ ratio\footnote{Throughout the paper the C/O ratio is calculated
using the abundances of carbon and oxygen expressed as number
fractions, i.e. C/O~$=\varepsilon_{\rm C}/\varepsilon_{\rm O}$ following the
definition given by Eq.~(\ref{eq_xicsi}).}
are also indicated. 
Scrolling Table~\ref{tab_sun} from top to bottom we note that $Z_{\odot}$ significantly
decreases, passing from  $\sim 0.019$ in AG89 down to  $\sim 0.012$ in GAS07. This implies that
opacity tables constructed assuming the same $Z$ may notably differ depending
on the adopted solar mixture.  Concerning C/O,  a key parameter affecting the opacities
for $\log(T) \la 3.5$, we see that it spans a rather narrow range ($0.43 \la$ C/O $\la 0.53$)
passing from one  compilation to the other, except for the H01 which corresponds to a higher value, 
C/O $\sim 0.72$. How much these differences in the reference solar mixtures may 
impact on the resulting opacities is discussed in Sect.~\ref{ssec_ksun}. 

\begin{table}
\centering
\begin{minipage}{0.5\textwidth}
\begin{tabular}{lccc}
\hline
\hline
Reference  & $Z_{\odot}$ & (C/O)$_{\odot}$ 
& (C/O)\footnote{This abundance ratio is defined by Eq.~(\ref{eq_crit1}).}$_{{\rm crit},1}$\\
\hline
Anders \& Grevesse 1989 (AG89) & 0.0194 & 0.427 & 0.958\\
Grevesse \& Noels 1993  (GN93) & 0.0173 & 0.479 & 0.952\\
Grevesse \& Sauval 1998 (GS98) & 0.0170 & 0.490 & 0.947 \\
Holweger 2001 (H01)\footnote{The elemental abundances are taken from Grevesse \& Sauval (1998), but 
for  C, N, O, Ne, Mg, Si, and Fe that are modified following the revision by Howeger (2001).} 
  & 0.0149 & 0.718 & 0.937 \\
Lodders 2003     (L03)       & 0.0132 & 0.501 & 0.929 \\ 
Grevesse et al. 2007  (GAS07) & 0.0122 & 0.537 & 0.929 \\
Caffau et al. 2009 (C09)\footnote{The elemental abundances are taken from Grevesse \& Sauval (1998), but 
for N, O, and Ne following the revision by Caffau et al. (2008, 2009).}
    & 0.0155 & 0.575 & 0.938 \\ 
\hline
\end{tabular}\par
   \vspace{-0.75\skip\footins}
   \renewcommand{\footnoterule}{}
\end{minipage}
\caption{Compilations of the solar chemical composition adopted in the computation 
of the EOS and gas opacities. For each  mixture
the solar total metallicity $Z_{\odot}$ (in mass fraction), 
the abundance ratios  (C/O)$_{\odot}$ and (C/O)$_{{\rm crit},1}$ are indicated for comparison.
The latter marks a critical boundary for the gas molecular chemistry in the range 
$3.2~\le~\log(T)~\le~3.6$. The C and O abundances are expressed as
number fractions.}
\label{tab_sun}
\end{table}

\begin{figure}
\resizebox{\hsize}{!}{\includegraphics{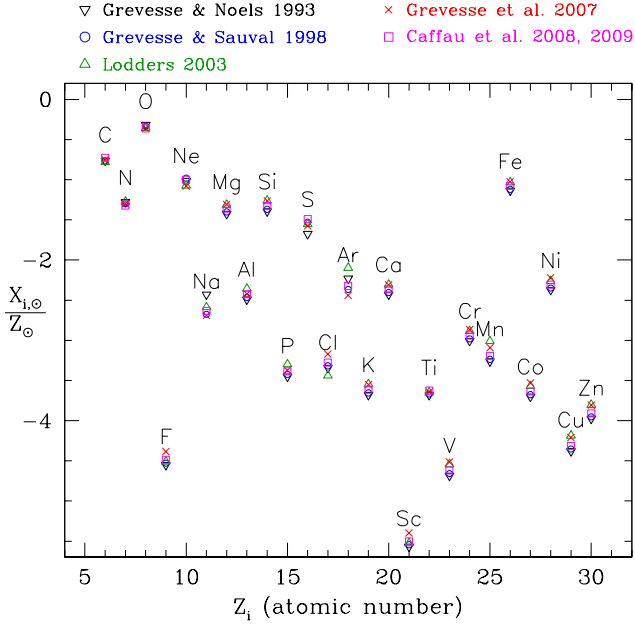}}
\caption{Fractional abundances of elements, 
with nuclear charge $Z_i=6-30$, normalised to the solar metallicity
according to various compilations, as indicated.
 }
\label{fig_fracsol}
\end{figure}

\begin{figure}
\resizebox{\hsize}{!}{\includegraphics{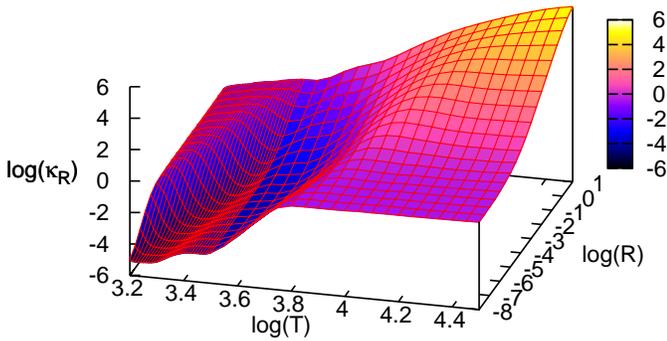}}
\caption{Rosseland mean opacity as a function of variables $T$ and $R$ over the
entire parameter space considered in our calculations.
The adopted composition
is assumed to have  $Z=Z_{\rm ref}=0.02$, $X=0.7$  and the metal abundances 
scaled-solar to the GS98 mixture.} 
\label{fig_opac_map}
\end{figure}

Let us indicate with $\mathcal{N}_Z$ the number of metals, i.e. the chemical elements heavier than helium,
with atomic number $Z_i \ge 3$.
Each metal is characterised by an abundance $X_i$ in mass fraction and, equivalently, an abundance $\varepsilon_i$ in number
fraction, respectively defined as:
\begin{equation}
\label{eq_xicsi}
X_i =  \displaystyle \frac{A_i N_i}{\sum_{j=1}^{\mathcal{N}_{\rm el}} A_j N_j} 
\,\,\,\,\,\,\,{\rm and}\,\,\,\,\,\,\,
\varepsilon_i  =  \displaystyle \frac{N_i}{N_{\rm a}}=\displaystyle \frac{N_i}{\sum_{j=1}^{\mathcal{N}_{\rm el}} N_j}\,,
\end{equation}
where $N_i$ is the number density of nuclei of type $i$ with atomic mass $A_i$, and $N_{\rm a}$ is the
total number density of all atomic species (with the same notation as in Sect.~\ref{sssec_consrel}).
In both cases the normalisation condition must hold, i.e. $\sum_{i=1}^{\mathcal{N}_{\rm el}}X_i=1$ and
$\sum_{i=1}^{\mathcal{N}_{\rm el}}\varepsilon_i=1$.
The total metal abundance is given by $Z= \sum_{i\ge 3} X_i$ in mass fraction, 
and $\varepsilon_{Z} = \sum_{i\ge 3} \varepsilon_i$ in number fraction.

We assign each metal species the variation factors, $f_i$ and $g_i$,
relative to the reference mixture:  
\begin{equation}
\label{eq_fi}
X_i=f_i X_{i,{\rm ref}}
\,\,\,\,\,\,\,{\rm and}\,\,\,\,\,\,\,
\varepsilon_i = g_i \varepsilon_{i,{\rm ref}}\, .
\end{equation} 

The reciprocal relations between $X_i$ and $\varepsilon_i$ derive straightforwardly:
\begin{equation}
\label{eq_xeps}
X_i  =  \displaystyle\frac{A_i \varepsilon_i}{\sum_{j=1}^{\mathcal{N}_{\rm el}} A_j \varepsilon_j}
\,\,\,\,\,\,\,{\rm and}\,\,\,\,\,\,\,
\varepsilon_i  = \displaystyle\ \frac{X_i/A_i}{\sum_{j=1}^{\mathcal{N}_{\rm el}}X_j/A_j }\,\, , 
\end{equation}
as well as those between $f_i$ and $g_i$ for metals:
\begin{equation}
\label{eq_figi}
f_i = g_i \, \displaystyle\frac{\sum_{j\ge 3} A_j \varepsilon_j^{\rm ref}}{\sum_{j\ge 3} A_j \varepsilon_j}
\,\,\,\,\,\,\,{\rm and}\,\,\,\,\,\,\,
g_i = f_i \, \displaystyle\frac{\sum_{j\ge 3} X_j^{\rm ref}/A_j}{\sum_{j\ge 3} X_j/A_j}\, .
\end{equation}
We have verified that $f_i \approx g_i$ as long as they are not too large and the ratios between
the two summations in the left-hand side members of Eq.~(\ref{eq_figi}) do not deviate significantly
from unity (see, for instance, Table~\ref{tab_alpha}).

In principle, the reference chemical mixture can be {\em any} given chemical composition. 
Frequent choices are, for instance,  mixtures with scaled-solar partitions of metals, or with 
enhanced abundances of $\alpha$-elements. The \AE SOPUS code is structured to allow large freedom
in specifying the reference mixture.
For simplicity, in the following we will adopt the solar mixture as the reference composition, so that 
the reference metal abundances are 
\begin{equation}
\label{eq_xref}
X_{i, {\rm ref}}= X_{i, {\odot}}\displaystyle\frac{Z}{Z_{\odot}}
\,\,\,\,\,\,\,{\rm and}\,\,\,\,\,\,\,
\varepsilon_{i, {\rm ref}}= 
\varepsilon_{i,{\odot}}\displaystyle\frac{\varepsilon_{Z}}{\varepsilon_{Z_{\odot}}}
\end{equation} 
with clear meaning of the symbols.
The partitions,  $X_{i, {\odot}}/Z_{\odot}$, of chemical elements from
C to Zn are shown in Fig.~\ref{fig_fracsol} for a few compilations of
the solar chemical composition.

According to the notation presented by Annibali et al. (2007), 
the chemical elements can be conveniently divided into three classes depending on the sign 
of $f_i$ (or $g_i$) , namely:
\begin{itemize}
\item {\em enhanced elements} with $f_{i}>1$  (or $g_{i}>1$);
\item {\em depressed elements} with $f_{i}<1$ (or $g_{i}<1$); 
\item {\em fixed elements} with $f_{i}=1$ (or $g_{i}=1$).
\end{itemize}
The latter correspond to the reference abundances, i.e. scaled-solar in the case discussed here.
Moreover, let us designate 
\begin{itemize}
\item {\em selected elements} with $f_{i}\neq 1$ (or $g_{i} \neq 1$)
\end{itemize}
 the group of elements which are assigned 
variation factors different from unity (either enhanced or depressed), as part of the input specification.
 We limit the discussion here to the case of the abundances $X_i$
expressed in mass  fraction, since exactly the same scheme, with the due substitutions, can be applied to the abundances $\varepsilon_i$ 
in number fraction. In this respect one should  bear in mind that the conversions $X_i \rightleftarrows \varepsilon_i$
are obtained with Eqs.~(\ref{eq_xeps}). 
Starting from the reference mixture, then the new mixture can be obtained in two distinct ways:

\begin{enumerate}
\item  Case $Z \neq Z_{\rm ref}$. The enhancement/depression factors $f_i$
of the {\em selected elements}
produce a net increase/depletion of total metal content 
relative to the reference metallicity
$Z_{\rm ref}$. The actual metallicity  is calculated directly 
with $Z=\sum_{i=1}^{\mathcal{N}_Z} f_i X_{i, {\rm ref}}$.
In this case all $\mathcal{N}_Z$ variation factors $f_i$ 
can be freely specified without any additional constrain.

\item Case $Z=Z_{\rm ref}$. The enhancement/depression factors produce  
non-scaled-solar partitions of metals, while the total reference
metallicity $Z_{\rm ref}$ is to be preserved.
This constraint can be fulfilled with various schemes, e.g. by properly varying the total abundance of 
{\em all other non-selected elements} so as to balance the abundance variation
of the {\em selected group}. For instance, if the selected elements have all $f_i >0$, so that
we refer to them as {\em enhanced group}, then the whole positive abundance variation 
should be compensated by the negative abundance variation of the complementary  {\em depressed group}.
Another possibility is to define a {\em fixed group} of elements 
whose abundances should not be varied, hence not involved in the balance procedure; 
in this case the preservation of the metallicity is obtained by 
properly changing the abundances of a lower number of atomic species among the non-selected ones.

In principle, the quantities
$f_i$ can be chosen independently for up 
to a maximum of $(\mathcal{N}_Z-1)$ elements, while the remaining factor
is bound by the $Z=Z_{\rm ref}$ condition. 
A simple practise is to assign  the same factor to all the elements
belonging to the {\em selected group}, either {\em enhanced} or {\em depressed},
 as frequently done for
$\alpha$-enhanced mixtures. In this respect more details can be found in Sect.~\ref{ssec_alpha}.
\end{enumerate}

The former  case ($Z \neq Z_{\rm ref}$) properly describes a chemical mixture in which  
the abundance variations are the product of nuclear burnings occurring in the stellar interiors.
This applies, for instance, to thermally-pulsing asymptotic giant
branch (TP-AGB) stars whose envelope chemical composition is 
enriched in C and O produced by He-shell flashes and convected 
to the surface by the third dredge-up, 
which leads to an effective increment of the  global metallicity
$(Z>Z_{\rm ref})$.

The latter case ($Z=Z_{\rm ref}$) corresponds, for instance, to chemical mixtures with 
a scaled-solar abundance of CNO elements $X_{\rm CNO}$, but different ratios e.g. 
$X_{\rm C}/X_{\rm CNO}$,  $X_{\rm N}/X_{\rm CNO}$, and $X_{\rm O}/X_{\rm CNO}$. 
Alternatively, if we consider the abundances in number fractions,
 the condition, $\varepsilon_{\rm CNO}=$ const., may describe 
the surface composition of an intermediate-mass star after 
the second dredge-up on the
early AGB, when products of complete CNO-cycle are brought up to the surface.
In this case the total number of CNO catalysts does not change, while
C and O have been partly converted to $^{14}$N.
Another example may refer to     
 $\alpha$- enhanced mixtures with different [$\alpha$/{\rm Fe}]$>0$
but the same metal content $Z$.

Finally, it should be noticed that, once the actual metallicity $Z$ is
determined,  in both cases 
the normalisation condition implies that 
the helium abundance is given by the relation  $Y=1-X-Z$.

\section{Results}
\label{sec_results}
In the following sections we will discuss a few applications
of the new opacity calculations, selecting those ones that may be particularly
relevant in the computation of stellar models.   
For completeness, our results are compared with other opacity
data available in the literature.   
\subsection{Scaled-solar mixtures}
\label{ssec_ksun}

\begin{figure*}
\begin{minipage}{0.33\textwidth}
\resizebox{1.\hsize}{!}{\includegraphics{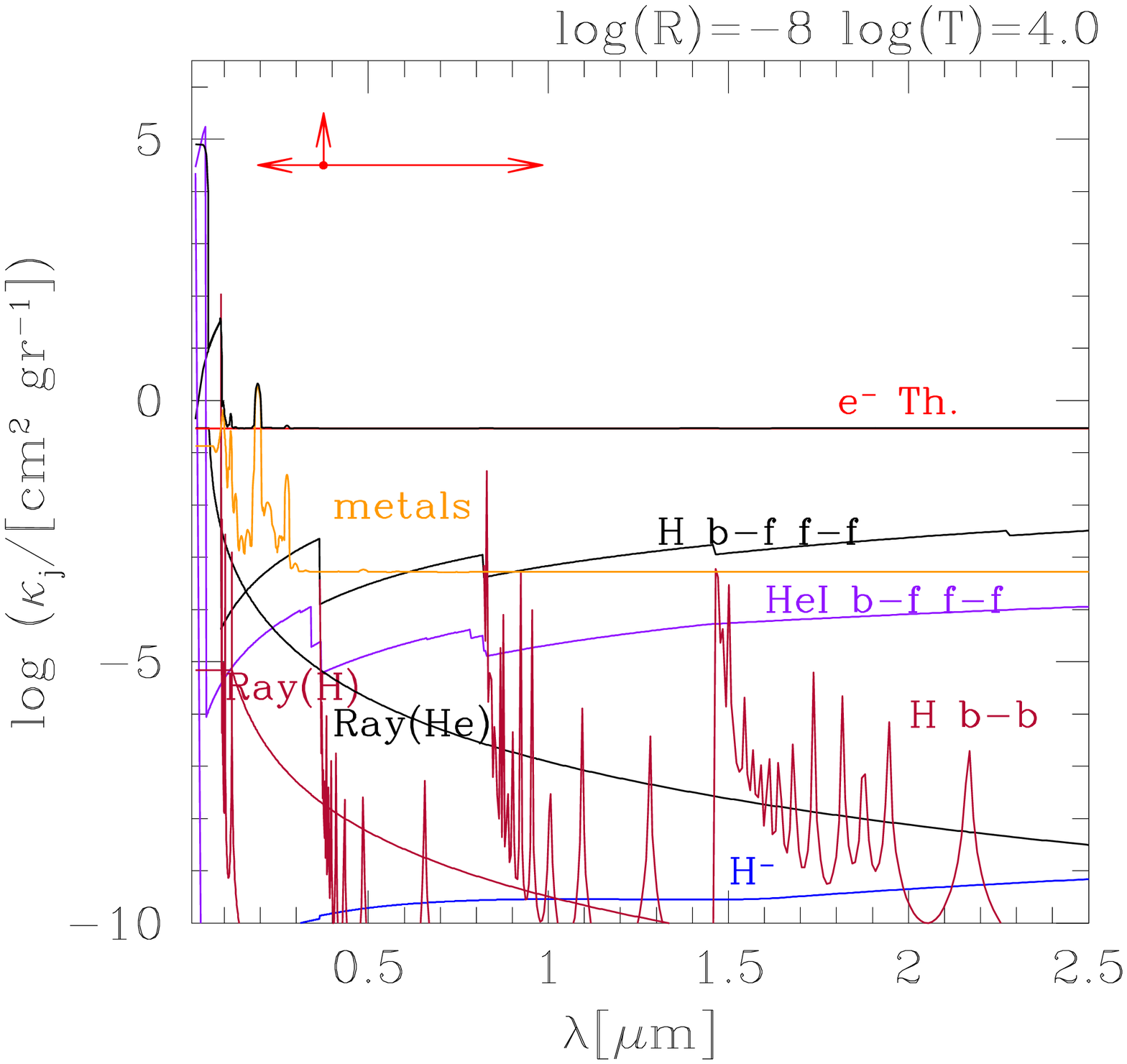}}
\end{minipage}
\begin{minipage}{0.33\textwidth}
\resizebox{1.\hsize}{!}{\includegraphics{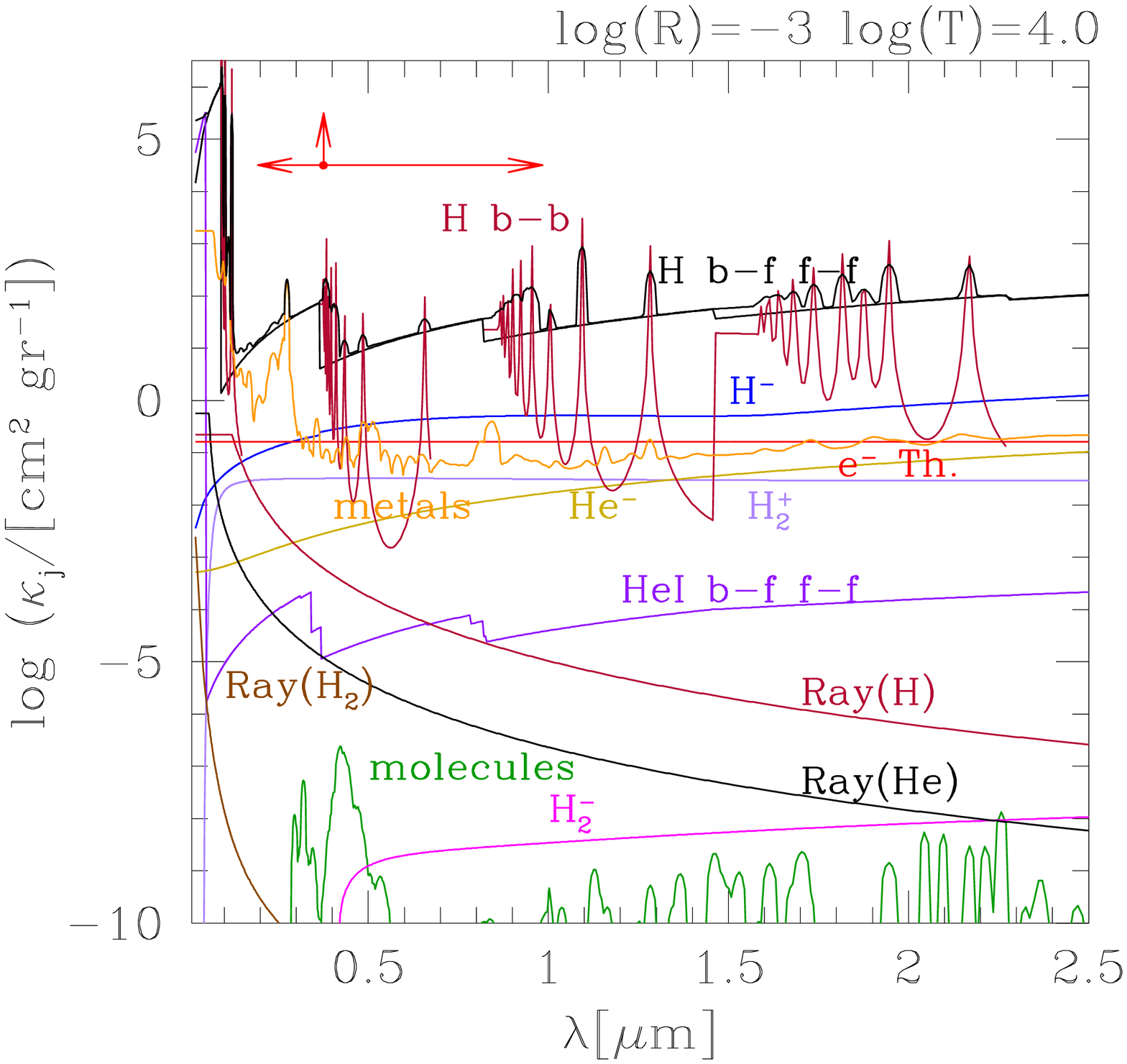}}
\end{minipage}
\begin{minipage}{0.33\textwidth}
\resizebox{1.\hsize}{!}{\includegraphics{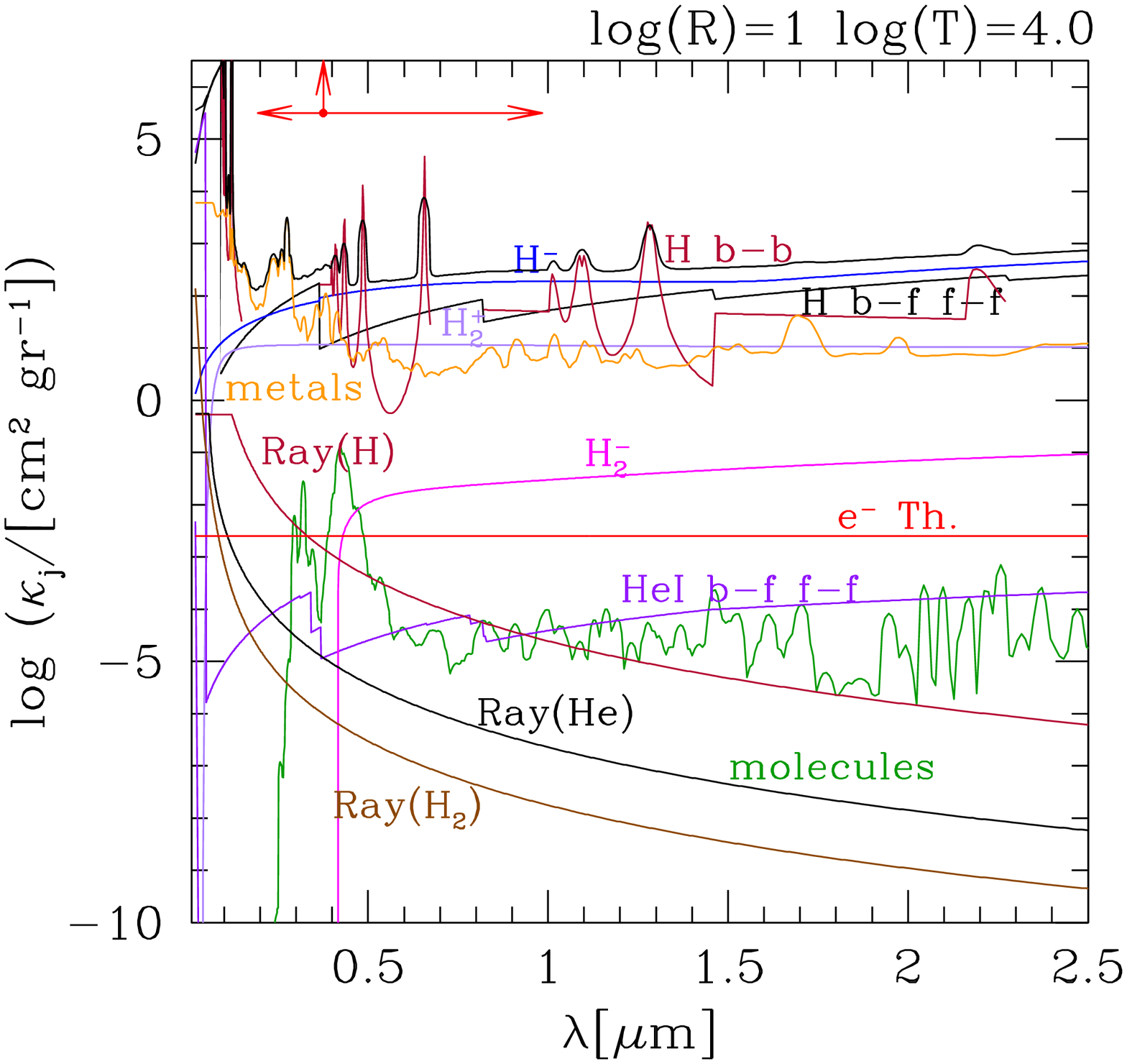}}
\end{minipage}
\begin{minipage}{0.33\textwidth}
\resizebox{1.\hsize}{!}{\includegraphics{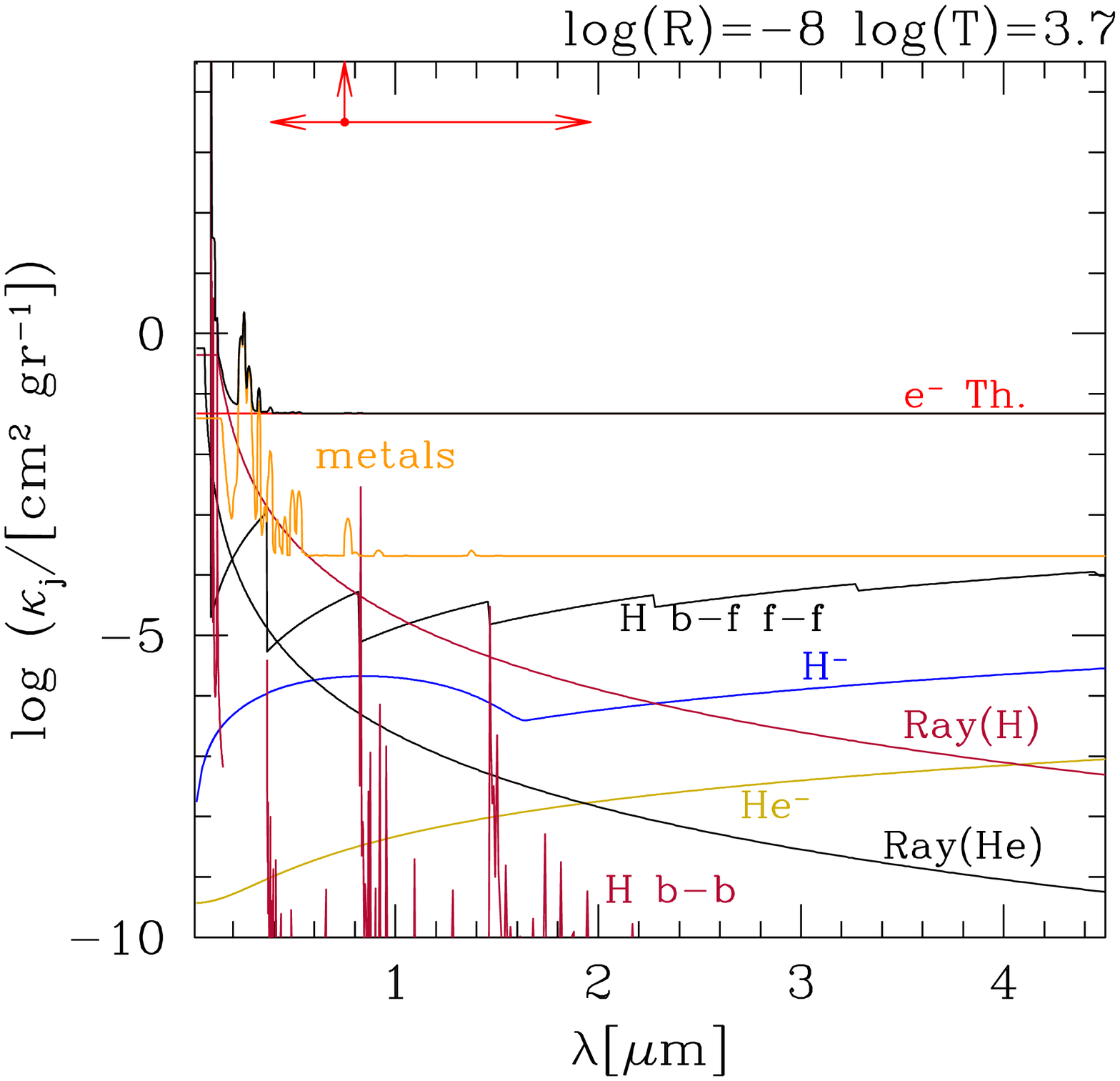}}
\end{minipage}
\begin{minipage}{0.33\textwidth}
\resizebox{1.\hsize}{!}{\includegraphics{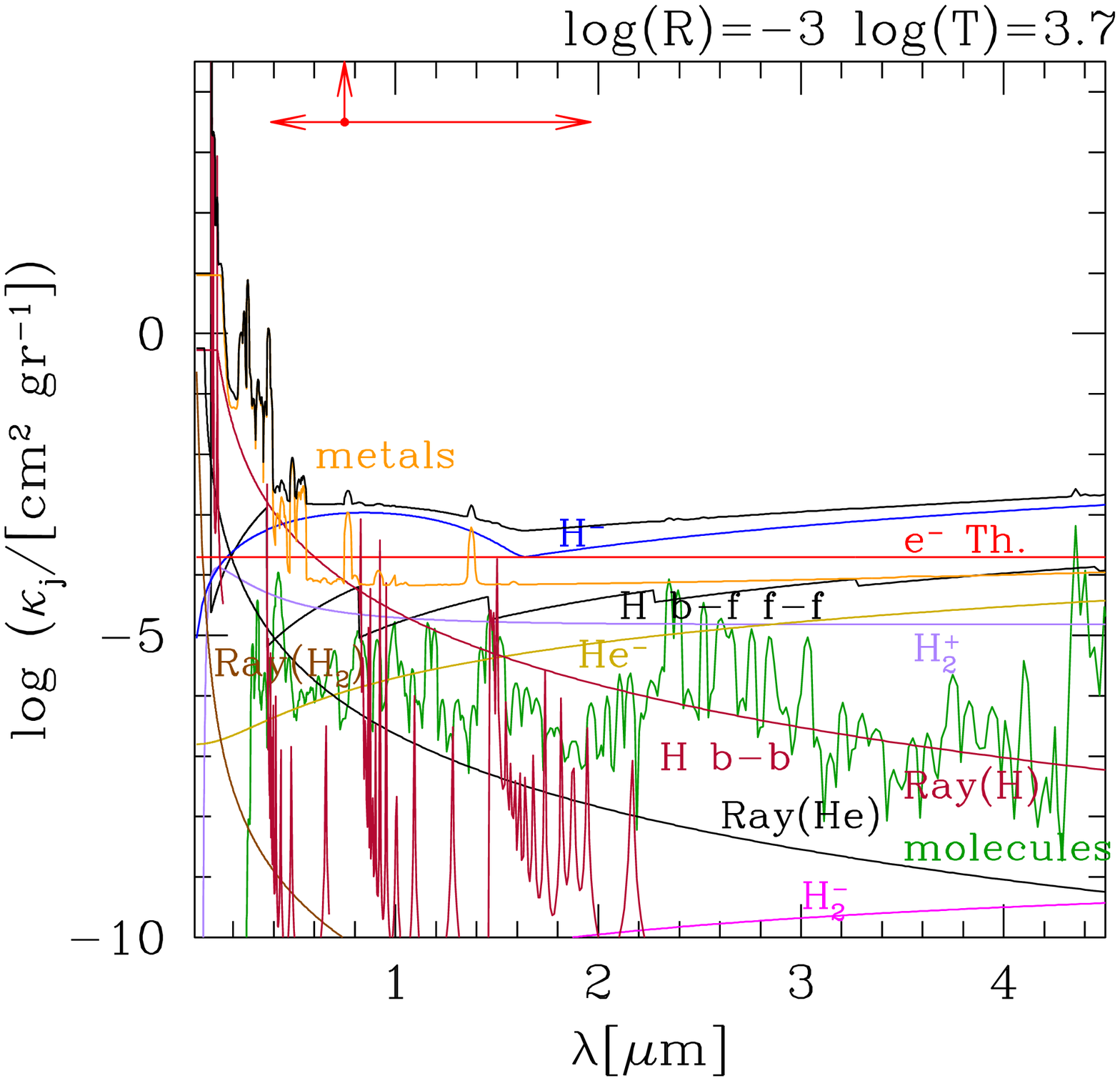}}
\end{minipage}
\begin{minipage}{0.33\textwidth}
\resizebox{1.\hsize}{!}{\includegraphics{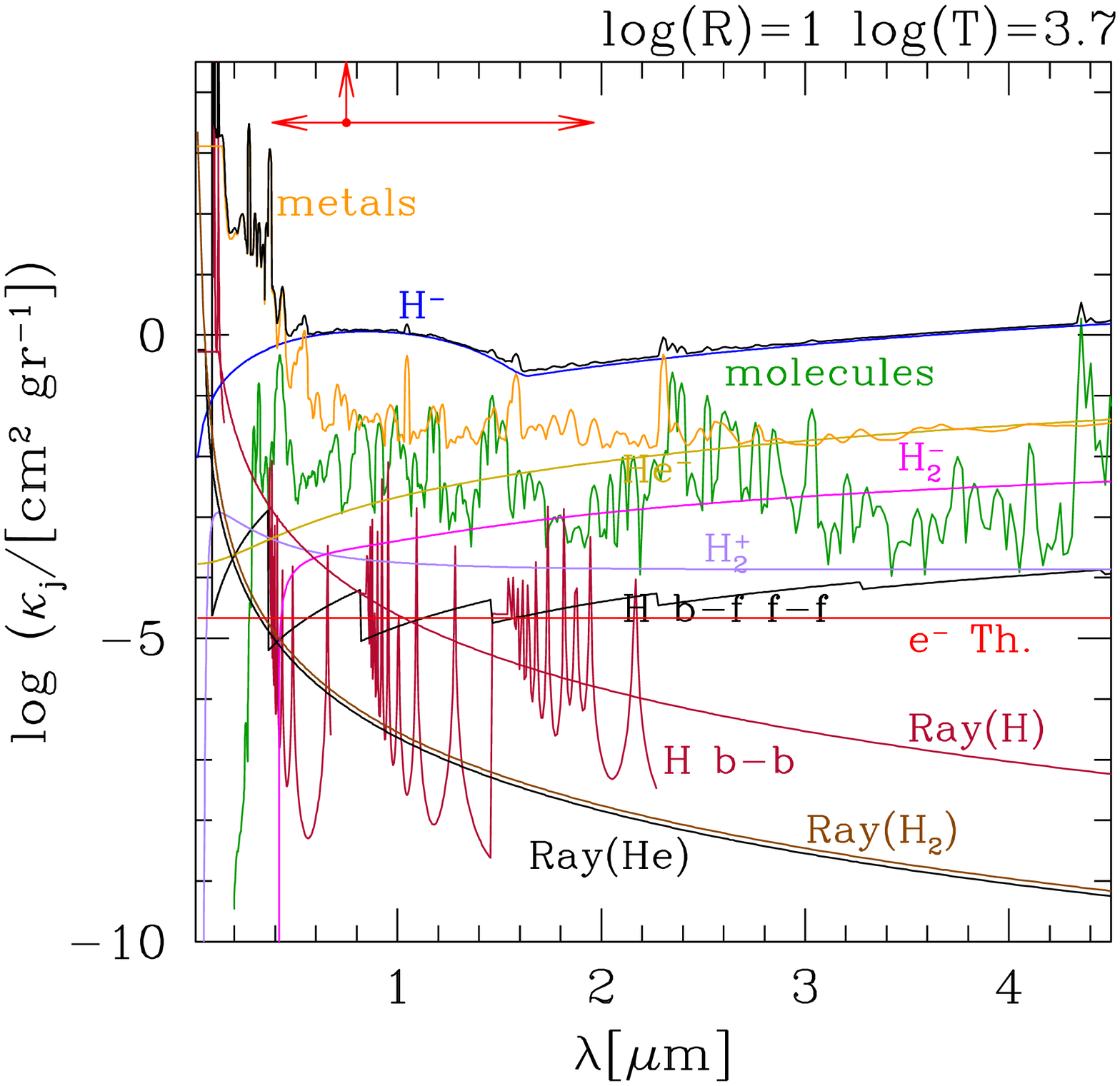}}
\end{minipage}
\begin{minipage}{0.33\textwidth}
\resizebox{1.\hsize}{!}{\includegraphics{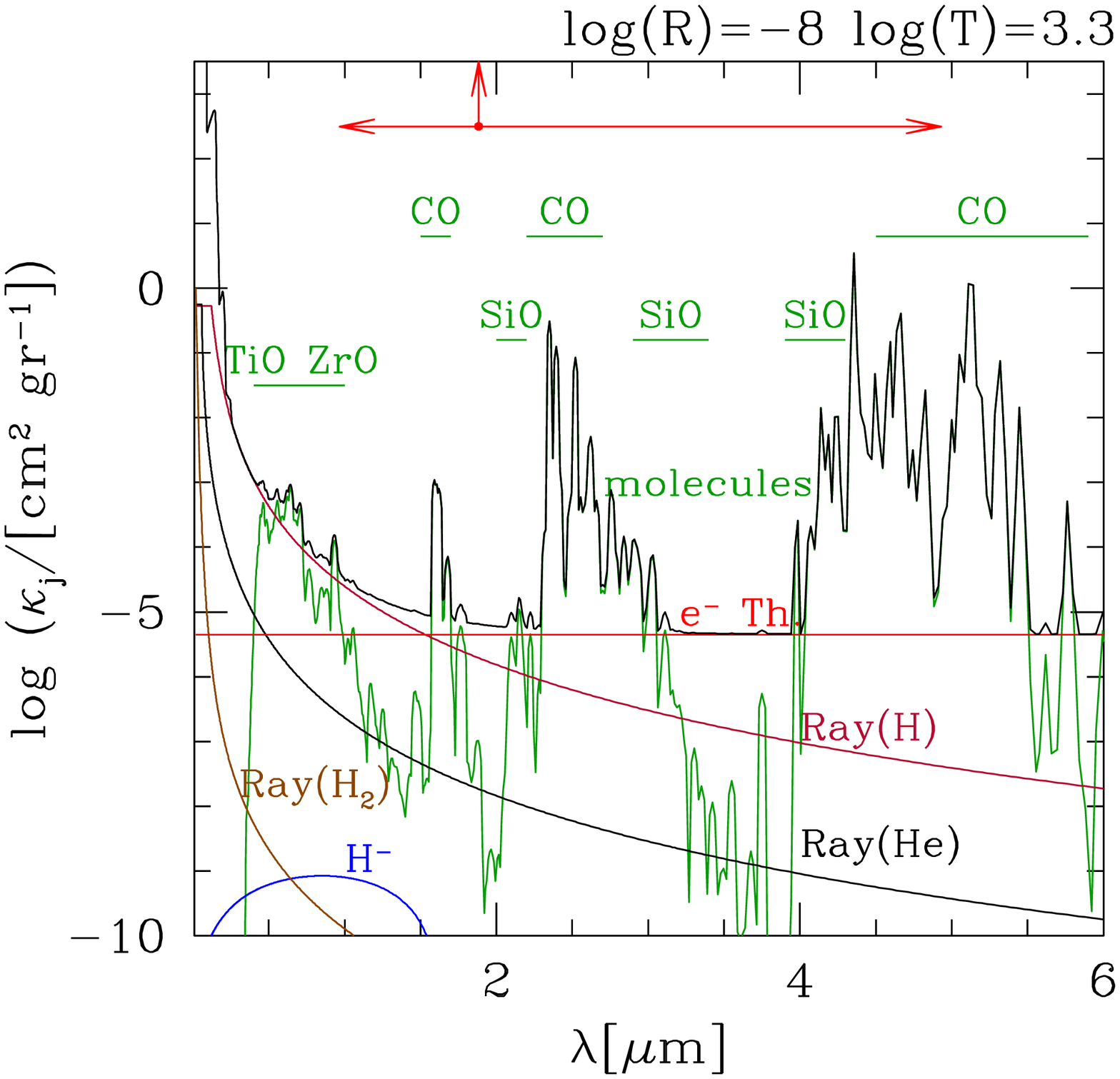}}
\end{minipage}
\begin{minipage}{0.33\textwidth}
\resizebox{1.\hsize}{!}{\includegraphics{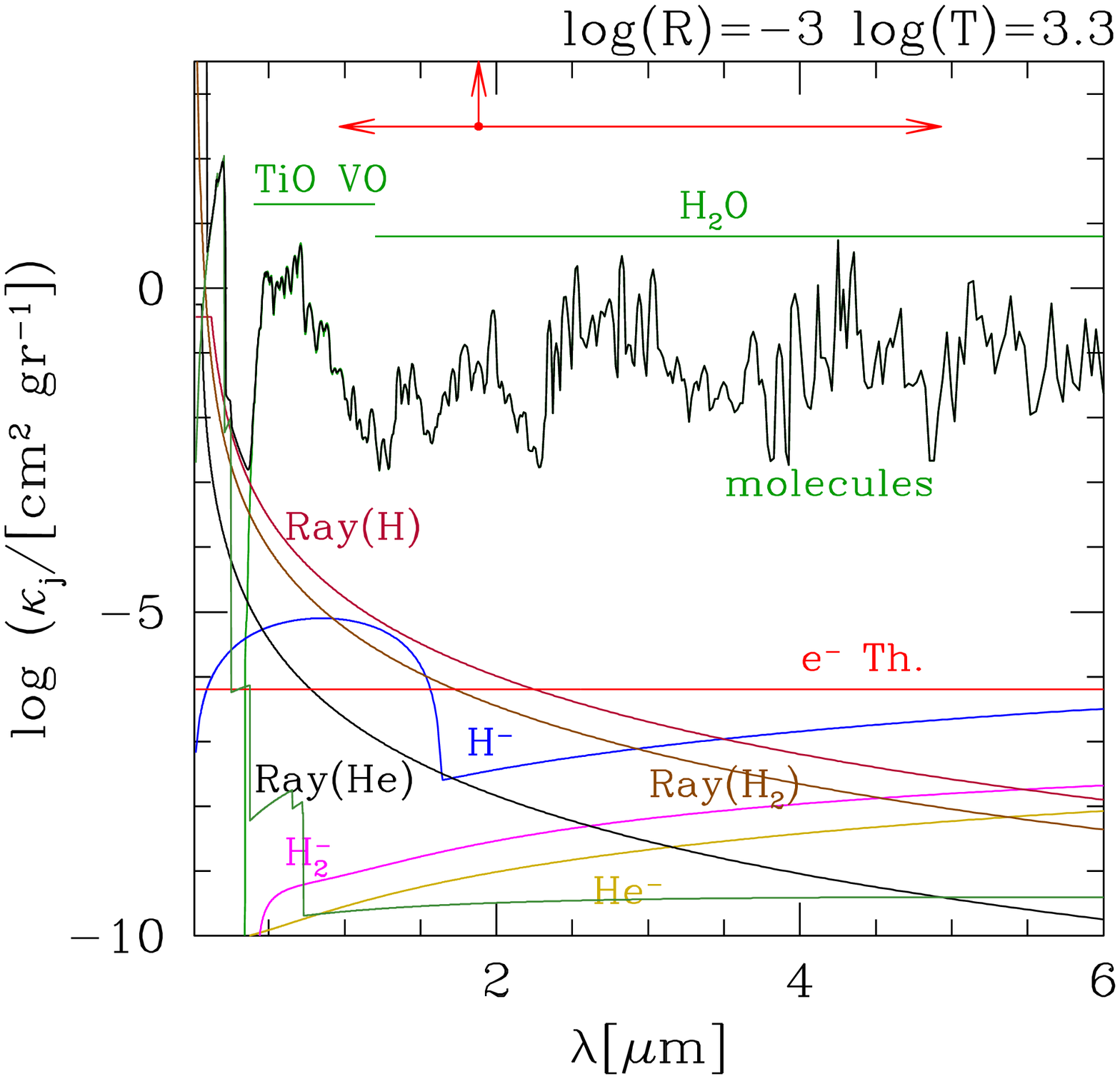}}
\end{minipage}
\begin{minipage}{0.33\textwidth}
\resizebox{1.\hsize}{!}{\includegraphics{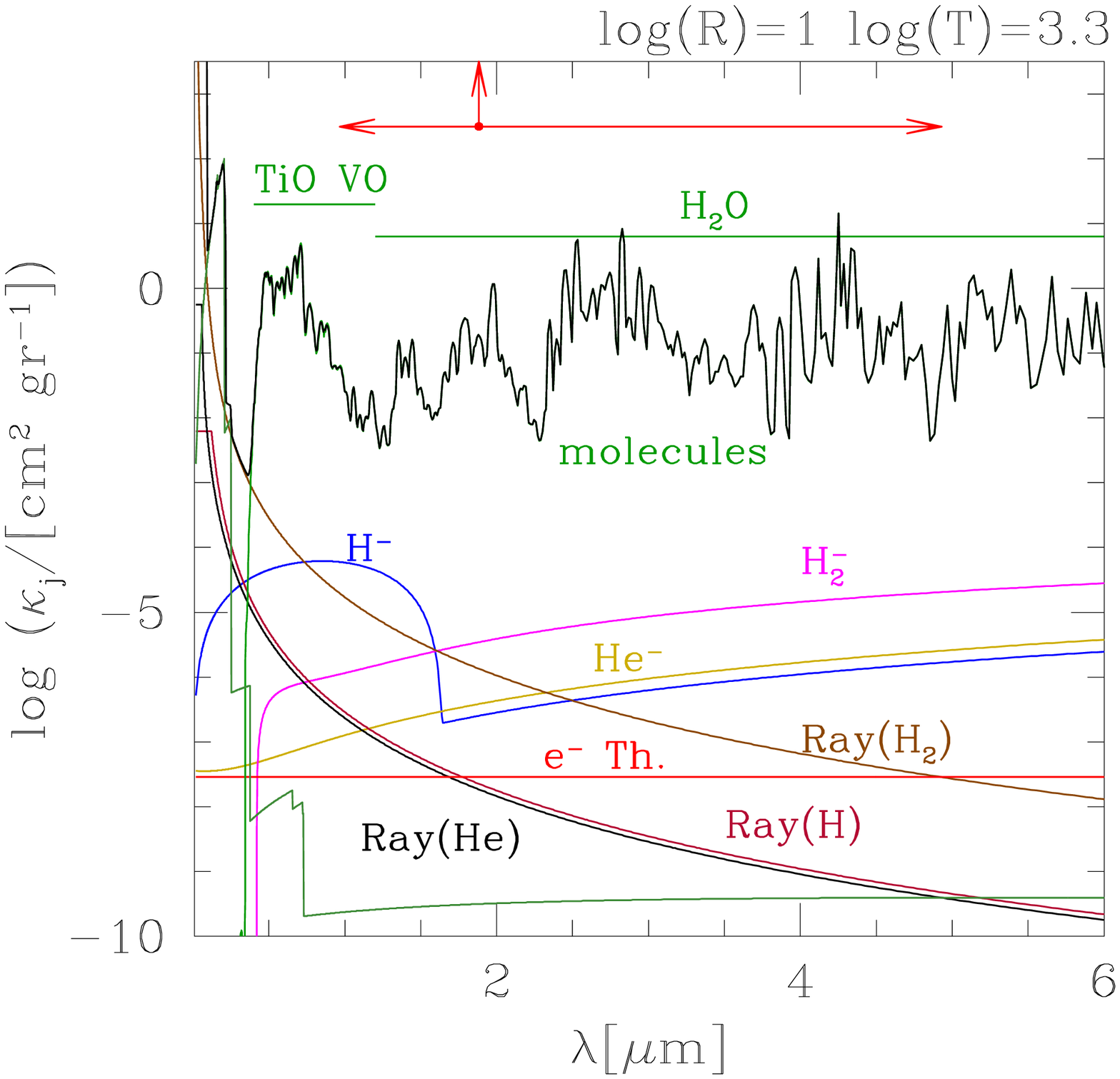}}
\end{minipage}
\caption{Monochromatic absorption coefficients for several opacity
sources as a function of the wavelength, for three values
of the temperature and  three values of the $R$ variable, as indicated. The chemical composition
is defined by $Z=Z_{\rm ref}=0.02$, $X=0.7$ with metal abundances 
scaled-solar to the GAS07 mixture. The total coefficient is depicted by the highest black line.
The vertical arrow marks the wavelength of the maximum of the Rosseland weighting function
(given by Eq.\ref{eq_lmax}),
while the horizontal arrows delimit the wavelength range within which the Rosseland weighting function
drops by a factor $1/e$. 
Where  molecular absorption bands are important, the corresponding spectral intervals 
are also indicated. For graphical purpose only,
line absorption coefficients for molecules and atoms are
smoothed by convolution with a Gaussian function. The variance is empirically chosen to depend 
on the wavelength so as to have a neat representation without missing important spectral details.
}
\label{fig_sig_z02}
\end{figure*}

\begin{figure*}
\begin{minipage}{0.33\textwidth}
\resizebox{1.\hsize}{!}{\includegraphics{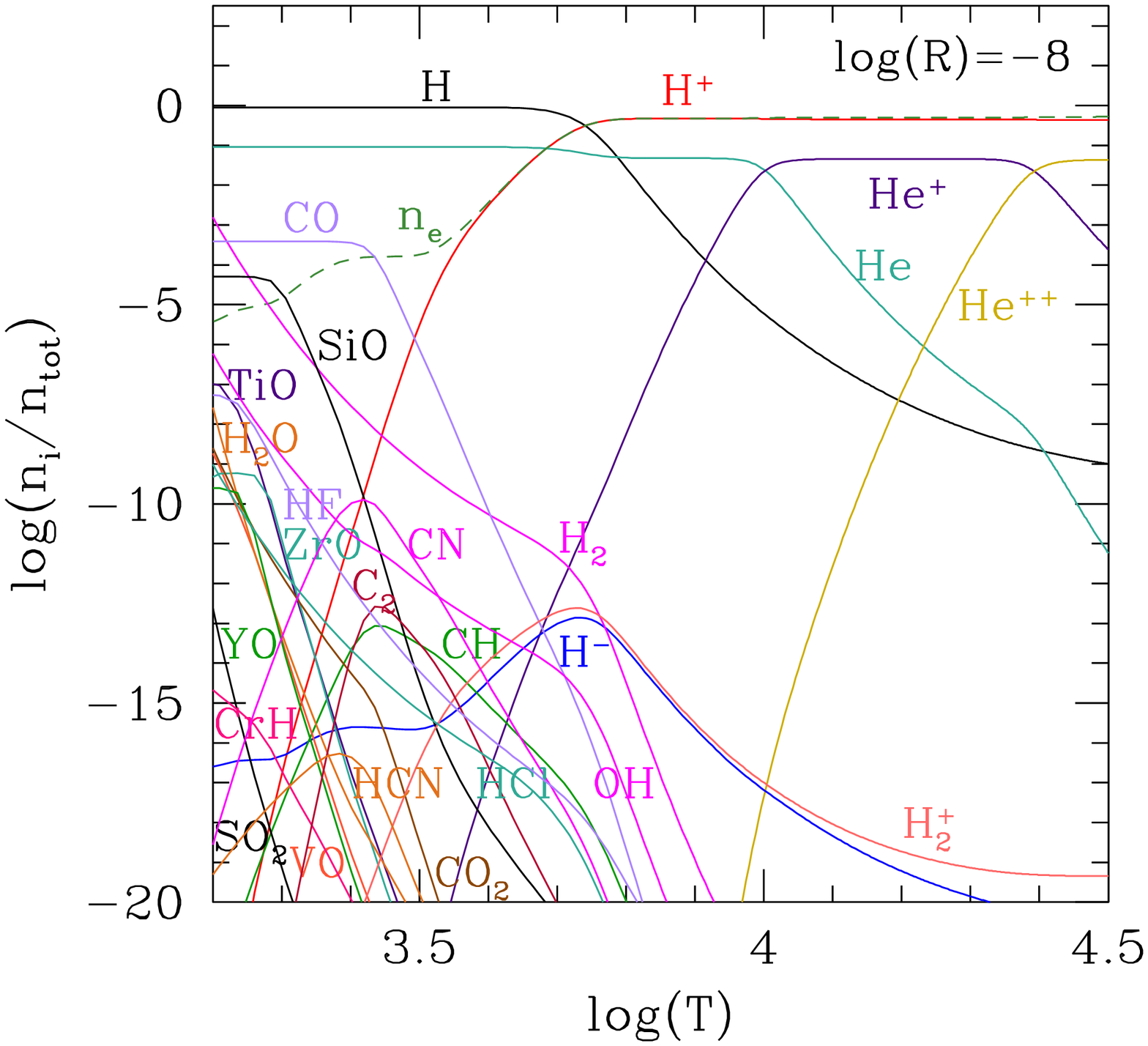}}
\end{minipage}
\begin{minipage}{0.33\textwidth}
\resizebox{1.\hsize}{!}{\includegraphics{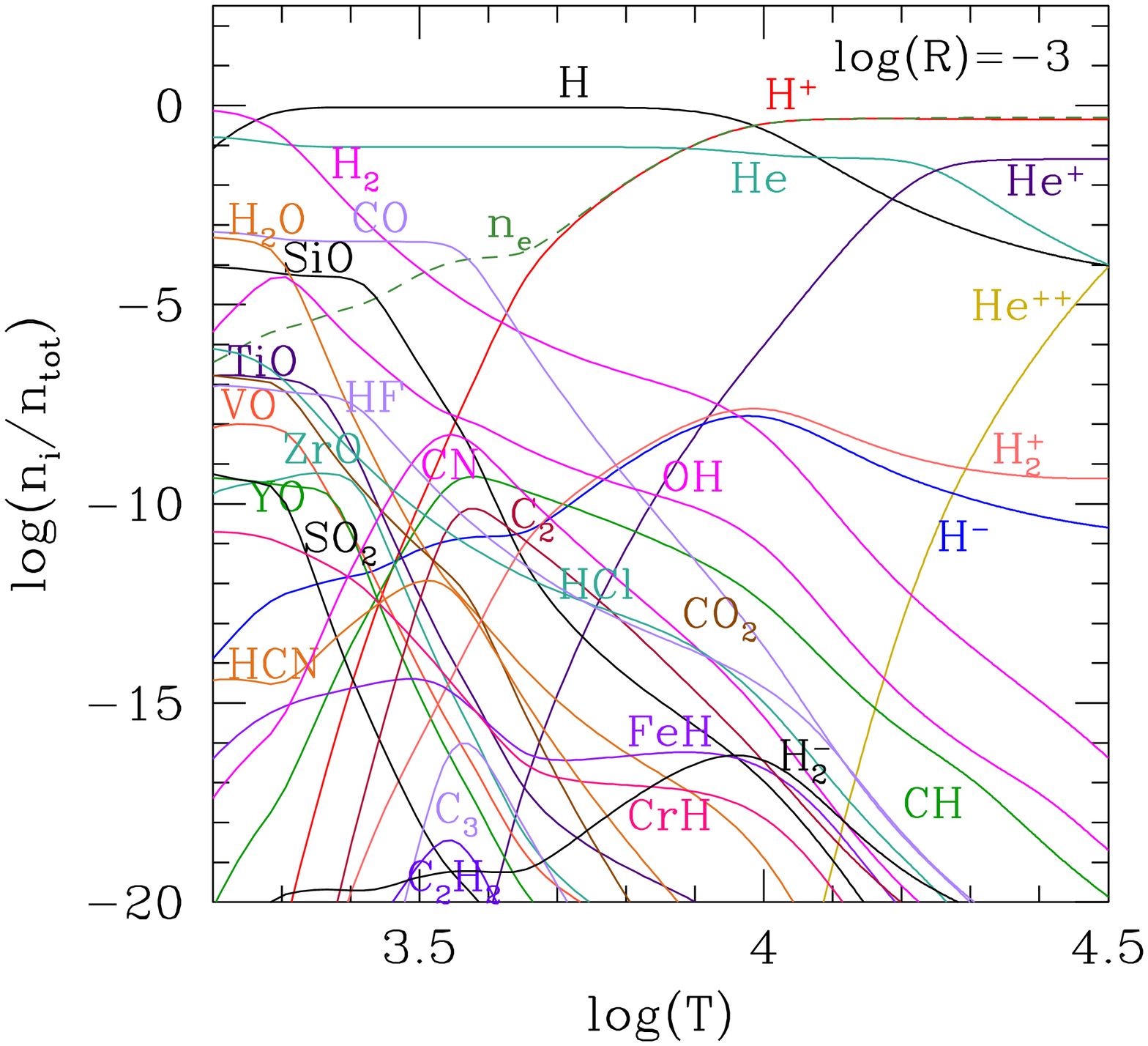}}
\end{minipage}
\begin{minipage}{0.33\textwidth}
\resizebox{1.\hsize}{!}{\includegraphics{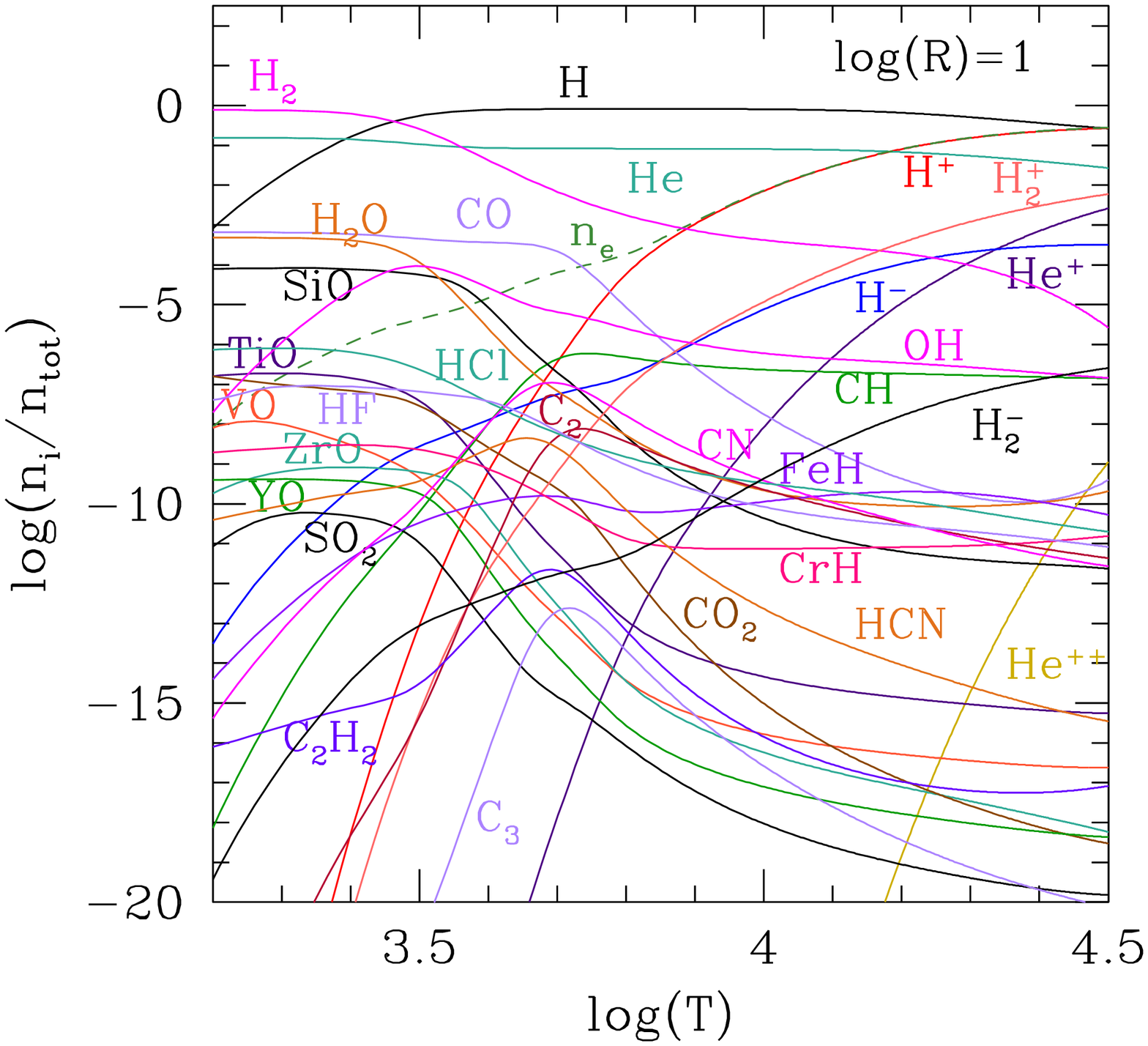}}
\end{minipage}
\begin{minipage}{0.33\textwidth}
\resizebox{1.\hsize}{!}{\includegraphics{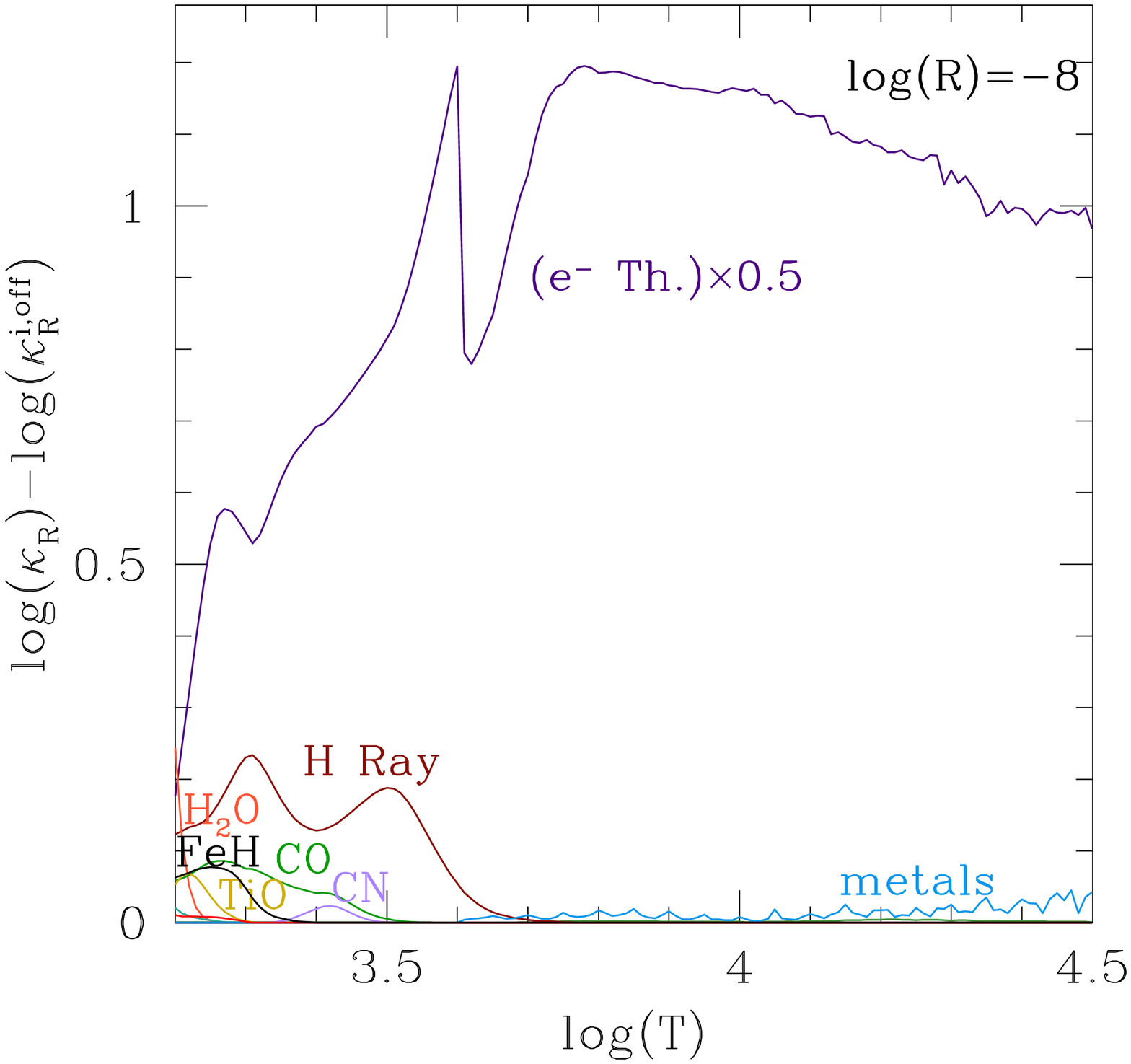}}
\end{minipage}
\begin{minipage}{0.33\textwidth}
\resizebox{1.\hsize}{!}{\includegraphics{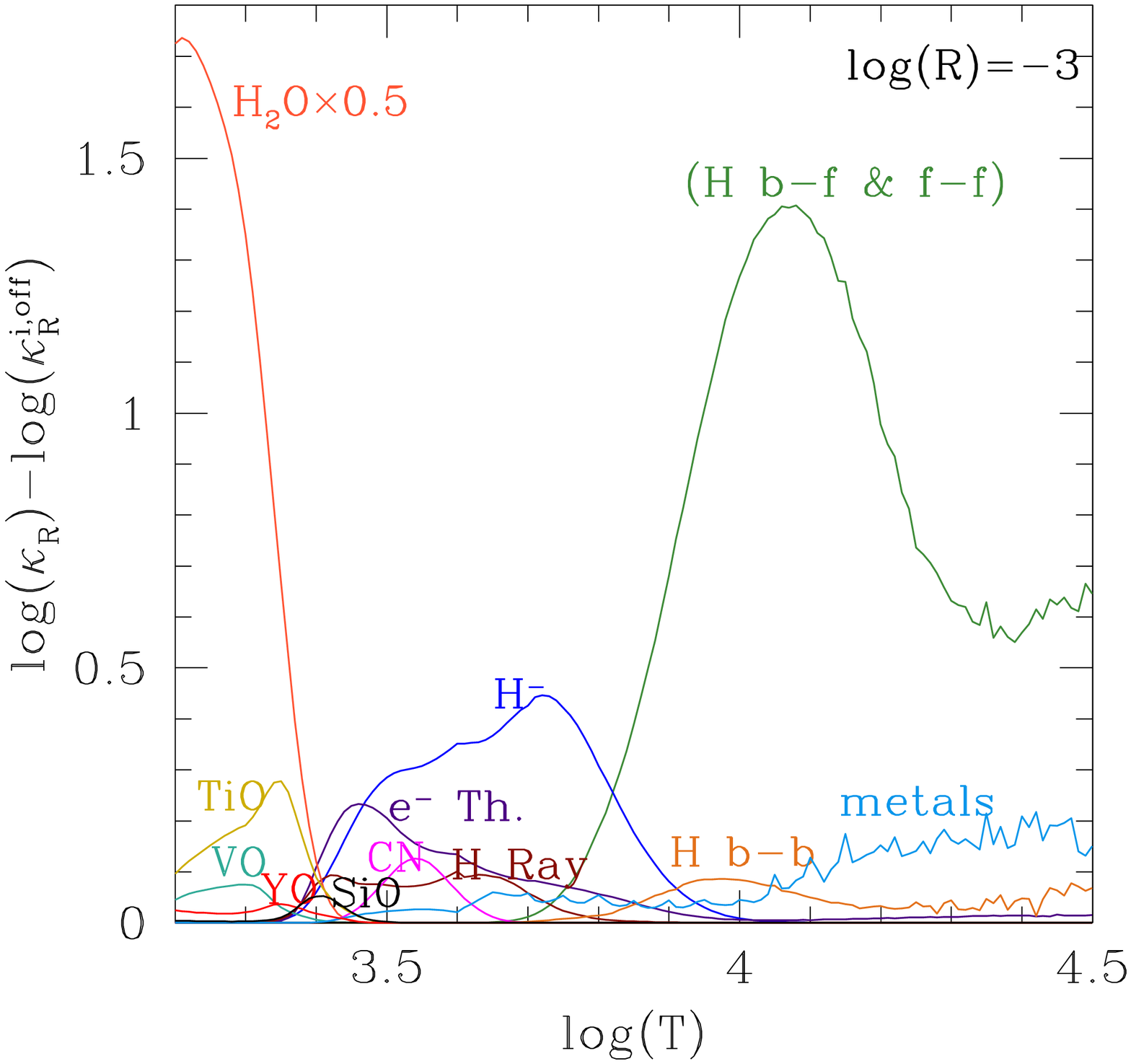}}
\end{minipage}
\begin{minipage}{0.33\textwidth}
\resizebox{1.\hsize}{!}{\includegraphics{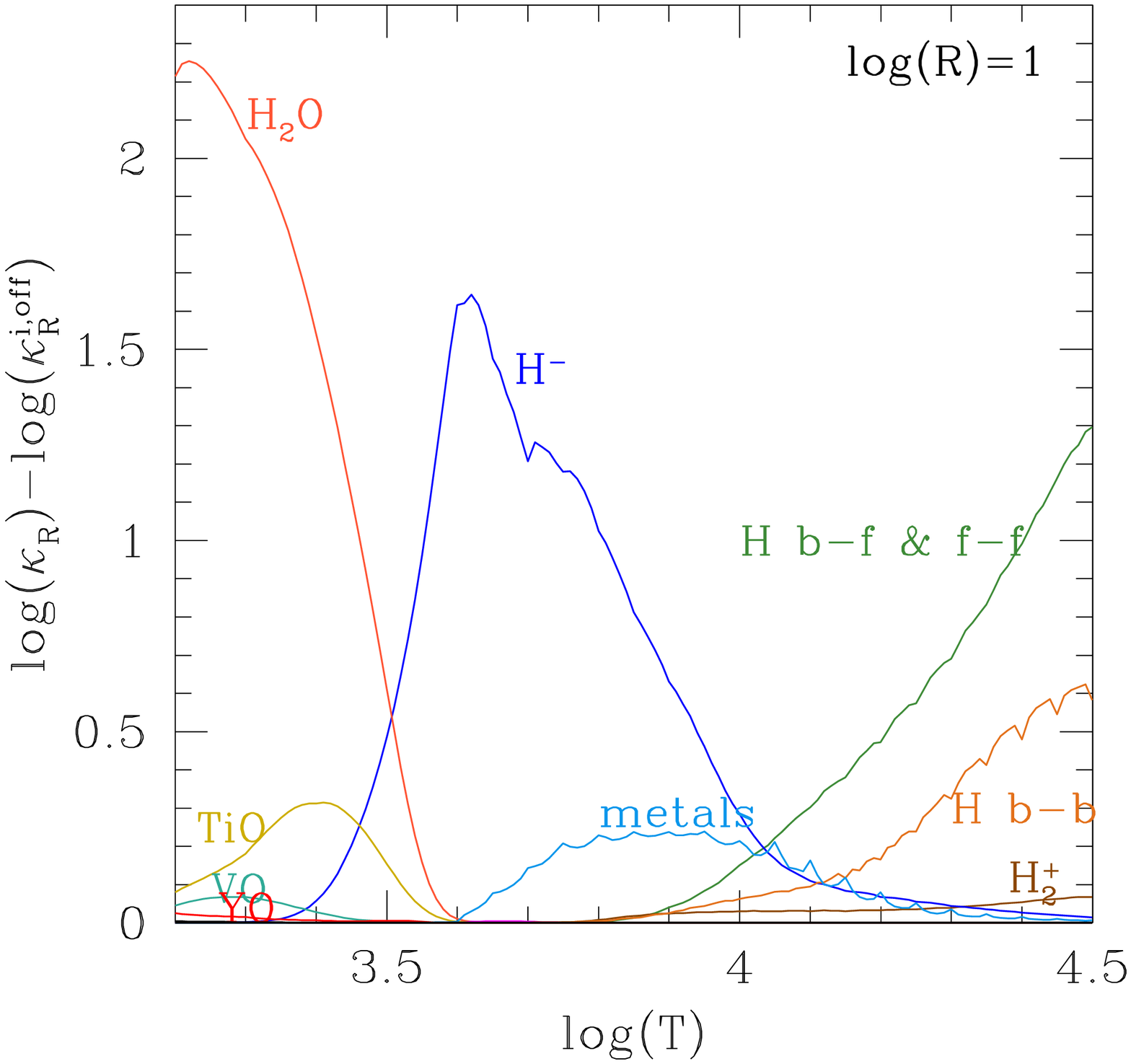}}
\end{minipage}
\caption{Top panels:
concentrations of various chemical species as a function of temperature,
for three values of the $R$ parameter, as indicated.
Bottom panels: contributions of different opacity sources (both continuous and line-absorption
processes) to the total RM opacity.
Each curve corresponds to $\log(\kappa)-\log(\kappa_j^{\rm off})$, where
$\kappa$ is the {\em full} opacity  including
{\em all} opacity sources here considered, whereas $\kappa_j^{\rm off}$ is
the {\em reduced} opacity obtained {\em omitting one} particular source
at once (labelled nearby). The logarithmic notation allows to highlight the
temperature domains which characterise the different opacity contributors.
The adopted chemical composition consists of $X=0.7$, $Z=Z_{\rm ref}=0.02$,
with elemental abundances scaled according to the GAS07 solar
mixture. }
\label{fig_chem_koff_z02}
\end{figure*}

Let us first illustrate the case of scaled-solar mixtures, which will serve
as reference for other compositions. 
As an example, Fig.~\ref{fig_opac_map} visualises the tri-dimensional plot of one
opacity  table  calculated over the whole $\log(T)-\log(R)$ parameter space for a
given chemical mixture. The latter is characterised by ($X=0.7$; $Z_{\rm ref}=0.02$; 
$Z=Z_{\rm ref}$; $f_{i}=1$, for $i=3\cdots N_{\rm el}$) according to the notation introduced
in Sect.~\ref{sec_tables},  meaning that all metal abundances
are scaled-solar.
One can see that the grid of the state variables (i.e. $\Delta\log(T)=0.01$ for 
$3.2 \le \log(T)\le 3.5$, and  $\Delta\log(T)=0.05$ 
for $3.5 < \log(T)\le 4.5$;  $\Delta\log(R)=0.5$)  
is sufficiently dense  to allow a smooth variation
of $\kappa_{\rm R}$ all over the parameters space,  
which is  a basic requirement for accurate interpolation. 

\begin{figure}
\resizebox{\hsize}{!}{\includegraphics{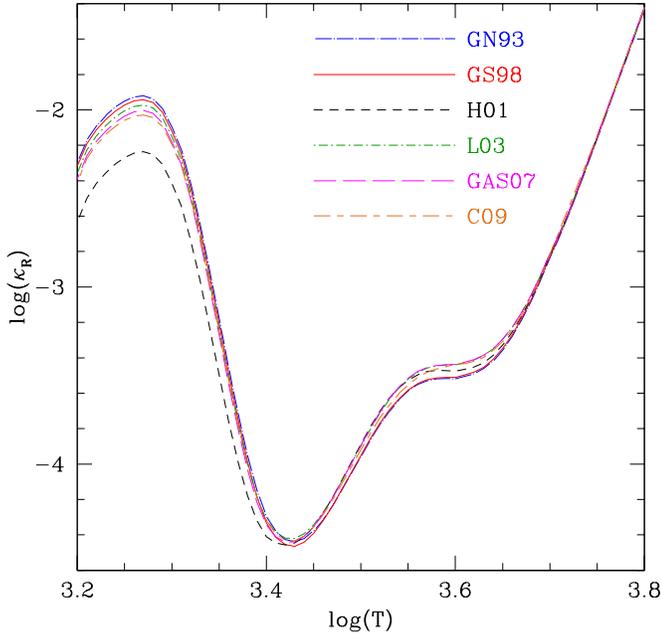}}
\caption{Rosseland mean opacity as a function of temperature and assuming $\log R=-3$.
The adopted chemical composition consists of $X=0.7$, $Z=Z_{\rm ref}=0.02$,
with elemental abundances scaled according to a few compilations of the solar mixture
abundances, namely: Grevesse \& Noels  1993;  Grevesse \& Sauval 1998; Holweger 2001; Lodders 2003;
 Grevesse, Asplund \& Sauval 2007; Caffau et al. (2008, 2009).
Note the significant depression of the H$_2$O bump in the Holweger (2001) case compared to the others,
due to the lower oxygen abundance, hence (C/O)$_{\odot}$ ratio.
}
\label{fig_ksun}
\end{figure}

Different opacity sources dominate the total $\kappa_{\rm R}$ in different regions
of the  $\log(T)-\log(R)$ plane. Roughly speaking,  we may say that the
continuous and atomic opacities prevail at higher temperatures, while molecular
absorption plays the major r\^ole for $\log(T) \la 3.5$. It has been known for long 
time (see e.g. Alexander 1975),
for instance, that the prominent opacity bump peaking
at $\log(T)=\sim 3.25$ in Fig.~\ref{fig_ksun} is mainly due to the strong absorption of 
H$_2$O molecular bands.
To delve deeper into the matter
it is instructive to look  at Fig.~\ref{fig_sig_z02}  and
Fig.~\ref{fig_chem_koff_z02}, which illustrate the basic ingredients 
affecting the RM opacity and their dependence on wavelength, temperature and density.

Figure~\ref{fig_sig_z02} displays the spectral behaviour of 
the monochromatic opacity coefficient per unit mass,
$\kappa_{j}(\nu)$,
of several absorption and scattering processes, as defined by Eqs.~(\ref{eq_kabs})--(\ref{eq_kscat}).
We consider three representative values of the temperature (i.e. $\log(T)=3.3,\,3.7,\ 4.0$) and
three choices of the $R$ variable (i.e. $\log(R)=-8,\,-3,\ 1$), for a total of nine panels
that should sample the main opacity domains.  For each temperature, 
we also indicate in Fig.~\ref{fig_sig_z02} the
spectral range most relevant for the Rosseland mean, by marking 
 the wavelength, $\lambda_{\rm max}$,
 at which the Rosseland weighting function reaches its maximum value (given by Eq.~\ref{eq_lmax}), and the
interval across which it decreases by a factor $1/e$.

At larger temperatures, i.e. $\log(T)=4.0$ and  $\lambda_{\rm max} \sim 0.38\,\mu$m (top panels),   
the total monochromatic coefficient is essentially determined
by the Thomson e$^-$~scattering at very low gas densities (see the top-left panel for $\log(R)=-8$), while 
the H opacity (bound-bound, bound-free, and free-free transitions) plays the major r\^ole
at large $\rho$. Next to hydrogen, some non-negligible contribution comes from atomic absorption
at shorter wavelengths.

At intermediate temperatures, i.e. $\log(T)=3.7$ and $\lambda_{\rm max}\sim 0.75\,\mu$m
(middle panels), Thomson e$^-$~scattering again controls the total 
absorption coefficient at the lowest densities, whereas at increasing $\rho$ 
the most significant opacity sources are due to metals and H$^-$ absorption  
(electron photo-detachment for $\lambda < 1.644\,\mu$m and free-free transitions).

At lower temperatures, i.e. $\log(T)=3.3$ and $\lambda_{\rm max}\sim 1.88\,\mu$m (bottom  panels),
the molecular absorption bands (mainly of H$_2$O, VO, TiO, ZrO, CO) dominate the total absorption coefficient 
at any gas density except for very low values,
where the spectral gaps between the molecular bands are filled in with the Thomson e$^-$~scattering
coefficient. Due to its harmonic character, the  Rosseland mean opacity emphasises
just these opacity holes, so that the total $\kappa_{\rm R}$ for $\log(T)=3.3$ and $\log(R)=-8$ 
will be mostly determined by the Thomson e$^-$~scattering, with a smaller
contributions from molecules. 

This fact becomes more evident
with the help of Fig.~\ref{fig_chem_koff_z02}, which provides
complementary information  
on both the chemistry of the gas, and the characteristic  
temperature windows of different opacity sources.
Results are presented as a function of temperature for three values of the parameter $R$.

As for the chemistry (top panels of Fig.~\ref{fig_chem_koff_z02}), we show the concentrations of
a few species, selecting them among those that are opacity contributors, while 
leaving out all other chemicals to avoid 
over-crowding in the plots (we recall that \AE SOPUS solves the chemistry for $\mathcal{N}_{\rm tot} \approx 800$ species).
It is useful to remark a few important features, namely:  
i) at lower temperatures molecular formation becomes more
efficient at increasing density, ii) the most abundant molecule is either carbon monoxide (CO) 
thanks to its high binding energy at low and intermediate densities, or molecular hydrogen (H$_2$) at higher
densities; iii)  the electron density $n_e$ is essentially supplied by  H
ionisation down to temperatures $\log(T)\simeq 3.8-3.6$, below which the
main electrons donors are nuclei with low-ionisation potentials, such
as: Mg, Al, Na, Si, Fe, etc. (see Fig.~\ref{fig_electron} and 
Sect.~\ref{ssec_alpha} for more discussion of this point).

The bottom panels display the contributions of several absorption/scattering processes to the total
RM opacity. This is done by considering, for a given source $j$,
 the ratio $\kappa_j^{\rm off}/\kappa_{\rm R}$, where $\kappa_j^{\rm off}$ is
the {\em reduced} RM opacity obtained by including all opacity sources but for the
$j^{\rm th}$ itself. 

At very low densities, i.e. $\log(R)=-8$ (left-hand side
panel of Fig.~\ref{fig_chem_koff_z02}) 
the most important opacity source,  all over the temperature
range under consideration, is  by far
Thomson scattering from free electrons. Note that at lower temperatures
a relatively important contribution is provided by Rayleigh scattering from neutral hydrogen, 
while the r\^ole of molecules is marginal since 
at these low densities molecular formation is inefficient.

Different is the case with $\log(R)=-3$ (middle panel of Fig.~\ref{fig_chem_koff_z02}).
We can distinguish three main opacity domains as a function of temperature.
At lower temperatures, say for  $3.2 \la \log(T)\la 3.6$, molecules completely rule the opacity,
 with H$_2$O being the dominant source
for $\log(T)\la 3.4$. Additional modest contributions come from metal oxides, such as TiO, VO, YO, 
and SiO. 
 Note that, though for C/O~$<1$ the chemistry is dominated by O-bearing
molecules, there is a small opacity bump due to CN at $\log(T) \approx  3.5$.  
At intermediate temperatures, $3.6\la  \log(T)\la 3.8$, 
the most important r\^ole is
played by the H$^{-}$ continuum opacity, which in turn depends on the availability of
free electrons supplied by ionised metals. Additional opacity 
contributions are provided by
Thomson scattering from electrons and Rayleigh scattering from 
neutral hydrogen.
At larger temperatures, $3.8\la  \log(T)\la 4.5$, the total RM opacity is
determined mostly by the b-f and f-f continuous absorption from hydrogen, with 
further contributions from b-b transitions of H  and atomic opacities.
  
In the high density case with  $\log(R)=1$ (right-hand side  panel of
Fig.~\ref{fig_chem_koff_z02}), the opacity pattern is similar to the
one just described, with a few differences.
The most noticeable ones are the sizable growth  
of the  H$^{-}$ opacity bump in the intermediate temperature window, and the 
increased importance of the H lines at higher temperatures.

Finally, we close this section by examining the sensitiveness of the
RM opacity to the underlying reference solar mixture. 
Figure~\ref{fig_ksun} shows an example of our opacity calculations 
made adopting a few solar abundances compilations available in the
literature. They are summarised in Table~\ref{tab_sun}.
The largest differences are expected for $\log T \la 3.4$, 
where the RM opacity is dominated by the opacity bump caused by the
H$_2$O molecule, whose amplitude is extremely sensitive to 
the excess of oxygen with respect to carbon, hence to the C/O ratio.
In fact, we notice that the opacity curves corresponding to
GN93, GS98, L03, GAS07, and C09 lie rather close one to each other, just
reflecting the proximity of their C/O ratios ($\approx 0.5-0.6$; see
Table~\ref{tab_sun}). For the same reason, the RM opacity predicted
at $\log T \sim 3.3$ with the H01 solar mixture is  roughly 50$\%$
lower, given the higher C/O ratio ($\approx 0.7$).

Some differences in RM opacity are also expected in the $3.5\la \log(T)\la
3.65$ interval, which is affected
mainly by the CN molecular bands and the negative hydrogen ion
H$^{-}$.
We see in Fig.~\ref{fig_ksun} that most of the results split into two curves:
the opacities based on L03 and GAS07 (and partly also C09) are higher than those
referring to GN93 and GS98 solar mixtures.
In this case the differences are not caused by the CN molecule, but
rather reflect the differences in the electron density. 
As one can notice in Fig.~\ref{fig_fracsol}, 
L03, GAS07 (and C09) compilations correspond to higher solar
partitions, $X_{i,\odot}/Z_{\odot}$, of those elemental species
that mostly provide the  budget of free electrons at these
temperatures, such as: Mg, Si, Ca, and Fe (see also Fig.~\ref{fig_electron}).
As a consequence, the H$^-$ opacity is strengthened in comparison to the
GN93 and GS98 cases.
On the other hand, 
the opacity curve corresponding to the  H01 mixture lies somewhere
in the middle. This is the indirect result of the larger C/O ratio
(i.e. more carbon is available)
which favours a larger concentration, hence opacity contribution,
of the CN molecule in this temperature window. 

The arguments developed here indicate that the
expression ``standard solar composition''  should be always  
specified explicitly together with its reference compilation and not taken  for
granted, since significant differences arise in the RM opacities
depending on the adopted solar mixture. 

\subsubsection{Comparison with other authors}
\label{sssec_kcomp}
\begin{figure*}
\begin{minipage}{0.50\textwidth}
\resizebox{\hsize}{!}{\includegraphics{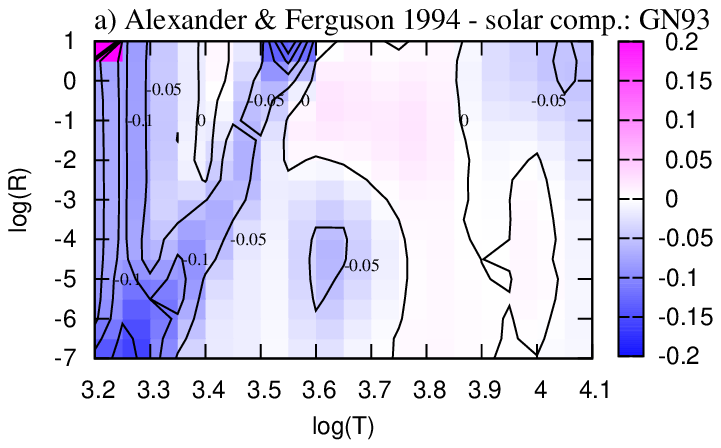}}
\end{minipage}
\hfill
\begin{minipage}{0.50\textwidth}
\resizebox{\hsize}{!}{\includegraphics{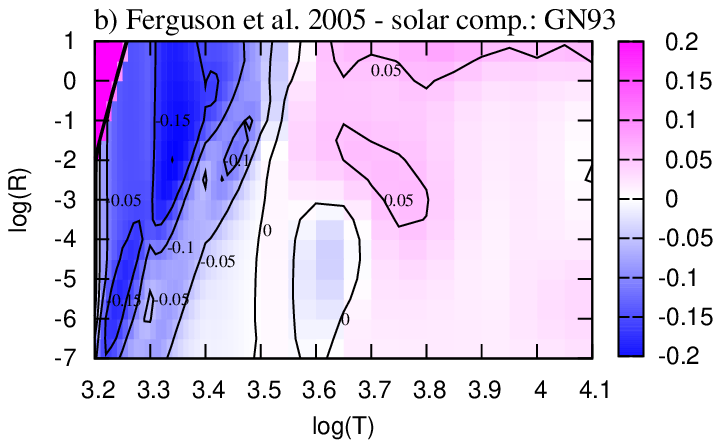}}
\end{minipage}
\begin{minipage}{0.50\textwidth}
\resizebox{\hsize}{!}{\includegraphics{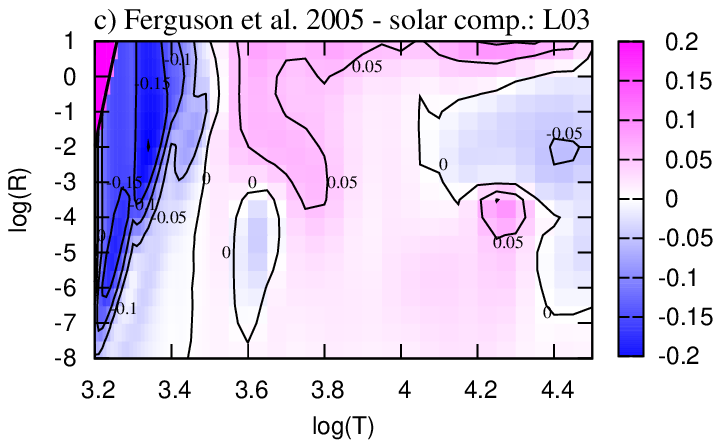}}
\end{minipage}
\hfill
\begin{minipage}{0.50\textwidth}
\resizebox{\hsize}{!}{\includegraphics{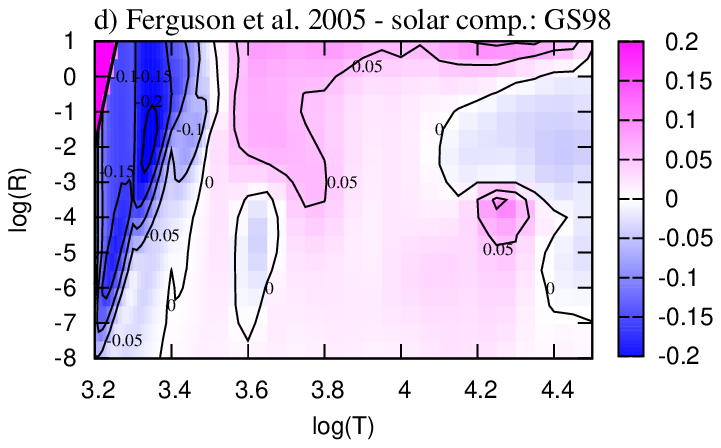}}
\end{minipage}
\begin{minipage}{0.50\textwidth}
\resizebox{\hsize}{!}{\includegraphics{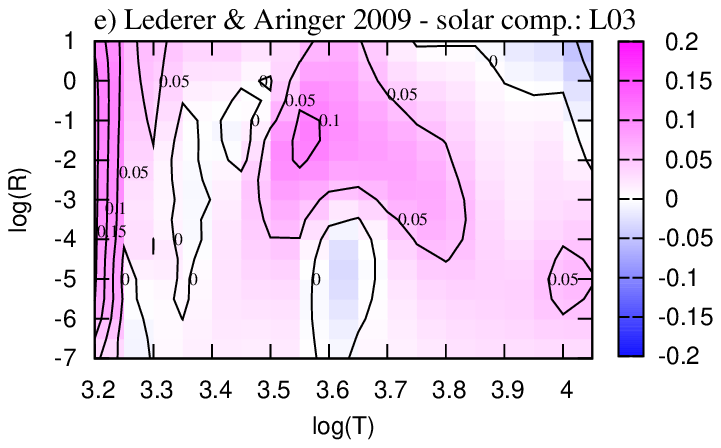}}
\end{minipage}
\hfill
\begin{minipage}{0.50\textwidth}
\resizebox{\hsize}{!}{\includegraphics{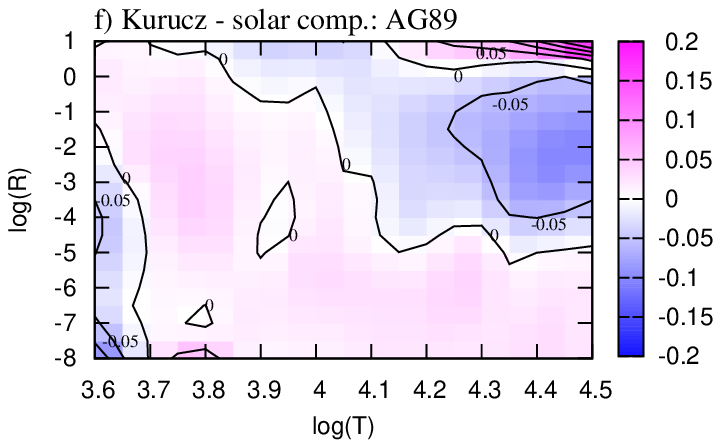}}
\end{minipage}
\begin{minipage}{0.50\textwidth}
\resizebox{\hsize}{!}{\includegraphics{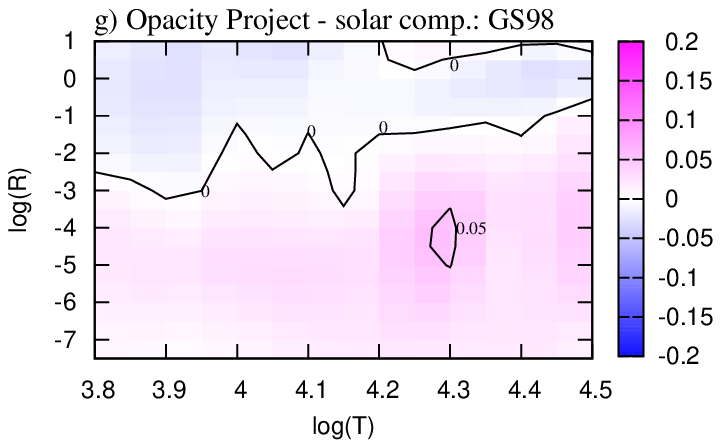}}
\end{minipage}
\hfill
\begin{minipage}{0.50\textwidth}
\resizebox{\hsize}{!}{\includegraphics{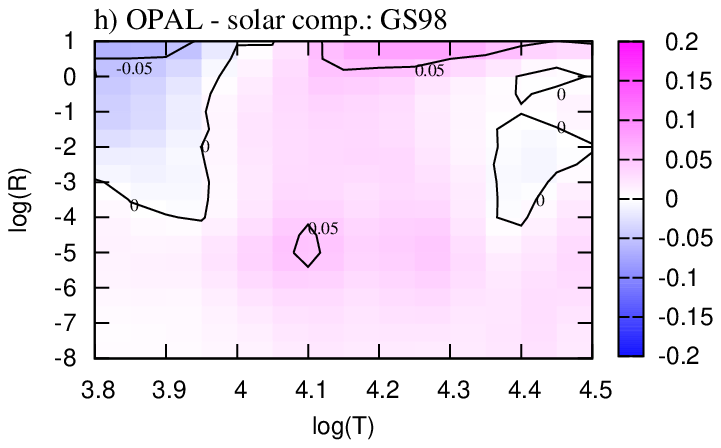}}
\end{minipage}
\caption{Comparison between our RM opacity results and those provided
by other authors, in terms of $\log(\kappa_{\rm R}^{\rm
author})-\log(\kappa_{\rm R}^{\rm \AE SOPUS})$. 
Contour lines, with an incremental step of $0.05$ dex, are superimposed to guide the eye.
In all cases, except for Kurucz , the adopted chemical composition corresponds to  $Z=Z_{\rm
ref}=0.02$, $X=0.7$.
External data are taken from: 
Alexander \& Ferguson (1994) and  Ferguson et al. (2005) adopting 
the Grevesse \& Noels (1993) solar mixture (panels a and b);
Ferguson et al. (2005) assuming  the solar abundances from Lodders (2003) 
(panel c) and Grevesse \& Sauval (1998) (panel d);
Lederer \& Aringer (2009) adopting the  Lodders (2003) solar mixture
(panel e); Kurucz' web database 
for a chemical composition with $Z=Z_{\odot}=0.0194$,
$X=X_{\odot}=0.7065$ according to the Anders \& Grevesse (1989) solar mixture (panel f);
OP and OPAL assuming the  Grevesse \& Sauval (1998) solar mixture (panels g and h).
}
\label{fig_kcomp_z02}
\end{figure*}

As a next step we checked our opacity results  
against tabulated RM data made publicly available from other authors.
In Fig.~\ref{fig_kcomp_z02} we show eight representative comparisons,
based on: the widely-used and well-tested database
set up by the Wichita State University group, i.e. Alexander \& Ferguson (1994), 
Ferguson et al. 2005 (hereafter also F05); 
the recent data by Lederer \& Aringer 2009 (hereafter also LA09) stored in the VizieR service; 
the RM data available in the Robert L. Kurucz' homepage, and
the OPAL and OP data computed via their interactive web-masks.
The $R$ and $T$ intervals are different depending on the source
considered. 
For instance, the comparisons with the  OPAL and OP opacities cover the range
from $3.8 \le \log(T) \le 4.5$, since no molecular contribution is
included in the OPAL and OP data.

In general we can conclude that the check is quite satisfactory in all cases
under examination, as
our opacity values agree with the reference data mostly within
$\pm 0.05$ dex, with the largest differences
reaching up to $\approx \pm 0.10-0.20$ only in narrow regions.
 
Let us start discussing the comparison with Alexander \& Ferguson (1994) and
Ferguson et al. (2005), illustrated in panels from a) to d) assuming various
reference solar compositions.
First we notice that the small magenta areas in the upper-left corners 
of the four panels are not included in the test, 
since at those densities and temperatures dust is expected
to condensate
\footnote{The inclusion of dust in pre-computed opacities is in any case 
problematic since in real stars it will hardly form under equilibrium conditions.}, 
whereas our EOS describes the matter in the gas phase.

Besides this, in all cases the agreement between the opacity data 
of the Wichita State University group and  
\AE SOPUS is very good for $3.4 \le \log(T) \le 4.5$, the differences $\Delta\log(\kappa_R)$ 
being mostly comprised within $\pm 0.05$ dex throughout the $R$
range. For $\log(T)<3.4$ the deviations between F05 and \AE SOPUS
appear to grow with a systematic trend, i.e. $\log(\kappa_R^{\rm\AE
SOPUS}) > \log(\kappa_R^{\rm F05})$, at increasing $R$. Anyhow, the variations
are not dramatic, the biggest values arriving at $\approx
-0.15/-0.20$. This result is not surprising since this is just
the region where molecular absorption dominates, so that the predicted
RM opacity is sensitive to differences in 
the treatment of the molecular line opacities (line
lists,  broadening, adopted frequency grid, etc.).

This applies also when comparing different releases of the same database 
as it is illustrated, for instance, by panels a) and b) relative to the
data of the Wichita State University group.
We notice that where \AE SOPUS exhibits the best agreement 
($< 0.05$ dex) with Alexander \& Ferguson (1994) at $\log(T)\approx 3.4$ and  $\log(R)\ga -3$, 
the largest differences ($0.15 - 0.20$ dex) show up instead in the comparison with F05
for the same set of abundances. 
In this respect, we expect that much of the discrepancy between
F05 and \AE SOPUS for $3.2 \le \log(T) \le 3.4$ is due to
the different molecular line data adopted for water vapour, i.e. 
Partridge \& Schwenke (1997) and Barber et al. (2006), respectively.

Support to the above interpretation is found when comparing  panel c) and e), 
the latter showing the
check of \AE SOPUS results against Lederer \& Aringer (2009) for the L03
solar mixture. 
As we see the agreement here is quite fair all over the
$\log(T)-\log(R)$ diagram, even in the low-$T$ corner dominated by
H$_2$O, VO, and TiO absorption, where larger differences with F05
(panel c) arise. As a matter of fact, in \AE SOPUS we adopt essentially the same
molecular data as in  LA09, so that a good
match is in principle expected.

Finally, let us briefly comment on the bottom panels (g and h) of
Fig.~\ref{fig_kcomp_z02}, relative to two data sets, OP and OPAL,
which  are widely used to describe the RM opacity of the gas 
in the high-T regions,
say for $\log(T) > 4.0$. The comparison with \AE SOPUS 
in the overlapping interval, $3.8 \le \log(T) \le 4.5$, is really
excellent, so that the  OP and OPAL opacity tables may be 
smoothly complemented in the low-$T$ regime with \AE SOPUS calculations.
\begin{figure*}
\begin{minipage}{0.48\textwidth}
\resizebox{\hsize}{!}{\includegraphics{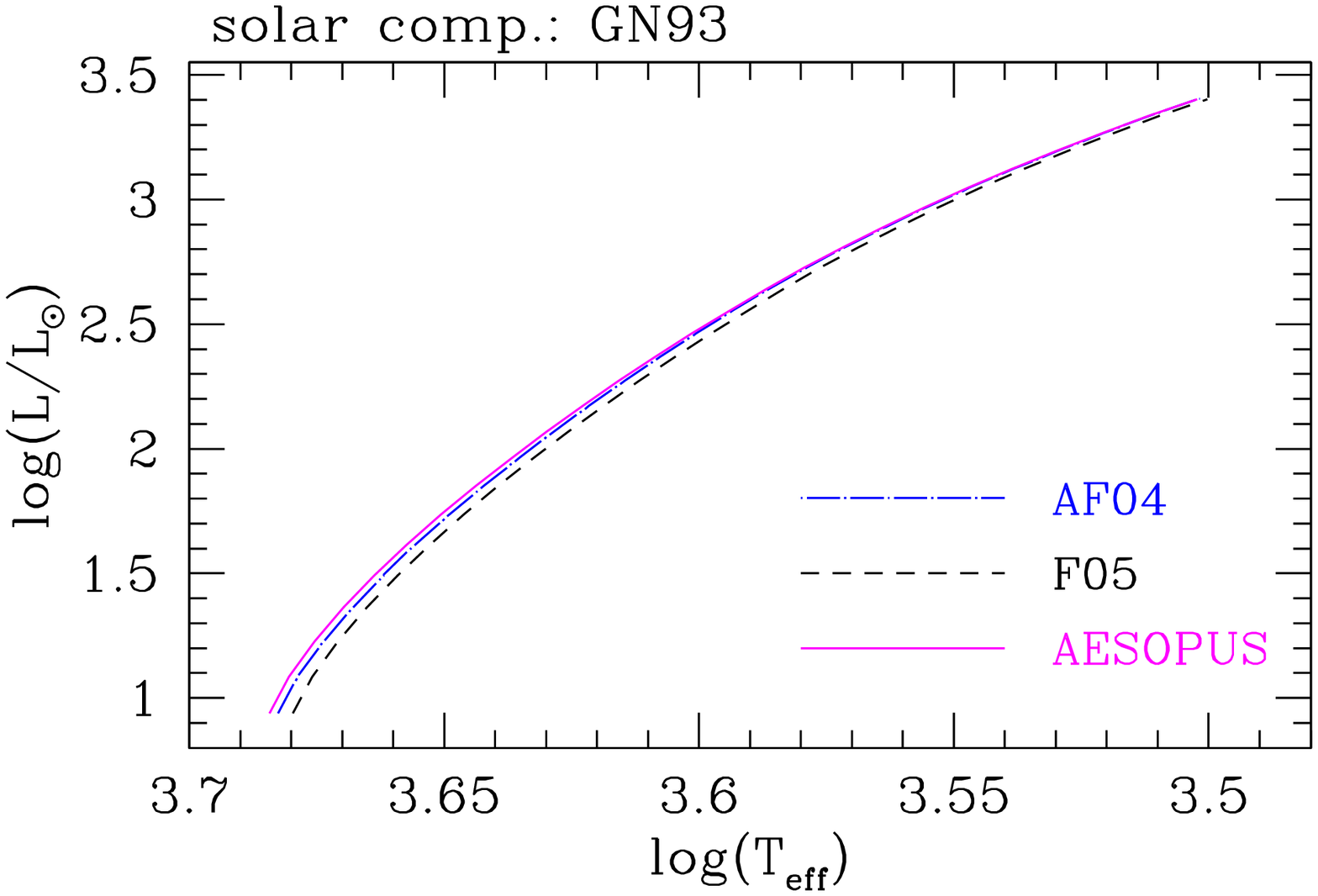}}
\end{minipage}
\hfill
\begin{minipage}{0.48\textwidth}
\resizebox{\hsize}{!}{\includegraphics{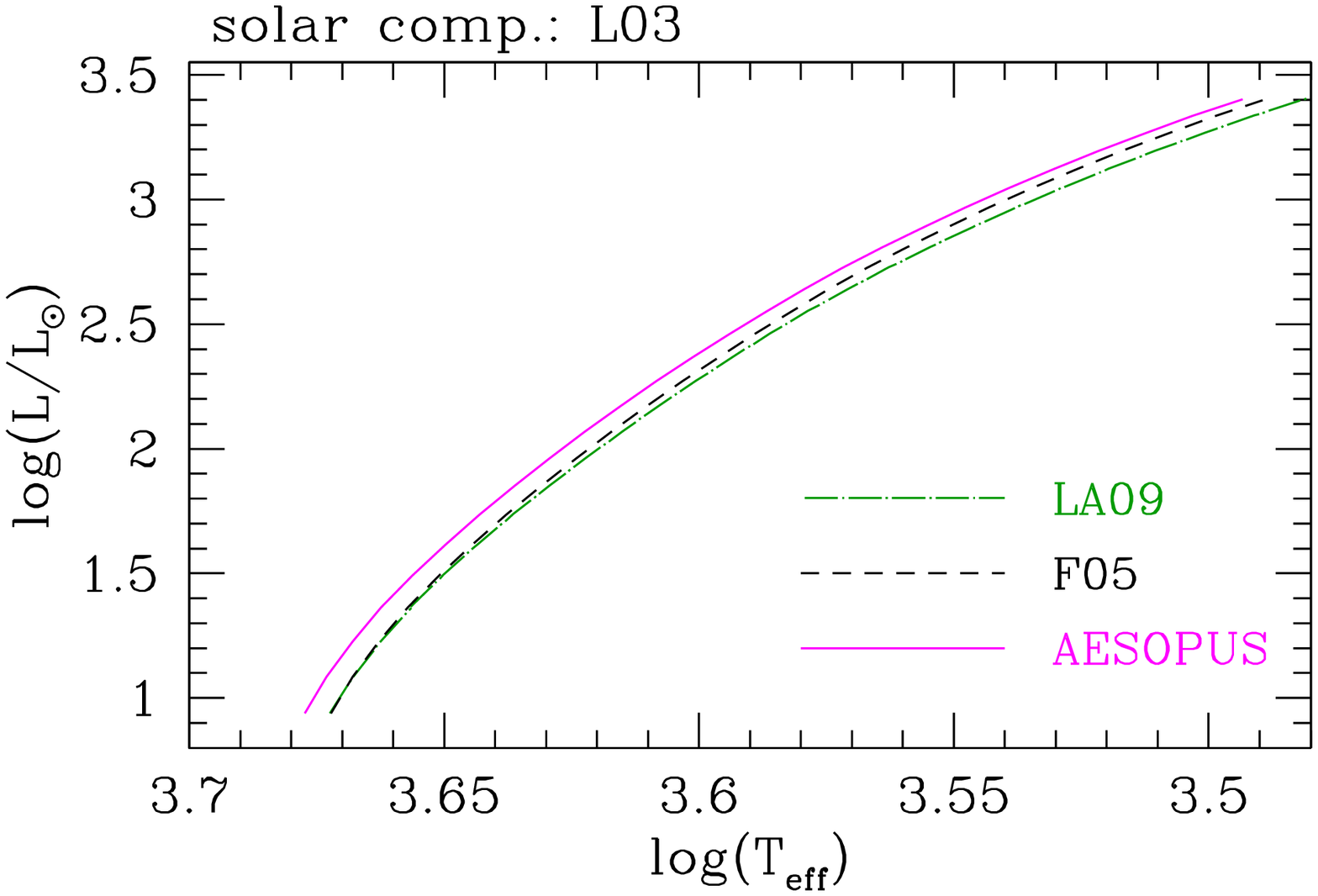}}
\end{minipage}
\caption{Predicted RGB tracks described by a $1.5\, M_{\odot}$ model 
with $Z=0.02,\,X=0.7$ and scaled-solar abundances of metals according to either
GN93 (left panel) or L03 (right panel). The luminosity is derived from the core-mass
luminosity relation given by Boothroyd \& Sackmann (1988), while increasing 
the core mass from $0.20\, M_{\odot}$ to $0.45\, M_{\odot}$.
The effective temperature is the result of envelope integrations (see the text for more
details). The different curves correspond to RM opacity tables computed by different authors, 
in the temperature range $3.2\,\le \log(T) \le 4.0$. }
\label{fig_rgb_opac}
\end{figure*}

\begin{figure*}
\begin{minipage}{0.48\textwidth}
\resizebox{\hsize}{!}{\includegraphics{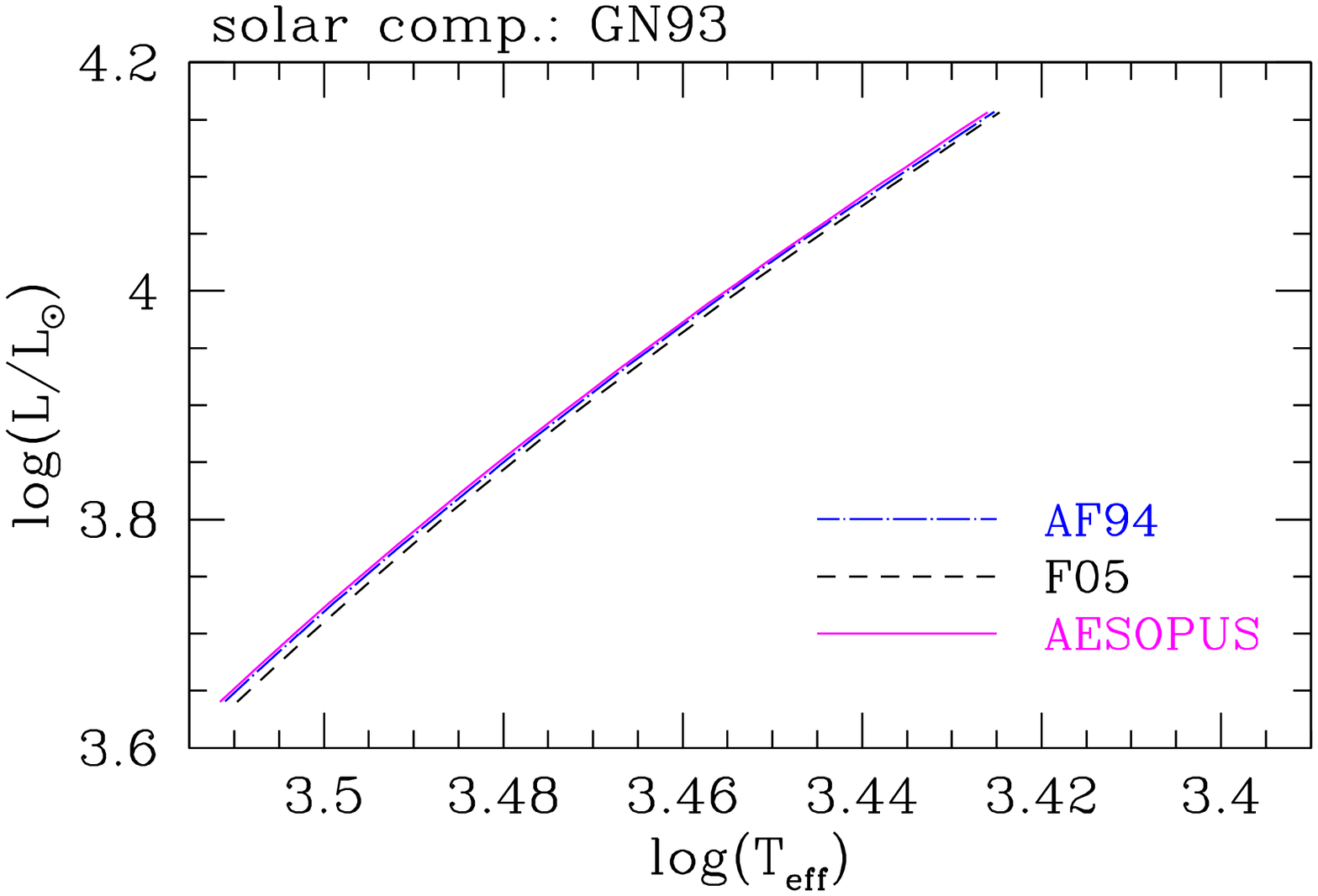}}
\end{minipage}
\hfill
\begin{minipage}{0.48\textwidth}
\resizebox{\hsize}{!}{\includegraphics{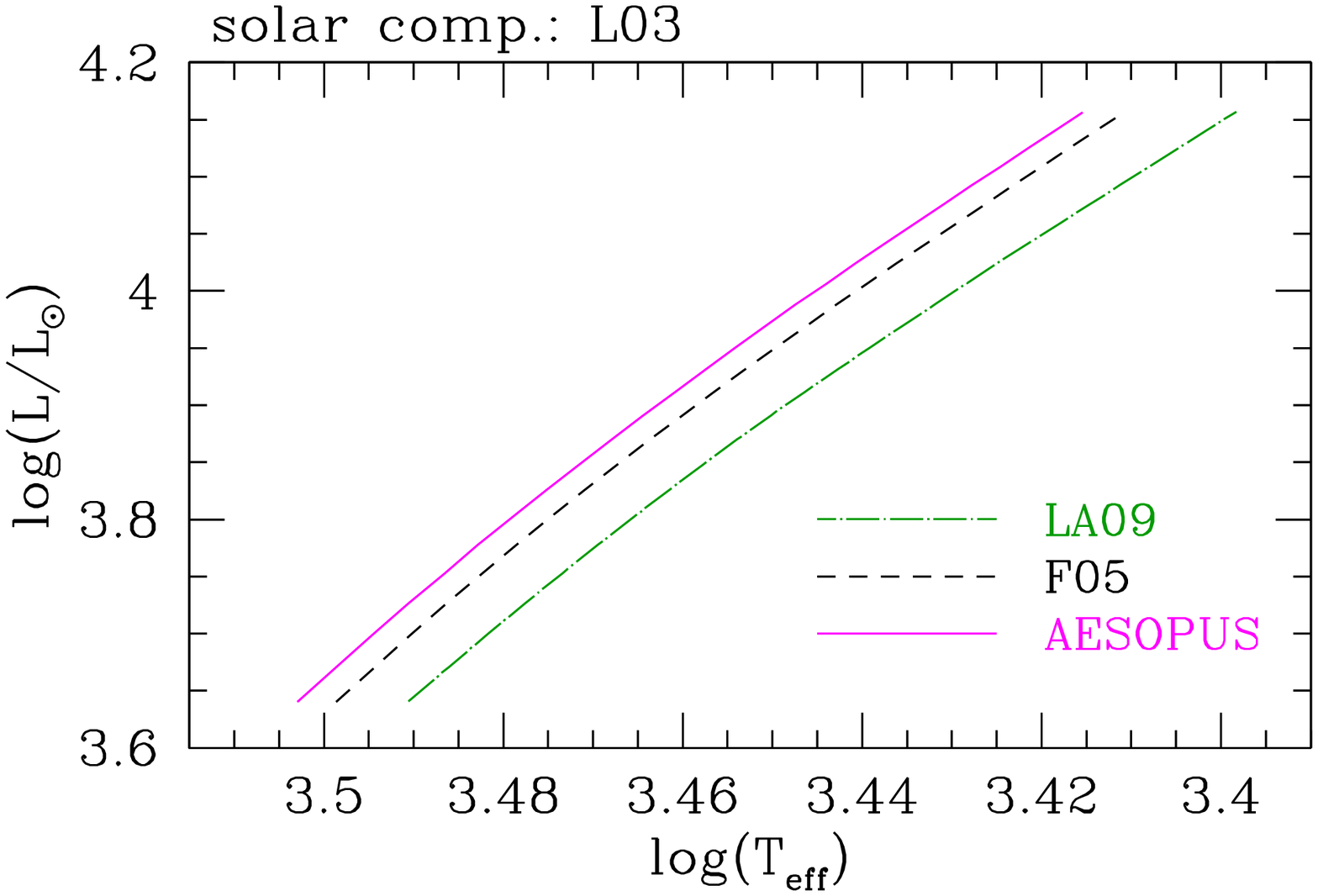}}
\end{minipage}
\caption{Predicted AGB tracks described by a $2.0\, M_{\odot}$ model 
with $Z=0.02,\,X=0.7$ and scaled-solar abundances of metals according to either
GN93 (left panel) or L03 (right panel). The luminosity is derived from the core-mass
luminosity relation, at the quiescent stage of the pre-flash maximum, 
given by  Wagenhuber \& Groenewegen (1998), while increasing 
the core mass from $0.50\, M_{\odot}$ to $0.75\, M_{\odot}$. The effective temperature is the result of
 envelope integrations (see the text for more
details). The different curves correspond to different RM opacity tables, 
in the temperature range $3.2\,\le \log(T) \le 4.0$.}
\label{fig_agb_opac}
\end{figure*}

\subsubsection{Tests with stellar models}
\label{sssec_starmod}
The numerical differences in $\kappa_{\rm R}$ between different authors, illustrated
in previous Sect.~\ref{sssec_kcomp},
assume a physical meaning when one analyses their impact  
on the models in which the Rosseland mean opacities are employed.
As already mentioned in Sect.~\ref{sec_intro}, the largest 
astrophysical use of pre-tabulated $\kappa_{\rm R}(\rho,\,T)$ 
is in the field of stellar evolution models to describe, in particular, the thermodynamic structure of
the most external layers including the atmosphere.  

While it is beyond the scope of this paper to perform a detailed analysis of the effects
of low-$T$ opacities on stellar structure and evolution, 
we consider here two illustrative cases, i.e. the predicted location in the H-R diagram of the 
Hayashi tracks described by low-mass stellar models while evolving through the RGB and AGB phases.
To investigate the differences in $T_{\rm eff}$ brought about by different choices of
low-$T$ opacity tables, we have carried out numerical integrations of a complete envelope model (basically 
the same as the one included in the Padova stellar evolution code) which extends from the atmosphere down to
surface of the degenerate core. The overall numerical procedure is fully described 
in Marigo et al. (1996, 1998), and Marigo \& Girardi (2007), so that it will not be repeated here.
The mixing-length parameter is assumed $\alpha=1.68$.  

As a matter of fact, it  has long been known that the atmospheric opacity is critical in determining
the position in the H-R diagram of a red-giant star (e.g. Keeley 1970; Scalo \& Ulrich 1975). 
We also recall that during the quiescent burning stages of both  RGB and AGB phases of a low-mass star 
the stellar luminosity is essentially controlled by the mass of the central core 
(and the chemical composition of the gas), being largely independent of the envelope mass.
Adopting suitable core-mass luminosity relations available in the literature, 
for given value of the core mass and chemical composition, envelope integrations yield 
the effective temperature at the corresponding luminosity. 
We have repeated this procedure 
increasing the core mass -- from $0.2\,M_{\odot}$ to $0.46\,M_{\odot}$ for the RGB and  from 
$0.5\,M_{\odot}$ to $0.75\,M_{\odot}$ for the AGB -- and 
adopting different opacity tables for $T \le 10\,000$ K. 

The results for  $1.5\,M_{\odot}$ and $2.0\,M_{\odot}$ models with $Z=0.02,\, X=0.7$ are  
shown in Figs.~\ref{fig_rgb_opac} and \ref{fig_agb_opac} for the RGB and AGB tracks respectively.
We have adopted low-$T$ opacities from AF94, F05, LA09, and \AE SOPUS, and two reference solar compositions, 
i.e. GN93 and L03.  
In all cases the computations with the opacities from \AE SOPUS 
and from the Wichita State University group are in close agreement, typically
being abs$(\log T_{\rm eff}^{\rm AF94}-\log T_{\rm eff}^{\small\rm \AE SOPUS}) \la 0.001$ dex
(ranging from $\sim 5$~K to $\sim 20$~K) and 
abs$(\log T_{\rm eff}^{\rm F05}-\log T_{\rm eff}^{\rm \AE SOPUS}) \la 0.005$ dex
(ranging from  $\sim 10$~K to $\sim 50$~K).
The deviations from the results with LA09 opacities are somewhat larger, 
$0.005 \la$ abs$(\log T_{\rm eff}^{\rm LA09}-\log T_{\rm eff}^{\rm \AE SOPUS}) \la 0.02$ dex
(ranging from  $\sim 50$~K to $\sim 100$~K).
In this respect it should be recalled that 
in the $T_{\rm eff}$-range considered here, $3.4 \la \log(T_{\rm eff}) \la 3.7$,  
the main opacity contributors are the absorption by
H$^{-}$ and Thompson e$^-$~scattering (the concentration of water vapour is still relatively low
even at the lowest temperatures; see Fig.~\ref{fig_chem_koff_z02}), 
so that differences in opacities are likely due to differences in the description of the 
H$^{-}$ opacity, and/or in the density of free electrons, which in turn may be affected 
by differences in the partition functions of the ions with low-ionisation potentials.
Anyhow, the temperature differences among the RGB and AGB 
tracks are in most cases lower than the current uncertainty affecting 
the semi-empirical $T_{\rm eff}$-scale of F-G-K-M giants  
($\sigma\sim 60-80$~K; e.g. Ram{\'{\i}}rez \& Mel{\'e}ndez 2005; Houdashelt et al. 2000).

\subsection{Varying C-N-O mixtures}
\label{ssec_kcno}

In several situations Rosseland mean opacities for non-scaled solar abundances should be
used. One of these cases applies, for instance, to stellar models 
in which the surface abundances of  C, N, and O
are altered via mixing and/or wind processes. A remarkable example corresponds
to the TP-AGB phase of low- and intermediate-mass stars, whose
envelope composition may be enriched with  primary carbon 
(and possibly oxygen) 
via the third dredge-up, or with newly synthesised nitrogen by 
hot-bottom burning.
As a net consequence, the abundances of C, N, and O as well as their
abundance ratios may be significantly changed
compared to their pre-TP-AGB values (Wood \& Lattanzio 2003).
 Most critical is the variation of 
the surface C/O ratio, which controls the chemistry of the gas at the
low temperatures typical of the atmospheres of AGB stars
(e.g. Marigo 2002).

Indeed, one of the aims of the present work is to provide a flexible 
computational tool to generate RM opacities for any value of 
combination of the C-N-O abundances, hence C/O ratio. 
 
\begin{figure}
\resizebox{1.05\hsize}{!}{\includegraphics{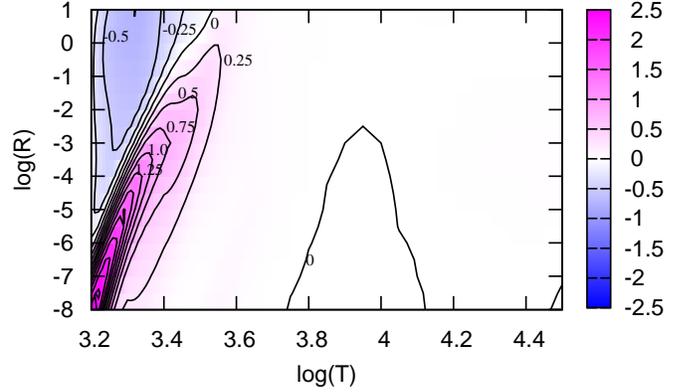}}
\caption{Comparison of RM opacities relative to two gas mixtures with $Z_{\rm ref}=0.02$, $X=0.7$ 
but different C/O ratios, 
namely C/O$=1.3$ and C/O$=$C/O$_{\odot}=0.49$ according to GS98 solar composition.
The colour map shows the difference $\log(\kappa_{\rm R}^{{\rm C/O}=1.3})- \log(\kappa_{\rm R}^{{\rm C/O}=0.49})$
throughout the standard location in the $\log(T)-\log(R)$ diagram of one opacity table computed with \AE SOPUS.
The contour lines corresponds to differences $\Delta\log(\kappa_{\rm R})$ multiple of $\pm 0.25$ dex.
Note the large deviations occurring in the low-$T$ region dominated by molecular absorption.} 
\label{fig_opac_co1p3_map}
\end{figure}

Figure~\ref{fig_opac_co1p3_map} shows clearly that big changes in 
$\kappa_{\rm R}$ are expected at low temperatures, say $\log(T)< 3.5$, 
when passing from an O-rich to a C-rich chemical mixture.
For instance, at $\log(T)=3.3$ RM opacities of a gas with  C/O$=1.3$ 
become much larger than in the case with  C/O$=0.49$ at lower densities, $-8 \la \log(R)\la -3$,
while the trend is reversed at increasing density, $\log(R)>-3$.
This fact is extremely important for the consequences it brings about 
to the evolutionary properties of C stars 
(see e.g. Marigo \& Girardi 2007; Cristallo et al. 2007; Marigo et al. 2008; 
Weiss \& Ferguson 2009; Ventura \& Marigo 2009).

In this context we will analyse in detail the impact of changing the
C/O ratio in a gas mixture, thus simulating the effect of the third
dredge-up in TP-AGB stars. 
\begin{figure}
\begin{minipage}{0.5\textwidth}
\resizebox{1\hsize}{!}{\includegraphics{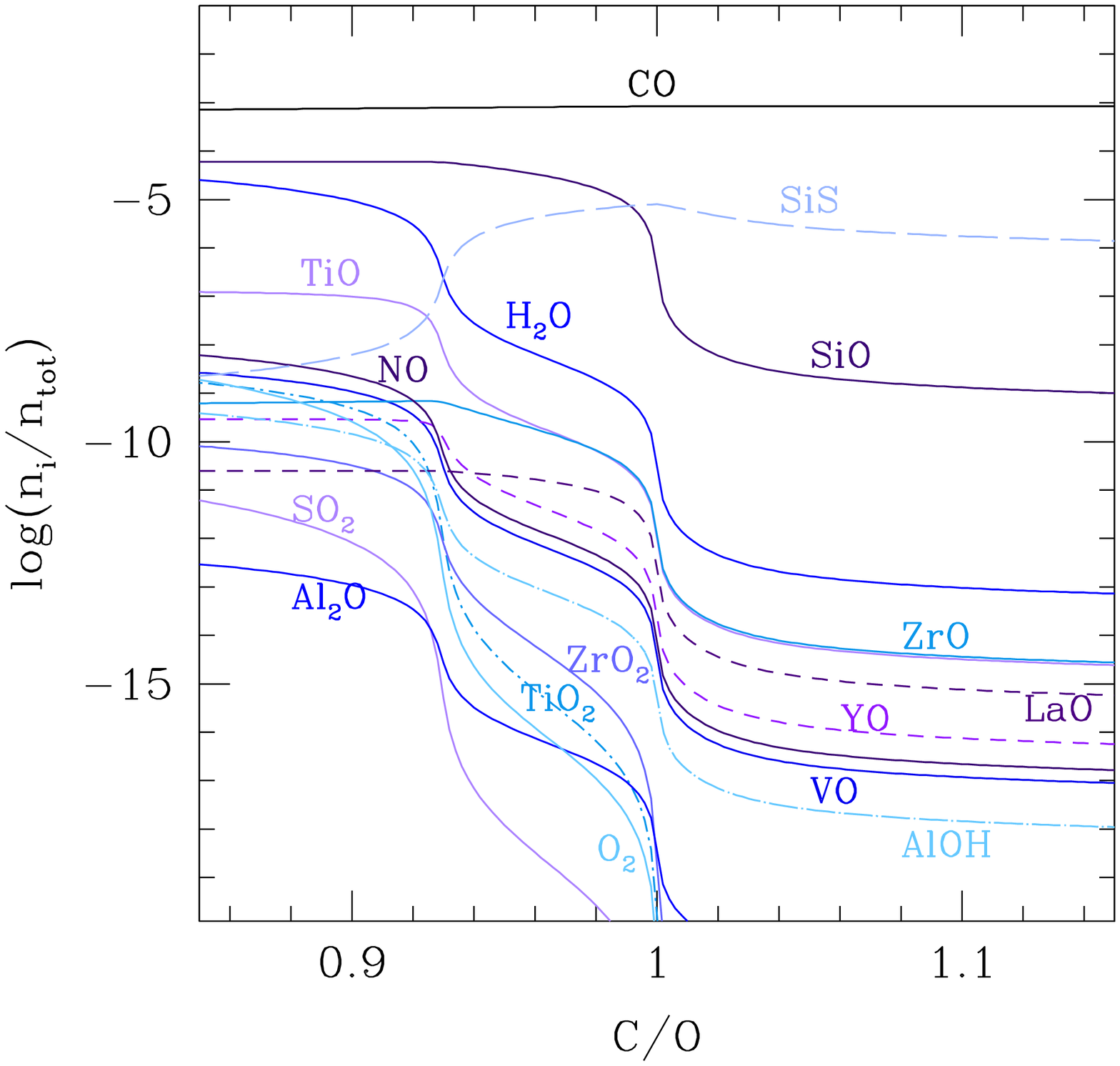}}
\end{minipage}
\hfill
\begin{minipage}{0.5\textwidth}
\resizebox{1\hsize}{!}{\includegraphics{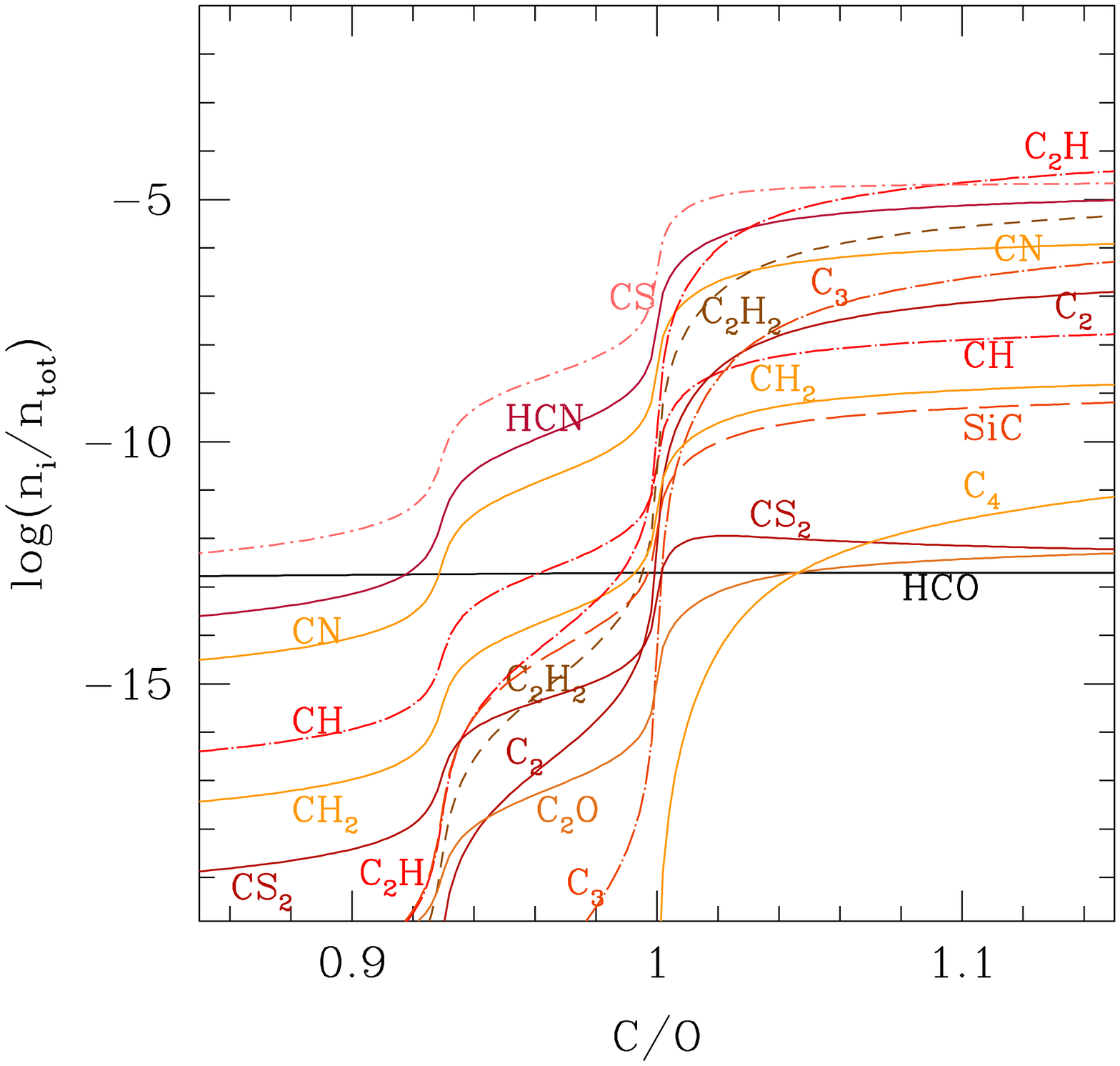}}
\end{minipage}
\caption{Concentrations of several gas species as a function of the
  C/O ratio, in a gas mixture with $\log(T)=3.3$, $\log(R)=-3$ (or equivalently $\log(\rho)=-11.1$),
$Z_{\rm ref}=0.02$, and $X=0.7$, 
and adopting the GAS07 reference solar partitions. 
The increase of C/O follows that of C, while O abundance is kept
unchanged. The actual metallicity $Z$ also increases with C. 
The molecules are divided into two groups, namely: O-bearing molecules (top panel) and C-bearing molecules
(bottom panel). 
Note the sharp change in molecular concentrations at C/O~$\approx 1$.}
\label{fig_chemco}
\end{figure}
\subsubsection{Molecular chemistry: the key r\^ole of the C/O ratio}
Figure~\ref{fig_chemco} illustrates the abrupt change in the chemical equilibria when the C/O ratio
passes from below to above unity, in a gas with $\log(T)= 3.3$ and $\log(R)=-3$
($\log(\rho)=-11.1$).
From a more careful inspection of Fig.~\ref{fig_chemco} we see that
the abundance curves of the O-bearing molecules (top panel) and the C-bearing 
molecules (bottom panel) follow mirror trends, exhibiting two sudden changes  of values at
C/O~$\approx 0.93$ and C/O~$\approx 1.0$. We may say that these two C/O
values bracket the transition
region between the O-dominated and the C-dominated chemistry. 
As discussed by Ferrarotti \& Gail (2002) the abrupt changes in the chemical equilibria 
at  C/O~$\approx 0.93$  
and C/O~$\approx 1.0$ respectively 
correspond to the critical values of the carbon abundance 
\begin{eqnarray}
\label{eq_crit1}
\varepsilon_{\rm C}^{{\rm crit},1}  =   \varepsilon_{\rm O} - \varepsilon_{\rm Si}\,\,\, \longrightarrow\,\,\, 
{\left(\frac{\rm C}{\rm O}\right)}_{{\rm crit},1}
& = & \frac{\varepsilon_{\rm C}^{{\rm crit},1}}{\varepsilon_{\rm O}} = 1 -  \displaystyle\frac{\varepsilon_{\rm Si}}{\varepsilon_{\rm O}}\\
\nonumber
\varepsilon_{\rm C}^{{\rm crit},2}  =  \varepsilon_{\rm O}\quad\quad\,\,\,\,\, \longrightarrow\,\,\, 
{\left(\frac{\rm C}{\rm O}\right)}_{{\rm crit},2} & = & 
\frac{\varepsilon_{\rm C}^{{\rm crit},2}}{\varepsilon_{\rm O}} = 1
\end{eqnarray}
The existence of $\varepsilon_{\rm C}^{{\rm crit},1}$ and $\varepsilon_{\rm C}^{{\rm crit},2}$
can be understood considering the extraordinary high bond energies of the 
two monoxide molecules CO and SiO, i.e.  $E_{\rm B}({\rm CO})=11.16$~eV and  $E_{\rm B}({\rm SiO})=8.29$~eV,
as well as the usually large concentrations of the involved species, 
i.e.  C, O, and to a less extent Si.
Following  Ferrarotti \& Gail (2002) 
for temperatures $T\la 1\,500$ K, at which dust is expected to condensate, one must also
consider the contribution of another strongly-bound molecule, SiS ( $E_{\rm B}({\rm SiS})=6.46$~eV), so that 
the first critical carbon abundance  should be redefined as
$\varepsilon_{\rm C}^{{\rm crit},1} =   \varepsilon_{\rm O} - \varepsilon_{\rm Si} + \varepsilon_{\rm S}$. 
Since this study deals with the gas chemistry for $\log(T) \ge 3.2$
(i.e. without dust formation)
in the following we limit our discussion to the case described by Eq.~(\ref{eq_crit1}).

In most cases the bond strength of CO mostly determines the chemical equilibria:
as long as  $\varepsilon_{\rm C} < \varepsilon_{\rm C}^{\rm crit,1}$, the excess of oxygen 
atoms, $\varepsilon_{\rm O}-\varepsilon_{\rm C}$, is available for the formation 
of O-bearing molecules -- such as
SiO, H$_2$O, TiO, VO, YO, etc. --, while as soon as 
 $\varepsilon_{\rm C} > \varepsilon_{\rm C}^{\rm crit,2}$, i.e.  C/O~$> 1$, the situation is reversed 
and the excess of of carbon 
atoms, $\varepsilon_{\rm C}-\varepsilon_{\rm O}$, takes part in C-bearing molecules 
such as CN, HCN, C$_2$, C$_2$H$_2$ , SiC, etc. 
This also explains why, unlike the others, the abundances of the
molecules involving the carbon monoxide, like CO itself and HCO, show
a flat behaviour with the C/O ratio.

The situation is somewhat different  
in the transition interval, $\varepsilon_{\rm C}^{\rm crit,1} \la \varepsilon_{\rm C} \la \varepsilon_{\rm C}^{\rm crit,2}$,
where the molecular pattern is controlled also by SiO, in addition to CO.
The C, O, and Si atoms are now almost completely absorbed in the CO and SiO monoxides,
which are the most abundant molecules, as shown in Fig.~\ref{fig_chemco}.
In other words, the excess of oxygen atoms over carbon is trapped  in
the molecular bond with silicon, which accounts for the first
abundance drop of 
the other O-bearing molecules at C/O~$\simeq 0.93$.

It is clear from Eq.~(\ref{eq_crit1}) that the value of (C/O)$_{\rm crit,1}$
 depends on the assumed oxygen and silicon abundances. In principle any change in the 
ratio ${\rm Si}/{\rm O}$ would correspond to a different (C/O)$_{\rm crit,1}$. 
As a reference case, it is instructive to compare the results for different choices of the solar
abundances. They are listed in Table~\ref{tab_sun}.
Passing from the AG89 to the most recent GAS07 compilation, the
 C/O$_{\rm crit,1}$ decreases from $\simeq 0.96$ to $\simeq 0.93$, implying 
that the transition  from the O- to the C-dominated chemistry takes place over a wider range
of the C/O ratio, i.e. $\sim 0.93-1$ for GAS07 in place of  $\sim 
0.96-1.00$ for AG89. 
As we will see later 
in this section, the knowledge of this critical ratio is of crucial importance since it
defines the onset of the transition between two chemical regimes, with consequent dramatic effects  
on the corresponding RM opacities of the gas 
(see for instance Figs.~\ref{fig_opac_varco_map} and \ref{fig_opac_varco_map_ag89_g07}).  
\begin{figure*}
\begin{minipage}{0.33\textwidth}
\resizebox{1.\hsize}{!}{\includegraphics{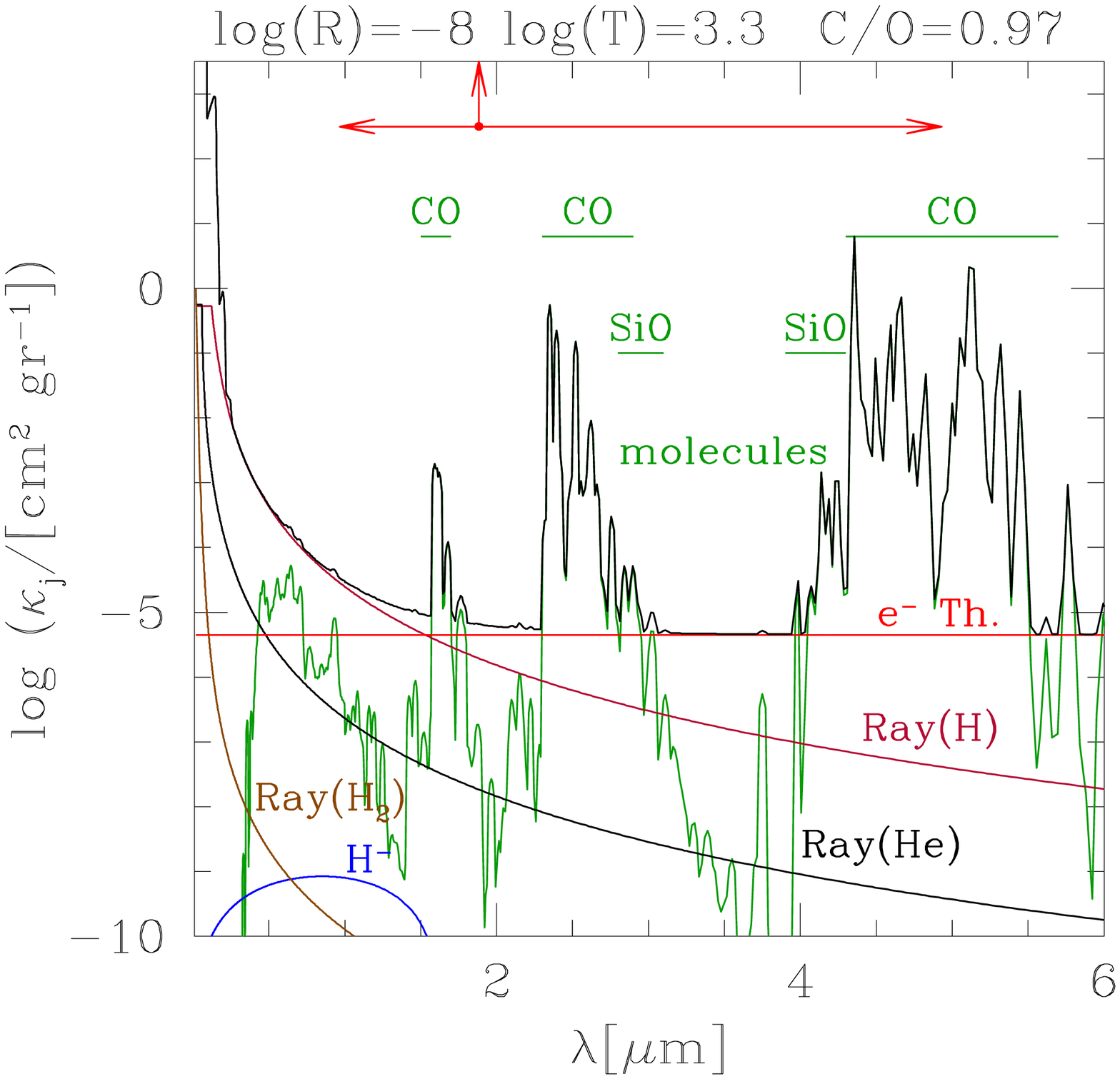}}
\end{minipage}
\begin{minipage}{0.33\textwidth}
\resizebox{1.\hsize}{!}{\includegraphics{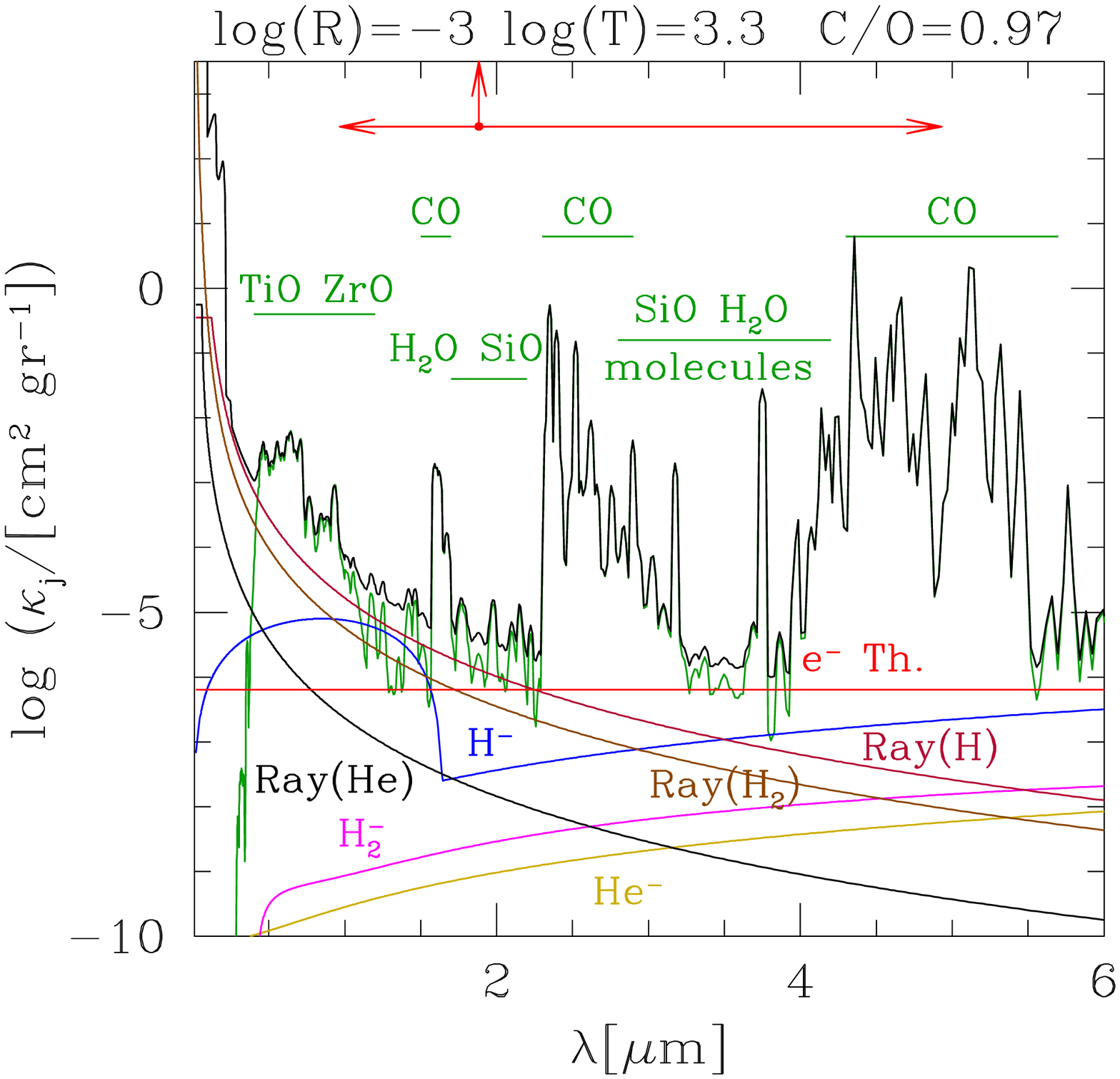}}
\end{minipage}
\begin{minipage}{0.33\textwidth}
\resizebox{1.\hsize}{!}{\includegraphics{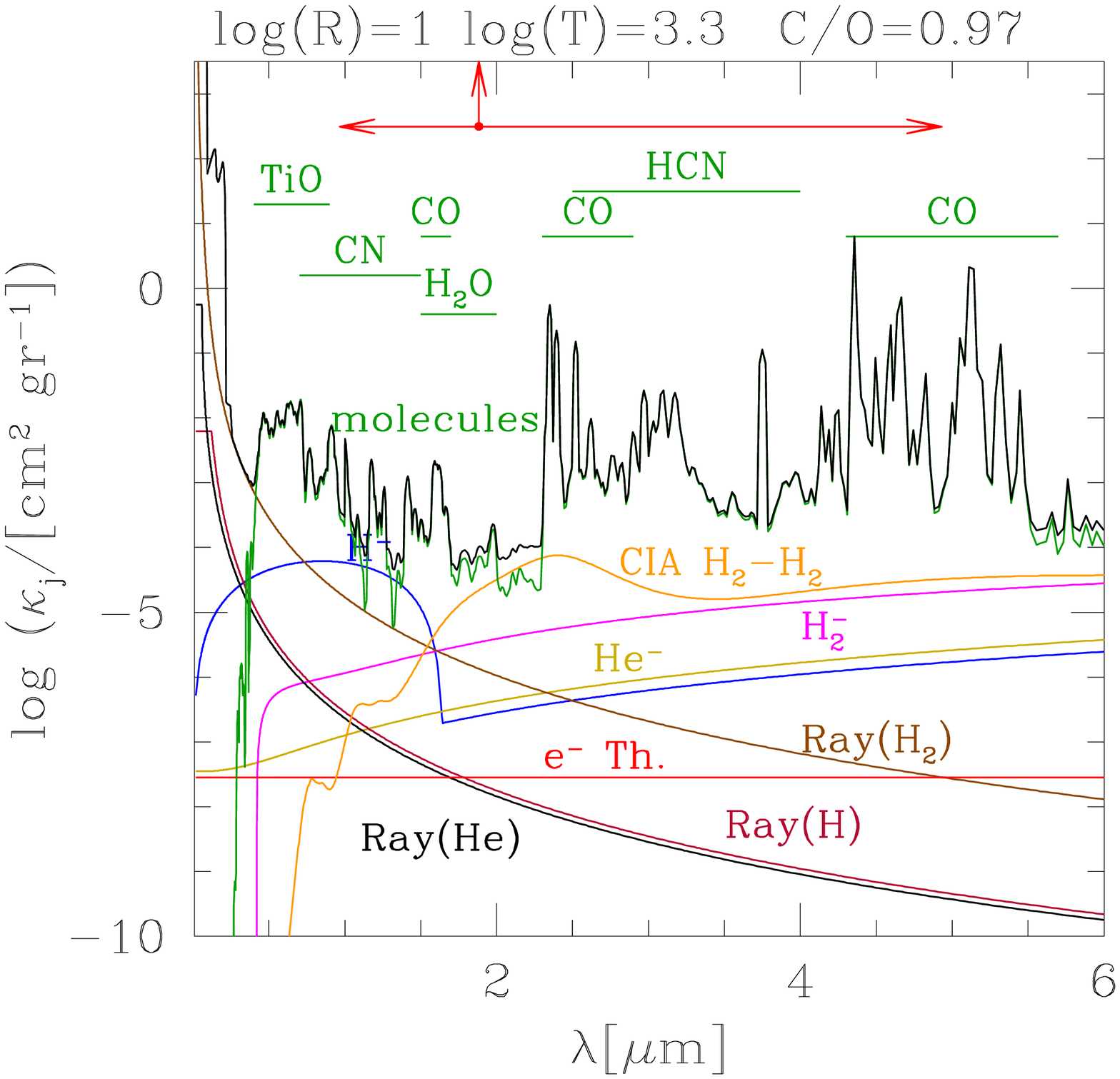}}
\end{minipage}
\begin{minipage}{0.33\textwidth}
\resizebox{1.\hsize}{!}{\includegraphics{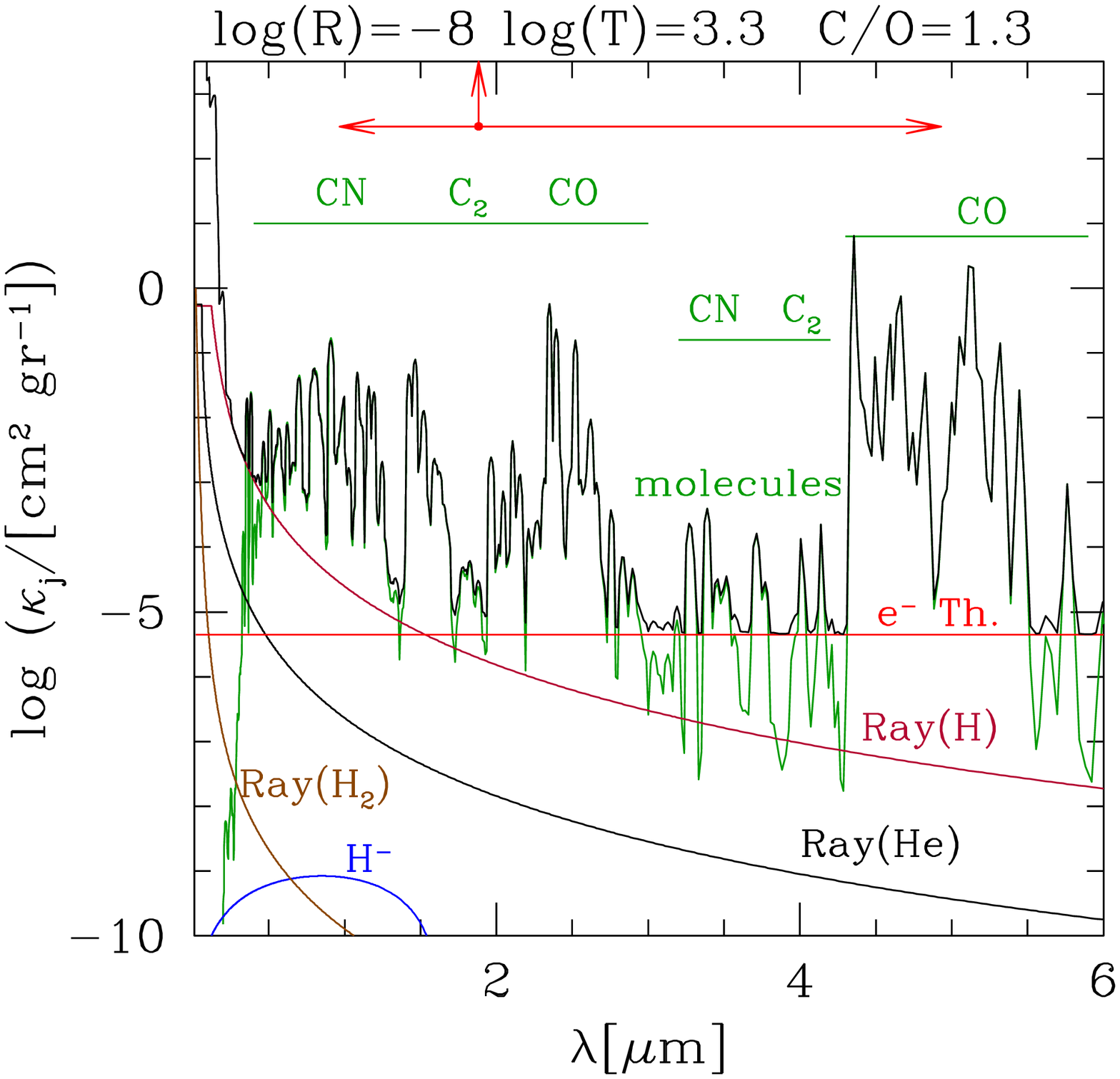}}
\end{minipage}
\begin{minipage}{0.33\textwidth}
\resizebox{1.\hsize}{!}{\includegraphics{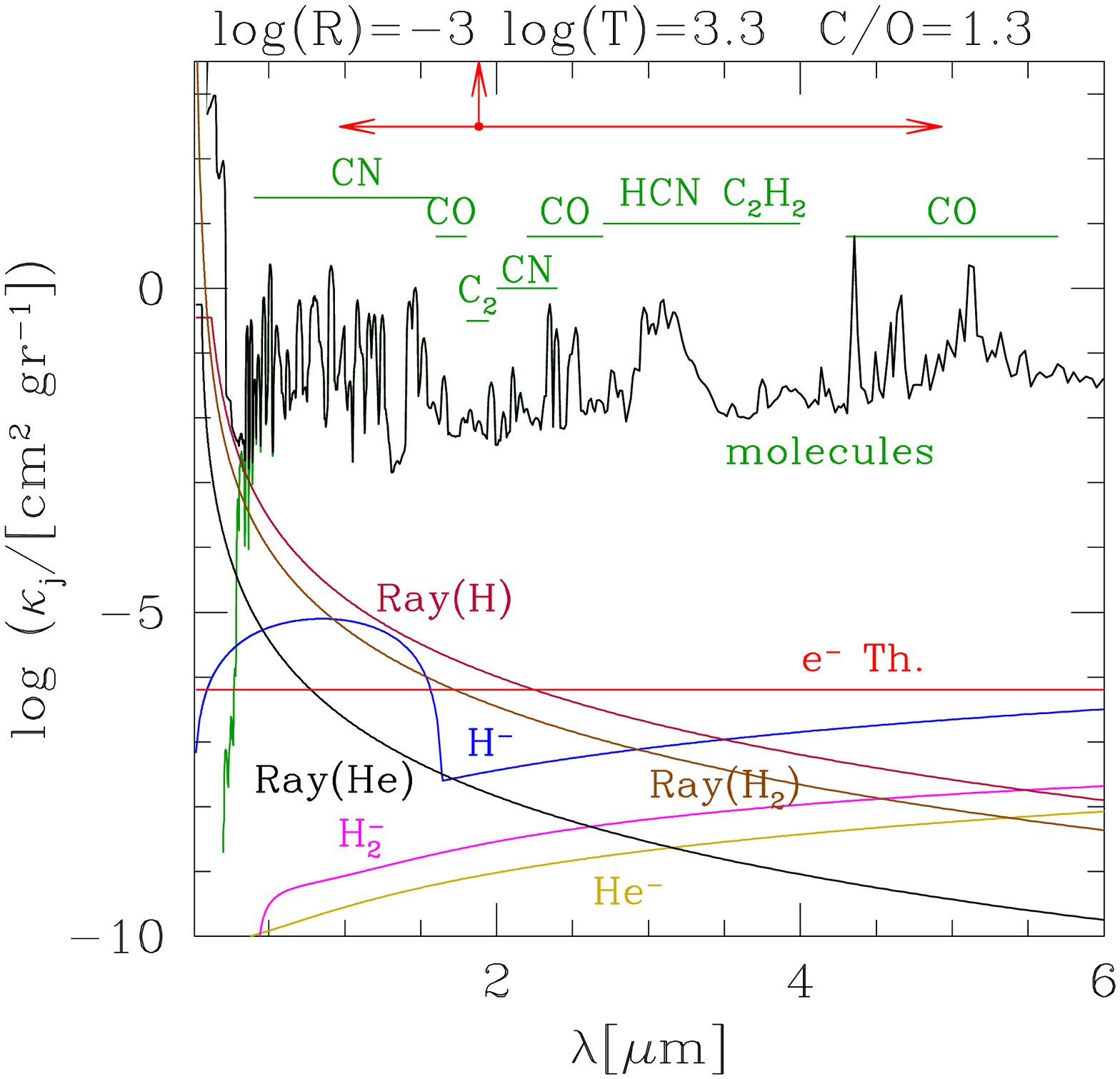}}
\end{minipage}
\begin{minipage}{0.33\textwidth}
\resizebox{1.\hsize}{!}{\includegraphics{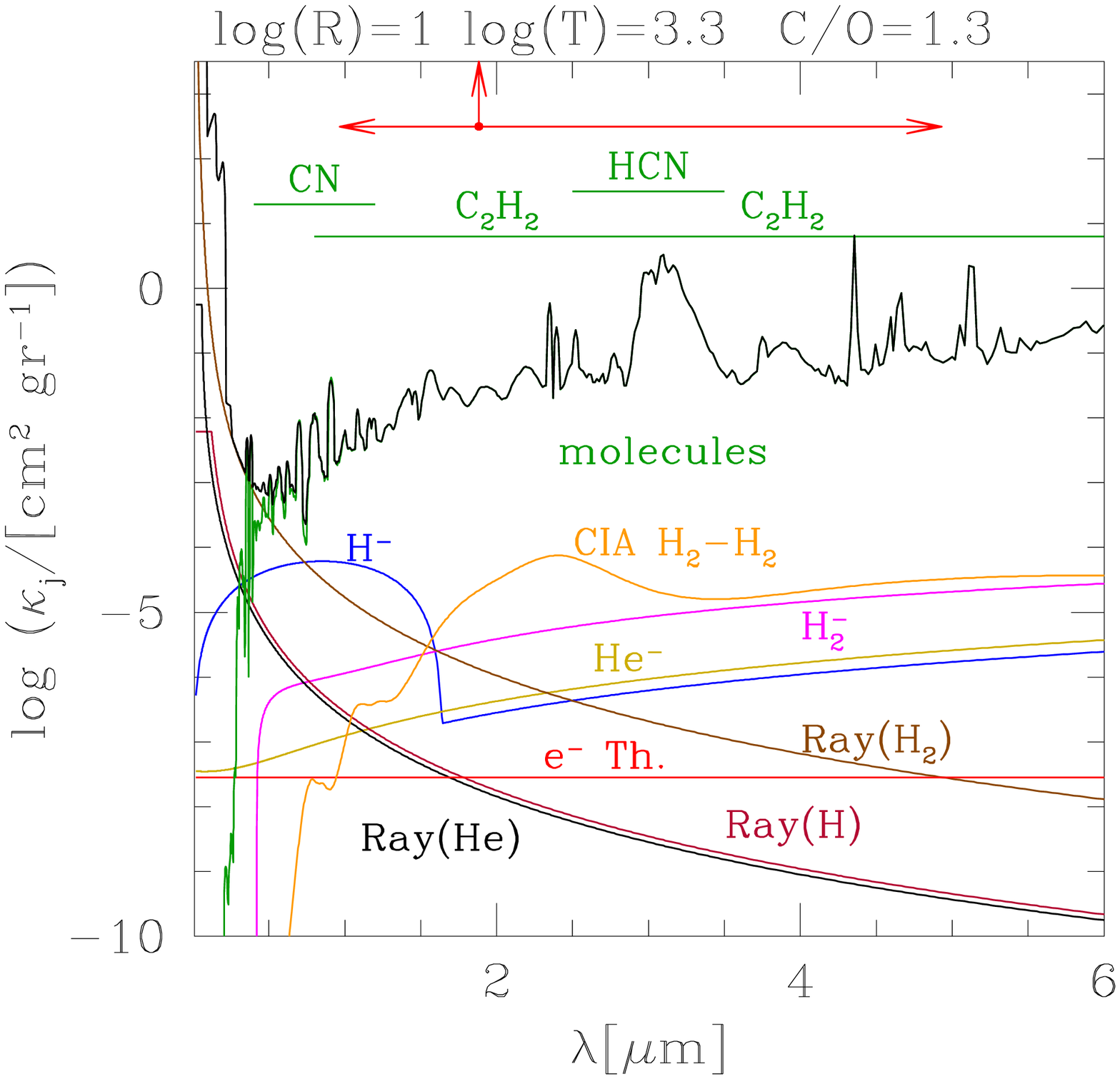}}
\end{minipage}
\caption{The same as in Fig.~\ref{fig_sig_z02}, but for gas mixtures with  C/O=$0.97$
(upper panels) and  C/O=$1.30$ (bottom panels) and $\log(T)=3.3$.
Note that in the spectral range relevant for the RM opacity, 
the total monochromatic coefficient is affected by heterogeneous sources 
(e.g. TiO, ZrO, CO, H$_2$O, CN, Thomson e$^-$~scattering) for C/O=$0.97$, 
while  absorption by C-bearing molecules dominate for C/O=$1.30$. }
\label{fig_sigc}
\end{figure*}

\begin{figure*}
\begin{minipage}{0.33\textwidth}
\resizebox{\hsize}{!}{\includegraphics{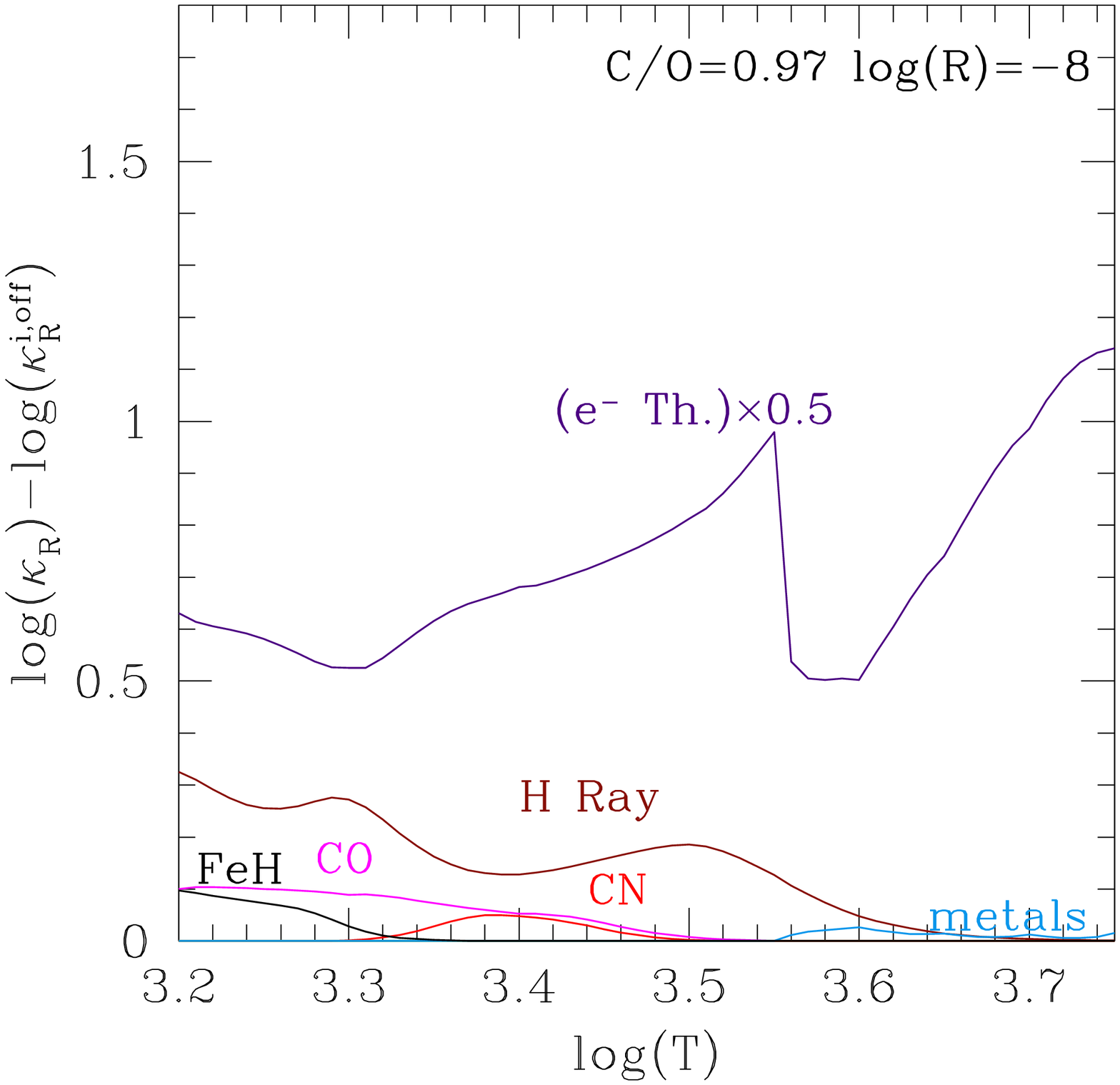}}
\end{minipage}
\begin{minipage}{0.33\textwidth}
\resizebox{\hsize}{!}{\includegraphics{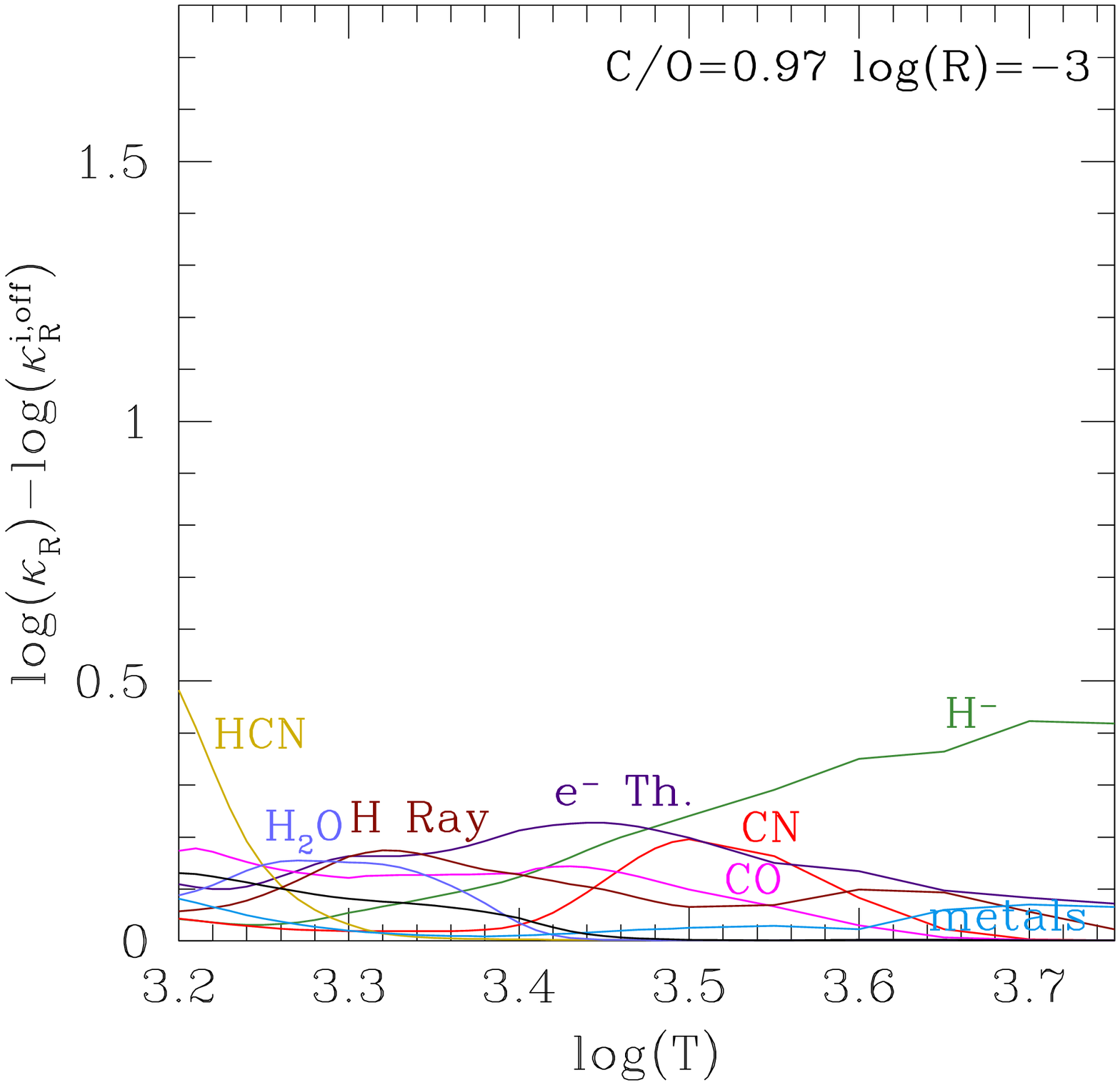}}
\end{minipage}
\begin{minipage}{0.33\textwidth}
\resizebox{\hsize}{!}{\includegraphics{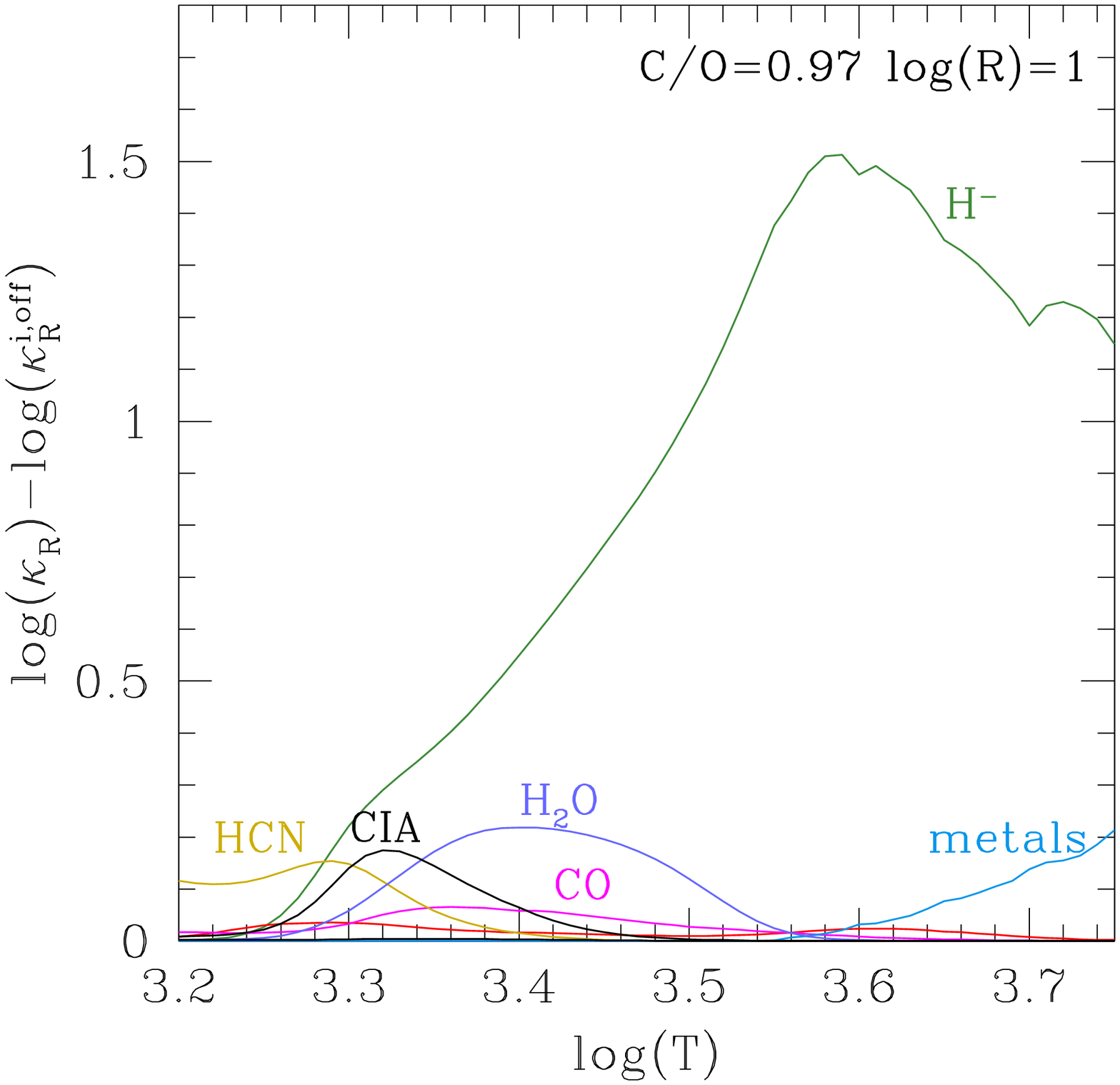}}
\end{minipage}
\begin{minipage}{0.33\textwidth}
\resizebox{\hsize}{!}{\includegraphics{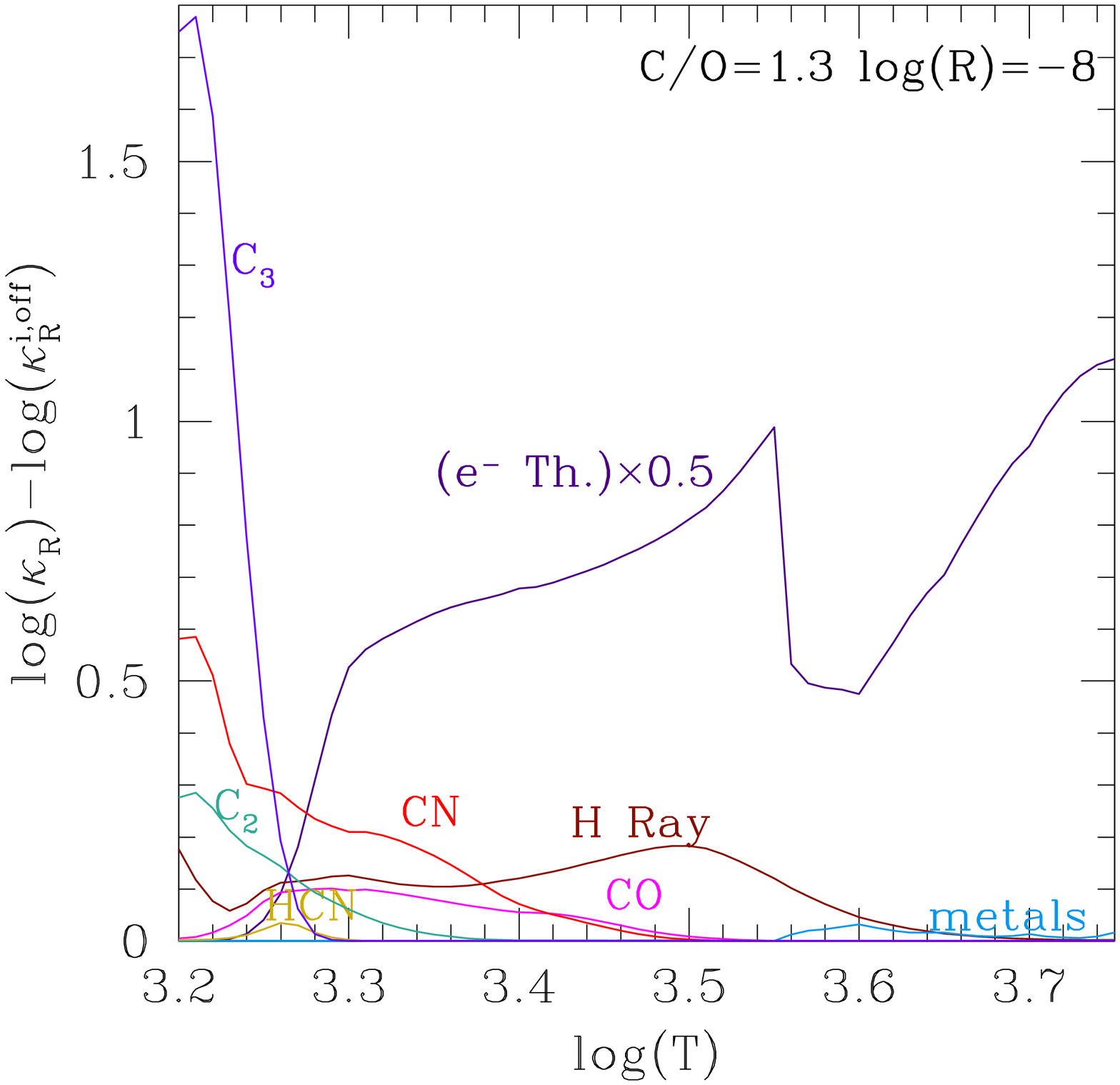}}
\end{minipage}
\begin{minipage}{0.33\textwidth}
\resizebox{\hsize}{!}{\includegraphics{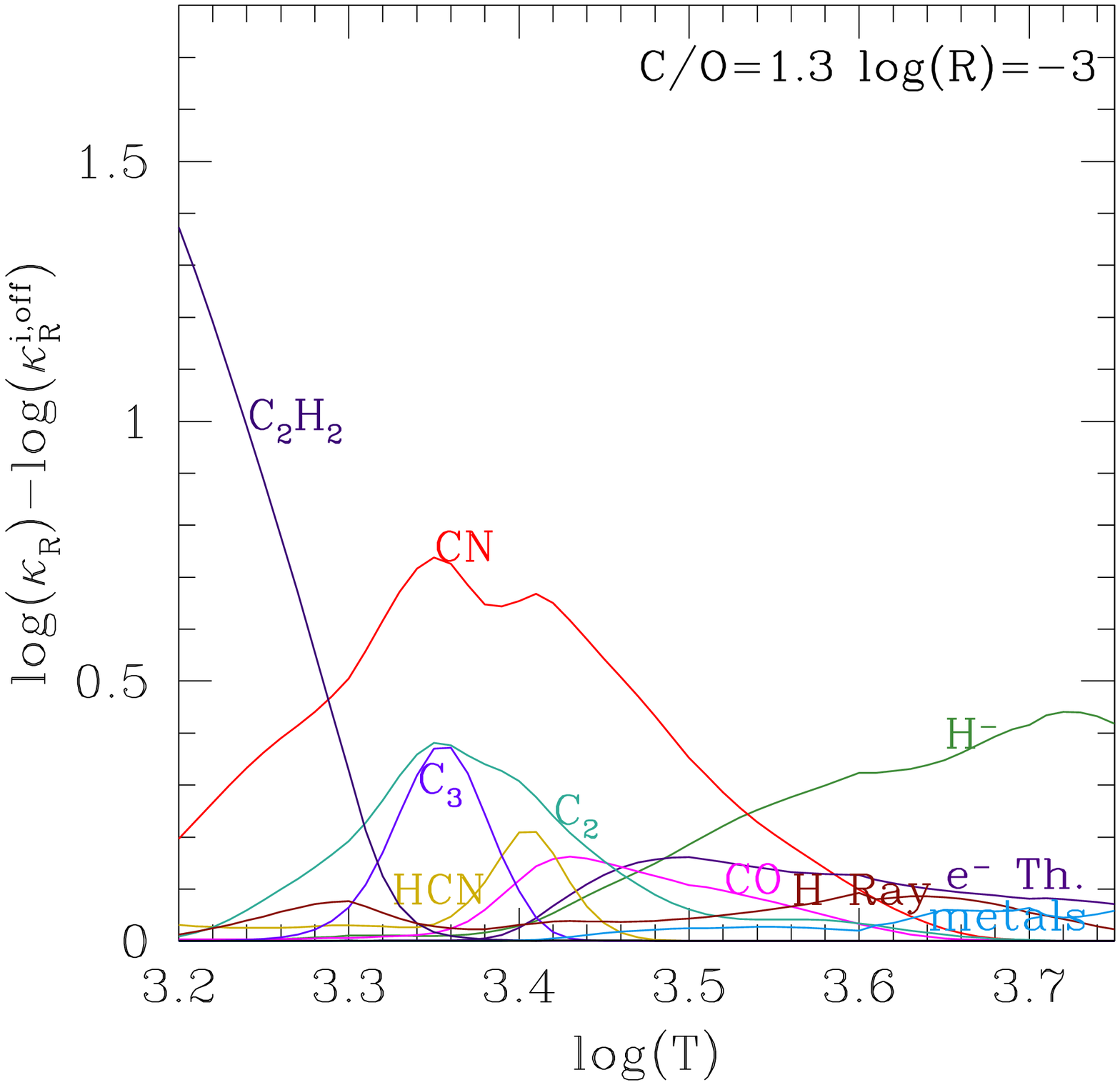}}
\end{minipage}
\begin{minipage}{0.33\textwidth}
\resizebox{\hsize}{!}{\includegraphics{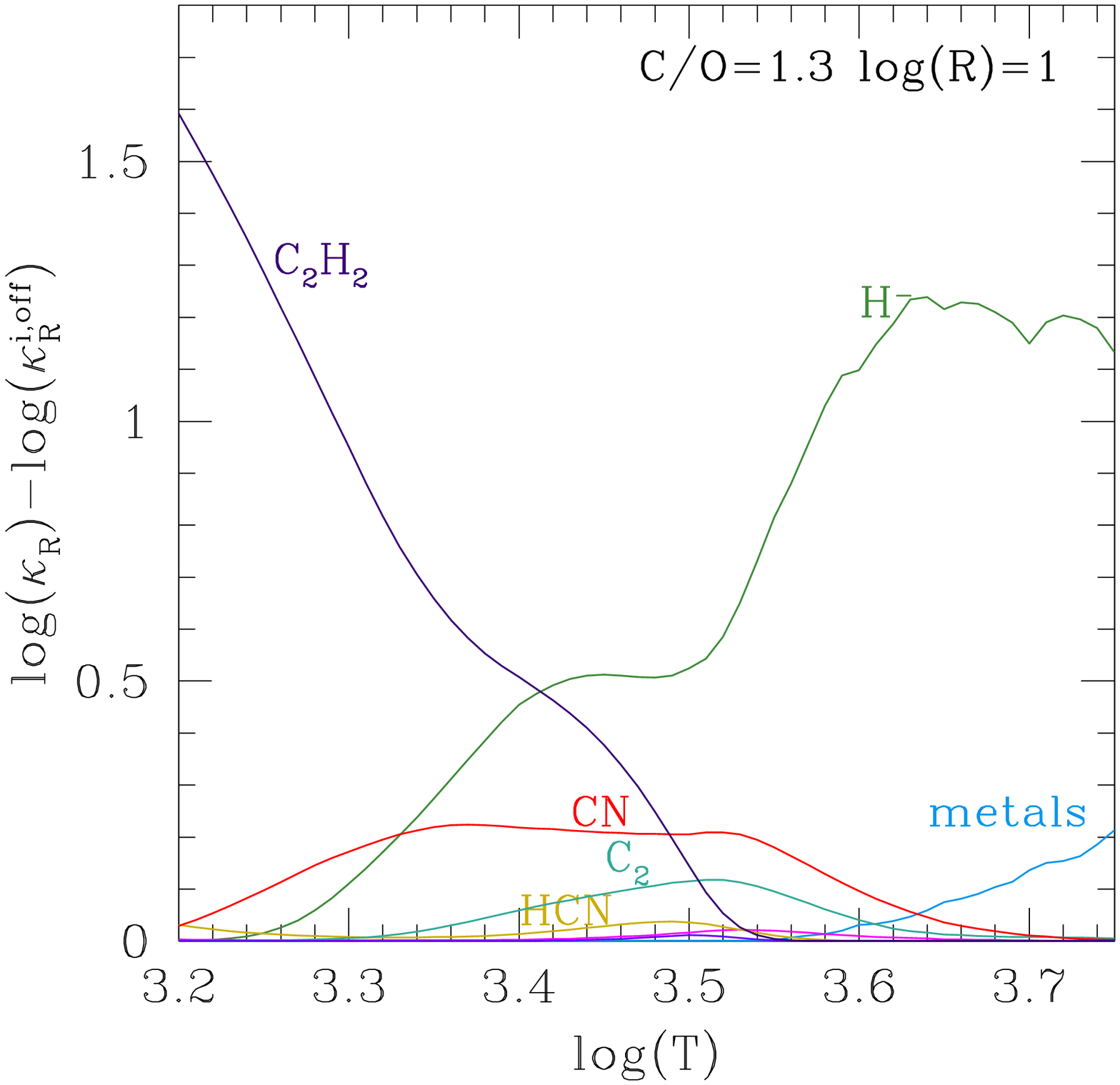}}
\end{minipage}
\caption{The same as in Fig.~\ref{fig_chem_koff_z02} but for gas mixtures 
with C/O=$0.97$ (upper panels) and C/O~$=1.3$ (bottom panels),  and zoomed into the 
molecule-dominated temperature region. Note the various opacity bumps
of the C-bearing molecules in the C/O=$1.30$ case, while comparable contributions from 
both O-rich and C-rich molecules are present in the  C/O=$0.97$ case.
}
\label{fig_koff_crich}
\end{figure*}

\subsubsection{Opacity sources at increasing C/O ratio}
The extreme sensitiveness of the molecular chemistry -- for $\log(T)\la
3.5-3.6$ depending on 
the density -- to the C/O parameter has striking consequences on the
low-temperature gas opacities, 
as shown in Fig.~\ref{fig_sigc}, relative to  $\log(T)=3.3$ and three
values of the $R$ parameter.
This figure can be interestingly compared with the bottom panels  of 
Fig.~\ref{fig_sig_z02}, describing the case of an oxygen-rich scaled-solar chemistry.
For instance, 
we see that at $\log(T)=3.3$ and $\log(R)=-3$ (bottom-mid panel of Fig.~\ref{fig_sigc}) 
the total monochromatic 
coefficient $\kappa(\nu)$ for C/O~$=1.3$ is mostly determined by the
absorption bands of molecules such 
as  HCN and CN,
while in a gas with the same thermodynamic conditions and solar 
C/O~$\simeq 0.5$, the dominating species are
H$_2$O, TiO, and VO (see bottom-mid panel of Fig.~\ref{fig_sig_z02}).

At the same temperature and density, and for C/O~$=0.97$  (upper-mid panel of Fig.~\ref{fig_sigc})
the total coefficient $\kappa(\nu)$ is, on
average, lower than in the other two cases, being mostly affected by
the absorption bands of CO,  while the gaps in between 
are populated by the weaker molecular bands of H$_2$O, SiO, ZrO, TiO, etc.
At lower densities ($\log(R)=-8$; upper-left panel of Fig.~\ref{fig_sigc}) Rayleigh scattering 
from neutral H and Thomson scattering from free electrons fill the spectral intervals
between the CO absorption bands, while at higher densities 
($\log(R)=1$; upper-right panel of Fig.~\ref{fig_sigc}) the total monochromatic coefficient is
completely dominated by molecular absorption, with a sizable contribution by CIA($\HH_2/\HH_2$)
at $\lambda \simeq 2$ $\mu$m, just in correspondence of the maximum $\lambda_{\rm max}$ 
of the weighting function of the Rosseland mean (see Eq.~\ref{eq_lmax}).
 
\begin{figure}
\resizebox{\hsize}{!}{\includegraphics{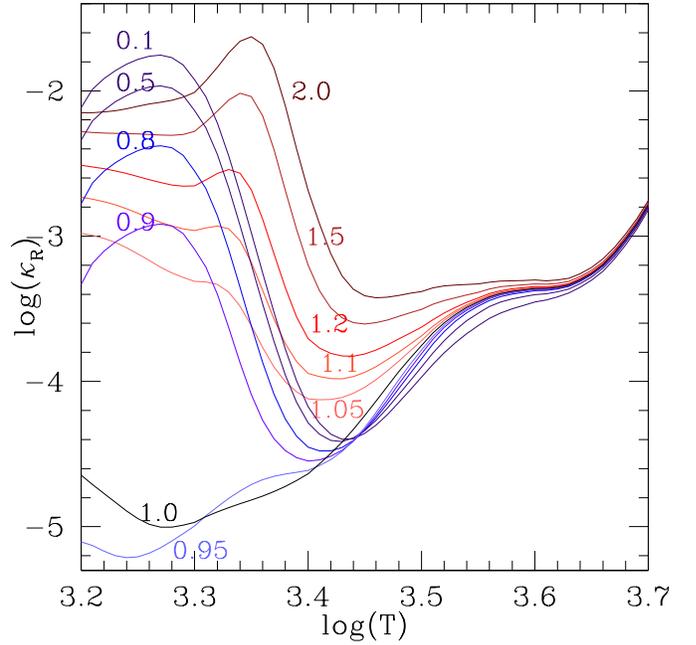}}
\caption{Rosseland mean opacity as a function of temperature, assuming $\log(R)=-3$, and at
  increasing C/O, from $0.1$ up to $2.0$. The reference composition
is defined by ($Z_{\rm ref}=0.02$, $X=0.7$) and assuming the metal abundances 
scaled-solar to the GAS07 mixture. 
The abundance of carbon is made vary (hence the actual $Z$), while keeping unchanged 
that of oxygen.} 
\label{fig_opac_varco}
\end{figure}

\begin{figure}[h]
\resizebox{\hsize}{!}{\includegraphics{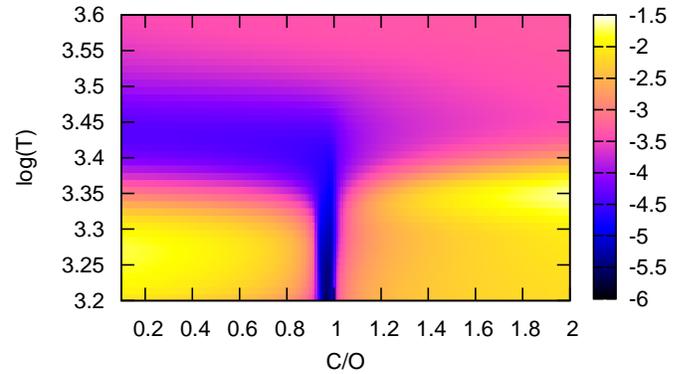}}
\caption{Rosseland mean opacity as a function of the temperature and increasing C/O.
adopting the GAS07 solar mixture, and assuming $Z_{\rm ref}=0.02$, $X=0.7$, and $\log(R)=-3$.   
The abundance of carbon is made vary accordingly to the current C/O ratio (so that the actual metallicity
varies as well),
while that of oxygen is kept fixed at its scaled-solar value.} 
\label{fig_opac_varco_map}
\end{figure}

\begin{figure}
\begin{minipage}{0.5\textwidth}
\resizebox{\hsize}{!}{\includegraphics{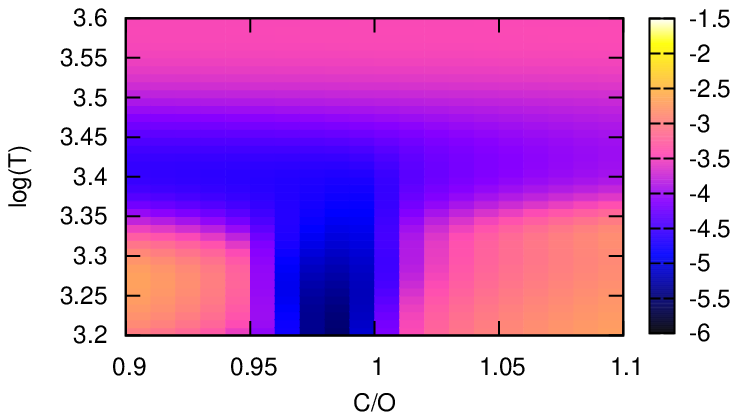}}
\end{minipage}
\vfill
\begin{minipage}{0.5\textwidth}
\resizebox{\hsize}{!}{\includegraphics{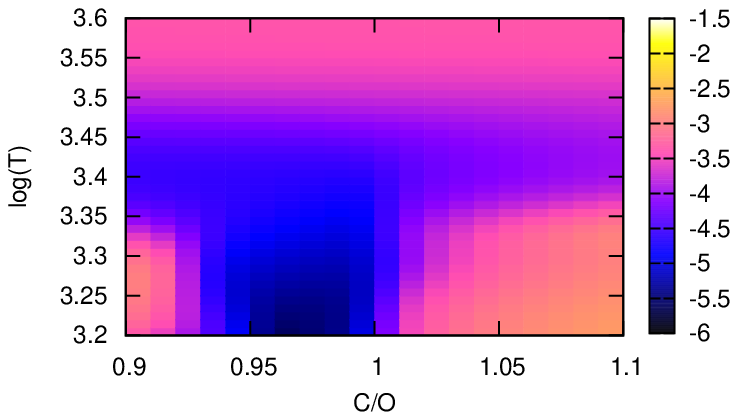}}
\end{minipage}
\caption{The same as in Fig.~\ref{fig_opac_varco_map}, but zoomed into
a narrower interval around C/O~$=1$. The reference solar compositions are
AG89 (top panel) and GAS07 (bottom panel). } 
\label{fig_opac_varco_map_ag89_g07}
\end{figure}

The sharp changes in the chemistry and monochromatic coefficient $\kappa(\nu)$ as a function of C/O 
impact as much strongly on the integrated RM opacity $\kappa_{\rm R}$,
which is evident in Figs.~\ref{fig_koff_crich} -- 
\ref{fig_opac_varco_map_ag89_g07}.

For the same two C/O values considered above, Fig.~\ref{fig_koff_crich} shows the
contributions of different opacity sources to the RM opacity
 as a function of the temperature (and assuming $\log(R)=-8,\,-3,\,1$).
An instructive comparison with the results for a scaled-solar chemistry 
can be done with the help of Fig.~\ref{fig_chem_koff_z02}. 
In the case with C/O~$=0.97$  (upper panels of Fig.~\ref{fig_koff_crich})
 Rayleigh scattering from hydrogen and Thomson
scattering from free electrons dominate for $\log(R)=-8$, becoming comparable 
with the molecular sources for $\log(R)=-3$. 
Moreover, we notice that at this C/O value, representing the
transition between different chemistry regimes, the opacity pattern is quite
heterogeneous as it includes the contributions from both O-bearing and
C-bearing molecules.
For instance, we see that  H$_2$O is important at lower temperatures, CN 
shows up at larger temperatures, while CO contributes over a larger
temperature interval.

In the case with C/O~$=1.3$  (bottom  panels of Fig.~\ref{fig_koff_crich}) 
the most noticeable features at different 
densities are the following.
At $\log(R)=-8$ and   $\log(T)\la 3.3$
the largest contribution come from C$_3$ (and CN, C$_2$), while at larger temperatures
the  electron scattering dominates. At $\log(R)=-3$
the high and broad opacity bump of CN that dominates the RM opacity over
a wide temperature interval, $3.30 \la \log(T) \la 3.55$, 
while the C$_2$H$_2$ contribution  is prominent for  $\log(T) \la
3.30$. In addition, other C-bearing molecules 
(C$_2$, C$_3$, HCN, CO) provide
non-negligible contributions to the RM opacity.
Finally, at $\log(R)=1$ the polyatomic molecule C$_2$H$_2$  is the most
efficient contributor to $\kappa_{\rm R}$ for $\log(T) \le 3.4$, while the
 hydrogen anion becomes  prominent at higher temperatures.

The complex behaviour of the  RM opacities 
as a function of the C/O ratio is exemplified with the aid of
Fig.~\ref{fig_opac_varco} for  
$3.2 \la \log(T)\la 3.6$, the temperature range in which molecules become the most efficient radiation 
absorbers. It turns out that  while 
the C/O ratio increases from $0.1$ to $0.9$ the opacity bump peaking at ($\log(T)\simeq 3.3$ for 
$\log(R)=3$) -- mostly due to H$_2$O -- becomes more and more depressed because of the smaller 
availability of O atoms. Then, passing from C/O~$=0.9$ down to  C/O~$=0.95$ the H$_2$O feature
actually disappears and $\kappa_{\rm R}$ drastically drops by more than two orders of magnitude.
In fact, at this C/O value the chemistry enters the transition region
already discussed (see Fig.~\ref{fig_chemco}), so that most of both O and C atoms are trapped in the CO
molecule at the expense of the other  molecular
species, belonging to both the O- and  C-bearing groups. 
 At C/O~$=1$ the RM opacity increases at the lowest temperatures,
$\log(T) \la 3.3$ , while a sudden upturn is
expected as soon as C/O slightly exceeds unity, as displayed by the
curve for C/O~$=1.05$ in Fig.~\ref{fig_opac_varco}. This fact reflects
the drastic change in the molecular equilibria from the O- to the
C-dominated regime.
Then, at increasing C/O ($1.1$, $1.2$, $1.5$, and $2.0$) the opacity curves
move upward following a more gradual trend, which is related with the
strengthening of the C-bearing molecular absorption bands.

\begin{figure}
\resizebox{\hsize}{!}{\includegraphics{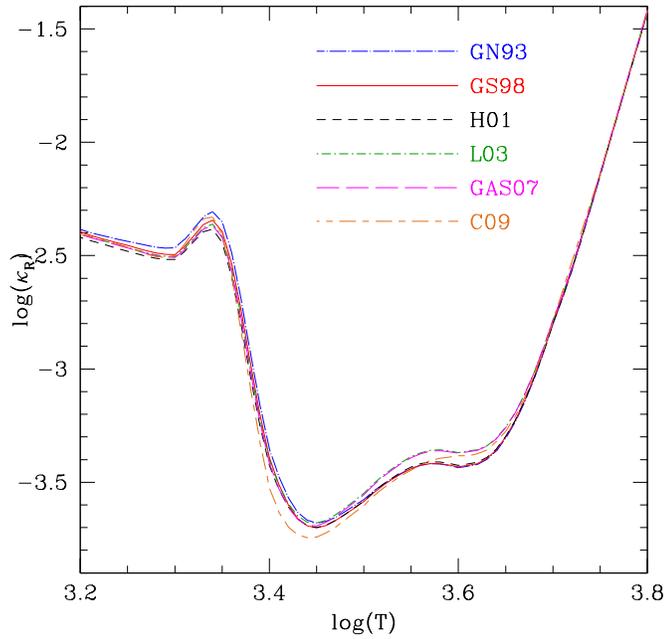}}
\caption{Rosseland mean opacity as a function of the temperature in a
gas with $Z_{\rm ref}=0.02$, $X=0.7$,  and $\log R=-3$. The adopted
chemical composition is characterised by C/O~$=1.3$,
for various compilations of the reference solar mixture, 
namely: GN93, GS98, H01, L03, GAS07,and C09. In each case the actual 
metallicity $Z> Z_{\rm ref}$ because of the increase in C abundance.
}
\label{fig_co1p3}
\end{figure}

\begin{figure}
\resizebox{\hsize}{!}{\includegraphics{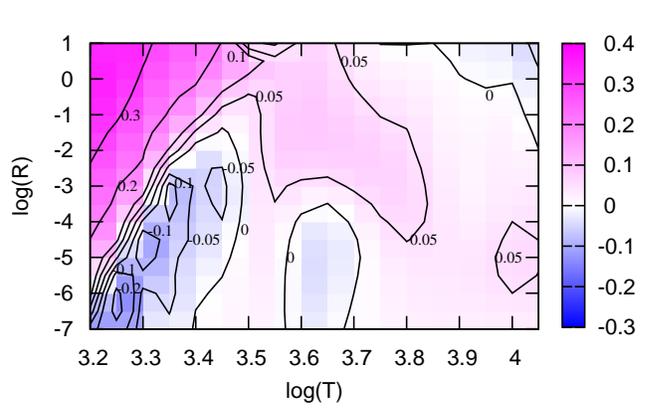}}
\caption{Comparison between our RM opacities and the data from 
Lederer \& Aringer (2009), in terms of $\log(\kappa_{\rm R}^{\rm
author})-\log(\kappa_{\rm R}^{\rm \AE SOPUS})$. A few Contour lines are plotted with
the corresponding values (in dex). 
The chemical mixture is defined by $Z_{\rm ref}=0.02$, $X=0.7$, and
C/O~$\simeq 1.49$. The reference solar composition is L03.}
\label{fig_led09}
\end{figure}
\begin{figure}
\resizebox{\hsize}{!}{\includegraphics{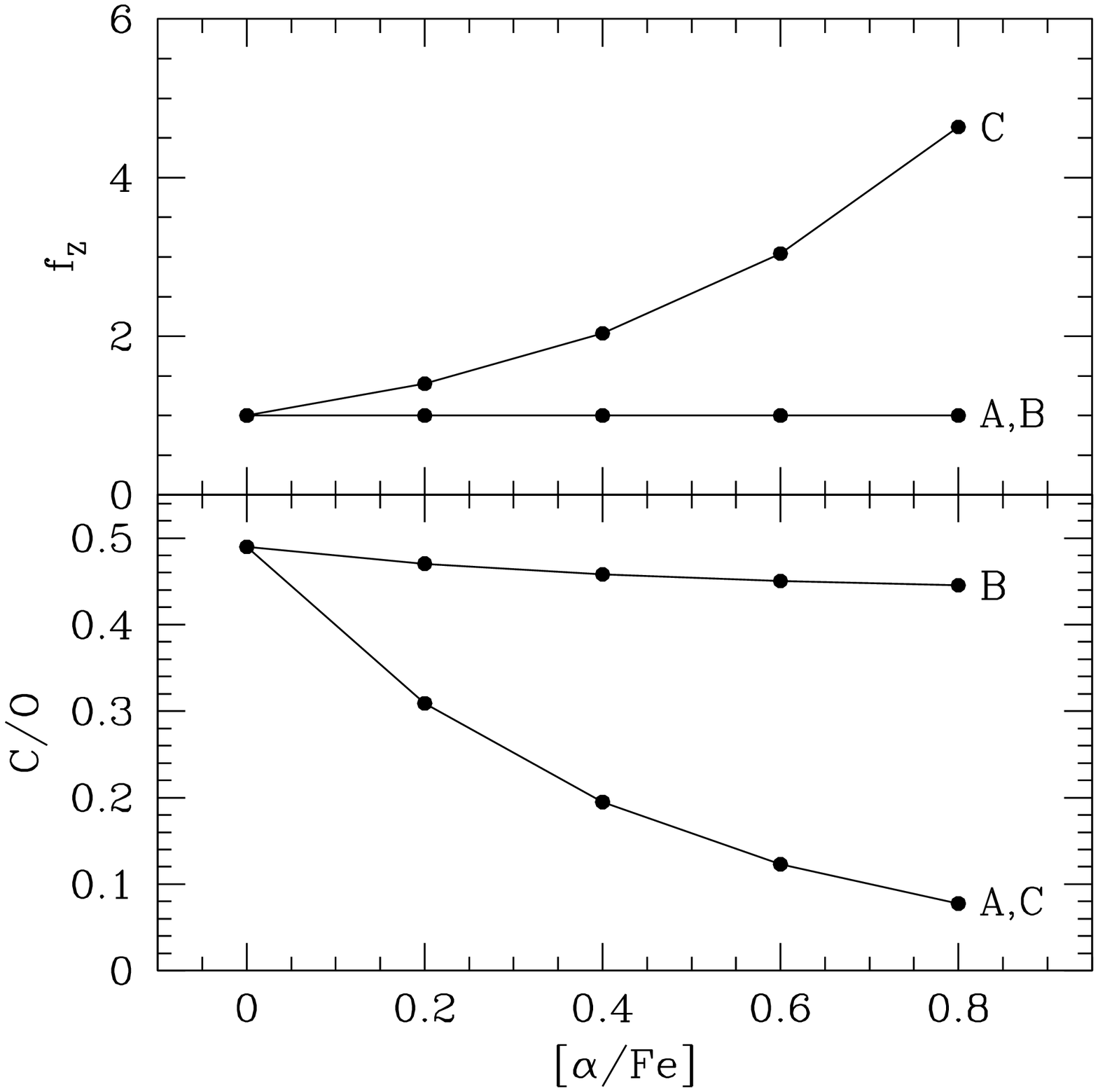}}
\caption{Relation between the total metallicity $Z=f_Z\, Z_{\rm ref}$, the C/O ratio
  and the degree of $\alpha$-enhancement $[\alpha/{\rm Fe}]$,
for the three chemical {\em mixtures} $A$, $B$, and $C$
defined in Sect.~\ref{ssec_alpha}, and adopting $Z_{\rm ref}=0.02$.
The reference solar mixture is GS98.}
\label{fig_alphacoz}
\end{figure}

An enlightening picture of the dependence of the RM opacity on the
C/O ratio is provided by Fig.~\ref{fig_opac_varco_map}, which displays
the map of $\log(\kappa_{\rm R})$ at varying temperature and C/O, for
fixed $\log(R)=-3$.
In this diagram the drop in opacity  
marking the transition region between the O-rich and C-dominated opacity 
is neatly visible as a narrow vertical strip of width 
$0.95 \la {\rm C/O} \la 1.00$ (assuming GS98 as reference solar mixture)
for temperatures $3.2\le \log(T)\la 3.35$. 
This C/O range exactly coincides with the transition interval,
(C/O)$_{\rm crit,1} \la$~C/O~$\la$ (C/O)$_{\rm crit,2}$, between the
O- and C-dominated chemistry.
As already mentioned, the lower limit C/O$_{\rm crit,1}$  
is particularly sensitive to the abundance of silicon relative to 
oxygen.  
In respect to this, Fig.~\ref{fig_opac_varco_map_ag89_g07} shows an 
enlargement of the opacity map over a narrow interval around C/O~$=1$, 
for two choices
of the reference solar composition, i.e. AG89 and GAS07.
It is evident that the opacity dip affects a larger C/O range in the
case of GAS07 as it corresponds to a higher ratio, 
(Si/O)$_{\odot}=7.079\, 10^{-2}$, 
compared  to AG89 with (Si/O)$_{\odot}=4.168\, 10^{-2}$.
Once chosen the reference solar mixture,  one should take this 
feature into account when computing RM opacity
tables at varying C/O ratio, in order to have a good sampling of
the critical region, and avoid inaccurate interpolations between 
grid points belonging to different regimes.
  
Going back to Fig.~\ref{fig_opac_varco_map} we also notice that in the $3.4 <
\log(T)\la 3.6$ the RM opacity increases with C/O. This fact is due to the
increasing contribution from the CN molecule, which 
is one of the relevant opacity sources in this temperature interval
(see bottom-middle panel of Fig.~\ref{fig_chem_koff_z02} for C/O~$=0.54$, and
Fig.~\ref{fig_koff_crich}
for C/O~$=0.95$ and  C/O~$=1.3$). 
It is worth remarking that the effect on the H$^{-}$ opacity 
due to the increased carbon abundance is quite modest and only affects
the opacity for $\log(T) > 3.6$, when ionised carbon is expected to provide 
some fraction of the available free electrons (see Fig.~\ref{fig_electron}).
A more exhaustive consideration of this point is given in Sect.~\ref{ssec_alpha},
when discussing the case of $\alpha$-enhanced mixtures.
For larger temperatures the differences in opacity at increasing C/O
progressively reduce and practically vanish
for $\log(T) > 3.7$, when the opacity is controlled by the hydrogen bound-free
and free-free transitions. 

Let us now briefly comment the sensitiveness of the
results to the reference solar mixture. 
To this aim Fig.~\ref{fig_co1p3} 
illustrates the trend of RM opacity as a function of the temperature in
a carbon-rich gas (C/O~$=1.3$) with the same $Z_{\rm ref}=0.02$, but
different choices of the solar composition.
The differences show up for $\log(T)\la 3.65$ and in most cases are
modest, thus confirming the key r\^ole of
the C/O ratio  
in determining the basic features of the 
molecular opacities.
Another point which deserves some attention is the behaviour of the
RM opacity in the $3.55\la \log(T)\la 3.65$ interval, which is affected
mainly by the CN molecular bands and the negative hydrogen ion
H$^{-}$. A detailed discussion of this point has been already
developed in Sect.~\ref{ssec_ksun}.

\subsubsection{Practical hints on interpolation}
At given metallicity $Z$ and partitions of the metal species $X_i/Z$, 
interpolation between pre-computed opacity tables is usually 
performed as a function of the state variables 
(e.g. $T$ and $R$) and the hydrogen abundance $X$.

When dealing with chemical mixtures with changing elemental abundances, 
as in the case of the atmospheres of TP-AGB stars, one has to
introduce additional independent parameters, in principle as many as 
the varying chemical species. 

Let us consider here the most interesting application, that is the case of
TP-AGB stars which experience 
significant changes in the surface abundances of CNO
elements, hence in the C/O ratio.
Suppose, for simplicity, to have a chemical mixture with C/O$>1$.
Correct interpolation requires that
not only the carbon abundance $X_{\rm C}$ is adopted 
as independent parameter, but also the C/O ratio given its crucial
r\^ole in the molecular chemistry and opacity 
(see Figs.~\ref{fig_chemco} and \ref{fig_opac_varco}).
In addition, one should pay attention to the
drastic changes in $\kappa_{\rm R}$ in the proximity of C/O$=1$.   
The narrow opacity dip, delimited by the boundaries
 C/O$_{\rm crit,1}=1-\varepsilon_{\rm Si}/\varepsilon_{\rm O}$ and  
C/O$_{\rm crit,2}=1$ 
(see Figs.~\ref{fig_opac_varco_map}-\ref{fig_opac_varco_map_ag89_g07}), 
should be sampled with at least $1$ or $2$ opacity tables, 
to avoid substantial mistakes in the interpolated values. 

A useful example of an interpolation scheme suitable to treat the 
 complex chemical evolution 
predicted at the surface of TP-AGB stars undergoing  both the third dredge-up and hot-bottom
burning can be found in Ventura \& Marigo (2009), where the grid
of pre-computed opacity tables covers wide ranges of  C-N-O
abundances (and C/O ratio).
Following the formalism introduced in Sect.~\ref{sec_tables}, the adopted
independent parameters (besides $T$, $R$ and $X$) 
are the variation factors $f_{\rm C}$, $f_{\rm C/O}$, and  $f_{\rm N}$
(defined by Eq.~\ref{eq_fi}), which are assigned values both $>1$
(i.e. enhancement) and  $<1$ (i.e. depletion) to account for the
composite effect on the surface composition produced by the third
dredge-up and hot-bottom burning.
In fact, the C/O ratio may initially increase due to the
the third dredge-up and then decrease when hot-bottom burning 
consumes carbon in favour of nitrogen.

Finally it should be remarked that, when dealing with 
C-rich mixtures, adopting 
both $f_{\rm C}$ and $f_{\rm C/O}$
(rather than either $f_{\rm C}$ or $f_{\rm C/O}$)
as independent parameters allows more robust results, since 
the interpolation is piloted by both the actual carbon abundance 
(mainly affecting the strength of the opacity curves)
and the actual C/O ratio (mainly influencing the morphology of the
opacity curves; see Fig.~\ref{fig_opac_varco}).

\subsubsection{Comparison with other authors}
Finally, we close our discussion on the RM opacities for  C-rich mixtures  
by comparing our results with the data calculated by Lederer \& Aringer
(2009). Figure~\ref{fig_led09} shows an example for a gas mixture
characterised
by $Z_{\rm ref}=0.02$, $X=0.7$, and C/O~$=1.49$.
In general, the agreement between the two calculations is reasonably
good, but worse than that for scaled-solar
mixtures (see Fig.~\ref{fig_kcomp_z02}, panel b).
The largest differences show up at the lower temperatures, where
the RM opacity is dominated by the CN, C$_2$H$_2$, C$_2$, HCN, C$_3$ molecular bands.
This migth appear a bit odd since  both sets of calculations
adopt basically the same molecular data (see Table~\ref{tab_opacmol}).

In the range $3.2 \la \log(T) \la 3.4$, compared to Lederer \& Aringer
(2009), \AE SOPUS predicts larger RM opacities (up to $0.1/0.2$ dex)
across a strip  with $-7 \la \log(R) \la 3$, and lower values  (up to $\simeq
0.3$ dex) for $\log(R) > -3$.

One likely motivation of the former difference is that the scaling introduced 
by LA09 to the original 
$gf$ values in the C$_2$  line list (Querci et al. 1974) is not included 
in our calculations. As discussed by LA09 (see their figure 10)
not applying this correction to the line strengths of C$_2$ causes an
increase of  $\log(\kappa_{\rm R})$ up to $\simeq 0.1$ dex, which is
just what we get in terms of $\log(\kappa_{\rm R}^{\rm LA09})-\log(\kappa_{\rm R}^{\rm \AE SOPUS})$ 
in that particular region of the diagram.
On the other hand, more recently Aringer et al. (2009) have shown
that omitting this scaling modification to the original C$_2$ line list
improves the comparison between synthetic and observed colours of
carbon stars (see their figure 15).

The latter discrepancy between LA09 and \AE SOPUS 
at larger densities has not a clear reason at present. 
We note that in this region of the $\log(T)-\log(R)$ diagram, 
the dominant contribution
to the RM opacity is provided by C$_2$H$_2$ (see bottom panels of Fig.~\ref{fig_sigc}).
We are currently investigating possible differences among the partition
function and/or dissociation energy of this molecule, adopted in the EOS
calculations by LA09 and \AE SOPUS.


\subsection{$\alpha$-enhanced mixtures}
\label{ssec_alpha}
We will analyse a few important aspects related to
RM opacities of $\alpha$-enhanced mixtures, i.e. characterised by having
$[\alpha/{\rm Fe}]>0$, according to the notation (in dex):
\begin{equation}
\label{eq_afem}
[\alpha/{\rm Fe}]=\log \left(\frac{X_{\alpha}}{X_{\rm Fe}}\right) -
\log \left(\frac{X_{\alpha,\odot}}{X_{\rm Fe,\odot}}\right)
\end{equation}
where $X_{\alpha,\odot}$ and $X_{\rm Fe,\odot}$ are the total mass fractions
of the $\alpha$-elements and Fe-group elements, respectively.
In the following we allocate O, Ne, Mg, Si, S, Ca, and Ti in 
the $\alpha$-group,  while V, Cr, Mn, Fe, Co, Ni, Cu, and Zn
are assigned to the Fe-group.
It should be noticed that, since Fe is by far the most abundant element of its
group, the ratio $[\alpha/{\rm Fe}]$ calculated with Eq.~(\ref{eq_afem}) coincides with 
the ratio computed using the abundances in number fraction:
\begin{equation}
\label{eq_afen}
[\alpha/{\rm Fe}]=\log \left(\frac{\varepsilon_{\alpha}}{\varepsilon_{\rm Fe}}\right) -
\log \left(\frac{\varepsilon_{\alpha,\odot}}{\varepsilon_{\rm Fe,\odot}}\right)
\end{equation}
For simplicity in our discussion we take as {\em selected elements} 
all  $\alpha$-elements which are
given the same $[\alpha/{\rm Fe}]> 0$. However, it should be remarked that 
any other prescription, concerning both the selected elements and  
the corresponding $[X_i/{\rm Fe}]$ (i.e. positive or negative),
 can be set by the user via the  \AE SOPUS interactive web page.

First of all,  we call attention to the fact that a given value of the ratio
$[\alpha/{\rm Fe}]$ is not sufficient to specify 
the chemical mixture unambiguously. The same degree of
$\alpha$-enhancement may correspond to quite different situations, as exemplified in the following.

Adopting the formalism introduced
in Sect.~\ref{sec_tables} and introducing the quantity $f_Z=Z/Z_{\rm ref}$ 
($g_Z=\varepsilon_Z/\varepsilon_{Z_{\rm ref}}$), 
we define three different 
$\alpha$-enhanced compositions that, in our opinion, may describe possibly 
frequent applications. 
They are characterised as follows (considering the metal abundances
expressed in mass fractions):
\begin{itemize}
\item  {\em Mixture $A$}: $Z=Z_{\rm ref}$ hence $f_Z=1$; $f_i > 0$ for $\alpha$-elements
({\em enhanced group}); $f_i < 0$ for any other element ({\em depressed group}).
In this case the {\em fixed group} (with $f_i = 0$) is empty.
\item {\em Mixture} $B$: $Z=Z_{\rm ref}$ hence $f_Z=1$; $f_i > 0$ for $\alpha$-elements
({\em enhanced group}); $f_i < 0$ for the Fe-group elements 
({\em depressed group}); $f_i = 0$ for
any other element ({\em fixed group}).
\item  {\em Mixture} $C$: $Z>Z_{\rm ref}$ hence $f_Z>1$;  $f_i > 0$ for $\alpha$-elements
({\em enhanced group});  $f_i < 0$ for any other element ({\em depressed group}).
In this case the {\em fixed group} (with $f_i = 0$) is empty, as for {\em mixture} $A$.
\end{itemize}

For each of the three mixtures considered here, 
Table~\ref{tab_alpha} lists the 
variation factors, $f_i$ and $g_i$, of the most relevant elements, i.e. C, N, O,
Fe-group elements, and the metallicity parameter, $f_Z$  and $g_Z$, as a function
of a few selected $[\alpha/{\rm Fe}]$ values.
The general analytical derivation of the abundance variation factors as a function of the selected 
$[X_i/{\rm Fe}]$ for the three kinds of mixtures is detailed in Appendix~\ref{apx_afe}.
Figure~\ref{fig_alphacoz} displays the expected trends of C/O and $f_Z$ 
at increasing $[\alpha/{\rm Fe}]$ assuming
$Z_{\rm ref}=0.02$ and the GS98 solar composition.

For  a given $[\alpha/{\rm Fe}]$ value, 
the three mixtures  have distinctive abundance features when
compared to the reference composition, i.e. with 
$Z=Z_{\rm ref}$ and scaled-solar partitions of metals. 
In particular, for their relevance to the resulting RM opacity, it
is worth considering the changes in the CNO abundances, and mostly in the
C/O ratio.
\begin{table*}
\label{tab_alpha}
\begin{center}
\begin{tabular}{ccccccccccccc}
\hline
\hline
$[\alpha/{\rm Fe}]$ & {\em Mixture} & $f_{\rm C}$ & $g_{\rm C}$ & $f_{\rm N}$ & $g_{\rm N}$ 
& $f_{\rm O}$ & $g_{\rm O}$ & $f_{\rm Fe-group}$ & $g_{\rm Fe-group}$ & $f_Z$  & $g_Z$ \\
\hline
0.2 & $A$ & 0.714 & 0.721 & 0.714 & 0.721 & 1.131 & 1.143 & 0.714 & 0.721 & 1.000 & 0.990 \\ 
    & $B$ & 1.000 & 0.982 & 1.000 & 0.982 & 1.041 & 1.022 & 0.657 & 0.645 & 1.000 & 1.018 \\
    & $C$ & 0.714 & 0.721 & 0.714 & 0.721 & 1.131 & 1.143 & 0.714 & 0.721 & 1.392 & 1.387 \\ 
0.4 & $A$ & 0.491 & 0.500 & 0.491 & 0.500 & 1.233 & 1.256 & 0.491 & 0.500 & 1.000 & 0.982 \\
    & $B$ & 1.000 & 0.970 & 1.000 & 0.970 & 1.060 & 1.037 & 0.426 & 0.413 & 1.000 & 1.031 \\
    & $C$ & 0.491 & 0.500 & 0.491 & 0.500 & 1.233 & 1.256 & 0.491 & 0.500 & 2.004 & 2.000 \\
0.6 & $A$ & 0.329 & 0.336 & 0.329 & 0.336 & 1.308 & 1.339 & 0.329 & 0.336 & 1.000 & 0.977 \\
    & $B$ & 1.000 & 0.962 & 1.000 & 0.962 & 1.087 & 1.046 & 0.273 & 0.263 & 1.000 & 1.039 \\
    & $C$ & 0.329 & 0.336 & 0.329 & 0.336 & 1.308 & 1.339 & 0.329 & 0.336 & 2.950 & 2.972 \\
0.8 & $A$ & 0.215 & 0.222 & 0.215 & 0.222 & 1.360 & 1.398 & 0.215 & 0.222 & 1.000 & 0.973 \\
    & $B$ & 1.000 & 0.957 & 1.000 & 0.967 & 1.099 & 1.052 & 0.174 & 0.167 & 1.000 & 1.045 \\
    & $C$ & 0.215 & 0.222 & 0.215 & 0.222 & 1.360 & 1.398 & 0.215 & 0.222 & 4.392 & 4.513 \\
\hline
\end{tabular} 
\caption{Main characteristics of the $\alpha$-enhanced mixtures described
  in text. The variation factors of C, N, O, and Fe-group elements, defined by Eqs.~(\protect{\ref{eq_figi}})
 for abundances either in mass fraction or in mass fraction, 
are indicated together with the quantities  $f_Z=Z/Z_{\rm ref}$ and $g_Z=\varepsilon_{Z}/\varepsilon_{Z_{\rm ref}}$. 
The reference solar composition is GS98.}
\label{tab_alpha}
\end{center}
\end{table*}

\begin{figure*}
\begin{minipage}{0.33\textwidth}
\resizebox{\hsize}{!}{\includegraphics{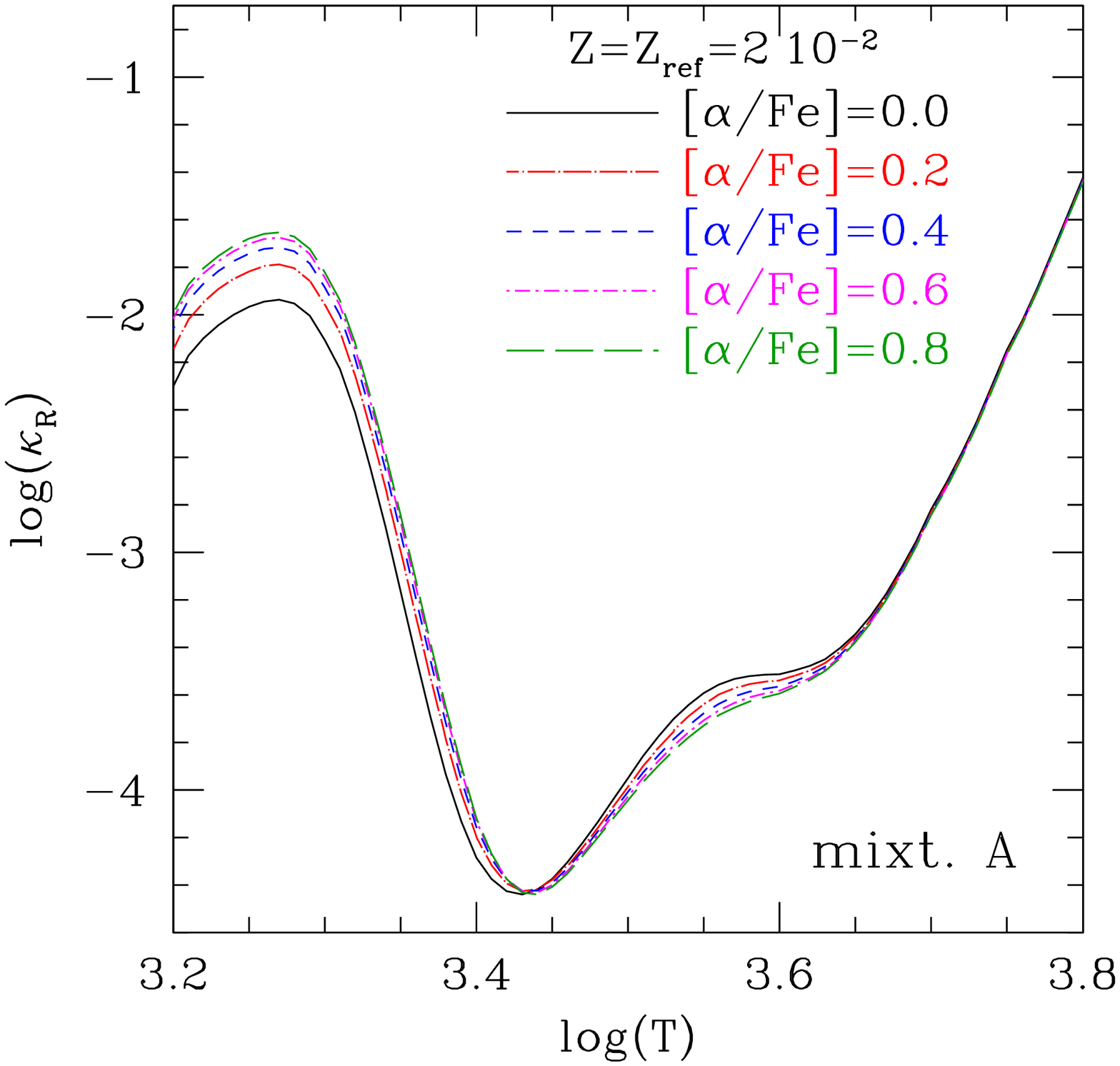}}
\end{minipage}
\begin{minipage}{0.33\textwidth}
\resizebox{\hsize}{!}{\includegraphics{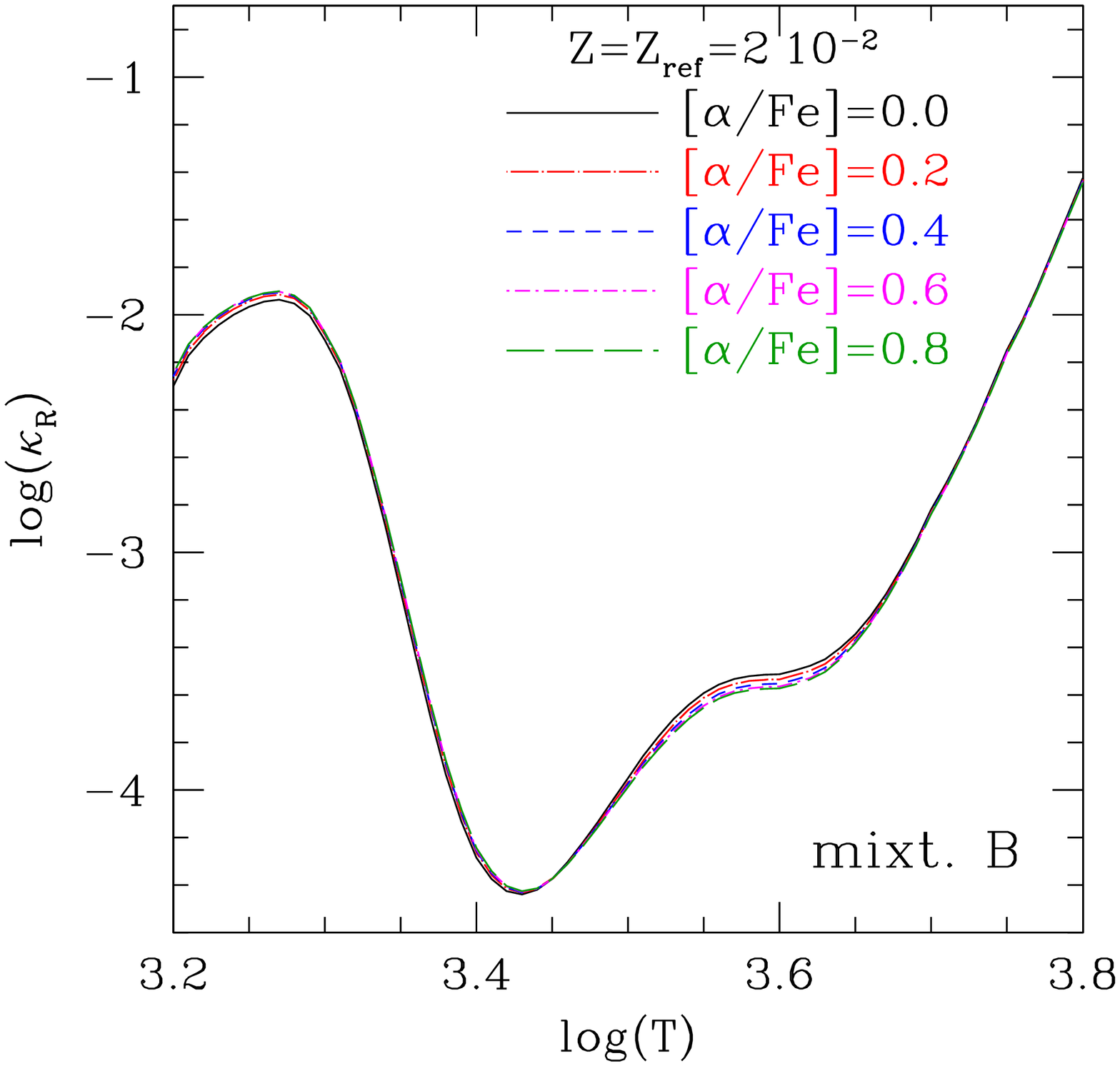}}
\end{minipage}
\begin{minipage}{0.33\textwidth}
\resizebox{\hsize}{!}{\includegraphics{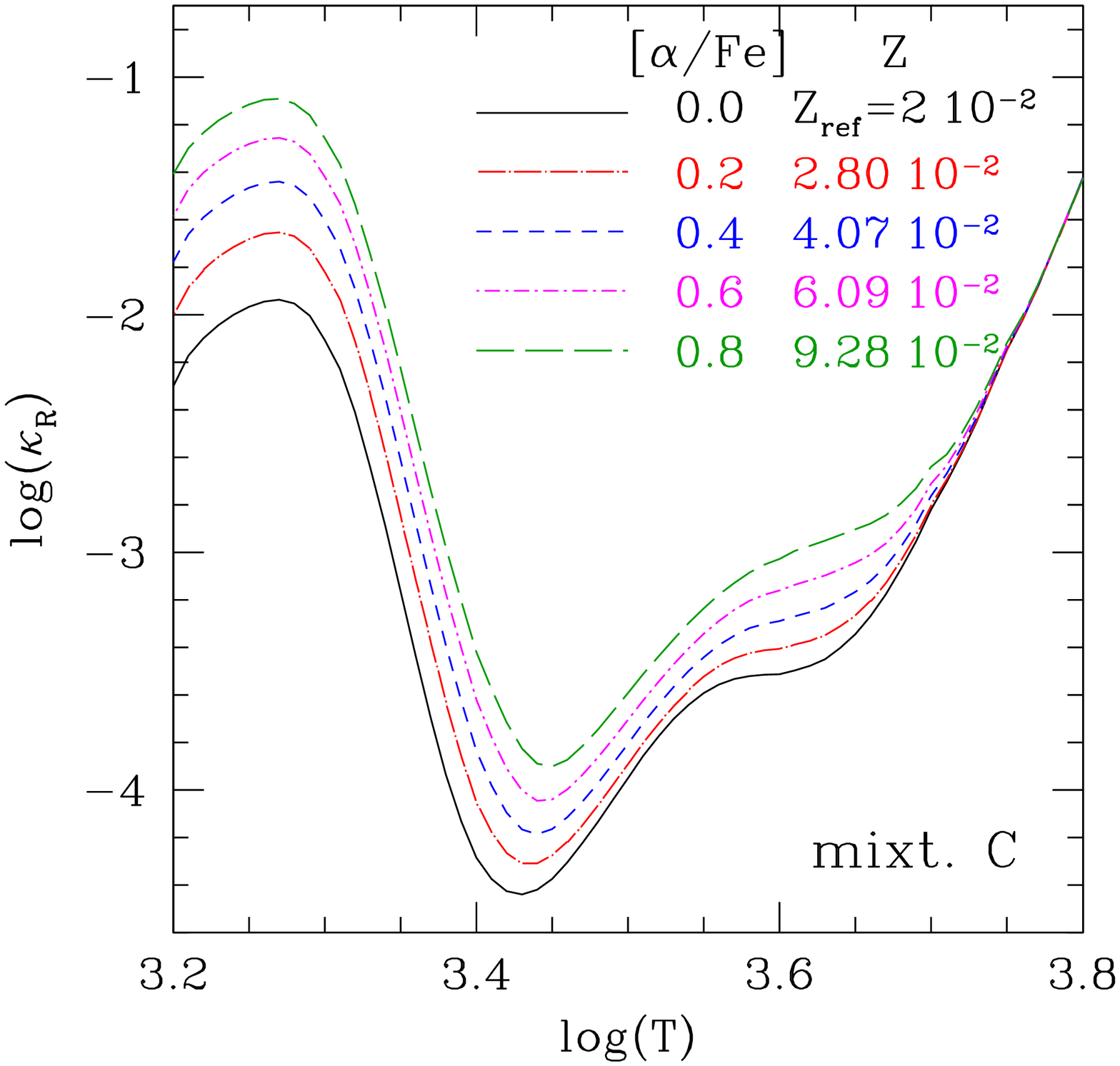}}
\end{minipage}
\caption{Rosseland mean opacity as a function of temperature and
  assuming  $\log R=-3$, for a gas with ($Z_{\rm ref}=0.02$, $X=0.7$)
 and various choices of the
  ratio  $[\alpha/{\rm Fe}]$, as indicated. The reference solar
  mixture is GS98. Results are shown for three choices of the
chemical composition, corresponding to {\em mixtures} $A$, $B$, $C$ 
described in the  text.
}
\label{fig_alphacomp}
\end{figure*}

\begin{figure*}
\begin{minipage}{0.33\textwidth}
\resizebox{\hsize}{!}{\includegraphics{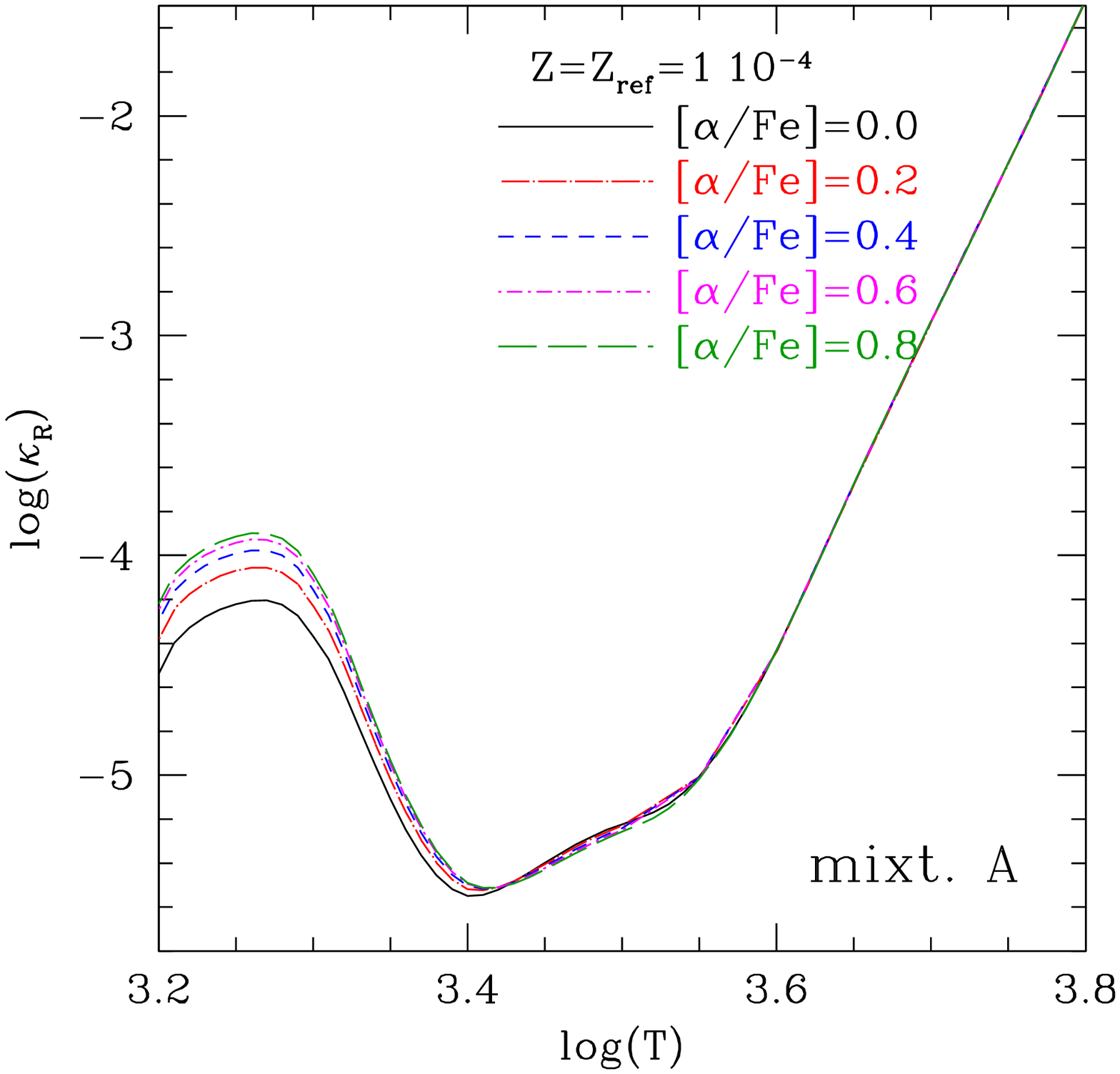}}
\end{minipage}
\begin{minipage}{0.33\textwidth}
\resizebox{\hsize}{!}{\includegraphics{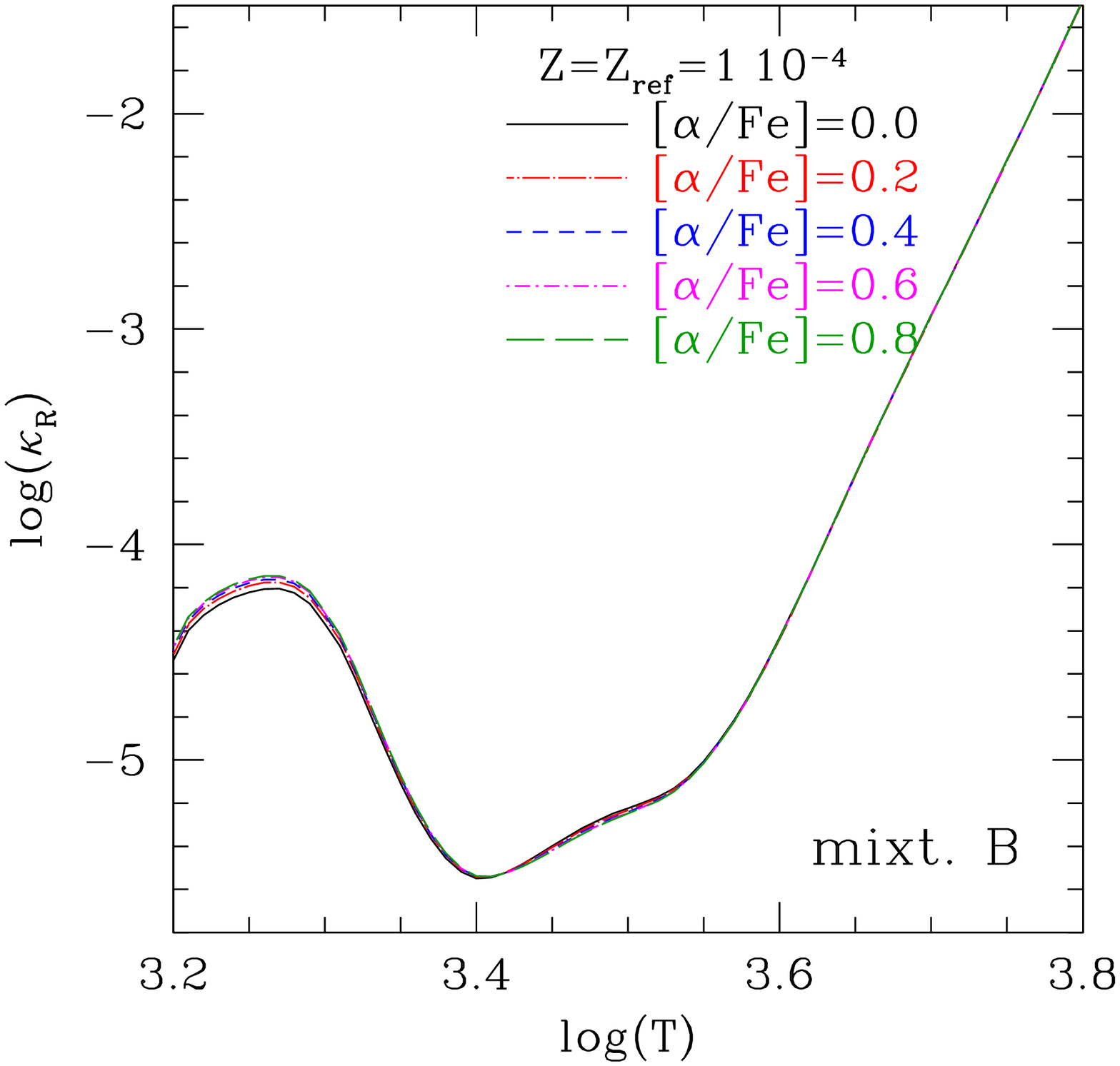}}
\end{minipage}
\begin{minipage}{0.33\textwidth}
\resizebox{\hsize}{!}{\includegraphics{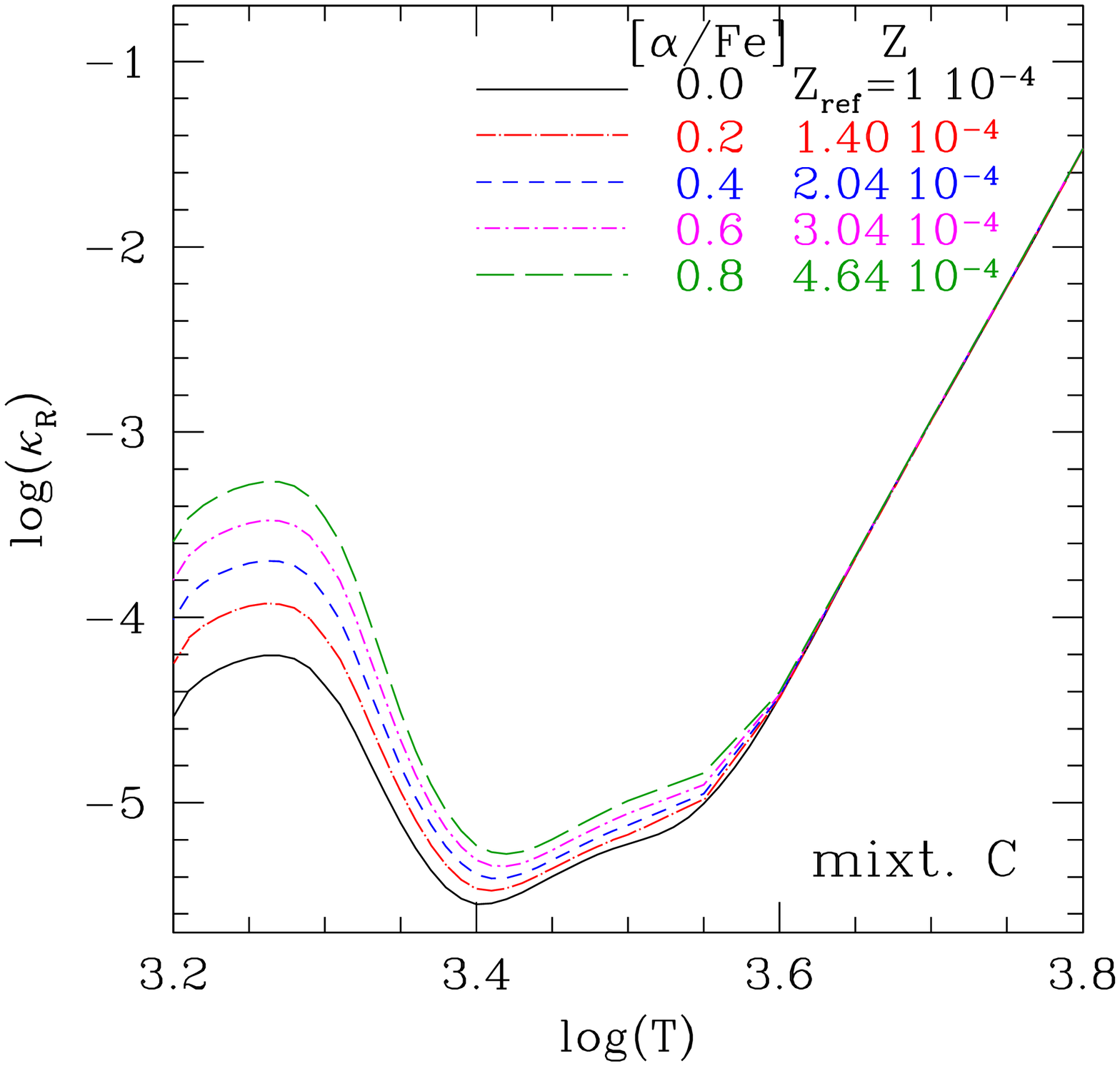}}
\end{minipage}
\caption{The same as in Fig.~\ref{fig_alphacomp}, but for $Z_{\rm ref}=0.0001$.}
\label{fig_alphacomp_z0001}
\end{figure*}
\begin{itemize}
\item
{\em Mixture} $A$ is depleted both in the iron-group elements as well as in
carbon and nitrogen. For instance, at  $[\alpha/{\rm Fe}]=0.4$, 
the abundances of the Fe-group elements are almost halved and the same
applies to C and N, while O is augmented by $\sim 23\%$. 
As a consequence the C/O ratio decreases significantly, passing from
(C/O)$_{\odot} \sim 0.49$ down to (C/O)$_{\odot} \sim 0.19$.
In general, the ratio C/O  lowers considerably at increasing $[\alpha/{\rm Fe}]$.
\item
{\em Mixture} $B$ is depleted in the iron-group elements, while C and N are
left unchanged.  At  $[\alpha/{\rm Fe}]=0.4$,  the abundances of the
Fe-group elements are depressed by $\sim 43\%$, while O is increased
by only $\sim 6\%$. In this case the  C/O ratio is just little
affected, changing from (C/O)$_{\odot} \sim 0.49$ to (C/O)$_{\odot} \sim 0.46$.
In general, the ratio C/O  slightly decreases at increasing $[\alpha/{\rm Fe}]$.
\item {\em Mixture} $C$ has the same characteristics of {\em mixture} $A$ in terms
of metal partitions, i.e. $(X_i/Z_{\rm ref})_A = (X_i/Z)_C$, 
but with a different metallicity.
It follows that the $C$ case shares with $A$ the same elemental ratios,
so that the C/O declines significantly
at increasing $[\alpha/{\rm Fe}]$, while the total metallicity
increases. For instance, at $[\alpha/{\rm Fe}]=0.4$  {\em mixture} $C$ 
corresponds to a metallicity $Z\simeq 2\, Z_{\rm ref}$ (see
Table~\ref{tab_alpha} and Fig.~\ref{fig_alphacoz}).
\end{itemize}

The aforementioned differences in the chemistry among the {\em mixture}s $A$, $B$,
and $C$ affect the resulting RM opacities, as displayed by
Fig.~\ref{fig_alphacomp} for $Z_{\rm ref}=0.02$ and Fig.~\ref{fig_alphacomp_z0001} 
for $Z=0.0001$, both assuming $X=0.7$ and $\log(R)=-3$.
Let us first discuss the results of the $A$ and $B$ cases with $Z_{\rm
ref}=0.02$.
 With respect to {\em mixture} $A$ (left panel of Fig.~\ref{fig_alphacomp}), we see that
at increasing $\alpha$-enhancement, the
 opacity variations show up with 
opposite trends in two temperature intervals, namely: 
at intermediate temperatures, $3.50\la \log(T) \la 3.65$, and 
at lower temperatures, $3.2\la \log(T) \la 3.4$.
Specifically, the opacity knee at $\log(T)\simeq 3.55$ slightly smooths,
 while the opacity bump at $\log(T)\simeq 3.3$
becomes more prominent with increasing $[\alpha/{\rm Fe}]$.

\begin{figure}
\resizebox{\hsize}{!}{\includegraphics{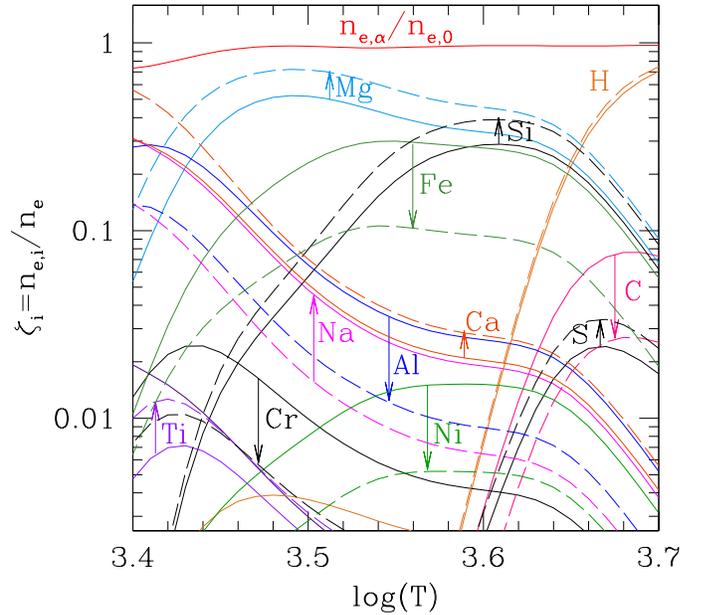}}
\caption{Contributions of free electrons $\zeta_i=n_{e,i}/n_e$, normalised to
the total electron density $n_e$, provided by different ions
as a function of temperature, in two gas mixtures with ($Z=Z_{\rm ref}=0.02$,
$X=0.7$, $\log(R)=-3$) and different
  partitions of the $\alpha$-elements, namely: $[\alpha/{\rm
  Fe}]=0.0$ (scaled-solar abundances; solid lines),  
and $[\alpha/{\rm Fe}]=+0.6$ ($\alpha$-enhanced {\em mixture} of type $A$; 
dashed lines). The arrows indicate the
increasing/decreasing trends when passing from $[\alpha/{\rm
  Fe}]=0.0$ to $[\alpha/{\rm Fe}]=+0.6$.
The highest curve (in red) displays the ratio $n_{e,\alpha}/n_{e,0}$, i.e.
the electron density of the $\alpha$-enhanced
composition relative to the scaled-solar case.
The reference solar mixture is GS98.}
\label{fig_electron}
\end{figure}

As already discussed in Sect.~\ref{ssec_ksun}, in the $3.5\la \log(T) \la 3.6$ interval 
the most effective opacity source is the 
negative hydrogen ion 
(see lower middle panel of Fig.~\ref{fig_chem_koff_z02}), which
positively correlates with the electron density, $n_e$.
 Figure~\ref{fig_electron} shows that
in this temperature range the principal electron donors are elements with
relatively low-ionisation potentials, mainly Mg, Si, Fe, Al, Ca, and
Na, which involve both the {\em enhanced group} and the {\em depressed group}.
For this reason, it turns out that in the $\alpha$-enhanced {\em mixture} of
type $A$ the decreased number of electrons contributed 
by Fe (together with C, Na, Al, Cr, Ni) 
is practically counter-balanced by the increased number of electrons
removed from the $\alpha$-atoms such as Mg, Si, and S. The net effect is just
a very little reduction in the electron density.
In the case exemplified in Fig.~\ref{fig_electron} even 
 a large $\alpha$-enhancement  $[\alpha/{\rm Fe}]=0.6$
corresponds to a reduction of $n_e$ by just $\sim 6\%$  at $\log(T)=3.55$.
In turn, this small variation in $n_e$ produces a minor reduction
of the H$^{-}$ opacity.
From a  careful inspections of the results we find that the
depression of the opacity knee at $\log(T)\simeq 3.55$ should be rather
ascribed to the weakening of the CN molecular absorption bands, which
reflects the depression of both carbon and nitrogen abundances in
{\em mixture} $A$.
In fact, at these temperatures and $\log(R=-3)$ the CN contribution to the RM opacity
is not negligible (see lower middle panel of Fig.~\ref{fig_chem_koff_z02}).

 \begin{figure}
\begin{minipage}{0.49\textwidth}
\resizebox{\hsize}{!}{\includegraphics{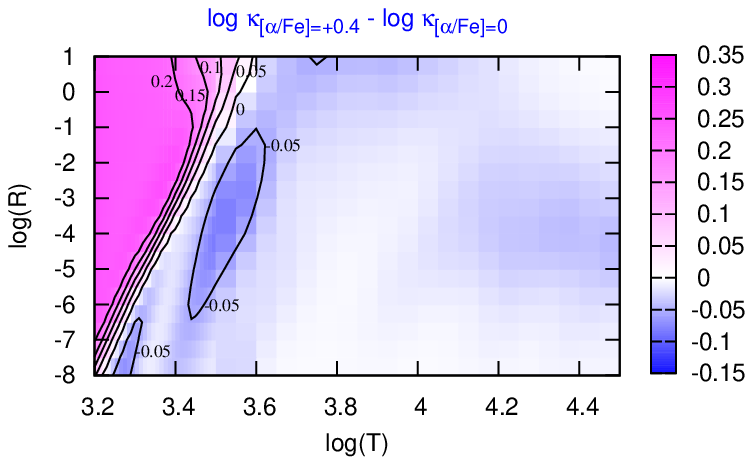}}
\end{minipage}
\vfill
\begin{minipage}{0.49\textwidth}
\resizebox{\hsize}{!}{\includegraphics{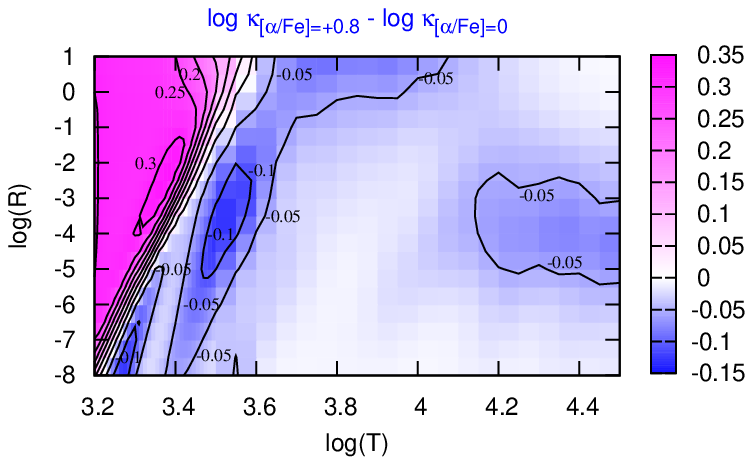}}
\end{minipage}
\caption{Differences in RM opacities between 
$\alpha$-enhanced mixtures,  with $[\alpha/{\rm Fe}]=+0.4$ (top panel) 
and $[\alpha/{\rm Fe}]=+0.8$ (bottom panel), 
and a scaled-solar composition according to GS98.  
The $\alpha$-enhanced mixtures are
constructed according to the $A$ scheme. In all case we assume 
($Z_{\rm ref}=0.02$, $X=0.7$). Contour lines, with an incremental step of $0.05$ dex, 
are over-plotted to guide the comparison.
}
\label{fig_alphadif}
\end{figure}

With respect to {\em mixture} $B$, we note that 
in the same temperature range, i.e. $3.50\la \log(T) \la 3.65$, 
the variations of the RM opacity at increasing $[\alpha/{\rm Fe}]$ are smaller than for {\em mixture} $A$, 
almost negligible. In fact, in {\em mixture} $B$ the carbon and nitrogen
abundances are left unchanged so that the opacity contribution from CN
is not expected to vary as well. Furthermore, the same arguments 
on the electron density, discussed for
{\em mixture} $A$, hold also in this case, and the H$^{-}$ opacity
contribution is predicted to change just slightly.

Let us now consider the  temperature interval $3.2\la \log(T) \la
3.4$, which is characterised by the opacity
bump due to the molecular absorption bands of H$_2$O, TiO, and ZrO.
We see from Figure~\ref{fig_alphacoz} (bottom panel) that 
the opacity peak grows at increasing $[\alpha/{\rm Fe}]$, 
reflecting the decrease of the C/O ratio. 
In fact, the concomitant enhancement of oxygen and the depression of
carbon favour the chemistry of the
O-bearing molecules, thus strengthening the opacity  contributions 
of H$_2$O, TiO, and ZrO at those temperature. The reader should refer
to  Sect.~\ref{ssec_kcno} for a broad
analysis of the dependence of the low-temperature opacity on the C/O
ratio. For the same reasons, in the case of {\em mixture} $B$ the opacity bump 
is practically insensitive to changes in $[\alpha/{\rm Fe}]$, 
since the decrease of C/O ratio is just marginal, as shown in
Fig.~\ref{fig_alphacoz}.

Figure~\ref{fig_alphadif} shows the differences in terms 
of $\Delta \log(\kappa_{\rm R})$ expected
when the chemical composition of the gas is enhanced in
$\alpha$-elements, according to {\em mixture} $A$. 
The same comments already spent for 
Fig.~\ref{fig_alphacomp} (left panel) 
hold here. At increasing $[\alpha/{\rm Fe}]$ 
negative deviations mostly take place in the region dominated 
by the absorption of H$^{-}$, while positive variations 
show up at lower temperatures, over a  
well-defined region in the $\log(T)-\log(R)$ diagram, the boundaries
of which are
determined by the thermodynamic conditions required to form H$_2$O
efficiently (see top panels of Fig.~\ref{fig_chem_koff_z02}), thus
becoming narrower at decreasing $R$.

The case of {\em mixture} $C$ deserves different remarks.
At increasing $[\alpha/{\rm Fe}]$ the RM opacity is predicted to 
be larger all over the temperature range  $3.2\la \log(T) \la 3.75$,
and the variations are always larger than for the other two mixtures.
This fact can be explained simply as a metallicity effect, since 
the global metal content increases with the $[\alpha/{\rm Fe}]$
as indicated by the $f_Z$ parameter (see Fig.~\ref{fig_alphacoz}). 
Therefore, {\em mixture} $C$ shares with {\em mixture} $A$ the same 
partition of metals (i.e. the same variation
factors $f_i$; see Table~\ref{tab_alpha}), but their abundances are
all higher, including those belonging to the {\em depressed group}.
The net effect is systematic increase of the RM opacity with $[\alpha/{\rm Fe}]$.

Finally, a  cautionary comment is worth being made.
It should be noticed that the while the $\alpha$-elements are the same
for the three mixtures here considered, the differences deal with i) which elements are 
assigned to the  {\em depressed group} and  to the {\em fixed group},
and ii) the total metallicity. The results discussed above show
clearly that this an important point which impacts on
the resulting RM opacities. Therefore, when using RM opacity tables 
one should be always aware of how the underlying $\alpha$-enhanced 
mixture has been 
constructed, since his/her results may be importantly affected. 
This aspect has been recently discussed 
by Dotter et al. (2007).
To our knowledge  available RM opacity tables adopt
$\alpha$-enhanced mixtures similar to our $A$ scheme 
(e.g. Ferguson  et al. 2005 and related website of the Wichita State University group).

\subsection{Other peculiar mixtures: C-N-O-Na-Mg-Al abundance anti-correlations}
\label{ssec_anomal}
\begin{figure}
\resizebox{\hsize}{!}{\includegraphics{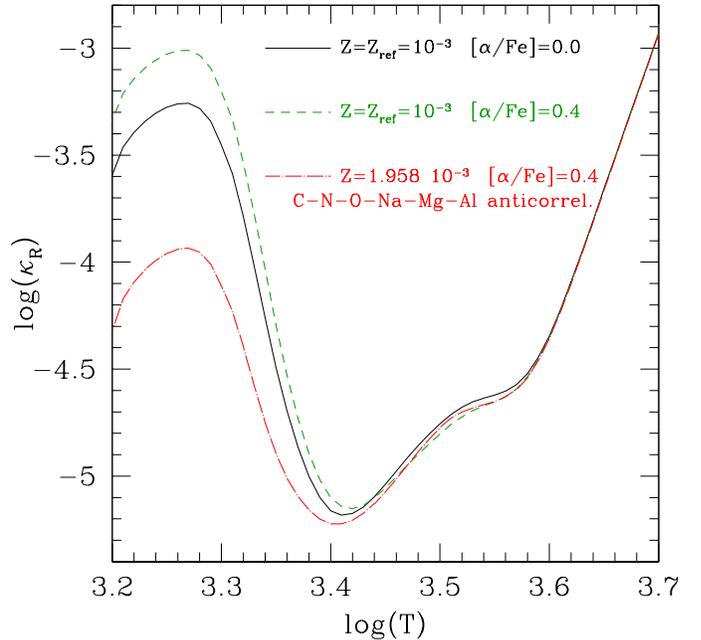}}
\caption{Rosseland mean opacity as a function of temperature and
  assuming  $\log R=-3$, for a gas with ($Z_{\rm ref}=0.001$,
$X=0.7$). The reference solar mixture is GS98.
Results are shown for three chemical mixtures, namely: i) scaled-solar
abundances of metals; enhanced abundances of $\alpha$-elements with $[\alpha/{\rm Fe}]=+0.4$
(according to {\em mixture} $A$; see Sect.~\ref{ssec_alpha});
peculiar chemical pattern characterised by additional C-N-O-Na-Mg-Al 
abundance variations superimposed to the $\alpha$-enhanced
mixture. See the text for details.
}
\label{fig_opac_anomal}
\end{figure}

Another relevant case is suggested by the peculiar chemical patterns
observed in stars of Galactic globular clusters (GGC), being characterised
by striking abundance anti-correlations between C-N and O-Na, and Mg-Al,
which are in turn superimposed on a typical $\alpha$-enhanced mixture 
(e.g. Gratton et al. 2001).
Stellar evolution models including low-temperature RM opacities
suitable for these particular compositions have been recently 
calculated (Salaris et al. 2006; Pietrinferni et al. 2009).

Figure~\ref{fig_opac_anomal} shows an example of RM opacities 
computed with \AE SOPUS for a gas mixture which would represent the pattern
of extreme C-N-O-Na-Mg-Al anti-correlations, as measured in 
GGC stars (Carretta et al. 2005).
The adopted abundance scheme is the following. We start with
our reference scaled-solar mixture, 
characterised by $Z_{\rm ref}=0.001,\, X=0.7$ and GS98 solar
composition. 
Then we construct a second composition with  $[\alpha/{\rm Fe}]=+0.4$ 
following the prescriptions for {\em mixture} $A$ (see
Sect.~\ref{ssec_alpha}). The C/O decreases from  (C/O)$_{\odot}\simeq 0.49$
to $\simeq 0.19$, while the total metallicity is preserved.
This fact explains the growth of the opacity peak due to H$_2$O at
$\log(T)\la 3.4$ in the $\alpha$-enhanced mixture.
The reader should go back to Sect.~\ref{ssec_alpha} for 
an extensive discussion on the  differences between the two 
RM opacity curves.

\begin{figure*}
\begin{minipage}{0.33\textwidth}
\resizebox{\hsize}{!}{\includegraphics{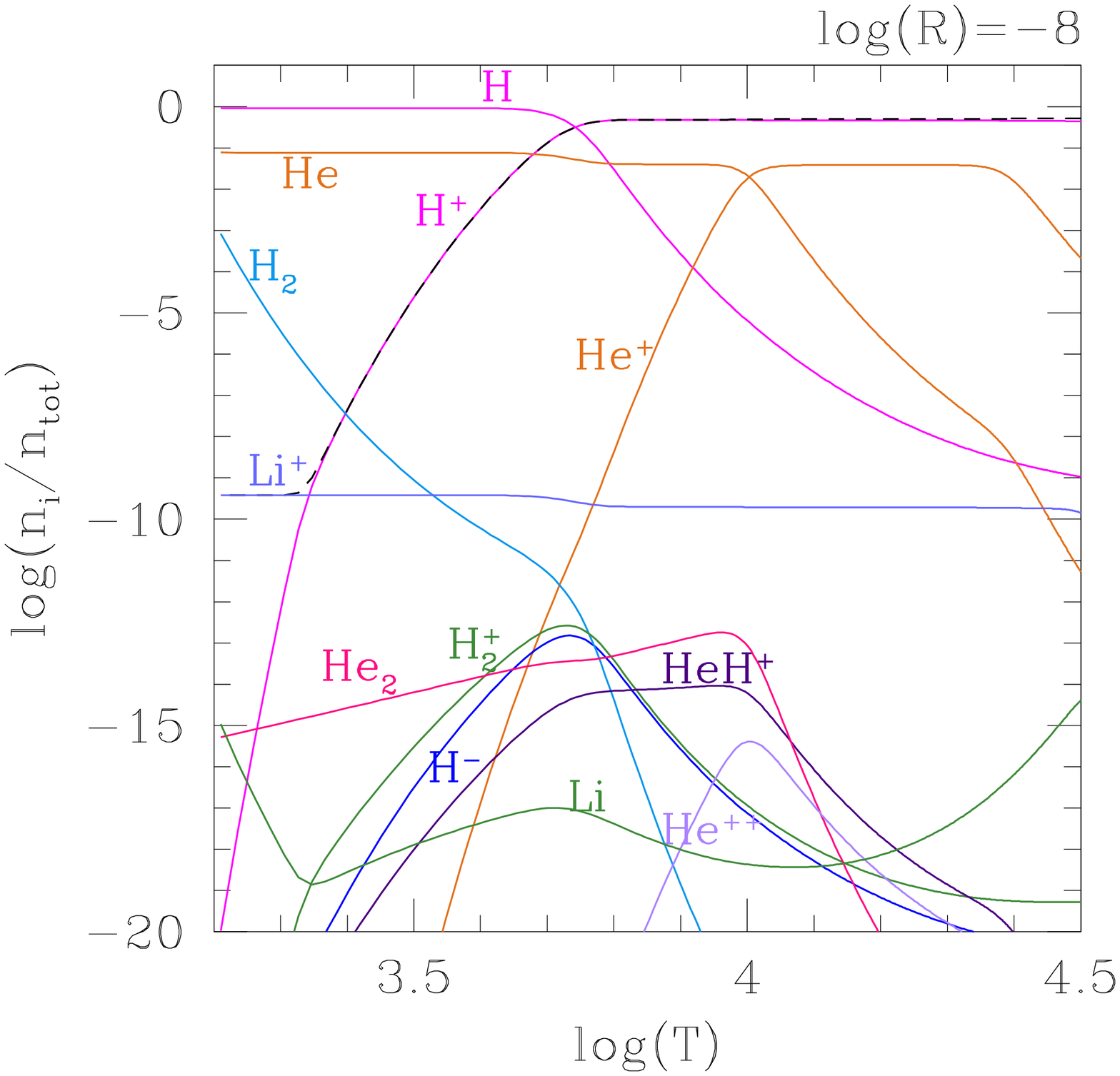}}
\end{minipage}
\begin{minipage}{0.33\textwidth}
\resizebox{\hsize}{!}{\includegraphics{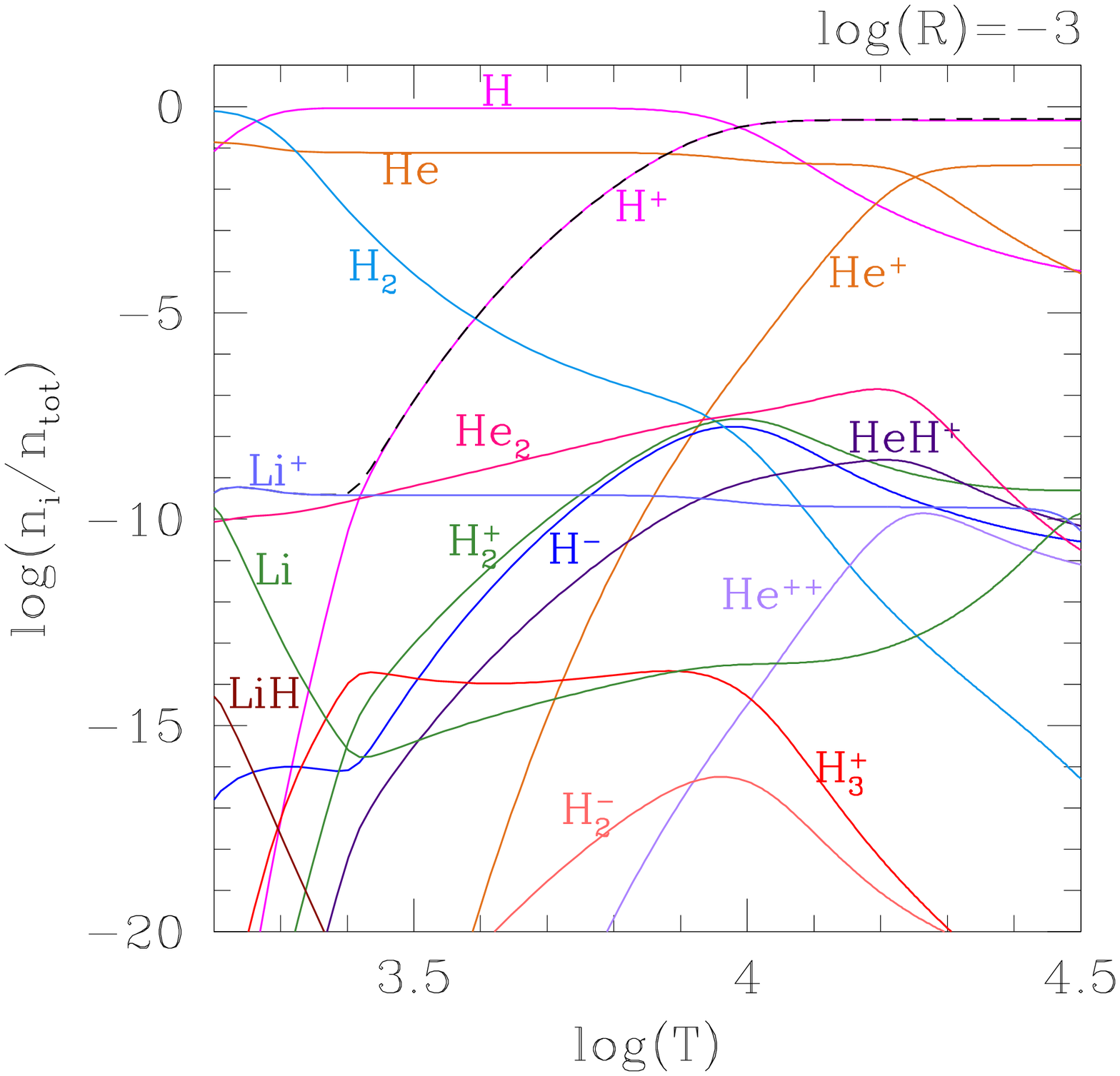}}
\end{minipage}
\begin{minipage}{0.33\textwidth}
\resizebox{\hsize}{!}{\includegraphics{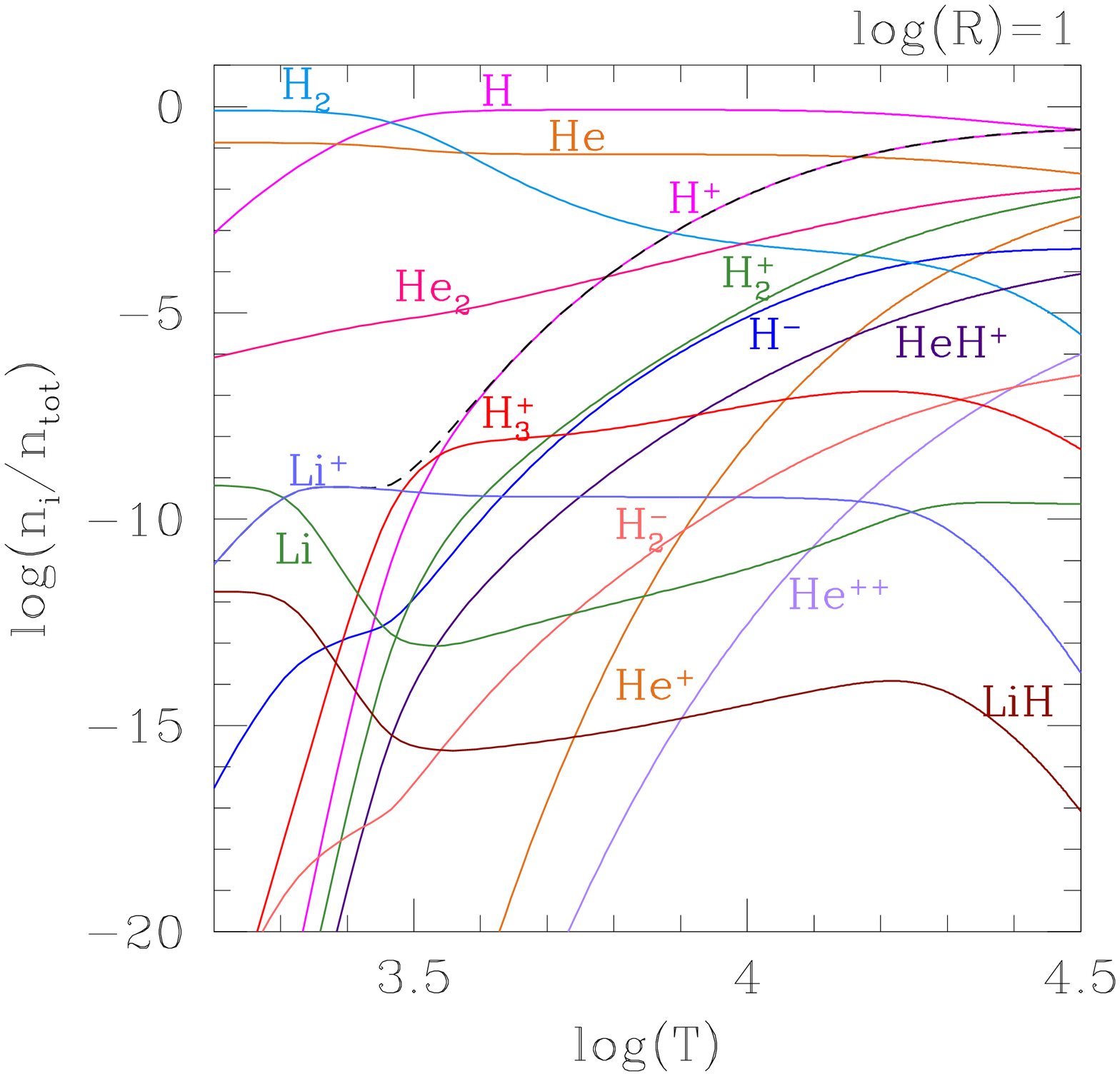}}
\end{minipage}
\caption{Concentrations of a few atomic and molecular species as a function of the temperature 
in a gas with primordial composition, adopting  $Z_{\rm ref}=0$, $X=0.7521$, and a lithium abundance 
of $\varepsilon_{\rm Li}/\varepsilon_{\rm H}=4.15\times 10^{-10}$, 
and assuming $\log(R)=-3$. The fraction of free electrons, $n_e$ is depicted by a dashed black line.
}
\label{fig_chem_z0}
\end{figure*}

\begin{figure*}
\begin{minipage}{0.33\textwidth}
\resizebox{\hsize}{!}{\includegraphics{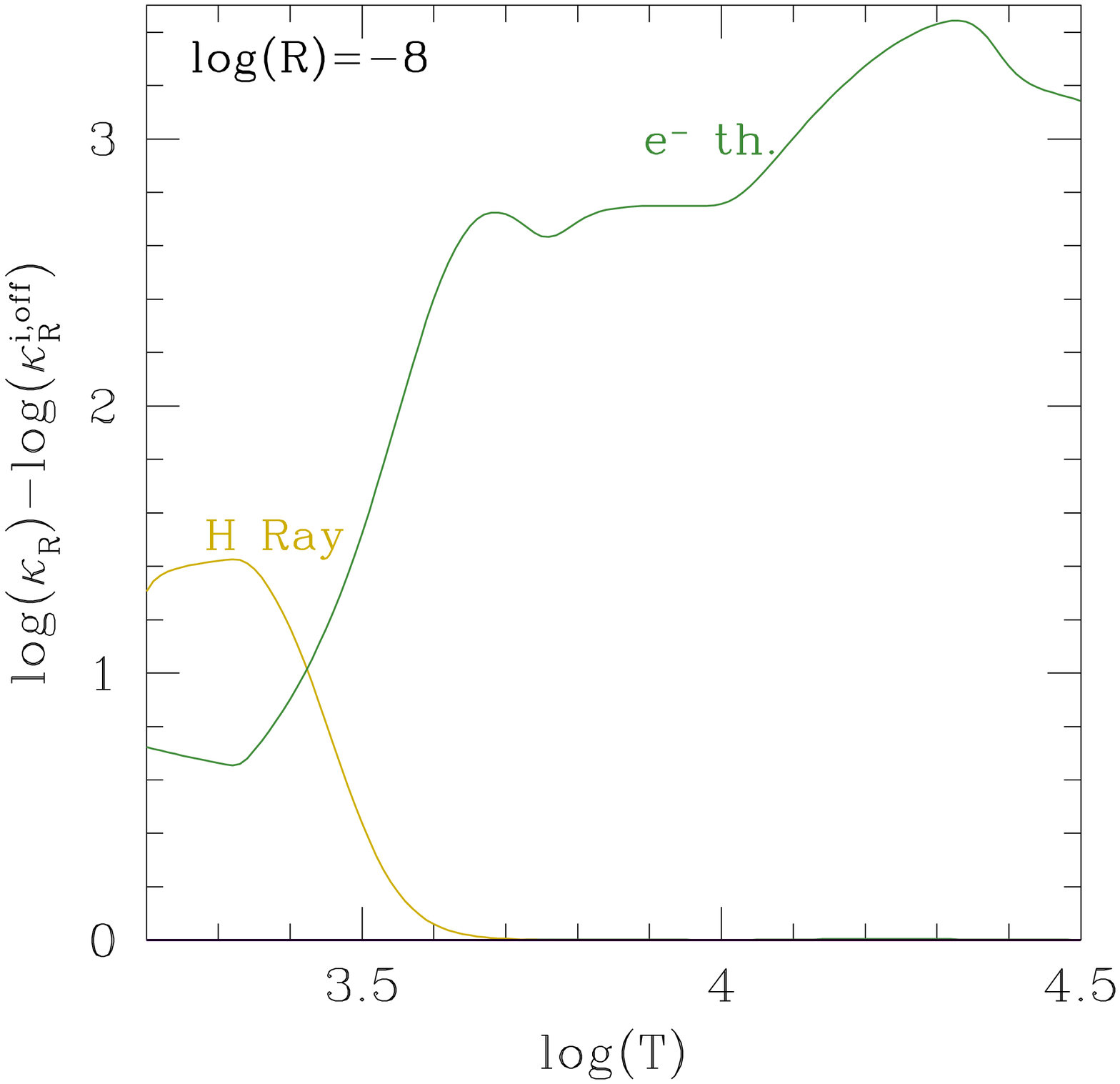}}
\end{minipage}
\begin{minipage}{0.33\textwidth}
\resizebox{\hsize}{!}{\includegraphics{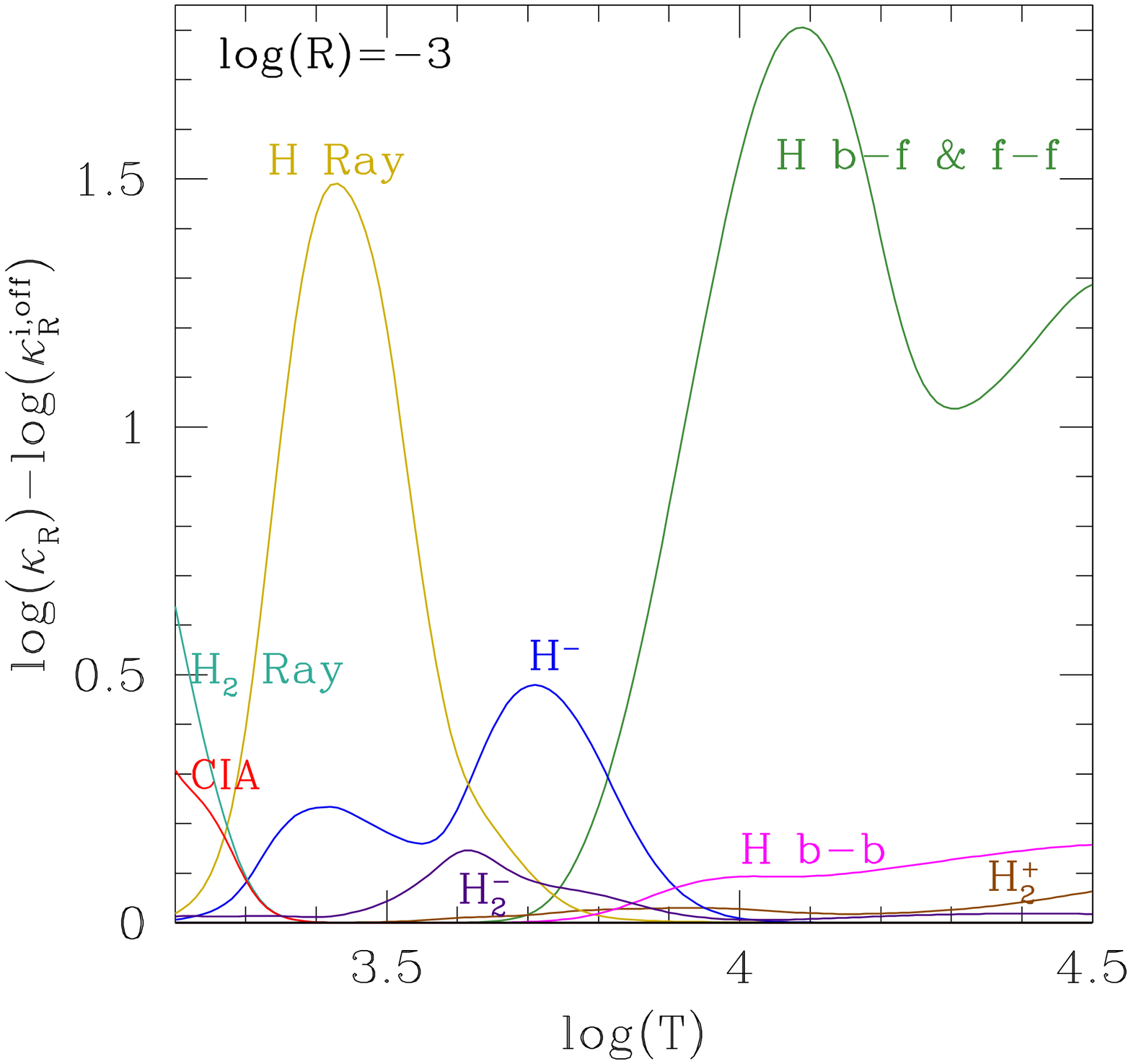}}
\end{minipage}
\begin{minipage}{0.33\textwidth}
\resizebox{\hsize}{!}{\includegraphics{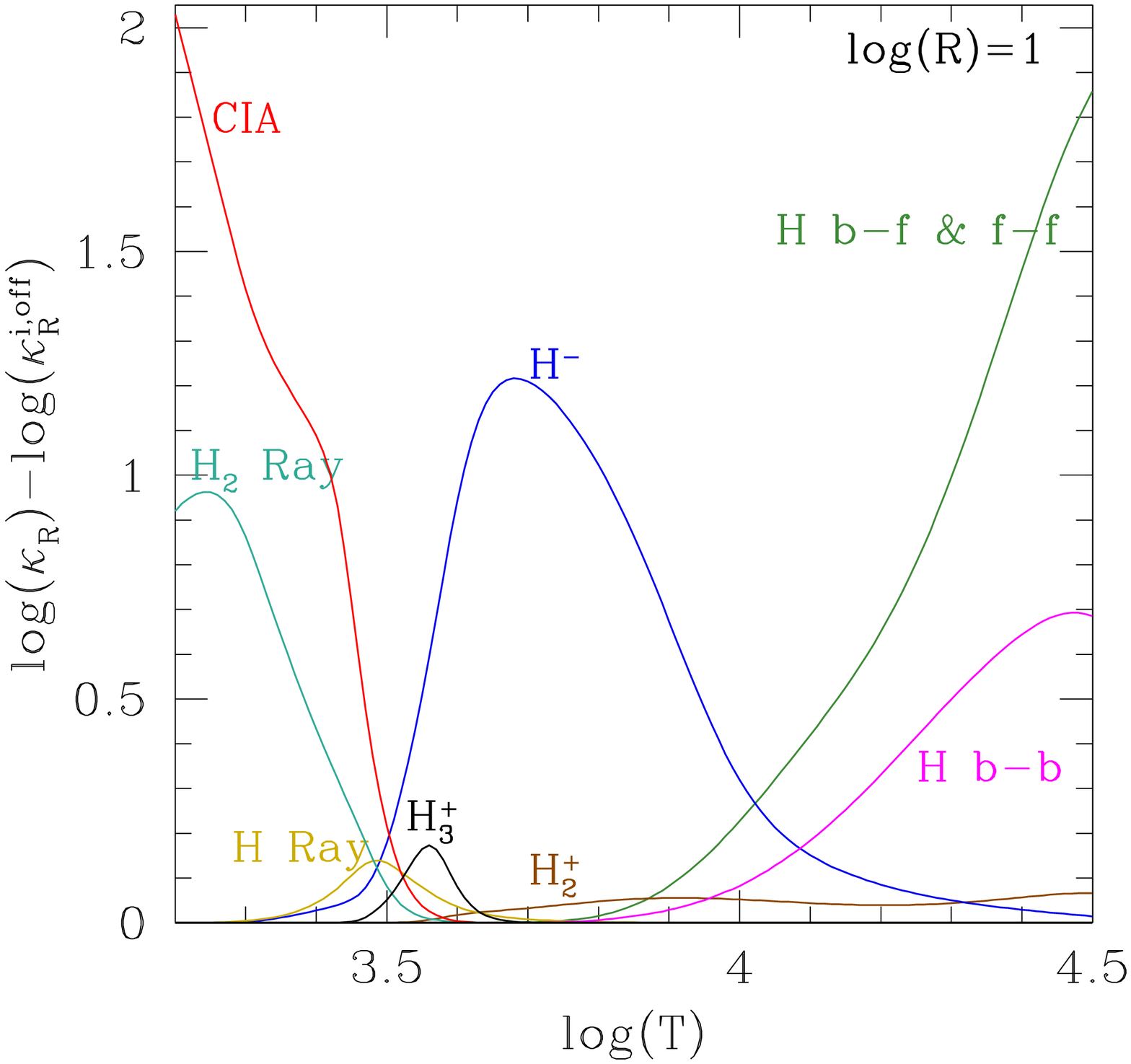}}
\end{minipage}
\caption{The same as in Fig.~\ref{fig_chem_koff_z02} but for a primordial composition with 
$Z_{\rm ref}=0$ and $X=0.7521$ and assuming three of the $R$ parameter, as indicated. 
Note the prominent bump of the CIA sources,
mainly due to H$_2$-H$_2$ collisions, at lower temperatures in the case for $\log(R)=1$.
}
\label{fig_z0_koff}
\end{figure*}

\begin{figure*}
\begin{minipage}{0.50\textwidth}
\resizebox{\hsize}{!}{\includegraphics{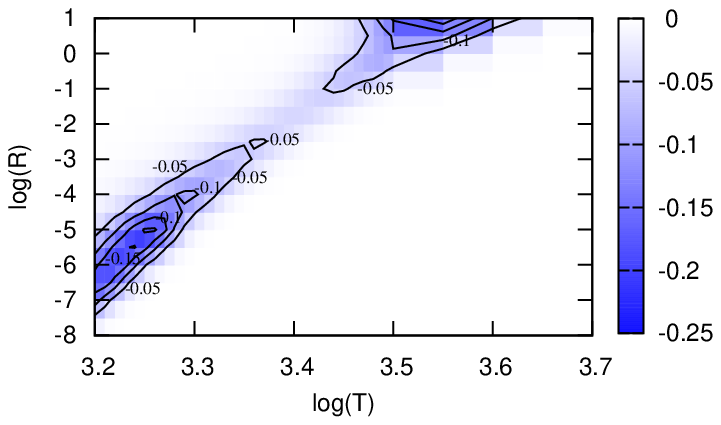}}
\end{minipage}
\begin{minipage}{0.50\textwidth}
\resizebox{\hsize}{!}{\includegraphics{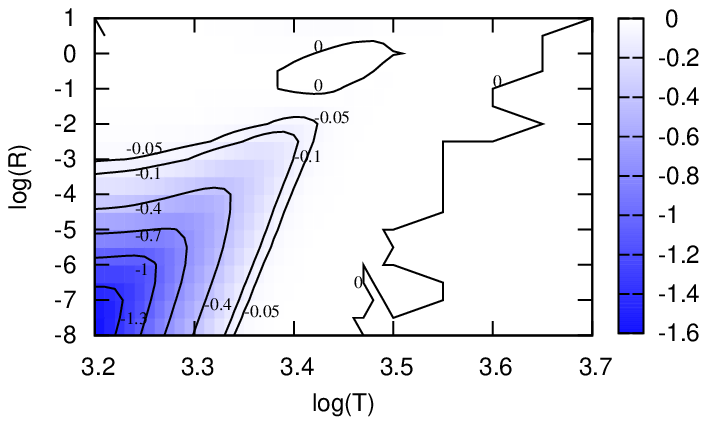}}
\end{minipage}
\caption{Difference 
$\Delta\log(\kappa_{R})=\log(\kappa_{\rm R})-\log(\kappa_{\rm R}^{\rm i, off})$ 
between the full RM opacity of our assumed
 primordial composition and the {\em reduced} opacity obtained either
leaving out  the chemistry 
of H$_3^{+}$ (left panel), or assuming a Li-free mixture (right panel). 
A few contour lines, labelled with the corresponding values (in dex), 
are superimposed to guide the eye.}
\label{fig_h3pli_dif}
\end{figure*}

\begin{figure*}
\begin{minipage}{0.49\textwidth}
\resizebox{\hsize}{!}{\includegraphics{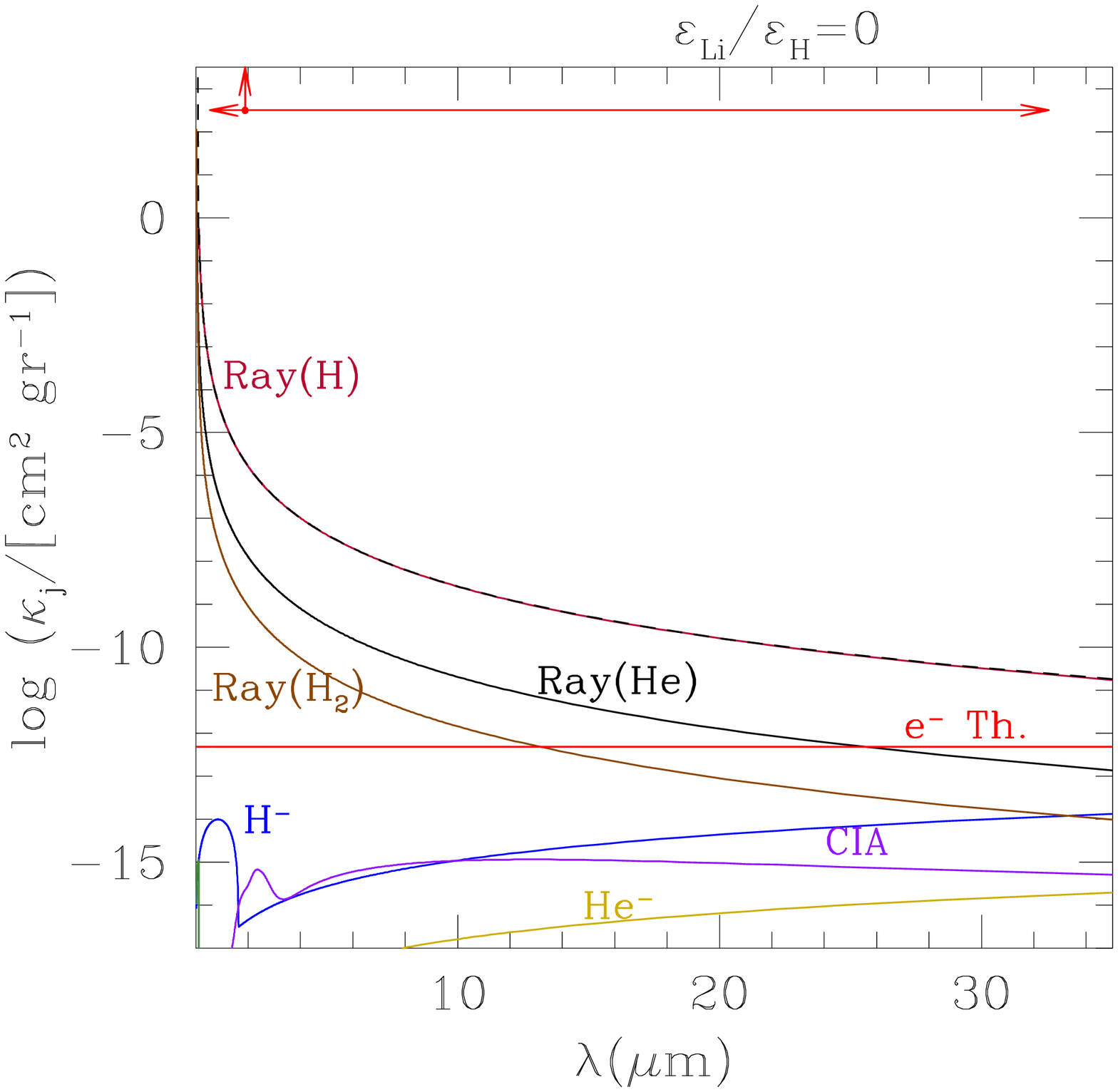}}
\end{minipage}
\begin{minipage}{0.49\textwidth}
\resizebox{\hsize}{!}{\includegraphics{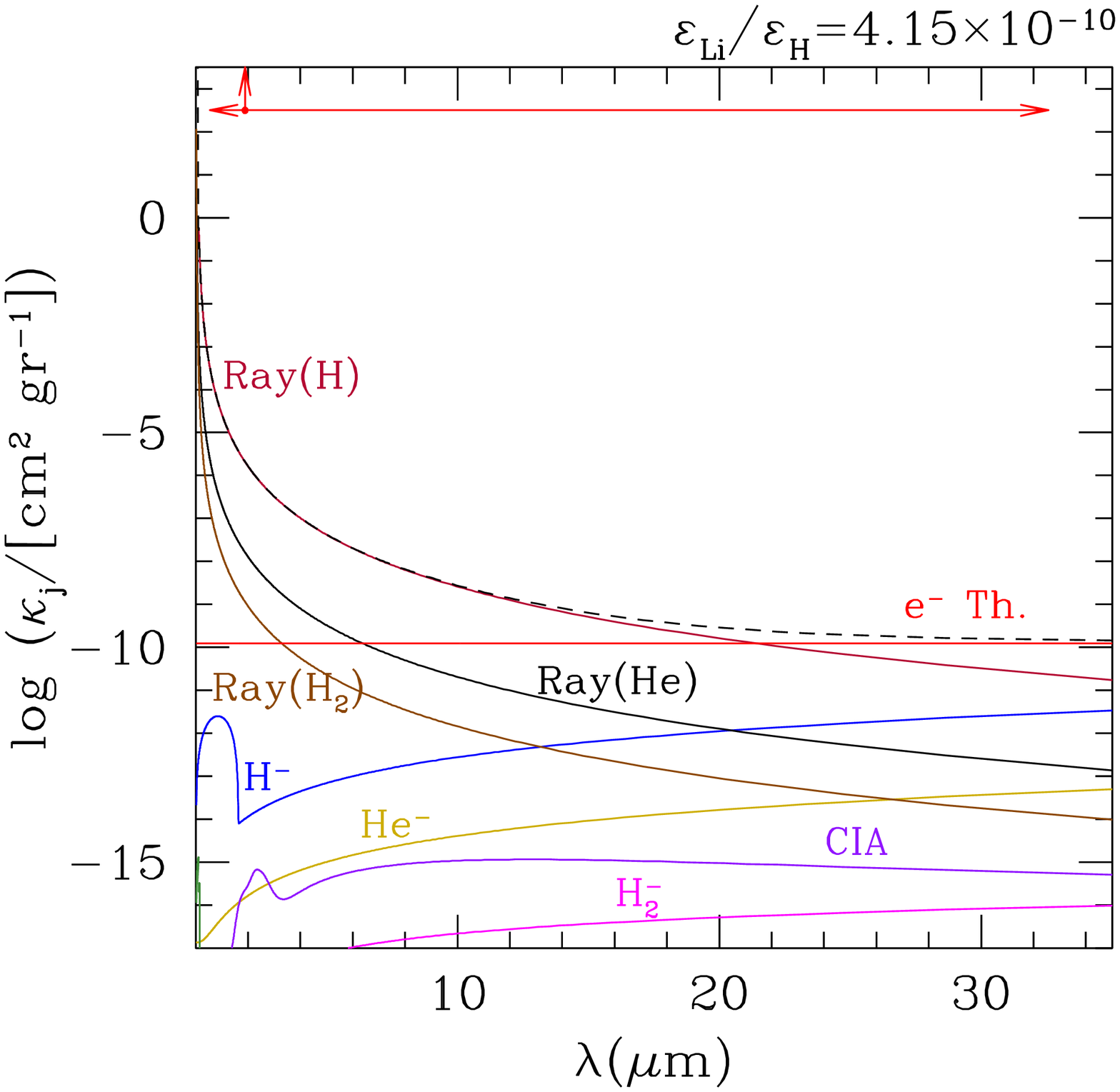}}
\end{minipage}
\caption{The same as in Fig.~\ref{fig_sig_z02} but for a primordial composition with 
$Z_{\rm ref}=0$ and $X=0.7521$, and assuming $\log(R)=-8$ and  $\log(T)=3.3$.
The arrows bracket the spectral range across
which the weighting function of the RM decays by a factor $1/100$. 
The left panel shows the results for
a lithium-free mixture, whereas the right panel illustrates the case for a primordial 
lithium abundance, as predicted by the SBBN in accordance with WMAP (Coc et al. 2004).}
\label{fig_linoli}
\end{figure*}

\begin{figure*}
\begin{minipage}{0.50\textwidth}
\resizebox{\hsize}{!}{\includegraphics{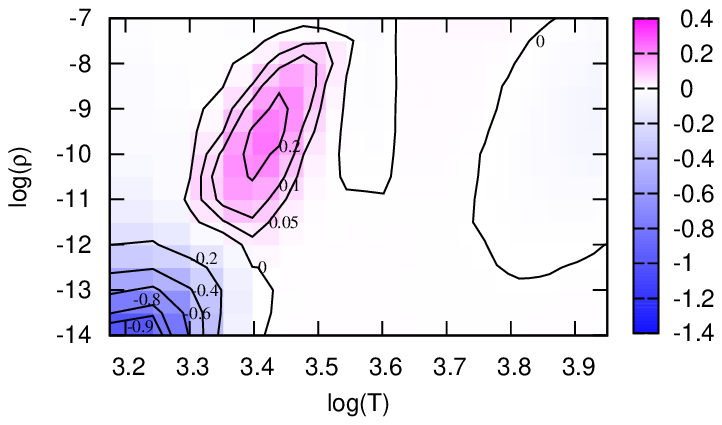}}
\end{minipage}
\begin{minipage}{0.50\textwidth}
\resizebox{\hsize}{!}{\includegraphics{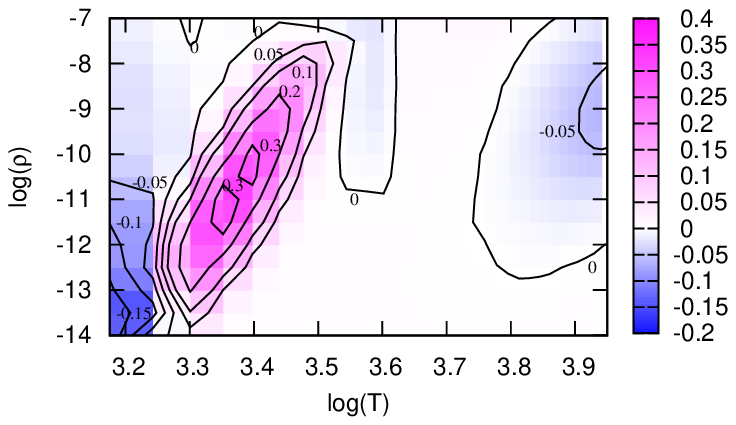}}
\end{minipage}
\caption{Comparison in terms of $\log(\kappa_{\rm R}^{\rm H04})-\log(\kappa_{\rm R}^{\rm \AE SOPUS})$ 
between our opacity results and the tabulated values by 
Harris (2004) for a metal-free mixture with  ($Z=0$, $X=0.7$).
A few contour lines, labelled with the corresponding values (in dex), 
are superimposed to help the comparison. Note how much the differences become significant at lower temperatures 
when assuming $\varepsilon_{\rm Li}/\varepsilon_{\rm H} =4.15\times 10^{-10}$  in our calculations (left
panel), while they drastically reduce adopting  $\varepsilon_{\rm Li}/\varepsilon_{\rm H}=0$ (right panel).
}
\label{fig_z0_harris04}
\end{figure*}
\begin{figure*}
\begin{minipage}{0.50\textwidth}
\resizebox{\hsize}{!}{\includegraphics{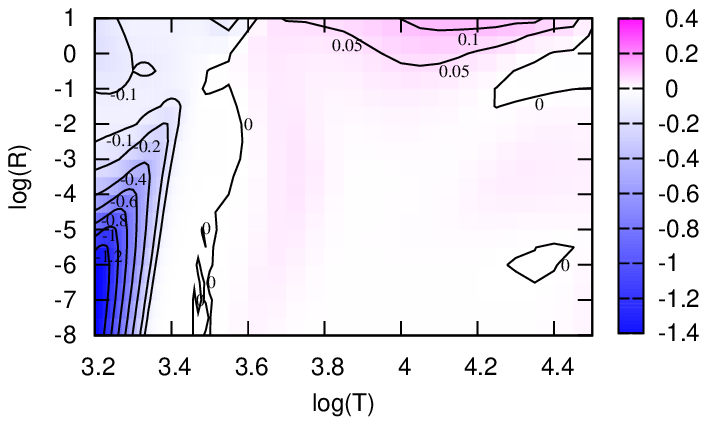}}
\end{minipage}
\begin{minipage}{0.50\textwidth}
\resizebox{\hsize}{!}{\includegraphics{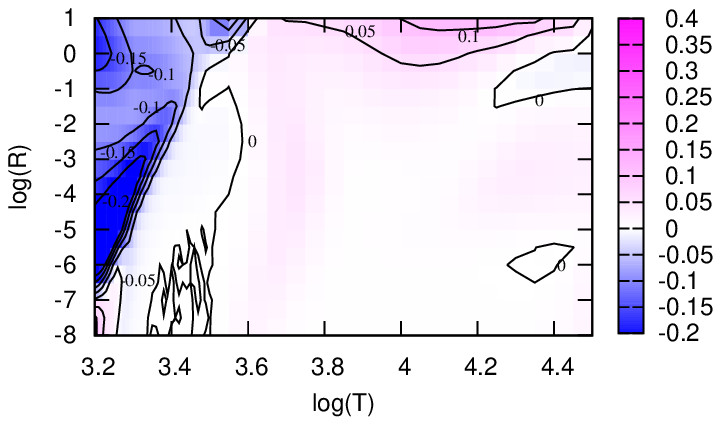}}
\end{minipage}
\caption{The same as in Fig.~\ref{fig_z0_harris04}, but in terms of 
$\log(\kappa_{\rm R}^{\rm F05})-\log(\kappa_{\rm R}^{\rm \AE SOPUS})$ 
between our opacity results and the tabulated values by 
Ferguson et al. (2005) for a metal-free mixture with  ($Z=0$, $X=0.7$).
}
\label{fig_z0_f05}
\end{figure*}

Finally we perturb the second mixture 
and add the C-N-O-Na-Mg-Al anti-correlation pattern
assuming the following abundance variations in dex (see Salaris et al. 2006):
$\log(f_{\rm C})=-0.6$;  $\log(f_{\rm N})=+1.8$;
$\log(f_{\rm O})=-0.8$; $\log(f_{\rm Na})=+0.8$;
$\log(f_{\rm Mg})=-0.4$; and
$\log(f_{\rm Al})=+1.0$.
By doing so the total metallicity almost doubles, $Z=1.97\,10^{-3}$, while the ratio
$[{\rm Fe/H}]\simeq -1.5 $ remains the same as in the genuine $\alpha$-enhanced mixture.

The increase in metallicity is mainly due to the augmented N
abundance, while those of C and O both drop considerably.
The resulting C/O ratio is now  $\simeq 0.31$, and the total 
$\varepsilon_{(\rm C)}+\varepsilon_{(\rm {O})}$ is decreased 
by $\simeq 83 \%$. 
This fact explains that, despite of the overall increase in $Z$, the
opacity curve of the peculiar mixture lies systematically lower
than the others in the temperature region dominated by the H$_2$O
bump. 

In the temperature interval
$3.4 \la \log(T) \la 3.6$ the differences in RM opacity among the
three curves in Fig.~\ref{fig_opac_anomal} are quite small and should be mainly
ascribed to differences in the abundances of electron donors, 
which in turn affect the strength of the
H$^{-}$ opacity.

\subsection{Metal-free mixtures}
\label{ssec_kz0}

The last important application we discuss here deals with
RM opacities suitable for zero-metallicity  gas with a primordial composition.
Following the standard Big Bang nucleosynthesis (SBBN), the most 
abundant elements to be synthesised first were H, He, with small quantities of
D and Li, and tiny (and negligible) traces of Be and B.
In this work we assume a primordial mixture made up of $X=0.7521$,
$\varepsilon_{\rm Li}/\varepsilon_{\rm H}=4.15\times 10^{-10}$ (ratio of abundances by number), and $Y=1-X-$Li (hence $Z=0$), 
these values being  predicted by the SSBN in accordance with the baryon-to-photon ratio  as derived by 
the Wilkinson Microwave Anisotropy Probe (WMAP; Coc et al. 2004).
The abundances of B ad Be are reasonably neglected, 
since 
$\varepsilon_{\rm Be}/\varepsilon_{\rm H}$ and
$\varepsilon_{\rm B}/\varepsilon_{\rm H} < 10^{-17}$ 
according to models of primordial nucleosynthesis (Thomas et al. 1993, 1994).

Figure~\ref{fig_chem_z0} shows the predicted chemistry of a primordial gas as a function of the 
temperature and three selected values of the $R$ parameter, and correspondingly Fig.~\ref{fig_z0_koff}  
illustrates the relative contributions of the most important opacity sources
to the total RM opacity.
It is worth noticing the following points.

At lower densities (e.g. left panels with $\log(R)=-8$) the abundance of the negative 
 hydrogen ion H$^{-}$ grows very little, the  H$_2$ molecule 
does not form efficiently even at the lowest temperatures, and the concentration 
of H$_3^+$ is negligible (reaching a maximum value $\log(n/n_{\rm tot})\sim -20.8$ 
at $\log(T)\sim 3.35$). The total RM opacity
is completely dominated by scattering processes, namely Thomson scattering from free electrons
at higher temperatures, and scattering from  hydrogen atoms at lower temperatures.

At increasing  densities (e.g. going from $\log(R)=-3$ to $\log(R)=1$) the abundances of
most relevant species like H$_2$ , H$^{-}$,  H$_3^{+}$ grow  higher and higher.
At intermediate densities (i.e. middle panel of  Fig.~\ref{fig_z0_koff}) we  may
distinguish three different temperature ranges, namely: $3.2 \la \log(T) \la 3.6$ dominated
by scattering from H atoms,  $3.6 < \log(T) \la 3.85$ characterised by the contribution
of  H$^{-}$, and $3.85 \la \log(T) \la 4.5$ controlled by the continuous absorption
of H (bound-free and free-free transitions). Free electrons are provided by H$^{+}$ and Li$^{+}$
as in the previous case.

Finally, at the highest densities (i.e. right panel of  Fig.~\ref{fig_z0_koff}) we notice that 
the RM opacity in the low-temperature region $3.2 \la \log(T) \la 3.5$ 
is determined by collision-induced absorptions (mainly CIA 
due to H$_2$-H$_2$ collisions); the H$^{-}$ opacity bump is prominent  in the range
 $3.2 < \log(T) \la 4.0$; and continuous and discrete processes due to H are dominant
at higher temperatures.

It should be remarked that Thomson scattering as well as absorption by negative ions
(i.e. H$^{-}$,  H$_2^{-}$, He$^{-}$) crucially depend on the amount of available free electrons.
 By looking at the curve of the electron concentration (dashed line)  in Fig.~\ref{fig_chem_z0} we see 
that, among the positive ions, three are the main electron donors in a primordial
gas,  i.e. H$^{+}$, Li$^{+}$, and H$_3^{+}$.  Ionisation of hydrogen
atoms accounts for $n_{e}$ at the higher temperatures down to $\log(T)\sim 3.6-3.3$ depending 
on the density, ionised lithium practically provides all free electrons at lower temperatures, while
H$_3^{+}$ contributes free electrons only over an intermediate temperature range depending on
the gas density.  

Let us first consider the case of H$_3^{+}$. The importance of this ion for the electron  
budget of a primordial gas has been extensively discussed by  Lenzuni et al. (1991) and Harris
et al. (2004, hereafter also H04). In this latter paper the authors have 
pointed out that the inclusion of  H$_3^{+}$, with the most recent partition function
of Neale \& Tennyson (1995), may increase the RM opacity mostly via an indirect effect on the chemistry, i.e.
by favouring larger concentrations of  H$^{-}$ and, to a less extent, via the direct absorption by H$_3^{+}$.
The authors have also analysed possible effects on the evolution of very low-mass stars of zero-metallicity.
In \AE SOPUS we have included the H$_3^{+}$ chemistry, its free-free opacity,
while neglecting the H$_3^{+}$ line opacity. However, as shown by H04, 
this latter provides a small contribution (few $\%$)
to the RM opacity in most cases, with a peak of $15 \%$ at certain temperatures and densities.
Figure~\ref{fig_h3pli_dif} (left panel) displays the region in the $\log(T)-\log(R)$ plane
which is affected by the H$_3^{+}$ via its inclusion/omission in the gas chemistry. 
The differences in $\log(\kappa_{\rm R})$ are always negative 
along a diagonal strip in the 
$\log(T)-\log(R)$ diagram, meaning that the neglecting 
H$_3^{+}$ would lead to underestimate the gas opacity because we omit 
its contribution to $n_e$ (hence weakening the H$^{-}$ opacity and the Thomson electron scattering),
as well as its contribution as a true absorber (the free-free continuum  in our computations).

The case of  Li is perhaps more interesting since the primordial abundance of this element is 
predicted by the SBBN and accurately constrained by WMAP. An extensive analysis on the  importance of
 Li for the opacity of the primordial  gas has been carried out by Mayer \& Duschl (2005), to whom 
the reader should refer for a detailed discussion. Our computations essentially agree with
the findings of Mayer \& Duschl (2005).
From the inspection of the right panel of Fig.~\ref{fig_h3pli_dif} one can see that even a low
concentration of Li notably impacts on the resulting RM opacity, the effect being more
pronounced at lower temperatures and lower densities. For $\log(T)=3.2$ and $\log(R)=-8$ the difference 
in opacity is sizable, reaching a value as high as $\Delta\log(\kappa_{\rm R})\simeq 1.6$! 
Figure~\ref{fig_linoli} helps to get a better insight of the r\^ole of Li: when including it
in the primordial chemistry the total monochromatic absorption coefficient rises for $\lambda > 15\,\mu$m
due to the increased contribution of the Thomson electron scattering.
In fact, a larger amount of free electrons is provided by the first ionisation of lithium, as shown
 in Fig.~\ref{fig_chem_z0} (left panel).

\begin{figure}
\resizebox{\hsize}{!}{\includegraphics{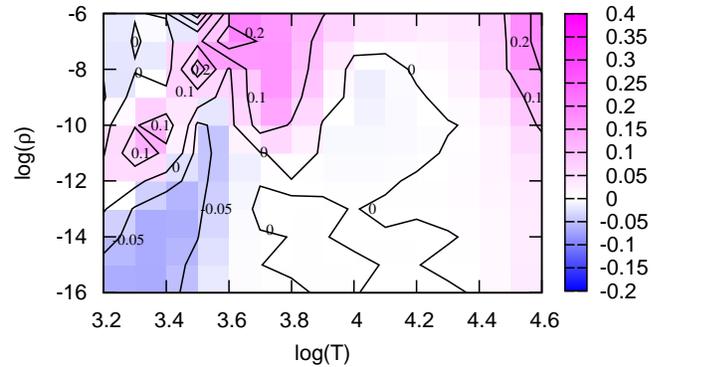}}
\caption{ The same as in Fig.~\ref{fig_z0_harris04} in terms of 
$\log(\kappa_{\rm R}^{\rm M05})-\log(\kappa_{\rm R}^{\rm \AE SOPUS})$ 
between our opacity results and the tabulated values by 
Mayer et al. (2005) for a metal-free mixture with  $Z=0$, $X=0.7521$,
and $\varepsilon_{\rm Li}/\varepsilon_{\rm H}=4.15\times 10^{-10}$.}
\label{fig_z0_meyer05}
\end{figure}

Finally, in Figs.~\ref{fig_z0_harris04}-\ref{fig_z0_meyer05} 
we present a few comparisons with recently published RM opacity data
for zero-metallicity gas, namely: Harris et al. (2004),  
Mayer \& Duschl (2005), and Ferguson et al. (2005).
In general the agreement is relatively good, mostly comprised within
$\pm 0.2$ dex, except for the large
differences (up to $-1.2-1.4$ dex) that arise in the comparison 
with H04 and F05 
at lower temperatures and densities. These discrepancies should be likely ascribed to
their neglecting of Li in the
chemical mixture, since they drastically reduce when
we omit Li from the equation of state.
We are not able to find clear reasons to the 
 remaining deviations for $\log(T) < 3.5$, temperatures at which 
Rayleigh scattering from H
and H$_2$, Thomson scattering from electrons, and CIA are
the dominant opacity contributors at varying density.
In general, differences in the thermodynamic data 
and input physics adopted to describe  
the processes listed in Table~\ref{tab_opacsource} 
might provide a reasonable explanation.

\section{Final remarks}
\label{sec_final}  
We have developed a new tool, \AE SOPUS, for computing Rosseland mean
opacities of an ideal gas in the low-temperature regime, 
$3.2 \le \log(T) \le 4.5$. 
The access to \AE SOPUS is made public via an
interactive web-interface (http://stev.oapd.inaf.it/aesopus), 
which enables the user to
specify with large freedom the input parameters, i.e.
the grid of the state variables $T$ and $R$,  
the reference solar composition, the total metallicity, 
and the abundance enhancement/depletion 
of all chemical elements, from H to U.

The Rosseland mean gas opacities, produced with a good accuracy 
(comparable to that of other opacity codes), are
delivered in a tabular form within a reasonably short time. 
At present, the typical computation time for one table at fixed
chemical composition, arranged with the default $T-R$ grid,
i.e. containing $N_T \times N_R= 67\times 19 = 1273$ opacity values, 
is less than  $50$ s with a 2.0 GHz processor.
Such a fast performance is attained thanks to 
the optimised use of the opacity sampling method 
to describe molecular line absorption, 
and the adoption of pre-tabulated absorption
 cross-sections for metals 
(from the Opacity Project database). In this way the line-opacity data  
is suitably arranged prior to the opacity computations, 
a process that, if otherwise performed on-the-fly, 
is in principle more accurate but at the cost of 
extremely long computing times (e.g. Ferguson et al. 2005).

On the other hand, several tests illustrated in the paper 
have proved that our procedure, besides being fast, 
is as well suitable to produce fairly accurate  
Rosseland mean opacities, to which the very fine spectral details 
are not critical as they are washed out, by construction, 
in the harmonic average of the monochromatic coefficient.

First applications of \AE SOPUS opacity tables in stellar
evolutionary calculations performed with the
Padova code for both scaled-solar 
(Bertelli et al. 2009),  and $\alpha$-enhanced mixtures 
(Bressan et al., in prep.), and with the ATON code for C-N-O varying
mixtures along the AGB (Ventura \& Marigo 2009) have yielded
promising results.
In particular, we find that the differences in the 
effective temperature of giant (RGB and AGB) models 
brought about by the adoption of different opacity 
data for the same chemical composition 
(e.g. \AE SOPUS, Ferguson et al. 2005, Lederer \& Aringer 2009) 
amount to a few tens of degrees, in most cases lower than (or comparable to) 
the typical uncertainty of the semi-empirical $T_{\rm eff}$-scale of red giants.
     
We wish all interested researchers may benefit from an easy access to
the low-temperature opacity data. Feedback and suggestions are
welcome.

\begin{acknowledgements}
We thank our referee, Jason~W. Ferguson, for his detailed and careful
examination of the paper.
This work was supported by the University of Padova
(60A02-2949/09), INAF/PRIN07 (CRA 1.06.10.03), 
and MIUR/PRIN07 (prot. 20075TP5K9).
B.A. acknowledges funding by the contract ASI-INAF I/016/07/0 and by the
Austrian Science Fund (FWF) projects P19503-N16 and P18939-N16.
We are grateful to  Alan W. Irwin for his contribution to the EOS part, 
based on source code available under the GPL from the SSynth project 
(http://sourceforge.net/projects/ssynth/) that is developed
by Alan W. Irwin and Ana M. Larson.
We thank L. Girardi for his valuable help in the development of   
the \AE SOPUS web-interface, Michael Lederer and Alessandro Bressan 
for useful discussions.
\end{acknowledgements}

\appendix

\section{EOS under ICE conditions: numerical details}
\label{apx_method}
The \AE SOPUS code solves the equation of state  assuming
instantaneous chemical
equilibrium by means of the Newtwon-Raphson technique. 
We consider the $\mathcal{N}_{\rm el}+2$ conservation equations 
(see Sect.~\ref{sssec_consrel}) formulated in the generic form: 
\begin{displaymath}
  \left \{ \begin{array}{ccc}
f_{1}[n_{\alpha},\,(\alpha=1,\cdots \mathcal{N}_{\rm el}),\, N_{\rm a}, n_e]  & = & 0\\
f_{2}[n_{\alpha},\,(\alpha=1,\cdots \mathcal{N}_{\rm el}),\, N_{\rm a}, n_e]  & = & 0\\
 \vdots  & &\\
f_{\mathcal{N}_{\rm el}}[n_{\alpha},\,(\alpha=1,\cdots \mathcal{N}_{\rm el}),\, N_{\rm a}, n_e] &=&0\\
\,\,\,f_{\rm e}[n_{\alpha},\,(\alpha=1,\cdots \mathcal{N}_{\rm el}),\, N_{\rm a}, n_e] &= &0\\
\,f_{\rm tot}[n_{\alpha},\,(\alpha=1,\cdots \mathcal{N}_{\rm el}),\, N_{\rm a}, n_e]& =&\,\,0\,,\\
\end{array}\right.
\end{displaymath}
which depend on the $\mathcal{N}_{\rm el}+2$ unknowns, namely: 
the number density of each neutral atom 
$n_{\rm alpha},\, \alpha=1,\dotsc N_{\rm el}$; the total number
density of atoms $N_{\rm a}$; and the electron density $n_e$.
Then we calculate the jacobian matrix ${\mathbf J}$ of the functions $f$'s with
respect to each unknown,
\begin{gather*}
 J_{ij}=\frac{ \partial f_i }{ \partial n_j }=\left(
    \begin{array}{ccccc}
\displaystyle\frac{ \partial f_1 }{ \partial n_{1} }\, & 
\dotsc\, &
\displaystyle\frac{ \partial f_1 }{ \partial n_{\mathcal{N}_{\rm el}}}\, &
\displaystyle\frac{ \partial f_1 }{ \partial N_{\rm a} }\, &
\displaystyle\frac{ \partial f_1 }{ \partial n_{e} } \\
\vdots & \ddots & \vdots & \vdots & \vdots \\
\displaystyle\frac{ \partial f_{\mathcal{N}_{\rm el}} }{ \partial n_{1} }\, & 
\dotsc\, &
\displaystyle\frac{ \partial f_{\mathcal{N}_{\rm el}} }{ \partial n_{\mathcal{N}_{\rm el}}}\, &
\displaystyle\frac{ \partial f_{\mathcal{N}_{\rm el}} }{ \partial N_{\rm a} }\, &
\displaystyle\frac{ \partial f_{\mathcal{N}_{\rm el}} }{ \partial n_{e} } \\
\displaystyle\frac{ \partial f_{e} }{ \partial n_{1} }\, & 
\dotsc\, &
\displaystyle\frac{ \partial f_{e} }{ \partial n_{\mathcal{N}_{\rm el}}}\, &
\displaystyle\frac{ \partial f_{e}}{ \partial N_{\rm a} }\, &
\displaystyle\frac{ \partial f_{e} }{ \partial n_{e} } \\
\displaystyle\frac{ \partial f_{\rm tot} }{ \partial n_{1} }\, & 
\dotsc\, &
\displaystyle\frac{ \partial f_{\rm tot} }{ \partial n_{\mathcal{N}_{\rm el}}}\, &
\displaystyle\frac{ \partial f_{\rm tot}}{ \partial N_{\rm a} }\, &
\displaystyle\frac{ \partial f_{\rm tot}}{ \partial n_{e} } 
 \end{array} 
\right) 
\end{gather*}

In practice, because the unknown quantities are all inherently non-negative
functions, their logarithmic forms are adopted. This prevents
physically unrealistic estimates from occurring during the iteration process.

For each chemical element $\alpha$, 
the abundance conservation equation (Eq.~\ref{eq_atom}) 
is conveniently written in the form:

\begin{equation}
f_{\alpha}=\ln \large\left(
\begin{array}{c}
\displaystyle\sum_{A=1}^{\mathcal{N}_{\rm tot}} n_A \nu_{A,\alpha} \\
\hline 
{ N_{\rm a} \varepsilon_{\alpha}}\\
\end{array}\right)
= 0\, ,
\end{equation}
where $n_A$ is the number density of particle $A$ which ranges over
all species, i.e. atoms, ions, and molecules;
$\nu_{A,\alpha}$ are stoichiometric coefficients that keep track of
how many times species of type $A$ contributes to the conservation
equation of type $\alpha$. In other words $\nu_{A,\alpha}$ 
represents the number of atoms of the $\alpha$ element contained in
species $A$.

The charge neutrality equation (Eq.~\ref{eq_charge}) is expressed in the form:
\begin{equation}
f_{e}=\ln \left(\frac{\displaystyle\sum_{i=1}^{\mathcal{N}_{\rm tot}} 
\displaystyle\sum_{r=1}^{pz}\frac{ r\, n_{A_i^{+r}}}{n_e}}
{1 + \displaystyle\sum_{j=1}^{\mathcal{N}_{\rm tot}} 
\frac{n_{A_j^{-}}}{n_e}}\right) = 0\, .
\end{equation}

Finally, the conservation equation of the total number density 
(Eq.~\ref{eq_ntot}) is rearranged in the
form:

\begin{equation}
f_{\rm tot}=\ln \large\left(
\begin{array}{c}
\displaystyle\sum_{\alpha=1}^{\mathcal{N}_{\rm el}}
\displaystyle\sum_{A=1}^{\mathcal{N}_{\rm mol}}
\nu_{A,\alpha} (n_{A} + n_{A^+} + n_{A^-}) \\
\hline 
\Delta N_{\rm a}^{\rm mol}\\
\end{array}\right)
= 0\, ,
\end{equation}
where $\sum_{A} n_A \nu_{A,\alpha}$ is extended over all molecules and
quantifies their contribution  to the conservation equation
of any given element $\alpha$; 
$\Delta N_{\rm a}^{\rm mol}=N_{\rm a} - (n_{\rm tot} - n_e)$
corresponds to the excess in $N_{\rm a}$  due to molecular formation.

In summary, 
after providing a first guess to the number densities, \AE SOPUS sets
them into the system and the jacobian matrix. In general, 
the guess will be inaccurate
so that the functions $f_i$ have finite values. 
Denoting with ${\mathbf F}$ and ${\mathbf n}$ the entire vectors 
of the values of $f_i$ and $n_j$, we deal with the matrix equation
\begin{equation}
{\mathbf f}({\mathbf n} + \delta{\mathbf n})={\mathbf f}({\mathbf n}) +
{\mathbf J} \, \delta {\mathbf n}=0\,, 
\end{equation}
which corresponds to a set of ${\mathcal N}_{\rm el} + 2$ linear
equations for the first-order corrections $ \delta {\mathbf n}$.
The matrix equation is solved with the LU decomposition method. 
The corrections are then
added to the solution vector of the number densities, 
${\mathbf n}^{\rm new} = {\mathbf n}^{\rm old}  + \delta {\mathbf n}$, 
and the process is iterated until the maximum
relative change in the densities 
becomes lower than a given accuracy $\delta_{n}$, i.e. typically
$10^{-5}$ in our computations.

\section{The frequency distribution}
\label{apx_fgrid}
\begin{figure}
\resizebox{\hsize}{!}{\includegraphics{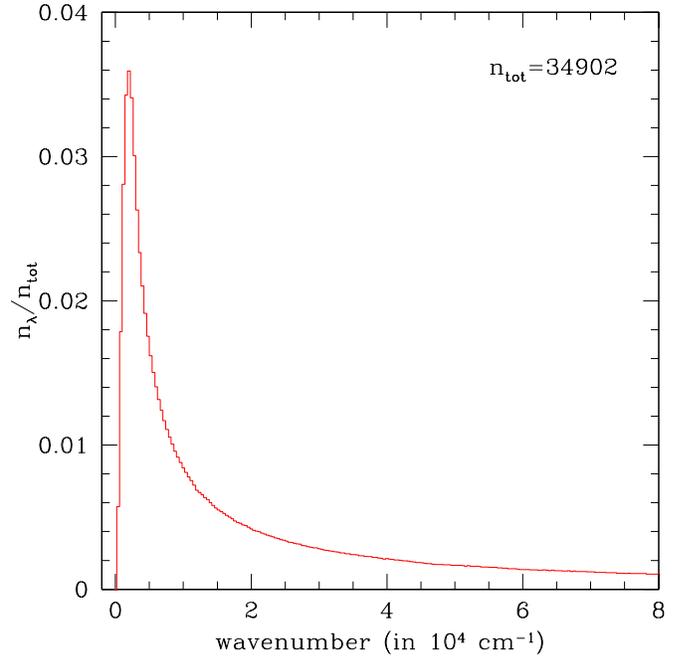}}
\caption{The histogram of the sampling frequency distribution with $n_{\rm tot} =
34902$ sampling points, selected following the scheme proposed by 
Helling \& J\o rgensen (1998). This represents the reference distribution
whence smaller frequency samples are extracted. See the text for more explanation.    
}
\label{fig_nugrid}
\end{figure}
Computing the RM opacity with Eq.~(\ref{eq_rosseland}) requires that
the total monochromatic absorption coefficient is evaluated at a 
finite number of frequency points. 
In principle the more the points, the more accurate the results should be.
However, since we are dealing with a mean quantity one can obtain still
good results using a relatively low number of frequency points, with
the advantage of speeding up the computations.

In respect to this some discussion can be found in 
Ferguson et al. (2005) who integrate over $24\,000$ points, and  
Lederer \& Aringer (2009) who adopt $5\,645$ points.
For the present work we have performed further tests 
to get useful indications on the relationship between the size of the 
frequency distribution and quality of the results, in terms
of accuracy (reliability)  and precision (reproducibility) of the results.

For this purpose we proceed as follows. First, we determine a seed frequency distribution 
by adopting the scheme proposed by
Helling \& J\o rgensen (1998), originally designed to optimize the 
selection of frequency points in the OS method.
In few words, a frequency distribution produces a correct spectral sampling 
if it obeys the condition
$E_{\tilde\nu}(T) \Delta\tilde\nu = const.$, i.e. expressing the constancy of the normalized energy
density of the Planckian, $E_{\tilde\nu}(T)$, over any arbitrary
interval $\Delta\tilde\nu$, where $\tilde\nu$ [cm$^{-1}$] is the wave-number.
Then, the seeked optimal distribution corresponds to the upper
envelope the entire sample of Planckian distributions evaluated at
different temperatures, so that we take
the maximum of the normalized energy density $E_{\tilde\nu}(T)$ at each  $\tilde\nu$.
The final distribution, shown  in Fig.~\ref{fig_nugrid}, is sharply peaked at lower frequencies and
declines exponentially at longer frequencies.

\begin{figure*}
\begin{minipage}{0.48\textwidth}
\resizebox{\hsize}{!}{\includegraphics{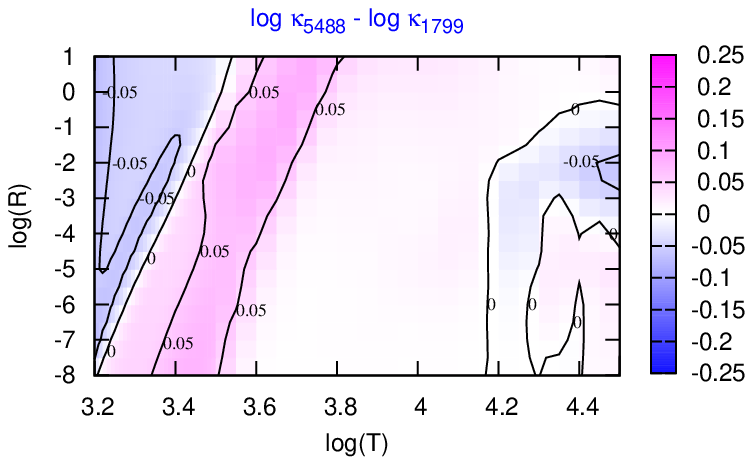}}
\end{minipage}
\hfill
\begin{minipage}{0.48\textwidth}
\resizebox{\hsize}{!}{\includegraphics{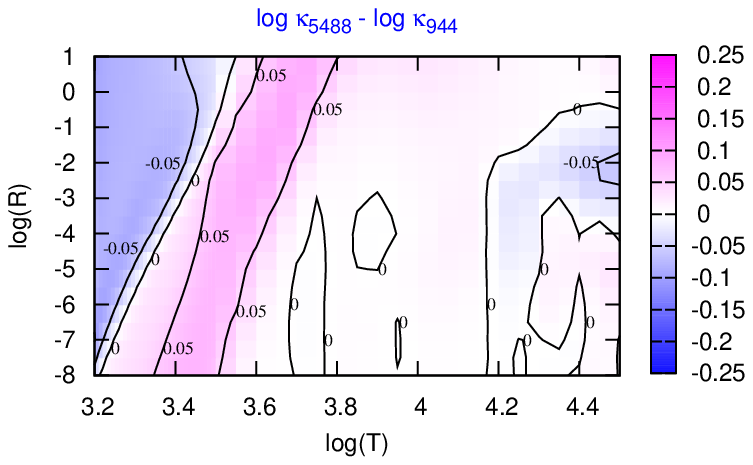}}
\end{minipage}
\hfill
\begin{minipage}{0.48\textwidth}
\resizebox{\hsize}{!}{\includegraphics{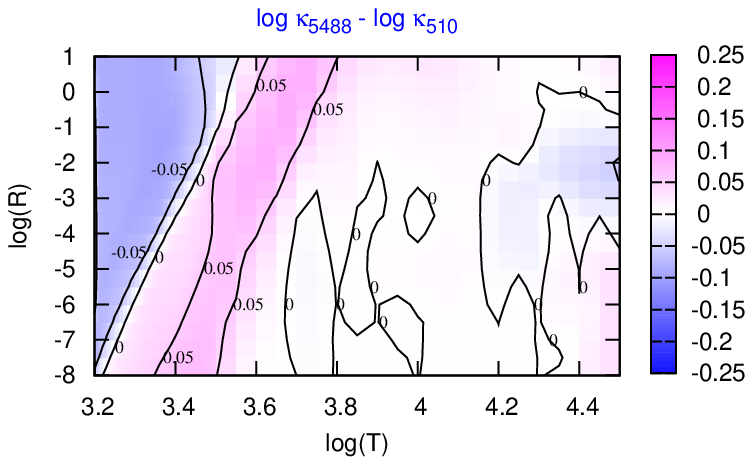}}
\end{minipage}
\hfill
\begin{minipage}{0.48\textwidth}
\resizebox{\hsize}{!}{\includegraphics{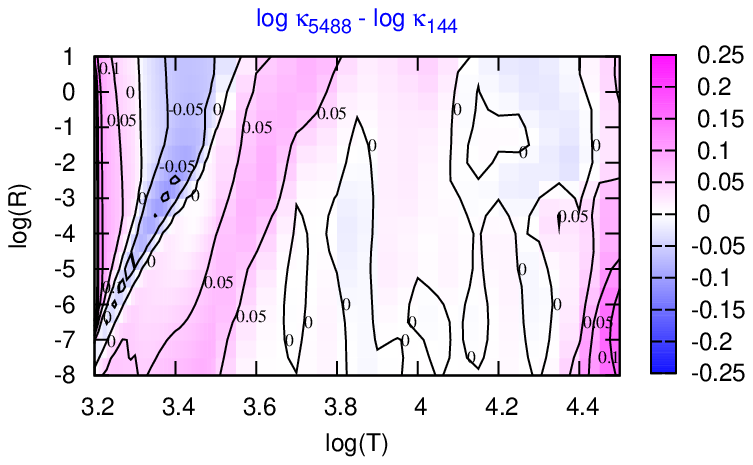}}
\end{minipage}
\caption{Differences in opacities between the reference frequency grid
  with $n_{\rm tot}=5488$  points, and other test cases with  $n_{\rm
  tot}=1799,\,944,\, 510,\,$ and $149$ points. The adopted chemical
  composition is characterized by $X=0.7$, $Z_{\rm ref}=0.02$, and scaled-solar
  abundances according to GS98.
}
\label{fig_gridvar1}
\end{figure*}

Once the seed distribution is constructed, any other  
frequency grid of given size is extracted from it
by using a Monte-Carlo technique. In our work we tested a few
cases adopting 5488, 1799, 944, 510, and 149 points.
Each grid is used to compute RM opacities 
for two chemical compositions characterized by:
$I$)  $X=0.7$, $Z=Z_{\rm ref}=0.020$ and scaled-solar abundances of metals, 
and $II$) a carbon-rich mixture with $X=0.7$, $Z_{\rm ref}=0.02$, 
$Z=0.026$ and C/O~$=1.5$. In the latter case carbon is made increase
relative to its scaled-solar value, producing a net increment
of the actual metallicity.
\begin{figure*}
\begin{minipage}{0.48\textwidth}
\resizebox{\hsize}{!}{\includegraphics{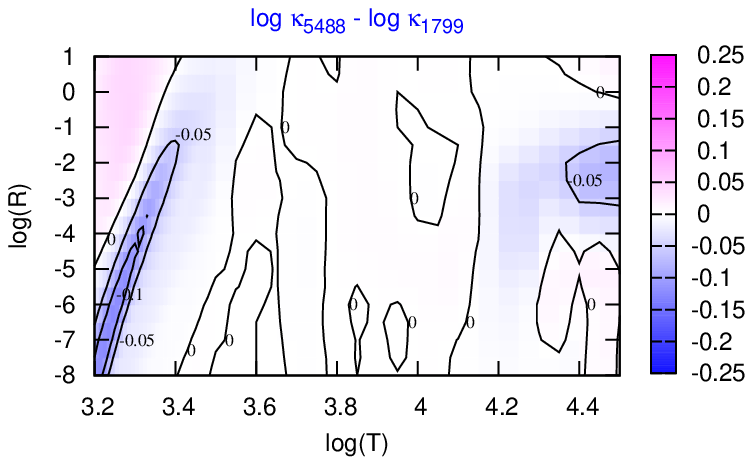}}
\end{minipage}
\hfill
\begin{minipage}{0.48\textwidth}
\resizebox{\hsize}{!}{\includegraphics{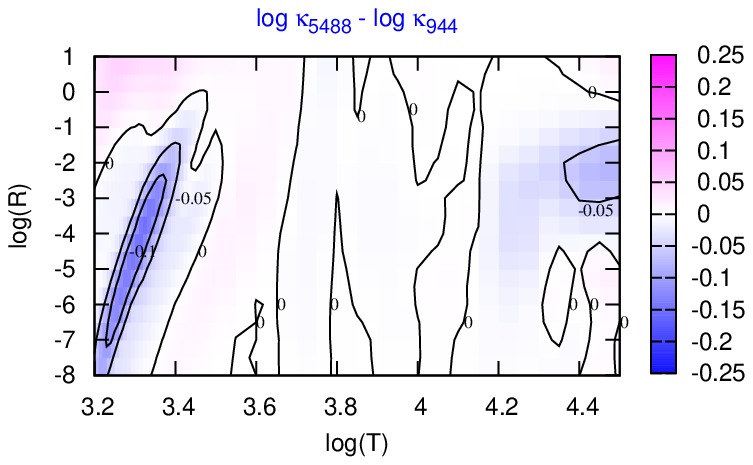}}
\end{minipage}
\hfill
\begin{minipage}{0.48\textwidth}
\resizebox{\hsize}{!}{\includegraphics{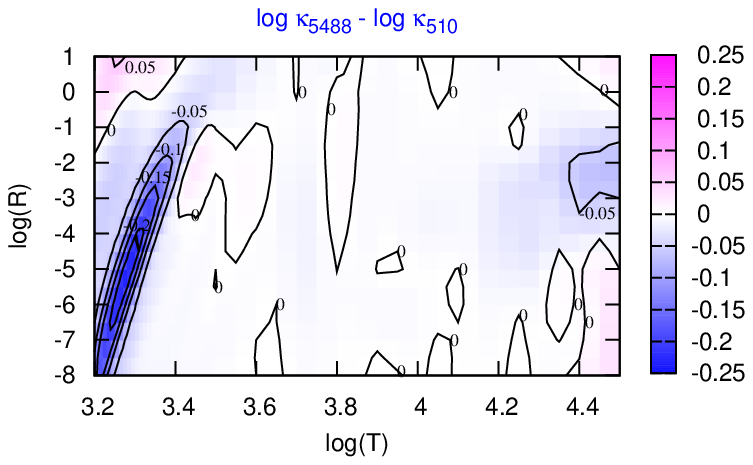}}
\end{minipage}
\hfill
\begin{minipage}{0.48\textwidth}
\resizebox{\hsize}{!}{\includegraphics{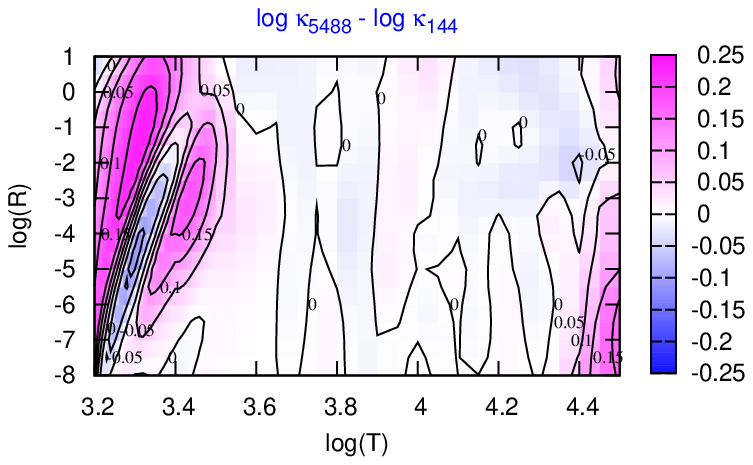}}
\end{minipage}
\caption{The same as in Fig.~\ref{fig_gridvar1}, but for the adopted chemical
  composition which is defined by $X=0.7$, $Z_{\rm ref}=0.02$, $Z=0.026$, and C/O~$=1.5$.
}
\label{fig_gridvar2}
\end{figure*}

Then, adopting as reference opacities those obtained with the densest frequency grid, i.e. $n_{\rm tot}=5488$,
we evaluate the differences, $\log(\kappa_{4588})-\log(\kappa_{n_j})$,  for each opacity subset 
computed with a lower frequency grid (i.e. $n_{j}=  1799, \,944,\, 510,\, {\rm and}\, 149$ points.
The results are shown in Fig.~\ref{fig_gridvar1} for mixture $I$
and Fig.~\ref{fig_gridvar2} for mixture $II$.

We see that in most cases the
differences remain small, within $\simeq \pm 0.05$ dex, 
over most of the $\log(T)-\log(R)$ space, and even with the 
smallest frequency set the loss in accuracy, though larger,
is not dramatic. As expected, the biggest deviations take place
at lower temperatures where the opacity contribution from 
molecular bands is more sensitive to the frequency sampling. 

In any case, it is worth noticing that the uncertainties 
brought about by the adopted frequency distribution are comparable,
if not lower, with the typical differences in RM opacities computed
with different codes (see, for instance, Figs.~\ref{fig_kcomp_z02} and \ref{fig_led09}).

\section{Chemical mixtures with non-solar $[X_i/{\rm Fe}]$ ratios: a general scheme}
\label{apx_afe}
Let us first consider non-scaled-solar mixtures in which the reference metallicity is preserved,
i.e. $Z=Z_{\rm ref}$.
Basing on the formalism introduced in Sects.~ \ref{ssec_chem} and \ref{ssec_alpha} we conveniently divide 
the metal species (with $Z_i \ge 3$) into three groups, namely:
\begin{itemize}
\item The {\em selected elements} with given
$\gamma_i=[X_i^s/X_{\rm Fe}]$ according to the input specification,
with  abundances $X_i^s=f_i^s \,X_{i,{\rm ref}}^s$;
\item The {\em fixed elements} with abundances $X_i^f= X_{i,{\rm ref}}^f$;
\item  The {\em balancing elements}, including all the other metals, with abundances
$X_i^b= f_b\,  X_{i,{\rm ref}}^b$.
\end{itemize}

We recall that the ratios $[X_i^s/X_{\rm Fe}]$ can be freely chosen
to be either positive or negative.
According to the adopted scheme, in order to preserve the metallicity the abundance variation 
of the {\em selected elements} should be compensated by the total
abundance variation of the {\em balancing elements}. It follows that,
by construction, all {\em balancing elements} share the same variation
factor $f^b$. 

Therefore, from the condition $Z=Z_{\rm ref}$, and the definition of
$[X_i^s/X_{\rm Fe}]$ for each of the $\mathcal{N}_{sel}$ selected elements,
we set up a system of $\mathcal{N}_{sel}+1$ equations:
\begin{equation*}
\left\{
\begin{array}{ccl}
Z_{\rm ref} & = & \displaystyle\sum_{i} X_i^s + \displaystyle\sum_{j} X_j^f + \displaystyle\sum_{k} X_k^b \\
\displaystyle\left[\frac{X_i^s}{X_{\rm Fe}}\right] & = & 
\log\displaystyle\left(\frac{X_i^s}{X_{\rm Fe}}\right)
-\log\displaystyle\left(\frac{X^s_{i,\odot}}{X_{\rm Fe, \odot}}\right)\,\,\,\,\,\,\quad\quad\quad
\forall\, i=1,\dotsc\mathcal{N}_{sel}\,\,,
\end{array}
\right.
\end{equation*}
which can be re-formulated with the aid of Eqs.~(\ref{eq_fi}) 
and (\ref{eq_xref}):
\begin{equation*}
\left\{
\begin{array}{ccl}
Z_{\odot} & = & \displaystyle\sum_{i} f_i^s X_{i,\odot}^s + \displaystyle\sum_{j} X_{j,\odot}^f + 
f^b \, \displaystyle\sum_{k} X_{k,\odot}^b \\
10^{\gamma_i} & = & \displaystyle \frac{f_i^s}{f^b}
\quad\quad\quad\quad\quad\quad\quad\quad\quad\quad\quad\quad\quad
\forall\, i=1,\dotsc\mathcal{N}_{sel}
\end{array}
\right.
\end{equation*}
for the unknowns $f_i^s$ and $f^b$.
Let us denote with 
\begin{equation}
\xi_{i,\odot} =\displaystyle\frac{X_{i, \odot}}{Z_{\odot}} 
\end{equation}
the partitions of metals in the solar mixture.
Eventually, from simple analytical passages we obtain the general solution:
\begin{equation}
\label{eq_fsfb}
\left\{
\begin{array}{ccl}
f^b & = & \displaystyle
\frac{\displaystyle\sum_{i}\xi_{i,\odot}^s+\displaystyle\sum_{k}\xi_{k,\odot}^b}
{\displaystyle\sum_{i}10^{\gamma_i}\xi_{i,\odot}^s + \displaystyle\sum_{k}\xi_{k,\odot}^b} \\
f_i^s & =  & f^b \, 10^{\gamma_i}
\quad\quad\quad\quad\quad\quad\quad\quad\quad
\forall\, i=1,\dotsc\mathcal{N}_{sel}
\end{array}
\right.
\end{equation}
which only depends on the specified ratios  $[X_i^s/X_{\rm Fe}]$ of
the {\em selected elements}, and the metal partitions in the reference
solar composition. 
It is useful to particularize Eq.~(\ref{eq_fsfb}) for the cases of
{\em mixture}s $A$ and $B$ introduced in Sect.~\ref{ssec_alpha}, and finally
derive the results for {\em mixture} $C$.

\noindent {\em Mixture $A$}\\
Since the {\em fixed group} is empty, we have 
$\sum_{i}\xi_{i,\odot}^s+\sum_{k}xi_{k,\odot}^b=1$, hence:
\begin{equation}
\label{eq_mixa}
f^b = \displaystyle\frac{1}{1 + 
{\displaystyle\sum_{i}(10^{\gamma_i}-1)\,\xi_{i,\odot}^s}}.
\end{equation}

\noindent {\em Mixture $B$}\\
Since the {\em balancing elements} are those belonging to the
Fe-group, i.e.  $\sum_{k}X_{k,\odot}^b = X_{{\rm Fe},\odot}$, we
get:
\begin{equation}
\label{eq_mixb}
f^b = \displaystyle\frac{\displaystyle\sum_{i}\xi_{i,\odot}^s+\xi_{{\rm Fe},\odot}}
{\displaystyle\sum_{i}10^{\gamma_i}\xi_{i,\odot}^s + \xi_{{\rm Fe},\odot}}
\end{equation}

\noindent{\em Mixture $C$}\\
Finally, we consider the case of {\em mixture} $C$, in which the reference
metallicity $Z_{\rm ref}$ should not be preserved, 
as the actual metallicity, $Z=f_Z Z_{\rm ref}$, follows
the total abundance variation of the {\em selected elements}.
In this case we consider the system of equations
\begin{equation*}
\left\{
\begin{array}{rcl}
Z= f_Z\, Z_{\rm ref} & = & \displaystyle\sum_{i} X_i^s + \displaystyle\sum_{j}
X_j^{non-s} \\
\displaystyle\left[\frac{X_i^s}{X_{\rm Fe}}\right] & = & 
\log\displaystyle\left(\frac{X_i^s}{X_{\rm Fe}}\right)
-\log\displaystyle\left(\frac{X^s_{i,\odot}}{X_{\rm Fe, \odot}}\right)
\,\,\,\quad\quad\quad
\forall\, i=1,\dotsc\mathcal{N}_{sel}\,\,,
\end{array}
\right.
\end{equation*}
where we only distinguish between {\em selected} and {\em non-selected} elements. 
From the definitions of the abundance variation factors, recalling that 
$\sum_{j} X_{j,\odot}^{non-s} =  Z_{\odot}-\sum_{i} X_{i,\odot}^{s}$,
and after some manipulation, we obtain the equations
\begin{equation*}
\left\{
\begin{array}{ccl}
Z_{\odot} & = & \displaystyle\sum_{i} f_i^s X_{i,\odot}^s +
\displaystyle\sum_{j} X_{j,\odot}^{non-s} \\
10^{\gamma_i} & = & \displaystyle \frac{f_i^s}{f^{non-s}}
\quad\quad\quad\quad\quad\quad\quad\quad\quad\quad\quad\quad\quad
\forall\, i=1,\dotsc\mathcal{N}_{sel}
\end{array}
\right.
\end{equation*}
for the unknowns $f_i^s$ and $f^{non-s}$. We notice that formally we deal
with exactly the same  equations as those for {\em mixture} $A$, once  
the {\em non-selected elements} are considered in place of the  {\em balancing elements}.
Hence, the seeked solution is 
given by Eq.~(\ref{eq_mixa}) where one substitutes $f^b$ with $f^{non-s}$. 
In other words, {\em mixture} $A$ and {\em mixture} $C$ share the same non-solar
metal partitions $(X_i/Z_{\rm ref})_A=(X_i/Z)_C$, 
but their metallicity is different by a factor $f_Z$.


\begin{thebibliography}{}
\bibitem[Alexander(1975)]{1975ApJS...29..363A} Alexander, D.~R.\ 1975, 
\apjs, 29, 363 
\bibitem[]{} Alexander D.~R.,  \& {Ferguson} J.~W.,  1994, \apj, 437,
  879
\bibitem[{{Alvarez} \& {Plez}(1998)}]{1998A&A...330.1109A}Alvarez, R. \& Plod, B. 1998, \aap, 330, 110
\bibitem[Anders \& Grevesse(1989)]{1989GeCoA..53..197A} Anders,~E., \& Grevesse,~ N. 1989,
 Geochim. Cosmochim. Acta, 53, 197 
\bibitem[Annibali et 
al.(2007)]{2007A&A...463..455A} Annibali, F., Bressan, A., Rampazzo, R., Zeilinger, W.~W., 
\& Danese, L.\ 2007, \aap, 463, 455
\bibitem[]{}Aringer, B. 2000, Ph.D. Thesis, University of Vienna
\bibitem[Aringer et al.(2009)]{2009arXiv0905.4415A} Aringer, B., Girardi, 
L., Nowotny, W., Marigo, P., \& Lederer, M.~T.\ 2009, arXiv:0905.4415 
\bibitem[Asensio Ramos 
\& Socas-Navarro(2005)]{2005A&A...438.1021A} Asensio Ramos, A., \& Socas-Navarro, H.\ 2005, \aap, 438, 1021 
\bibitem[{{Barber} {et~al.}(2006){Barber}, {Tennyson}, {Harris}, \&
  {Tolchenov}}]{2006MNRAS.368.1087B}
{Barber}, R.~J., {Tennyson}, J., {Harris}, G.~J., \& {Tolchenov}, R.~N. 2006,
  \mnras, 368, 1087
\bibitem[{{Bauschlicher} {et~al.}(2001){Bauschlicher}, {Ram}, {Bernath},
  {Parsons}, \& {Galehouse}}]{BauschlicherRam2001}
{Bauschlicher}, C.~W., J., {Ram}, R.~S., {Bernath}, P.~F., {Parsons}, C.~G., \&
  {Galehouse}, D. 2001, J. Chem. Phys., 115, 1312
\bibitem[]{}Beers  T. C., Christlieb  N., 2005, \araa, 43, 531
\bibitem[]{}Bertelli, G., Nasi, E., Girardi, L., \&  Marigo, P. \ 2009, \aap, submitted
\bibitem[Boothroyd 
\& Sackmann(1988)]{1988ApJ...328..641B} Boothroyd, A.~I., \& Sackmann, I.-J.\ 1988, \apj, 328, 641 
\bibitem[{{Borysow}, {J\o rgensen} \& {Fu}}{{Borysow}
  et~al.}{2001}]{2001JQSRT..68..235B}
{Borysow} A.,  {J\o rgensen} U.~G.,    {Fu} Y.,  2001, \jqsrt, 68, 235
\bibitem[{{Borysow}, {J\ orgensen} \& {Zheng}}{{Borysow}
  et~al.}{1997}]{1997A&A...324..185B}
{Borysow} A.,  {J\o rgensen} U.~G.,    {Zheng} C.,  1997, \aap, 324,
185
\bibitem[Caffau et 
al.(2009)]{2009A&A...498..877C} Caffau, E., Maiorca, E., Bonifacio, P., Faraggiana, R., Steffen, M., Ludwig, H.-G., Kamp, I., \& Busso, M.\ 2009, \aap, 498, 877 
\bibitem[Caffau et 
al.(2008)]{2008A&A...488.1031C} Caffau, E., Ludwig, H.-G., Steffen, M., Ayres, T.~R., Bonifacio, P., Cayrel, R., Freytag, B., \& Plez, B.\ 2008, \aap, 488, 1031 
\bibitem[Carbon et al.(1969)]{1969lls..symp..435C} Carbon, D., Gingerich, 
O.~J., \& Latham, D.~W.\ 1969, Low-Luminosity Stars, 435 
\bibitem[Carretta et 
al.(2005)]{2005A&A...433..597C} Carretta, E., Gratton, R.~G.,
Lucatello, S., Bragaglia, A., \& Bonifacio, P.\ 2005, \aap, 433, 597
 \bibitem[Clemens et al.(2009)]{2009MNRAS.392L..35C} Clemens, M.~S., 
Bressan, A., Nikolic, B., \& Rampazzo, R.\ 2009, \mnras, 392, L35 
\bibitem[Clemens et al.(2006)]{2006MNRAS.370..702C} Clemens, M.~S., 
Bressan, A., Nikolic, B., Alexander, P., Annibali, F., 
\& Rampazzo, R.\ 2006, \mnras, 370, 702 
\bibitem[\protect\citeauthoryear{{Coc}, {Vangioni-Flam}, {Descouvemont},
  {Adahchour} \& {Angulo}}{{Coc} et~al.}{2004}]{2004ApJ...600..544C}
{Coc}, A.,  {Vangioni-Flam}, E.,  {Descouvemont}, P.,  {Adahchour}, A.,  \&  {Angulo},
  C.,  2004, \apj, 600, 544
\bibitem[Cowley 
\& Barisciano(1994)]{1994Obs...114..308C} Cowley, C.~R., \& Barisciano, L.~P., Jr.\ 1994, 
The Observatory, 114, 308 
\bibitem[Cristallo et al.(2007)]{2007ApJ...667..489C} Cristallo, S., 
Straniero, O., Lederer, M.~T., \& Aringer, B.\ 2007, \apj, 667, 489 
\bibitem[]{} Dalgarno A., 1962, Spectral Reflectivity of the Earth
  Atmosphere III; The Scattering of light by Atomic
  Systems. Geophys. Corp. of America, GCA Tech Rep. No.~62-28-A. 
\bibitem[{{Dalgarno} \& {Williams}}{{Dalgarno} \&
  {Williams}}{1962}]{1962ApJ...136..690D}
{Dalgarno} A.,  {Williams} D.~A.,  1962, \apj, 136, 690
\bibitem[Dotter et al.(2007)]{2007ApJ...666..403D} Dotter, A., Chaboyer, 
B., Ferguson, J.~W., Lee, H.-c., Worthey, G., Jevremovi{\'c}, D., 
\& Baron, E.\ 2007, \apj, 666, 403 
\bibitem[]{} Dulick, M. Bauschlicher, C.~W. Jr, Burrows, A., 
Sharp, C.~M., Ram, R.~S. \& Bernath, P.~F. 2003, ApJ, 594, 651
\bibitem[Eddington(1922)]{1922MNRAS..83...32E} Eddington, A.~S.\ 1922, 
\mnras, 83, 32 
\bibitem[{{Ferguson} {et~al.}(2005){Ferguson}, {Alexander}, {Allard}, {Barman},
  {Bodnarik}, {Hauschildt}, {Heffner-Wong}, \& {Tamanai}}]{2005ApJ...623..585F}
{Ferguson}, J.~W., {Alexander}, D.~R., {Allard}, F., {et~al.} 2005, \apj, 623,
  585
\bibitem[Ferrarotti 
\& Gail(2002)]{2002A&A...382..256F} Ferrarotti, A.~S., \& Gail, H.-P.\
2002, \aap, 382, 256 
\bibitem[Gingerich(1969)]{1969tons.conf.....G} Gingerich, O.\ 1969, Theory 
and Observation of Normal Stellar Atmospheres,  
\bibitem[Gingerich(1964)]{1964SAOSR.167...17G} Gingerich, O.\ 1964, SAO 
Special Report, 167, 17 
\bibitem[{{Goorvitch} \& {Chackerian}(1994)}]{1994ApJS...91..483G}
{Goorvitch}, D. \& {Chackerian}, Jr., C. 1994, \apjs, 91, 483
\bibitem[Gratton et 
al.(2001)]{2001A&A...369...87G} Gratton, R.~G., Bonifacio, P.,
Bragaglia, A., et al.\ 2001, \aap, 369, 87 
\bibitem[Gratton et 
al.(2004)]{2004ARA&A..42..385G} Gratton, R., Sneden, C., \& Carretta, E.\ 2004, \araa, 42, 385 
\bibitem[{{Grevesse} {et~al.}(2007){Grevesse}, {Asplund}, \&
  {Sauval}}]{2007SSRv..130..105G}
{Grevesse}, N., {Asplund}, M., \& {Sauval}, A.~J. 2007, Space Science Reviews,
  130, 105
\bibitem[{{Grevesse} \& {Sauval}(1998)}]{1998SSRv...85..161G}
{Grevesse}, N. \& {Sauval}, A.~J. 1998, Space Science Reviews, 85, 161
\bibitem[Grevesse 
\& Noels(1993)]{1993oee..conf...14G} Grevesse, N., \& Noels, A.\ 1993, Origin and Evolution of the Elements, 14 
\bibitem[{{Harris}, {Lynas-Gray}, {Miller} \&
  {Tennyson}}{{Harris} et~al.}{2004}]{2004ApJ...600.1025H}
{Harris} G.~J.,  {Lynas-Gray} A.~E.,  {Miller} S.,    {Tennyson} J.,  2004,
  \apj, 600, 1025 
\bibitem[{{Harris} {et~al.}(2006){Harris}, {Tennyson}, {Kaminsky}, {Pavlenko},
  \& {Jones}}]{2006MNRAS.367..400H}
{Harris}, G.~J., {Tennyson}, J., {Kaminsky}, B.~M., {Pavlenko}, Y.~V., \&
  {Jones}, H.~R.~A. 2006, \mnras, 367, 400
\bibitem[Helling 
\& Lucas(2009)]{2009arXiv0906.0296H} Helling, C., \& Lucas, W.\ 2009, arXiv:0906.0296 
\bibitem[Helling et 
al.(2000)]{2000A&A...358..651H} Helling, C., Winters, J.~M., \& Sedlmayr, E.\ 2000, \aap, 358, 651 
\bibitem[{{Helling} \& {J{\o}rgensen}(1998)}]{1998A&A...337..477H}
{Helling}, C. \& {J{\o}rgensen}, U.~G. 1998, \aap, 337, 477
\bibitem[Holweger(2001)]{2001AIPC..598...23H} Holweger, H.\ 2001, Joint 
SOHO/ACE workshop ''Solar and Galactic Composition'', 598, 23 
\bibitem[Houdashelt et al.(2000)]{2000AJ....119.1448H} Houdashelt, M.~L., 
Bell, R.~A., \& Sweigart, A.~V.\ 2000, \aj, 119, 1448 
\bibitem[{{Hunger} \& {van Blerkom}}{{Hunger} \& {van
  Blerkom}}{1967}]{1967ZA.....66..185H}{Hunger} K.,  {van Blerkom} D.,
1967, Zeitschrift f\"ur Astrophysik, 66, 185
\bibitem[{{Iglesias} \& {Rogers}}{{Iglesias} \&
  {Rogers}}{1996}]{1996ApJ...464..943I}
{Iglesias} C.~A.,  {Rogers} F.~J.,  1996, \apj, 464, 943
\bibitem[]{} Irwin, A.~W. 1981, ApJS, 45, 621
\bibitem[]{} Irwin, A.~W. 1988, A\&AS, 74, 145
\bibitem[]{} John T.~L.,  1988, \aap, 193, 189
\bibitem[John(1975)]{1975MNRAS.172..305J} John, T.~L.\ 1975, \mnras, 172, 
305 
\bibitem[{{J{\o}rgensen}(1997)}]{Jorgensen1997}
{J{\o}rgensen}, U.~G. 1997, in {Molecules in Astrophysics: Probes and
  Processes}, ed. E.~F. {van Dishoeck}, {IAU Symp. 178} (Kluwer),
441--456
\bibitem[J{\o}rgensen et 
al.(2000)]{2000A&A...361..283J} J{\o}rgensen, U.~G., Hammer, D., Borysow, A., \& Falkesgaard, J.\ 2000, \aap, 361, 283 
\bibitem[{{J{\o}rgensen} {et~al.}(1989){J{\o}rgensen}, {Almlof}, \&
  {Siegbahn}}]{1989ApJ...343..554J}
{J{\o}rgensen}, U.~G., {Almlof}, J., \& {Siegbahn}, P.~E.~M. 1989, \apj, 343,
  554
\bibitem[{{Karzas} \& {Latter}}{{Karzas} \&
  {Latter}}{1961}]{1961ApJS....6..167K}
{Karzas} W.~J.,  {Latter} R.,  1961, \apjs, 6, 167
\bibitem[Keeley(1970)]{1970ApJ...161..643K} Keeley, D.~A.\ 1970, \apj, 161, 
643 
\bibitem[Kurucz(1970)]{1970SAOSR.309.....K} Kurucz, R.~L.\ 1970, SAO 
Special Report No.~309
\bibitem[]{} Kurucz, R.~L. 1993a, Atomic data for opacity calculations. 
             Kurucz CD-ROM No.1.
\bibitem[]{} Kurucz, R.~L. 1993b, Opacities for Stellar Atmospheres: Abundance Sampler.
             Kurucz CD-ROM No.14.
\bibitem[]{} Kurucz, R.~L. 1993c, Diatomic molecular data for opacity calculations. 
             Kurucz CD-ROM No.15.
\bibitem[{{Langhoff} \& {Bauschlicher}(1993)}]{LanghoffBauschlicher1993}
{Langhoff}, S.~R. \& {Bauschlicher}, C.~W. 1993, Chem. Phys. Lett., 211, 305
\bibitem[]{} Lattanzio, J.C., \& Wood, P.P. 2003, in 
	Asymptotic Giant Branch Stars, chapter 2, 
	eds. H.J. Habing \& Hans Olofsson, 
	\aap Library
\bibitem[]{} Lebedev, V.~S., Presnyakov, L.~P., \& Sobel\'man,
I.~I. 2003, Physics-Uspekhi, 46, 473 
\bibitem[Lederer 
\& Aringer(2009)]{2009A&A...494..403L} Lederer, M.~T., \& Aringer, B.\ 2009, \aap, 494, 403
 \bibitem[{{Lenzuni}, {Chernoff} \& {Salpeter}}{{Lenzuni}
  et~al.}{1991}]{1991ApJS...76..759L}
{Lenzuni} P.,  {Chernoff} D.~F.,    {Salpeter} E.~E.,  1991, \apjs, 76, 759 
\bibitem[{{Littleton}(2007)}]{Littleton1987}
{Littleton}, J.~E. 2007, private communication
\bibitem[Luo(1997)]{1997ApJ...491..366L} Luo, G.~Q.\ 1997, \apj, 491, 366 
\bibitem[{{Lodders}(2003)}]{2003ApJ...591.1220L}
{Lodders}, K. 2003, \apj, 591, 1220
\bibitem[Marigo(2002)]{2002A&A...387..507M} Marigo, P.\ 2002, \aap,
387, 507 
\bibitem[Marigo et 
al.(2008)]{2008A&A...482..883M} Marigo, P., Girardi, L., Bressan, A., Groenewegen, M.~A.~T., Silva, L., \& Granato, G.~L.\ 2008, \aap, 482, 883 
\bibitem[Marigo 
\& Girardi(2007)]{2007A&A...469..239M} Marigo, P., \& Girardi, L.\ 2007, \aap, 469, 239
\bibitem[Marigo et 
al.(1998)]{1998A&A...331..564M} Marigo, P., Bressan, A., \& Chiosi, C.\ 1998, \aap, 331, 564 
\bibitem[Marigo et 
al.(1996)]{1996A&A...313..545M} Marigo, P., Bressan, A., \& Chiosi, C.\ 1996, \aap, 313, 545 
\bibitem[]{} Mayer M., \& Duschl W.~J. 2005, MNRAS, 358, 614
\bibitem[McSaveney et al.(2007)]{2007MNRAS.378.1089M} McSaveney, J.~A., 
Wood, P.~R., Scholz, M., Lattanzio, J.~C., 
\& Hinkle, K.~H.\ 2007, \mnras, 378, 1089 
\bibitem[Mihalas et al.(1988)]{1988ApJ...331..815M} Mihalas, D., D\"appen, 
W., \& Hummer, D.~G.\ 1988, \apj, 331, 815 
\bibitem[]{} Neale, L. \&  Tennyson, J. 1995, ApJ, 454, L169
\bibitem[]{}Partridge, H., \& Schwenke, D.~W. 1997, J. Chem. Phys., 106, 4618
\bibitem[Pietrinferni et al.(2009)]{2009ApJ...697..275P} Pietrinferni, A., 
Cassisi, S., Salaris, M., Percival, S., 
\& Ferguson, J.~W.\ 2009, \apj, 697, 275 
\bibitem[{{Plez}(2007)}]{Plez2007}
{Plez}, B. 2007, private communication
\bibitem[]{} Press, W.~H., Teukolsky, S.~A., Vetterling, W.~T., \&
  Flannery, B.~P. 1986, Numerical Recipes 
(Cambridge: Cambridge University Press) 
\bibitem[{{Querci} {et~al.}(1974){Querci}, {Querci}, \&
  {Tsuji}}]{1974A&A....31..265Q}
{Querci}, F., {Querci}, M., \& {Tsuji}, T. 1974, \aap, 31, 265
\bibitem[Ram{\'{\i}}rez 
\& Mel{\'e}ndez(2005)]{2005ApJ...626..446R} Ram{\'{\i}}rez, I., \& Mel{\'e}ndez, J.\ 2005, \apj, 626, 446 
\bibitem[Rosseland(1924)]{1924MNRAS..84..525R} Rosseland, S.\ 1924, \mnras, 
84, 525 
\bibitem[{{Rothman} {et~al.}(2005){Rothman}, {Jacquemart}, {Barbe}, {Chris
  Benner}, {Birk}, {Brown}, {Carleer}, {Chackerian}, {Chance}, {Coudert},
  {Dana}, {Devi}, {Flaud}, {Gamache}, {Goldman}, {Hartmann}, {Jucks}, {Maki},
  {Mandin}, {Massie}, {Orphal}, {Perrin}, {Rinsland}, {Smith}, {Tennyson},
  {Tolchenov}, {Toth}, {Vander Auwera}, {Varanasi}, \&
  {Wagner}}]{2005JQSRT..96..139R}
{Rothman}, L.~S., {Jacquemart}, D., {Barbe}, A., {et~al.} 2005, Journal of
  Quantitative Spectroscopy and Radiative Transfer, 96, 139
\bibitem[Salaris et al.(2006)]{2006ApJ...645.1131S} Salaris, M., Weiss, A., 
Ferguson, J.~W., \& Fusilier, D.~J.\ 2006, \apj, 645, 1131 
\bibitem[]{} Sauval, A.~J., \& Tatum, J.~B. 1984, ApJS, 56, 193 
\bibitem[Scalo 
\& Ulrich(1975)]{1975ApJ...200..682S} Scalo, J.~M., \& Ulrich, R.~K.\ 1975, \apj, 200, 682 
\bibitem[{{Schwenke}(1997)}]{Schwenke1997}
{Schwenke}, D.~W. 1997, private communication
\bibitem[{{Schwenke}(1998)}]{1998cpmg.conf..321S}
{Schwenke}, D.~W. 1998, in Chemistry and Physics of Molecules and Grains in
  Space. Faraday Discussions No. 109, 321
\bibitem[{{Seaton}(2005)}]{2005MNRAS.362L...1S}
{Seaton}, M.~J. 2005, \mnras, 362, L1
\bibitem[]{} Seaton M.~J.,  {Yan} Y.,  {Mihalas} D., \& {Pradhan}
  A.~K.,  1994, \mnras, 266, 805
\bibitem[Sharp 
\& Burrows(2007)]{2007ApJS..168..140S} Sharp, C.~M., \& Burrows, A.\
2007, \apjs, 168, 140 
\bibitem[Smith 
\& Dworetsky(1988)]{1988eaa..conf...32S} Smith, K.~C., \& Dworetsky, M.~M.\ 1988, Elemental Abundance Analyses, 32 
\bibitem[Smith et al.(2002)]{2002AJ....124.3241S} Smith, V.~V., 
Hinkle, K.~H., Cunha, K., Plez, B., Lambert, D.~L. et~al.\ 
2002, \aj, 124, 3241 
\bibitem[Thomas et al.(1994)]{1994ApJ...430..291T} Thomas, D., Schramm, 
D.~N., Olive, K.~A., Mathews, G.~J., Meyer, B.~S., 
\& Fields, B.~D.\ 1994, \apj, 430, 291 
\bibitem[Thomas et al.(1993)]{1993ApJ...406..569T} Thomas, D., Schramm, 
D.~N., Olive, K.~A., \& Fields, B.~D.\ 1993, \apj, 406, 569 
\bibitem[{{Tipping}(2007)}]{Tipping2007}
{Tipping}, R.~H. 2007, private communication
\bibitem[Tsuji(1973)]{1973A&A....23..411T} Tsuji, T.\ 1973, \aap, 23, 411 
\bibitem[Tsuji(1964)]{1964AnTok...9.....T} Tsuji, T.\ 1964, Annals of the 
Tokyo Astronomical Observatory, 9, no.~1
\bibitem[Uttenthaler et al.(2008)]{2008ApJ...682..509U} Uttenthaler, S., 
Aringer, B., Lebzelter, T., K{\"a}ufl, H.~U., Siebenmorgen, R., 
\& Smette, A.\ 2008, \apj, 682, 509 
\bibitem[]{}Ventura, P., \& Marigo, P. 2009, \mnras, L294 
\bibitem[Wagenhuber 
\& Groenewegen(1998)]{1998A&A...340..183W} Wagenhuber, J., \& Groenewegen, M.~A.~T.\ 1998, \aap, 340, 183 
\bibitem[Weiss 
\& Ferguson(2009)]{2009arXiv0903.2155W} Weiss, A., \& Ferguson, J.~W.\ 2009, arXiv:0903.2155 
\end{thebibliography}
\end{document}